\documentclass[article]{elsarticle}

\usepackage{bm}
\usepackage{amsmath}
\usepackage{subfigure}
\usepackage{booktabs}
\usepackage{threeparttable}
\usepackage{multirow}
\usepackage{makecell}
\usepackage{color}
\begin{document}

\begin{frontmatter}

\title{A conservative implicit scheme for three-dimensional steady flows of diatomic gases in all flow regimes using unstructured meshes in the physical and velocity spaces}

\author[a]{Rui Zhang}
\ead[Rui Zhang]{zhangruinwpu@mail.nwpu.edu.cn}

\author[a,b,c]{Sha Liu\corref{mycorrespondingauthor}}
\cortext[mycorrespondingauthor]{Corresponding author}
\ead[Sha Liu]{shaliu@nwpu.edu.cn}

\author[a]{Jianfeng Chen}
\ead[Jianfeng Chen]{chenjf@mail.nwpu.edu.cn}

\author[a,b]{Chengwen Zhong}
\ead[Chengwen Zhong]{zhongcw@nwpu.edu.cn}

\author[a,b]{Congshan Zhuo}
\ead[Congshan Zhuo]{zhuocs@nwpu.edu.cn}

\address[a]{School of Aeronautics, Northwestern Polytechnical University, Xi’an, Shaanxi 710072, China}
\address[b]{Institute of Extreme Mechanics, Northwestern Polytechnical University, Xi’an, Shaanxi 710072, China}
\address[c]{National Key Laboratory of Science and Technology on Aerodynamic Design and Research, Northwestern Polytechnical University, Xi’an, Shaanxi 710072, China}

\begin{abstract}
A computationally accurate and efficient numerical method under a unified framework is crucial to various multi-scale scientific and engineering problems.
So far, many numerical methods have encountered various challenges in efficiently solving multi-scale non-equilibrium flows that cover a wide range of Knudsen numbers, especially the three-dimensional hypersonic flows.
In this study, a conservative implicit scheme is further presented for three-dimensional steady flows of diatomic gases in all flow regimes, where both the implicit microscopic kinetic equations based on Rykov model and the corresponding implicit macroscopic governing equations are solved synchronously.
Furthermore, a simplified multi-scale numerical flux inspired by the strategy of discrete unified gas kinetic scheme (DUGKS) is proposed to relieve the limitation of grid size and time step in all flow regimes.
The flux is constructed through a backward Euler difference scheme with the consideration of collision effect in the physical reconstruction of the gas distribution function on the cell interface.
Meanwhile, its asymptotic preserving property in the continuum limit is analyzed.
In order to pursue high computational efficiency in three-dimensional flow simulations, the unstructured discrete velocity space (DVS) and MPI parallel in DVS are adopted to further speed up the calculation.
Additionally, based on numerical experiments of Apollo 6 command module, an empirical generation criterion for three-dimensional unstructured DVS is proposed.
Numerical results indicate that, the present method is about one to two orders of magnitude faster than the explicit conserved DUGKS method.
The present method is proved to accurately and efficiently predict aerothermodynamic properties of hypersonic rarefied gas flows.
\end{abstract}

\begin{keyword}
\texttt Implicit unified scheme\sep 
Diatomic molecules\sep 
Hypersonic rarefied flows\sep 
Non-equilibrium flows
\end{keyword}

\end{frontmatter}


\section{Introduction}\label{sec:introduction}
Rarefied gas flows are prevalent in various scientific and engineering fields, including the micro-electro-mechanical systems (MEMS)~\citep{senturia_simulating_1997,mahdavi_study_2022,wang_investigation_2022}, hypersonic vehicles and suborbital or orbital spacecraft~\citep{lebeau_application_2001,walpot_base_2012,li_rarefied_2015,schouler_survey_2020}.
Take the space shuttle reentry problem for example, during the entire descending trajectory, vehicles will encounter different flow regimes from the free molecular, transitional, slip to continuum flows.
Another situation is that the complex multi-scale flows with both rarefied and continuum flows exist simultaneously in a single flow field, which are also known as local continuum breakdown problems that often appear in the flow field around spacecraft~\citep{ivanov_computational_1998}.
Until now scientists and engineers are faced with the daunting task of solving such multi-scale non-equilibrium flow problems under a unified framework.
In the rarefied regimes, traditional computational fluid dynamics (CFD) methods for solving Euler and Navier-Stokes (N-S) equations lose their validity.
The direct simulation Monte Carlo (DSMC) method~\citep{bird_molecular_1994} based on probabilistic modeling is one of the most popular methods.
It was first proposed by Bird~\citep{bird_approach_1963} in 1963, and has been widely applied and verified through comparisons with experiment measurements.
DSMC method plays a key role in providing physical solutions in the rarefied environment.
However, in near-continuum regime, the computational cost of DSMC is even larger because the cell size and time step have to be less than the molecular mean free path and collision time, respectively.

The deterministic method is another type of methodology, it employs the velocity distribution function to describe the distribution of particles in a gas, and solves the Boltzmann equation or its model equations~\citep{chu_kinetic_1965,li_study_2004,xu_unified_2010,guo_discrete_2013}.
In the framework of deterministic approximation, the discrete velocity method (DVM), also known as the discrete ordinate method (DOM)~\citep{chu_kinetic_1965,li_study_2004}, has been extensively used in the numerical simulation of rarefied gas flows over the past few decades.
In this category of methods, the transportation term and the collision term are completely decoupled by using operator splitting method, which result in a quite concise and very high efficiency scheme for high Knudsen (Kn) number flow simulation.
However, in the near-continuum and continuum flow regimes, it is prohibitively expensive to obtain an accurate result due to the restriction of the cell size and time step.
Furthermore, for the conventional DVM in the finite volume framework, the gas distribution function at the cell interface is simply reconstructed based on the initial data at the cell center.
Its intrinsic numerical dissipation is proportional to the time step~\citep{xu_dissipative_2001}.
Therefore, if the cell size is excessively larger than the molecular mean free path, the gas distribution function at the cell interface can markedly deviate from the Maxwellian equilibrium distribution due to the numerical error in the case of small Kn number.
Consequently, the excessive numerical viscosity will yield more dissipating result.

In order to develop a kinetic scheme for multi-scale non-equilibrium flow simulation, Xu and Huang proposed the unified gas kinetic scheme (UGKS)~\citep{xu_unified_2010,huang_unified_2012} for gas flows in all flow regimes.
In the framework of UGKS, both the equations of gas distribution functions and macroscopic flow variables will be updated alternately.
Then a local time-dependent analytical solution of model equations is carefully designed to construct multi-scale numerical flux at the cell interface, which couples the molecule transport and collision effects in a local time step.
Hence, the cell size restriction is first overcome and can be determined by the requirement for accuracy and resolution in the numerical simulation. Meanwhile, the time step is only constrained by the CFL condition.
Based on the similar consideration, Guo et al. proposed the discrete unified gas kinetic scheme (DUGKS)~\citep{guo_discrete_2013,guo_discrete_2015}, in which the gas distribution function at the cell interface is constructed through a characteristic difference solution of the model equations.
The particle transport and collision effects are coupled and evaluated in a time step.
As a result, the cell size and time step are also not passively limited by the molecular mean free path and collision time.
Recently, some simplified algorithms of UGKS~\citep{chen_simplification_2016} and DUGKS~\citep{zhong_simplified_2020}, conserved DUGKS~\citep{liu_conserved_2018,chen_compressible_2020} have been developed.
Although UGKS and DUGKS provide a unified framework for capturing the flow behaviors in all flow regimes~\citep{xu_unified_2021,guo_progress_2021}, the model equations need to be discretized in time, physical space and particle velocity space same as the DVM method.
Consequently, it still takes a lot of resources to solve the three-dimensional (3D) cases, especially the hypersonic flows, even with the rapid advancement of computer infrastructure.

In recent years, continuing efforts have been devoted to develop computationally efficient implicit methods for multi-scale flow simulations.
In order to solve the kinetic model equations implicitly, the discretization of equilibrium distribution function ${f^*}$ in the collision term should be elaborately designed.
By approximating the equilibrium state ${f^{*,n+1}}$ with ${f^{*,n}}$, Yang and Huang~\citep{yang_rarefied_1995}, Mao et al.~\citep{mao_study_2015} proposed the semi-implicit DVM and UGKS methods, respectively.
Then, Jiang et al.~\citep{jiang_implicit_2019} developed a 3D semi-implicit parallel UGKS solver for all flow regimes based on the Mao's method.
Peng et al.~\citep{peng_implicit_2016} proposed a implicit gas kinetic unified algorithm (GKUA) for multi-body reentry flow simulations.
However, it will slow down the convergence considerably since the gas distribution function $f$ is handled implicitly in the collision term while the equilibrium state ${f^*}$ is still treated explicitly~\citep{mieussens_discrete_2000,mieussens_discrete_velocity_2000}.
Mieussens~\citep{mieussens_discrete_2000} introduced a linear mapping between the equilibrium state ${f^*}$ and the gas distribution function $f$, namely ${f^{*,n+1}} = {f^{*,n}} + {\bf{M}}\left( {{f^{n + 1}} - {f^n}} \right)$, to establish a fully implicit DVM method.
Apparently, ${\bf{M}}$ is a large Jacobian matrix in the discrete velocity space which results in remarkable computation complexity.
Considering the strategy of simultaneously solving the gas distribution function and macroscopic variables in explicit UGKS method, Zhu et al.~\citep{zhu_implicit_2016,zhu_unified_2017} proposed a fully implicit UGKS method by developing an implicit macroscopic prediction technique to deal with the equilibrium state ${f^*}$.
In addition to solving the implicit gas distribution function equation, the implicit macroscopic governing equation is also solved to provide predicted macroscopic variables ${{\bm{\tilde W}}^{n + 1}}$.
Based on this predicted macroscopic variables ${{\bm{\tilde W}}^{n + 1}}$, a predicted equilibrium state ${{\tilde f}^{*,n+1}}$ can be obtained then the gas distribution function equation can be solved fully implicitly.
The use of the macroscopic prediction technique increases the efficiency of implicit UGKS greatly, especially for the near-continuum flows and highly non-equilibrium flows.
Subsequently, Zhu et al.~\citep{zhu_implicit_2019} further proposed an implicit UGKS for unsteady flow in all Knudsen regimes.
Inspired by the macroscopic prediction technique, Yang et al.~\citep{yang_improved_2018,yang_improved_2019} proposed an improved implicit DVM method.
Yuan et al.~\citep{yuan_conservative_2020} developed a conservative implicit scheme for diatomic gases in all flow regimes.
Furthermore, Su et al.~\citep{su_can_2020,su_multiscale_2021} put forward a general synthetic iterative scheme (GSIS) to solve rarefied gas flows within dozens of iterations at any Knudsen number.
Yuan et al.~\citep{yuan_multi_2021} proposed a multi-prediction implicit scheme for steady flows in all flow regimes to achieve a faster convergence rate with an inner iteration.
Besides those works, there are many other methods have been developed based on similar ideas, such as implicit DVM with inner iteration~\citep{yang_efficient_2022}, two-step implicit UGKS method~\citep{xu_ugks_based_2022}.

Although implicit algorithms for all flow regimes have advanced quickly in recent years, they have not been widely used to solve three-dimensional cases, especially hypersonic rarefied flows.
In this work, a conservative implicit scheme is further developed for three-dimensional steady flows of diatomic gases in all flow regimes based on the Rykov model equation~\citep{rykov_model_1975}.
Both the implicit governing equations of gas distribution functions and macroscopic variables will be solved alternately by using the macroscopic prediction technique.
In order to maintain multi-scale property for all flow regimes and pursue high computational efficiency in 3D flow simulations, a simplified multi-scale numerical flux at the cell interface is constructed through a backward Euler difference scheme inspired by the strategy of DUGKS.
Then, the 3D unstructured discrete velocity space (DVS)~\citep{yuan_conservative_2020,titarev_numerical_2017} is adopted to refine and coarsen the grid points flexibly according to the specific flows.
Furthermore, the discrete velocity space decomposition parallelization method and MPI parallel strategy are used.

The remainder of this paper is organized as follows.
In Sec.~\ref{sec:model}, the kinetic model equation for diatomic gases is briefly introduced.
In Sec.~\ref{sec:implicit}, the basic algorithm of conservative implicit scheme with simplified multi-scale numerical flux for diatomic gases is described in detail.
A series of test cases are performed and discussed to validate the efficiency and accuracy of the proposed method in Sec.~\ref{sec:cases}.
Finally, the summary and remark are given in Sec.~\ref{sec:conclusion}.

\section{The Rykov Kinetic Model}\label{sec:model}
\subsection{Gas kinetic model}\label{sec:rykov}
In this paper, we consider the kinetic description of diatomic gases in which the vibrational degrees of freedom are not excited and the rotational degrees of freedom can be treated classically.
In this case, besides three translational degrees of freedom $K_{tr}$, two rotational degrees of freedom $K_{rot}$ have to be included as well in a gas distribution function $f({\bm{x}},{\bm{u}},\varepsilon_{rot},t)$, where ${\bm{x}}$ and ${\bm{u}}$ are three dimensional physical space and particle velocity space, respectively.
The continuous variable $\varepsilon_{rot}$ ($\varepsilon_{rot}>0$) is molecular rotational energy, and $t$ is the time.

The macroscopic conserved variables, such as density $\rho$, momentum $\rho{\bm{U}}$ and energy $\rho E$ are defined as the moments of distribution function in the phase space $d{\bf{\Xi }}= d{\bm{u}}d{\varepsilon_{rot}}$ by
\begin{equation}\label{equ:f_solve_rho}
	\rho = \int{mf({\bm{x}},{\bm{u}},\varepsilon_{rot},t)d{\bf{\Xi }}},
\end{equation}
\begin{equation}\label{equ:f_solve_rhoU}
	\rho {\bm{U}} = \int {{\bm{u}}mf({\bm{x}},{\bm{u}},\varepsilon_{rot},t)d{\bf{\Xi }}},
\end{equation}
\begin{equation}\label{equ:f_solve_Energy}
	\begin{aligned}
		\rho E =\frac{1}{2}{\rho{|{\bm{U}}|^2}}+\frac{{{K_{tr}}+{K_{rot}}}}{2}\rho RT
		=\int {\left({\frac{1}{2}{m}{{|{\bm{u}}|}^2}+\varepsilon_{rot}}
			\right) f({\bm{x}},{\bm{u}},\varepsilon_{rot},t)d{\bf{\Xi }}}.
	\end{aligned}
\end{equation}
Here $R$ is the specific gas constant, $m$ is the molecular mass and $T$ is the equilibrium temperature.
The total energy $\rho E$ is the sum of translational energy ${\rho E}_{tr}$ and rotational energy ${\rho E}_{rot}$, the translational and rotational energies are defined as follows:
\begin{equation}\label{equ:f_solve_EnergyTr}
	\begin{aligned}
		\rho E_{tr} = \frac{1}{2}{\rho {|{\bm{U}}|^2}} + \frac{{K_{tr}}}{2}\rho RT_{tr}
		= \int {{\frac{1}{2}{m}{{|{\bm{u}}|}^2}}f({\bm{x}},{\bm{u}},\varepsilon_{rot},t)d{\bf{\Xi }}},
	\end{aligned}
\end{equation}
\begin{equation}\label{equ:f_solve_EnergyRot}
	\begin{aligned}
		\rho E_{rot} = \frac{K_{rot}}{2}\rho RT_{rot}
		= \int {{\varepsilon_{rot}}f({\bm{x}},{\bm{u}},\varepsilon_{rot},t)
			d{\bf{\Xi }}}.
	\end{aligned}
\end{equation}
Here $T_{tr}$ and $T_{rot}$ are translational and rotational temperatures, respectively.
The heat flux $\bm{q}$ is the sum of translational heat flux $\bm{q}_{tr}$ and rotational heat flux $\bm{q}_{rot}$, which are defined as:
\begin{equation}\label{equ:f_solve_qtra}
	{{\bm{q}}_{tr}} = \frac{1}{2}\int {{\bm{c}} {{{\left| {\bm{c}} \right|}^2} } 
		mf({\bm{x}},{\bm{u}},\varepsilon_{rot},t)
		d{\bf{\Xi }}},
\end{equation}
\begin{equation}\label{equ:f_solve_qrot}
	{{\bm{q}}_{rot}} = \int {{\bm{c}}{\varepsilon_{rot}}f({\bm{x}},{\bm{u}},\varepsilon_{rot},t)
		d{\bf{\Xi }}},
\end{equation}
where $\bm{c}=\bm{u}-\bm{U}$ is the peculiar velocity.
The stress tension ${{\bf P}}$ is defined from the second-order moment of the distribution function,
\begin{equation}\label{equ:f_solve_stress}
	{{\bf P}} = \int {{\bm c}{\bm c}mf({\bm{x}},{\bm{u}},\varepsilon_{rot},t)
		d{\bf{\Xi }}}.
\end{equation}

The collision integral of Rykov model~\citep{rykov_model_1975} is approximated by the sum of two integrals which correspond to the elastic and inelastic collisions.
In the absence of an external force, the equation in three dimensional space has the following form:
\begin{equation}\label{equ:model_equ_f}
	\frac{{\partial f}}{{\partial t}} + {\bm{u}} \cdot \frac{{\partial f}}{{\partial {\bm{x}}}} 
	= \frac{{{f^{tr}} - f}}{\tau } 
	+ \frac{{{f^{rot}} - {f^{tr}}}}{{{Z_{rot}}\tau }}
	= \frac{{{f^*} - f}}{\tau } : = \Omega \left( {{f^*},f} \right),
\end{equation}
where the BGK like collision operator $\Omega ({f^*},f)$ on the right side of Eq.~\ref{equ:model_equ_f} is consisted of two terms corresponding to translational and rotational relaxation, respectively.
$\tau$ is the relaxation time and $Z_{rot}$ is the rotational collision number.
The equilibrium distribution function $f^*$ is defined as:
\begin{equation}\label{equ:define_f*}
	{f^*} = \left( {1 - \frac{1}{{{Z_{rot}}}}} \right){f^{tr}} + \frac{1}{{{Z_{rot}}}}{f^{rot}},
\end{equation}
where distribution functions ${f^{tr}}$ and ${f^{rot}}$ are expressed as:
\begin{equation}\label{equ:ftra}
	\begin{aligned}
		{f^{tr}} &= n{\left( {\frac{1}{{2\pi R{T_{tr}}}}} \right)^{\frac{{3}}{2}}}\exp 
		\left({-\frac{{{{\left| {\bm{c}}\right|}^2}}}{{2R{T_{tr}}}}}\right)
		\frac{1}{{mR{T_{rot}}}}\exp \left( { - \frac{{{\varepsilon _{rot}}}}{{mR{T_{rot}}}}} \right)\\
		&\times \left\{ {1 + \frac{{{\bm{c}} \cdot {{\bm{q}}_{tr}}}}{{15R{T_{tr}}{p_{tr}}}}
			\left( {\frac{{ {{{\left| {\bm{c}} \right|}^2}} 
				}}{{R{T_{tr}}}} - 5} \right) + \left( {1 - \delta } \right)\frac{{{\bm{c}} \cdot 
					{{\bm{q}}_{rot}}}}{{R{T_{tr}}{p_{rot}}}}\left( {\frac{{{\varepsilon _{rot}}}}{{mR{T_{rot}}}} - 1} 
			\right)} \right\},
	\end{aligned}
\end{equation}
\begin{equation}\label{equ:frot}
	\begin{aligned}
		{f^{rot}} &= n{\left( {\frac{1}{{2\pi R{T}}}} \right)^{\frac{{3}}{2}}}\exp \left( { - \frac{{{{\left| 
							{\bm{c}} \right|}^2}}}{{2R{T}}}} \right)\frac{1}{{mR{T}}}\exp 
		\left( { - \frac{{{\varepsilon _{rot}}}}{{mR{T}}}} \right)\\
		&\times \left\{ {1 + {\omega _0}\frac{{{\bm{c}} \cdot {{\bm{q}}_{tr}}}}{{15R{T}{p_2}}}
			\left( {\frac{{ {{{\left| {\bm{c}} \right|}^2}}}}{{R{T}}}
				- 5} \right) + {\omega _1}\left( {1 - \delta } \right)\frac{{{\bm{c}} \cdot 
					{{\bm{q}}_{rot}}}}{{R{T}{p_2}}}\left( {\frac{{{\varepsilon _{rot}}}}{{mR{T}}} - 1} 
			\right)} \right\},
	\end{aligned}
\end{equation}
where $n$ is the molecular number density.

The relaxation time $\tau$ is determined by dynamic viscosity $\mu$ and translational pressure $p_{tr}$ with $\tau={\mu(T_{tr})/{{p_{tr}}}}$.
The dynamic viscosity $\mu(T_{tr})$ is related to inter-molecular interactions.
For variable hard-sphere (VHS)~\citep{bird_monte-carlo_1981} molecules, the dynamic viscosity is calculated by
\begin{equation}\label{equ:T_solve_mu}
	\mu  = {\mu _{ref}}{\left( {\frac{{{T_{tr}}}}{{{T_{ref}}}}} \right)^\omega },
\end{equation}
where $\omega$ is the viscosity index, which is 0.74 for nitrogen and 0.77 for oxygen~\citep{bird_molecular_1994}.
Besides, Sutherland formula~\citep{toro_riemann_2009} is frequently used to describe the dependence of viscosity on temperature rather well for a wide range of temperatures,
\begin{equation}\label{equ:T_solve_mu_Sut}
	\mu  = {\mu _{ref}}{\left( {\frac{{{T_{tr}}}}{{{T_{ref}}}}} \right)^{{3 \mathord{\left/
					{\vphantom {3 2}} \right.
					\kern-\nulldelimiterspace} 2}}}\frac{{{T_{ref}} + {C_s}}}{{{T_{tr}} + {C_s}}},
\end{equation}
where $C_s$ is 111K for nitrogen and 124K for air, $\mu_{ref}$ is the reference viscosity at the reference temperature $T_{ref}$.

\subsection{Reduced distribution functions}\label{sec:reduced}
The transport process of distribution function depends only on three dimensional particle velocity $\bm {u}$ and is irrelevant to continuous variable $\varepsilon_{rot}$.
Therefore, reduced distribution functions~\citep{chu_kinetic_1965} $G({\bm{x}},{\bm {u}},t)$ and $R({\bm{x}},{\bm {u}},t)$ are introduced in the numerical computations to save computational memory and cost.
\begin{equation}\label{equ:define_phi}
	\left( {\begin{array}{*{20}{c}}
			G \\
			R 
	\end{array}} \right) 
	= \int {\bm {\vartheta} f({\bm{x}},{\bm{u}},\varepsilon_{rot},t)d\varepsilon_{rot}},
\end{equation}
where the vector $\bm {\vartheta} ={\left({m,\varepsilon_{rot}}\right)^T}$.
Multiplying Eq.~\ref{equ:model_equ_f} by vector $\bm {\vartheta}$ and integrating the resulting equations with respect to $\varepsilon_{rot}$ from zero to infinity, Eq.~\ref{equ:model_equ_f} can be transformed into the following two equations:
\begin{equation}\label{equ:model_reduced_equ_phi}
	\begin{aligned}
		&\frac{{\partial G }}{{\partial t}} + {\bm{u}} \cdot \frac{{\partial G }}{{\partial {\bm{x}}}} 
		= \frac{{{G ^*} - G }}{\tau }: = \Omega \left( {{G ^*},G } \right),	\\
		&\frac{{\partial H }}{{\partial t}} + {\bm{u}} \cdot \frac{{\partial H }}{{\partial {\bm{x}}}} 
		= \frac{{{H ^*} - H }}{\tau }: = \Omega \left( {{H ^*},H } \right),
	\end{aligned}
\end{equation}
where reduced equilibrium distribution functions ${G^*}$ and ${R^*}$ are given by:
\begin{equation}\label{equ:eqdf_G*}
	{G^*}({\bm{x}},{\bm{u}},t) = \left( 1 - \frac{1}{{{Z_{rot}}}} \right){G^{tr}}({\bm{x}},{\bm{u}},t) 
	+ \frac{1}{{{Z_{rot}}}}{G^{rot}}({\bm{x}},{\bm{u}},t),
\end{equation}
\begin{equation}\label{equ:eqdf_R*}
	{R^*}({\bm{x}},{\bm{u}},t) = \left( 1 - \frac{1}{{{Z_{rot}}}} \right){R^{tr}}({\bm{x}},{\bm{u}},t) 
	+ \frac{1}{{{Z_{rot}}}}{R^{rot}}({\bm{x}},{\bm{u}},t),
\end{equation}
with
\begin{equation}\label{equ:eqdf_Gtra}
	{G^{tr}}({\bm{x}},{\bm{u}},t) = {g^{eq}}\left( {{T_{tr}}} \right)\left[ {1 + \frac{{{\bm{c}} 
				\cdot {{\bm{q}}_{tr}}}}{{15R{T_{tr}}{p_{tr}}}}\left( {\frac{{{{\left| {\bm{c}} \right|}^2}}}
			{{R{T_{tr}}}}- 5} \right)} \right],
\end{equation}
\begin{equation}\label{equ:eqdf_Grot}
	{G^{rot}}({\bm{x}},{\bm{u}},t) = {g^{eq}}\left( {{T}} \right)\left[ {1 + {\omega _0}\frac{{{\bm{c}} 
				\cdot{{\bm{q}}_{tr}}}}{{15R{T}{p}}}\left( {\frac{{{{\left| {\bm{c}} \right|}^2}}}{{R{T}}} - 5}
		\right)} \right],
\end{equation}
\begin{equation}\label{equ:eqdf_Rtra}
	{R^{tr}}({\bm{x}},{\bm{u}},t) = R{T_{rot}}\left[ {{G^{tr}} + \left( {1 - \delta } 
		\right)\frac{{{\bm{c}} 
				\cdot {{\bm{q}}_{rot}}}}{{R{T_{tr}}{p_{rot}}}}{g^{eq}}\left( {{T_{tr}}} \right)} \right],
\end{equation}
\begin{equation}\label{equ:eqdf_Rrot}
	{R^{rot}}({\bm{x}},{\bm{u}},t) = R{T}\left[ {{G^{rot}} + {\omega _1}\left( {1 - \delta } \right)
		\frac{{{\bm{c}}\cdot {{\bm{q}}_{rot}}}}{{R{T}{p}}}{g^{eq}}\left( {{T}} \right)} \right].
\end{equation}
In the above equations, the $g^{eq}$ is the Maxwellian equilibrium distribution function,
\begin{equation}\label{equ:eqdf_Maxwellian}
	{g^{eq}}\left(\bm {u}; {\rho ,{\bm{U}},T} \right) = \rho {\left( {\frac{1}{{2\pi RT}}} 
		\right)^{{3 \mathord{\left/
					{\vphantom {3 2}} \right.\kern-\nulldelimiterspace} 2}}}\exp \left( 
	{ - \frac{{{{\left| {{\bm{u}} -	{\bm{U}}} \right|}^2}}}{{2RT}}} \right).
\end{equation}

As a result, the relationship between the macroscopic flow variable $\bm {W}=(\rho,{\rho {\bm{U}}},\rho E,\rho E_{rot})^T$ and distribution function can be written in terms of the moments of reduced distribution functions,
\begin{equation}\label{equ:phi_solve_mac}
	\bm {W} = \int {{\bf {\Phi}}\left( {\bm{u}} \right) \cdot \left( {\begin{array}{*{20}{c}}
				G \\
				R
		\end{array}} \right) d{\bm{u}}},
\end{equation}
with
\begin{equation}\label{equ:PHI_matrix}
	{\bf {\Phi}}\left( {\bm{u}} \right) = \left[ {\begin{array}{*{20}{c}}
			1&0\\
			{\bm{u}}&0\\
			{\frac{1}{2}{{\left| {\bm{u}} \right|}^2}}&1\\
			0&1
	\end{array}} \right].
\end{equation}
The translational heat flux ${{\bm{q}}_{tr}}$, rotational heat flux ${{\bm{q}}_{rot}}$, and stress tension ${\bf P}$ are calculated by:
\begin{equation}\label{equ:phi_solve_qtra}
	{{\bm{q}}_{tr}} = \int {\frac{1}{2}{\bm{c}} {{{\left| {\bm{c}} \right|}^2}G}d{\bm{u}}},
\end{equation}
\begin{equation}\label{equ:phi_solve_qrot}
	{{\bm{q}}_{rot}} = \int {{\bm{c}}Rd{\bm{u}}},
\end{equation}
\begin{equation}\label{equ:phi_solve_stress}
	{{\bf P}} = \int {\bm {c} \bm {c}Gd{\bm{u}}}.
\end{equation}

\section{Conservative Implicit Method}\label{sec:implicit}
\subsection{Solution of microscopic governing equations}\label{sec:microequ}
The construction of conservative implicit scheme for diatomic gases is based on reduced model equations in the classical finite volume framework. In this work, the implicit method is developed by discretizing Eq.~\ref{equ:model_reduced_equ_phi} at time $t_{n+1}$ by a backward Euler method,
\begin{equation}\label{equ:implicit_mic_equ}
	\frac{{\left| {{V_i}} \right|}}{{\Delta t}}\left( {{\rm{ }}\phi _{i,k}^{n + 1} - \phi _{i,k}^n} \right) 
	+ \sum\limits_{j \in N\left( i \right)} {{A_{ij}}{\bm{u_k}} \cdot {{\bm{n}}_{ij}}\phi _{ij,k}^{n + 1}}  
	= \left| {{V_i}} \right|\frac{{\tilde \phi _{i,k}^{*,n + 1} - \phi _{i,k}^{n + 1}}}
	{{\tilde \tau _i^{n + 1}}},\phi = G, H.
\end{equation}
The $V_i$ is the volume of cell $i$, the sign $j$ denotes the neighboring cells of cell $i$ and $N(i)$ is the set of all of the neighbors of cell $i$.
$ij$ denotes the variable at the cell interface between the cell $i$ and $j$.
$A_{ij}$ is the interface area, and ${\bm {n}}_{ij}$ is the outward unit vector normal to the interface $ij$ from cell $i$ to cell $j$.
$\bm{u}_k$ is the discrete particle velocity space, and the symbol $\sim$ denotes the predicted variables for the next iteration step.
${\Delta t}$ is the numerical time step in the implicit system, which is not limited by the CFL condition.
To solve Eq.~\ref{equ:implicit_mic_equ}, it can be rewritten in a incremental form,
\begin{equation}\label{equ:implicit_mic_equ_delta_form}
	\left( {\frac{1}{{\Delta t}} + \frac{1}{{\tilde \tau _i^{n + 1}}}} \right){\rm{ }}\Delta \phi _{i,k}^
	{n + 1} + \frac{1}{{\left| {{V_i}} \right|}}\sum\limits_{j \in N\left( i \right)} {{A_{ij}}{\bm{u}_k} 
		\cdot {{\bm{n}}_{ij}}\Delta \phi _{ij,k}^{n + 1}}  = \frac{{\tilde \phi _{i,k}^{*,n + 1} - 
			\phi _{i,k}^n}}{{\tilde \tau _i^{n + 1}}} - F_{ij,k}^n,
\end{equation}
where $\Delta \phi _{i,k}^{n + 1} = \phi _{i,k}^{n + 1} - \phi _{i,k}^{n}$, $\Delta \phi_{ij,k}^{n + 1}= \phi _{ij,k}^{n + 1} - \phi _{ij,k}^{n}$, and the micro-flux $F_{ij,k}^n$ is
\begin{equation}\label{equ:implicit_mic_flux}
	F_{ij,k}^n = \frac{1}{{\left| {{V_i}} \right|}}\sum\limits_{j \in N\left( i \right)} {{A_{ij}}{\bm{u}_k} 
		\cdot {{\bm{n}}_{ij}}\phi _{ij,k}^n}.
\end{equation}
Since using alternative algorithms to compute the implicit fluxes in delta-form won't have an impact on the final convergent solution, the increment of the distribution function on the cell interface in the preceding equation will be constructed by the first-order upwind scheme,
\begin{equation}\label{equ:implicit_mic_face_delta_phi}
	\Delta \phi _{ij,k}^{n + 1} = \frac{1}{2}\left( {\Delta \phi _{i,k}^{n + 1} + \Delta \phi _{j,k}^{n + 1}} 
	\right) + \frac{1}{2}{\rm{sign}}\left( {{\bm{u}_k} \cdot {{\bm{n}}_{ij}}} 
	\right)\left( {\Delta \phi _{i,k}^{n + 1} - \Delta \phi _{j,k}^{n + 1}} \right).
\end{equation}
As a result, the final expression of Eq.~\ref{equ:implicit_mic_equ_delta_form} has a form of
\begin{equation}\label{equ:implicit_mic_final_equ}
	\begin{aligned}
		&{{\rm{D}}_{i,k}}\Delta \phi _{i,k}^{n + 1} + \sum\limits_{j \in N\left( i \right)} 
		{{{\rm{D}}_{j,k}}\Delta 
			\phi _{j,k}^{n + 1}}  = {\mathop{\rm Res}\nolimits} _{i,k}^n.\\
		&{{\rm{D}}_{i,k}} = \frac{1}{{\Delta t}} + \frac{1}{{\tilde \tau _i^{n + 1}}} + \frac{1}{2}\sum\limits_{j 
			\in N\left( i \right)} {\frac{{{A_{ij}}}}{{\left| {{V_i}} \right|}}{{\bm{u}}_k} \cdot {{\bm{n}}_{ij}}
			\left[ {1 + {\rm{sign}}\left( {{{\bm{u}}_k} \cdot {{\bm{n}}_{ij}}} \right)} \right]} {\rm{ }},\\
		&{{\rm{D}}_{j,k}} = \frac{1}{2}\frac{{{A_{ij}}}}{{\left| {{V_i}} \right|}}{{\bm{u}}_k} \cdot 
		{{\bm{n}}_{ij}}\left[ {1 - {\rm{sign}}\left( {{{\bm{u}}_k} \cdot {{\bm{n}}_{ij}}} \right)} \right],\\
		&{\mathop{\rm Res}\nolimits} _{i,k}^n = \frac{{\tilde \phi _{i,k}^{*,n + 1} - \phi _{i,k}^n}}
		{{\tilde \tau _i^{n + 1}}} - F_{ij,k}^n.
	\end{aligned}
\end{equation}
From the fully implicit equation Eq.~\ref{equ:implicit_mic_final_equ}, we know that once the residual ${\mathop{\rm Res}\nolimits} _{i,k}^n$ is obtained, it can be solved easily by using the classical Lower-Upper Symmetric Gauss-Seidel (LU-SGS) method~\citep{yoon_lu-ssor_1987,yoon_lower-upper_1988} or point relaxation Symmetric Gauss-Seidel (PR-SGS) method~\citep{rogers_comparison_1995,yuan_comparison_2002}.
Given that the SGS iteration for the microscopic equation is implemented in the whole DVS, and it is very time-consuming for the simulation of three dimensional hypersonic flows because a large amount of discrete velocity points is required to capture the non-equilibrium distributions.
In the current work, the Eq.~\ref{equ:implicit_mic_final_equ} are solved using two times' iteration of PR-SGS~\citep{zhu_implicit_2016,yuan_conservative_2020}, each of which consists of a forward sweep step and a backward sweep step.

The calculation of residual ${\mathop{\rm Res}\nolimits} _{i,k}^n$ requires the determination of micro-flux $F_{ij,k}^n$, as well as the equilibrium distribution function $\tilde \phi_{i,k}^{*,n + 1}$ and the relaxation time $\tilde \tau _i^{n+1}$.
It is crucial for determining the micro-flux $F_{ij,k}^n$ and the equilibrium distribution function $\tilde \phi _{i,k}^{*,n + 1}$ because there are about whether the scheme is accurate and efficient in all flow regimes, especially in the continuum flow regime.
The construction of $\tilde \phi _{i,k}^{*,n + 1}$ which is related to the predicted macroscopic variable vector $\tilde {\bm{W}}_i^{n+1}$ will be introduced in the next subsection~\ref{sec:macroequ}.
Later in the subsection~\ref{sec:flux}, we will detail the calculation of the simplified multi-scale micro-flux $F_{ij,k}^n$.

\subsection{Solution of macroscopic governing equations}\label{sec:macroequ}
In this work, the implicit macroscopic prediction technique proposed by Zhu et al.~\citep{zhu_implicit_2016} is adopted to deal with the equilibrium state $\tilde \phi _{i,k}^{*,n + 1}$.
From Eqs.~\ref{equ:eqdf_G*} to~\ref{equ:eqdf_Maxwellian}, we know that the determination of $\tilde \phi _{i,k}^{*,n + 1}$ requires the solution of the predicted macroscopic flow variables $\tilde {\bm{W}}_i^{n+1}$ which is further related to $\phi _{i,k}^{n + 1}$.
Therefore, $\tilde {\bm{W}}_i^{n+1}$ will be predicted firstly by solving the macroscopic governing equations implicitly.
Take the moment of Eq.~\ref{equ:implicit_mic_equ} in the continuous velocity space for ${\bf {\Phi}}\left( {\bm{u}} \right)$ (Eq.~\ref{equ:PHI_matrix}) with respect to $\bm u$ from $-\infty$ to $+\infty$, we can derive the fully implicit macroscopic governing equation,
\begin{equation}\label{equ:implicit_mac_equ}
	\frac{1}{{\Delta t}}\left( {{\rm{ }}{\bm{\tilde W}}_i^{n + 1} - {\bm{W}}_i^n} \right) + \frac{1}
	{{\left| {{V_i}} \right|}}\sum\limits_{j \in N\left( i \right)} {{A_{ij}}{\bm{F}}_{ij}^{n + 1}} 
	= {\bm{S}}_i^{n + 1},
\end{equation}
which could be rewritten in a delta-form,
\begin{equation}\label{equ:implicit_mac_equ_delta_form}
	\frac{1}{{\Delta t}}\Delta {\bm{\tilde W}}_i^{n + 1} + \frac{1}{{\left| {{V_i}} \right|}}\sum
	\limits_{j \in N\left( i \right)} {{A_{ij}}\Delta {\bm{F}}_{ij}^{n + 1}}  = {\bm{S}}_i^{n + 1} - 
	\frac{1}{{\left| {{V_i}} \right|}}\sum\limits_{j \in N\left( i \right)} {{A_{ij}}{\bm{F}}_{ij}^n},
\end{equation}
where $\Delta {\bm{\tilde W}}_i^{n + 1} = {\bm{\tilde W}}_i^{n + 1} - {\bm{W}}_i^{n}$, 
$\Delta {\bm{F}}_{ij}^{n + 1} = {\bm{F}}_{ij}^{n + 1} - {\bm{F}}_{ij}^{n}$. The source term 
${\bm{S}}_i^{n + 1}$ is expressed as
\begin{equation}\label{equ:implicit_mac_equ_source_term}
	{\bm{S}}_i^{n + 1} = \left( {\begin{array}{*{20}{c}}
			0\\
			{\bm{0}}\\
			0\\
			{\frac{{\tilde \rho _i^{n + 1}R\tilde T_i^{n + 1} - \left( {{W_{rot}}} \right)_i^{n + 1}}}
				{{{Z_{rot}}\tilde \tau _i^{n + 1}}}}
	\end{array}} \right) = \left( {\begin{array}{*{20}{c}}
			0\\
			{\bm{0}}\\
			0\\
			{\frac{{\tilde \rho _i^{n + 1}R\tilde T_i^{n + 1} - \Delta \left( {{{\tilde W}_{rot}}} 
						\right)_i^{n + 1} - \left( {{W_{rot}}} \right)_i^n}}{{{Z_{rot}}\tilde \tau _i^{n + 1}}}}
	\end{array}} \right),
\end{equation}
where $W_{rot}=\rho E_{rot}$ is the rotational energy.
The macro-flux ${\bm{F}}_{ij}^{n}$ is calculated from the distribution function $\phi _{ij,k}^{n}$ at the cell interface by numerical integrals in the DVS as follows:
\begin{equation}\label{equ:implicit_mac_equ_mac_flux}
	{\bm{F}}_{ij}^n = \frac{1}{{\left| {{V_i}} \right|}}\sum\limits_{j \in N\left( i \right)}
	{{A_{ij}}\left[ {\sum\limits_k {{\omega _k}\left( {{{\bm{u}}_k} \cdot {{\bm{n}}_{ij}}} 
				\right)\left( \begin{array}{l}
					G_{ij,k}^n\\
					{{\bm{u}}_k}G_{ij,k}^n\\
					\frac{1}{2}{\left| {{{\bm{u}}_k}} \right|^2}G_{ij,k}^n + R_{ij,k}^n\\
					R_{ij,k}^n
				\end{array} \right)} } \right]},
\end{equation}
where ${\omega _k}$ is the associated quadrature weight (the cell volume in the unstructured DVS mesh) at the discrete velocity point ${\bm u}_k$ (the centroid of a cell in the unstructured DVS mesh).

In order to develop a matrix-free algorithm, the implicit flux $\Delta {\bm{F}}_{ij}^{n + 1}$ is approximated by the Euler equation-based flux splitting method,
\begin{equation}\label{equ:implicit_mac_equ_delta_F}
	\Delta {\bm{F}}_{ij}^{n + 1} = {\bm{R}}_{ij}^{n + 1} - {\bm{R}}_{ij}^n,
\end{equation}
where ${\bm{R}}_{ij}$ is the Roe's flux function~\citep{blazek_computational_2015},
\begin{equation}\label{equ:implicit_mac_equ_roe_flux}
	{{\bm{R}}_{ij}} 
	= \frac{1}{2}\left[ {{{\bm{G}}_{ij}}\left( {{{\bm{W}}_i}} \right) + {{\bm{G}}_{ij}}\left( {{{\bm{W}}_j}} 
		\right) + {r_{ij}}\left( {{{\bm{W}}_i} - {{\bm{W}}_j}} \right)} \right].
\end{equation}
The ${\bm{G}}_{ij}$ is the Euler flux~\citep{blazek_computational_2015,yuan_conservative_2020}, which is a function of macroscopic flow variables
\begin{equation}\label{equ:implicit_mac_equ_euler_flux}
	{{\bm{G}}_{ij}}\left( {\bm{W}} \right) = \left( \begin{array}{l}
		\rho {\bm{U}} \cdot {{\bm{n}}_{ij}}\\
		\left( {\rho {\bm{UU}} + p{\bf{I}}} \right) \cdot {{\bm{n}}_{ij}}\\
		\left( {\rho E + p} \right){\bm{U}} \cdot {{\bm{n}}_{ij}}\\
		{\rho{E}_{rot}}{\bm{U}} \cdot {{\bm{n}}_{ij}}
	\end{array} \right),
\end{equation}
and $\bf{I}$ is the unit tensor, $r_{ij}$ is the spectral radius of the Euler flux Jacobian at the cell interface between the cell $i$ and the cell $j$,
\begin{equation}\label{equ:implicit_mac_equ_spectral_radius}
	{r_{ij}} = \left| {{{\bm{U}}_{ij}} \cdot {{\bm{n}}_{ij}}} \right| + {a_{ij}} + 
	2\frac{{{\mu _{ij}}}}{{{\rho _{ij}}\Delta {l_{ij}}}},{\rm{    }}\Delta {l_{ij}} = 
	\left| {{{\bm{x}}_i} - {{\bm{x}}_j}} \right|,
\end{equation}
where $a_{ij}$ is the acoustic speed at the cell interface.

Substituting Eqs.~\ref{equ:implicit_mac_equ_delta_F}$\sim$\ref{equ:implicit_mac_equ_euler_flux} into Eq.~\ref{equ:implicit_mac_equ_delta_form}, and since the relationship $\sum\nolimits_{j \in N\left( i \right)} {{A_{ij}}\Delta {{\bm{G}}_{ij}}\left( {{{\bm{W}}_i}} \right)}$ is always satisfied for any closed finite volume in a steady state solution, the governing equation for the macroscopic flow variables can be simplified as
\begin{equation}\label{equ:implicit_mac_equ_lusgs}
	\left( {\frac{1}{{\Delta t}} + \frac{1}{2}\sum\limits_{j \in N\left( i \right)} 
		{\frac{{{A_{ij}}}}{{\left| {{V_i}} \right|}}{r_{ij}}} } \right)\Delta {\bm{\tilde W}}_i^{n + 1} 
	+ \sum\limits_{j \in N\left( i \right)} {\frac{1}{2}\frac{{{A_{ij}}}}{{\left| {{V_i}} 
				\right|}}\left[ {\frac{{\partial {{\bm{G}}_{ij}}\left( {{\bm{W}}_j^n} \right)}}{{\partial 
					{\bm{W}}_j^n}} - {\bf{I}}{r_{ij}}} \right]\Delta {\bm{\tilde W}}_j^{n + 1}}  
	= {\bm{S}}_i^{n + 1} - {\bm{F}}_{ij}^n,
\end{equation}
where ${{\partial {\bm{G}}} \mathord{\left/{\vphantom {{\partial {\bm{G}}} {\partial {\bm{W}}}}} \right.\kern-\nulldelimiterspace} {\partial {\bm{W}}}}$ is the flux Jacobian.
Given that the source term $\bm S_i^{n+1}$ related to the rotational energy is not equal to zero.
Combining Eq.~\ref{equ:implicit_mac_equ_source_term}, we can get the expression as follows:
\begin{equation}\label{equ:implicit_mac_equ_fin}
	\begin{aligned}
		&{{\mathop{\rm D}\nolimits} _i}\Delta {\bm{\tilde W}}_i^{n + 1} + \sum\limits_{j \in N\left( i \right)} 
		{{{\mathop{\rm D}\nolimits} _j}\Delta {\bm{\tilde W}}_j^{n + 1}}  = {\mathop{\bm{Res}}\nolimits} _i^n.\\
		&{{\mathop{\rm D}\nolimits} _i} = \left\{ {\begin{array}{*{20}{l}}
				{\frac{1}{{\Delta t}} + \frac{1}{2}\sum\limits_{j \in N\left( i \right)} {\frac{{{A_{ij}}}}{{\left|
								{{V_i}} \right|}}{r_{ij}}} ,}&{W = \rho ,\rho {\bm{U}},\rho E}.\\
				{\frac{1}{{\Delta t}} + \frac{1}{2}\sum\limits_{j \in N\left( i \right)} {\frac{{{A_{ij}}}}{{\left| 
								{{V_i}} \right|}}{r_{ij}}}  + \frac{1}{{{Z_{rot}}\tilde \tau _i^{n + 1}}},}&{W = \rho {E_{rot}}}.
		\end{array}} \right.\\
		&{{\mathop{\rm D}\nolimits} _j} = \frac{1}{2}\frac{{{A_{ij}}}}{{\left| {{V_i}} \right|}}\left[ 
		{\frac{{\partial {{\bm{G}}_{ij}}\left( {{\bm{W}}_j^n} \right)}}{{\partial {\bm{W}}_j^n}} - 
			{\bf{I}}{r_{ij}}} \right].\\	
		&{\mathop{\bm{Res}}\nolimits} _i^n = \left\{ {\begin{array}{*{20}{l}}
				{ - {\bm{F}}_{ij}^n,}&{W = \rho ,\rho {\bm{U}},\rho E}.\\
				{\frac{{\tilde \rho _i^{n + 1}R\tilde T_i^{n + 1} - W_i^n}}{{{Z_{rot}}\tilde \tau _i^{n + 1}}} 
					- {{F}}_{ij}^n,}&{W = \rho {E_{rot}}}.
		\end{array}} \right.
	\end{aligned}
\end{equation}

The implicit macroscopic governing equation Eq.~\ref{equ:implicit_mac_equ_fin} can also be solved by using the PR-SGS method.
Since the calculation of macroscopic equations in the discrete physical space is less than that of microscopic equations in the discrete physical space and discrete velocity space.
In the present work, forty times' iteration for the macroscopic equations is performed per time step, each of which consists of a forward sweep step and a backward sweep step.
The Eq.~\ref{equ:implicit_mic_final_equ} and Eq.~\ref{equ:implicit_mac_equ_fin} are the update rules for the microscopic distribution functions and the predicted macroscopic flow variables, respectively.
Finally, once the micro-flux $F_{ij,k}^n$ across the cell interface is obtained, the macro-flux ${\bm{F}}_{ij}^{n}$ can be calculated from the micro-flux, and the implicit macroscopic and microscopic equations are uniquely determined.

\subsection{Simplified multi-scale numerical flux}\label{sec:flux}
At the cell interface, building a multi-scale numerical flux is crucial since it determines whether the scheme is multi-scale and suitable for all flow regimes.
In this paper, the idea of DUGKS is adopted and a simplified multi-scale numerical flux will be used.
We evolve the initial distribution function inside the cell to the interface taking into account the particle transport and collision process through a temporal difference scheme of the model equation.
Integrating the Eq.~\ref{equ:model_reduced_equ_phi} within a local physical time step $s_{ij}$ along the characteristic line ${\bm x}+{\bm u}_k{}t$ whose end point ${\bm x}_{ij}$ is the middle point of the cell interface $ij$, we derive the characteristic line solution of Rykov model equation,
\begin{equation}\label{equ:flux_dis_equ}
	\begin{aligned}
		\phi \left( {{{\bm{x}}_{ij}},{{\bm{u}}_k},{t_n} + {s_{ij}}} \right) - \phi \left( {{{\bm{x}}_{ij}} -
			{{\bm{u}}_k}{s_{ij}},{{\bm{u}}_k},{t_n}} \right) 
		= {s_{ij}}\frac{{{\phi ^*}\left( {{{\bm{x}}_{ij}},{{\bm{u}}_k},{t_n} + {s_{ij}}} \right) - 
				\phi \left( {{{\bm{x}}_{ij}},
					{{\bm{u}}_k},{t_n} + {s_{ij}}} \right)}}{{\tau {\left( {{{\bm{x}}_{ij}},{t_n} + {s_{ij}}} 
					\right)}}}.
	\end{aligned}
\end{equation} 
The local time step $s_{ij}={\bf{min}}({\Delta t}_{i}, {\Delta t}_{j})$, ${\Delta t}_{i}$ and ${\Delta t}_{j}$ are determined according to CFL condition.
Therefore, the distribution function $\phi \left( {{{\bm{x}}_{ij}}, {{\bm{u}}_k},{t_n} + {s_{ij}}} \right)$ at the cell interface is calculated as
\begin{equation}\label{equ:flux_solve_phi}
	\begin{aligned}
		\phi \left( {{{\bm{x}}_{ij}},{{\bm{u}}_k},{t_n} + {s_{ij}}} \right) &= \frac{{\tau {\left( 
					{{{\bm{x}}_{ij}},{t_n} + {s_{ij}}} \right)}}}{{\tau {\left( {{{\bm{x}}_{ij}},{t_n} + {s_{ij}}} 
					\right)} + {s_{ij}}}}\phi \left( {{{\bm{x}}_{ij}} - {{\bm{u}}_k}{s_{ij}},
			{{\bm{u}}_k},{t_n}} \right) \\
		&+ \frac{{s_{ij}}}{{\tau {\left( {{{\bm{x}}_{ij}},{t_n} + {s_{ij}}} \right)} + {s_{ij}}}}{\phi ^*}
		\left( {{{\bm{x}}_{ij}},{{\bm{u}}_k},{t_n} + {s_{ij}}} \right).
	\end{aligned}
\end{equation} 
From Eq.~\ref{equ:flux_solve_phi}, we know that the particle transport and collision effects are coupled in the construction of distribution function at the cell interface.
In the free molecule limit, $\tau \left( {{{\bm{x}}_{ij}},{t_n} + {s_{ij}}} \right) \gg {s_{ij}}$, the distribution function $\phi \left( {{{\bm{x}}_{ij}} - {{\bm{u}}_k}{s_{ij}},{{\bm{u}}_k},{t_n}} \right)$ directly constructed from initial distribution function at cell will be the main part owing to the inadequate particle collisions.
The physical process of particle free transport in the free molecule flow regime is accurately described.
In the continuum flow regime, $\tau \left( {{{\bm{x}}_{ij}},{t_n} + {s_{ij}}} \right) \ll {s_{ij}}$, the system tends to equilibrium state because of the intensive particle collision.
Therefore, the part of equilibrium state ${\phi ^*}\left( {{{\bm{x}}_{ij}},{{\bm{u}}_k},{t_n} + {s_{ij}}} \right)$ at the cell interface will automatically take a dominant role and the scheme captures the flow physics in the continuum flow limit as well.
Moreover, the local physical local time step $s_{ij}$ is implemented to preserve the multi-scale property in the non-uniform unstructured mesh.
Consequently, Eq.~\ref{equ:flux_solve_phi} is a self-adaptive multi-scale numerical flux in the local physical cell size and time step.

\begin{figure}[!t]
	\centering
	\includegraphics[scale=0.45]{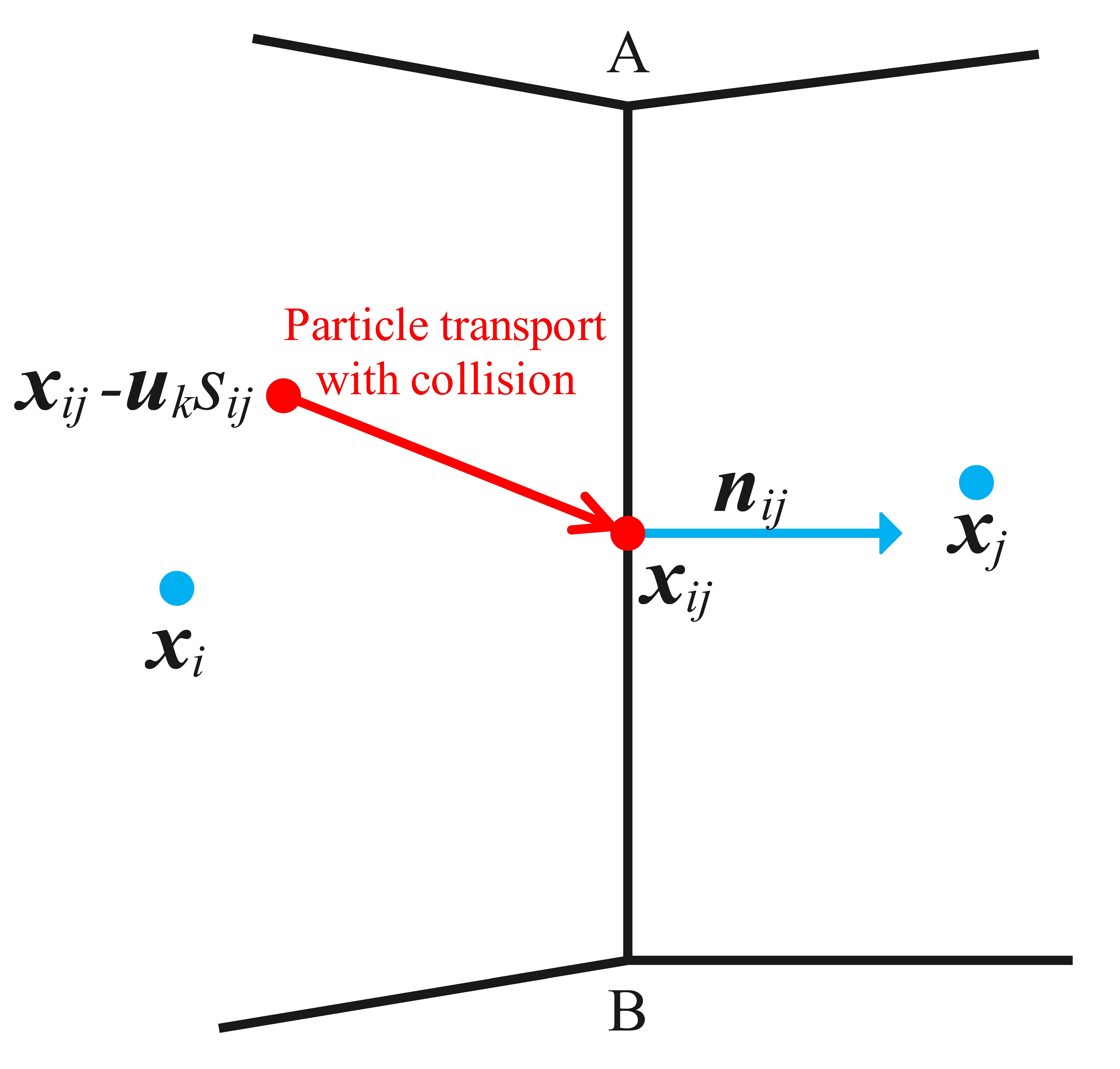}
	\caption{Sketch of two neighboring cells and the particle trajectories on a general unstructured mesh.}
	\label{Fig:Surface_Reconstruction}
\end{figure}

According to Eq.~\ref{equ:flux_solve_phi}, once the distribution function ${\phi}\left({{{\bm{x}}_{ij}}-{{\bm{u}}_k}{s_{ij}},{{\bm{u}}_k},{t_n}}\right)$ and the equilibrium distribution function $\phi^*({{{\bm{x}}_{ij}},{{\bm{u}}_k},{t_n} + {s_{ij}}})$ at the cell interface $\bm x_{ij}$ are obtained, then the distribution function $\phi \left( {{{\bm{x}}_{ij}},{{\bm{u}}_k},{t_n} + {s_{ij}}} \right)$ can be recovered.
The distribution function ${\phi}\left({{{\bm{x}}_{ij}}-{{\bm{u}}_k}{s_{ij}},{{\bm{u}}_k},{t_n}}\right)$ is obtained through the reconstruction of the initial distribution function data,
\begin{equation}\label{equ:flux_solve_phi_phi}
	{\phi}\left( {{{\bm{x}}_{ij}} - {{\bm{u}}_k}{s_{ij}},{{\bm{u}}_k},{t_n}} \right) = {\phi}\left( 
	{{{\bm{x}}_c},{{\bm{u}}_k},{t_n}} \right) + \left( {{{\bm{x}}_{ij}} - {{\bm{u}}_k}{s_{ij}} -
		{{\bm{x}}_c}} \right) \cdot L\left( {\nabla {{\phi}},{{\bm{x}}_c}} \right)\nabla 
	{\phi}\left( {{{\bm{x}}_c},{{\bm{u}}_k},{t_n}} \right),
\end{equation}
where ${\bm x}_c$ represent the central coordinates of the cell which the particles migrate from.
As shown in Fig.~\ref{Fig:Surface_Reconstruction}, the ${\bm x}_c$ equals to ${\bm x}_i$ if ${{{\bm{u}}_k}\cdot{{\bm{n}}_{ij}}} > 0$, or ${\bm x}_j$ otherwise.
The gradient $\nabla {\phi}({{{\bm{x}}_c},{{\bm{u}}_k},{t_n}})$ at the cell center is calculated using the least square method.
The function $L\left( {\nabla {{\phi}},{{\bm{x}}_c}} \right)$ in Eq.~\ref{equ:flux_solve_phi_phi} denotes the gradient limiter which is used to suppress numerical oscillations, and the Venkatakrishnan limiter~\citep{venkatakrishnan_convergence_1995} for flow simulations on unstructured mesh is adopted.

By taking the moment of Eq.~\ref{equ:flux_dis_equ} in the velocity space for ${\bf {\Phi}}\left( {\bm{u}} \right)$, the macroscopic flow variables $\bm {W}(\bm x_{ij},{t_n} + s_{ij})$ used to evaluate the equilibrium state $\phi^*({{{\bm{x}}_{ij}},{{\bm{u}}_k},{t_n} + s_{ij}})$ can also be derived directly.
\begin{equation}\label{equ:halftime_solve_rho}
	\rho \left( {{{\bm{x}}_{ij}},{t_n} + {s_{ij}}} \right) = \sum\limits_k 
	{{\omega _k}G\left( {{{\bm{x}}_{ij}} - {{\bm{u}}_k}{s_{ij}},{{\bm{u}}_k},{t_n}} \right)},
\end{equation}
\begin{equation}\label{equ:halftime_solve_U}
	\left( {\rho {\bm{U}}} \right)\left( {{{\bm{x}}_{ij}},{t_n} + {s_{ij}}} \right) = \sum\limits_k 
	{{\omega _k}{{\bm{u}}_k}G\left( {{{\bm{x}}_{ij}} - {{\bm{u}}_k}{s_{ij}},{{\bm{u}}_k},{t_n}} \right)},
\end{equation}
\begin{equation}\label{equ:halftime_solve_E}
	\left( {\rho E} \right)\left( {{{\bm{x}}_{ij}},{t_n} + {s_{ij}}} \right) = \sum\limits_k 
	{{\omega _k}\left[ {\frac{1}{2}{{\left| {{{\bm{u}}_k}} \right|}^2}G\left( {{{\bm{x}}_{ij}} 
				- {{\bm{u}}_k}{s_{ij}},{{\bm{u}}_k},{t_n}} \right) + R\left( {{{\bm{x}}_{ij}} 
				- {{\bm{u}}_k}{s_{ij}},{{\bm{u}}_k},{t_n}} \right)} \right]},
\end{equation}
\begin{equation}\label{equ:halftime_solve_Erot}
	\begin{aligned}
		\left( {\rho {E_{rot}}} \right)\left( {{{\bm{x}}_{ij}},{t_n} + {s_{ij}}} \right) &= 
		\frac{{{Z_{rot}}\tau \left( {{{\bm{x}}_{ij}},{t_n} + {s_{ij}}} \right)}}{{{Z_{rot}}\tau 
				\left( {{{\bm{x}}_{ij}},{t_n} + {s_{ij}}} \right) + {s_{ij}}}}\sum\limits_k {{\omega _k}
			R\left( {{{\bm{x}}_{ij}} - {{\bm{u}}_k}{s_{ij}},{{\bm{u}}_k},{t_n}} \right)} \\
		&+ \frac{{{s_{ij}}\rho \left( {{{\bm{x}}_{ij}},{t_n} + {s_{ij}}} \right)RT\left( {{{\bm{x}}_{ij}},
					{t_n} + {s_{ij}}} \right)}}{{{Z_{rot}}\tau \left( {{{\bm{x}}_{ij}},{t_n} + {s_{ij}}} 
				\right) + {s_{ij}}}}.
	\end{aligned}
\end{equation}
Furthermore, the translational and rotational heat fluxes $\bm{q}_{tr}(\bm x_{ij},t_n+s_{ij})$, $\bm{q}_{rot}(\bm x_{ij},t_n+s_{ij})$ can also be determined from the distribution function ${\phi}\left({{{\bm{x}}_{ij}}-{{\bm{u}}_k}s_{ij},{{\bm{u}}_k},{t_n}}\right)$ as:
\begin{equation}\label{equ:halftime_solve_qtra}
	{{\bm{q}}_{tr}}\left( {{{\bm{x}}_{ij}},{t_n} + {s_{ij}}} \right) = \frac{{\tau 
			\left( {{{\bm{x}}_{ij}},{t_n} + {s_{ij}}} \right)\frac{1}{2}\sum\limits_k {{\omega _k}
				{{\bm{c}}_k}{{\left| {{{\bm{c}}_k}} \right|}^2}G\left( {{{\bm{x}}_{ij}} -{{\bm{u}}_k}{s_{ij}},
					{{\bm{u}}_k},{t_n}} \right)} }}{{\tau \left( {{{\bm{x}}_{ij}},{t_n} + {s_{ij}}} \right) +
			{s_{ij}} - \frac{1}{3}{s_{ij}}\left( {1 - {1 \mathord{\left/
						{\vphantom {1 {{Z_{rot}}}}} \right.\kern-\nulldelimiterspace} {{Z_{rot}}}} + {{{\omega _0}} 
					\mathord{\left/{\vphantom {{{\omega _0}} {{Z_{rot}}}}} \right.
						\kern-\nulldelimiterspace} {{Z_{rot}}}}} \right)}},
\end{equation}
\begin{equation}\label{equ:halftime_solve_qrot}
	{{\bm{q}}_{_{rot}}}\left( {{{\bm{x}}_{ij}},{t_n} + {s_{ij}}} \right) = \frac{{\tau \left( 
			{{{\bm{x}}_{ij}},{t_n} + {s_{ij}}} \right)\sum\limits_k {{\omega _k}{{\bm{c}}_k}R\left( 
				{{{\bm{x}}_{ij}} - {{\bm{u}}_k}{s_{ij}},{{\bm{u}}_k},{t_n}} \right)} }}{{\tau \left( 
			{{{\bm{x}}_{ij}},{t_n} + {s_{ij}}} \right) + {s_{ij}} - {s_{ij}}\left( {1 - \delta } \right)
			\left( {1 - {1 \mathord{\left/{\vphantom {1 {{Z_{rot}}}}} \right.
						\kern-\nulldelimiterspace} {{Z_{rot}}}} + {{{\omega _1}} \mathord{\left/
						{\vphantom {{{\omega _1}} {{Z_{rot}}}}} \right.
						\kern-\nulldelimiterspace} {{Z_{rot}}}}} \right)}}.
\end{equation}
Up to now, the equilibrium distribution function $\phi^*({{{\bm{x}}_{ij}},{{\bm{u}}_k},{t_n} + s_{ij}})$ can be obtained from the macroscopic flow variables.
As a result, the distribution function $\phi({{{\bm{x}}_{ij}},{{\bm{u}}_k},{t_n} + s_{ij}})$ is solved by Eq.~\ref{equ:flux_solve_phi}.
The micro-flux $F_{ij,k}^n$ can be obtained using Eq.~\ref{equ:implicit_mic_flux} and the macro-flux ${\bm{F}}_{ij}^{n}$ is calculated from the micro-flux by numerical integrals in the DVS according to Eq.~\ref{equ:implicit_mac_equ_mac_flux}.

\subsection{Asymptotic preserving property of flux in the continuum limit}\label{sec:apflux}
The limiting transition to the continuum flow regime of Rykov model equation has been accomplished in the work of Rykov and Skobelkin on the basis of an asymptotic analysis~\citep{rykov_macroscopic_1978}.
In this work, we focus on investigating the distribution function used for the flux evaluation in the continuum limit, i.e., $\phi({{{\bm{x}}_{ij}},{{\bm{u}}_k},{t_n} + s_{ij}})$ given by Eq.~\ref{equ:flux_solve_phi}.
In the continuum limit, the gas distribution function $\phi$ can be approximated by the first-order Chapman-Enskog expansion,
\begin{equation}\label{equ:ce_phi}
	\phi  = {\phi ^*} - \tau {D_t}{\phi ^*} + O\left( {D_t^2} \right),
\end{equation}
where ${D_t} = {\partial _t} + {\bm{u}} \cdot \nabla$. According to Eq.~\ref{equ:flux_solve_phi_phi} (ignore the Venkatakrishnan limiter) and Eq.~\ref{equ:ce_phi}, the free transport term $\phi \left( {{{\bm{x}}_{ij}} - {{\bm{u}}_k}{s_{ij}},{{\bm{u}}_k},{t_n}} \right)$ can be approximated as follows:
\begin{equation}\label{equ:ce_phi_f_transport}
	\begin{aligned}
		&\frac{\tau }{{\tau  + {s_{ij}}}}\phi \left( {{{\bm{x}}_{ij}} - {{\bm{u}}_k}{s_{ij}},{{\bm{u}}_k},{t_n}} \right) = \frac{\tau }{{\tau  + {s_{ij}}}}\left[ {\phi \left( {{{\bm{x}}_c},{{\bm{u}}_k},{t_n}} \right) + \left( {{{\bm{x}}_{ij}} - {{\bm{u}}_k}{s_{ij}} - {{\bm{x}}_c}} \right) \cdot \nabla \phi \left( {{{\bm{x}}_c},{{\bm{u}}_k},{t_n}} \right)} \right]\\
		&\approx \frac{\tau }{{\tau  + {s_{ij}}}}\left[ {{\phi ^*}\left( {{{\bm{x}}_c},{{\bm{u}}_k},{t_n}} \right) + \left( {{{\bm{x}}_{ij}} - {{\bm{x}}_c}} \right) \cdot \nabla {\phi ^*}\left( {{{\bm{x}}_c},{{\bm{u}}_k},{t_n}} \right) + {s_{ij}}{\partial _t}{\phi ^*}\left( {{{\bm{x}}_c},{{\bm{u}}_k},{t_n}} \right)} \right] - \tau {D_t}{\phi ^*}\left( {{{\bm{x}}_c},{{\bm{u}}_k},{t_n}} \right),
	\end{aligned}
\end{equation}
where the high order derivative term is ignored.

Then we deal with the equilibrium state ${\phi ^*}\left( {{{\bm{x}}_{ij}},{{\bm{u}}_k},{t_n} + {s_{ij}}} \right)$ at the cell interface, take the moments of Eq.~\ref{equ:flux_dis_equ} and Eq.~\ref{equ:flux_solve_phi_phi} in the continuous velocity space for ${\bf {\Phi}}\left( {\bm{u}} \right)$ (Eq.~\ref{equ:PHI_matrix}) with respect to $\bm u$ from $-\infty$ to $+\infty$, respectively, one can obtain
\begin{equation}\label{equ:ce_w_f_1}
	{\bm{W}}\left( {{{\bm{x}}_{ij}},{t_n} + {s_{ij}}} \right) = \int {{\bf{\Phi }}
		\left( {\bm{u}} \right) \cdot \left( {\begin{array}{*{20}{c}}
				G\\
				R
		\end{array}} \right)\left( {{{\bm{x}}_{ij}} - {\bm{u}}{s_{ij}},{\bm{u}},{t_n}} 
		\right)d{\bm{u}}}  + {s_{ij}}\int {{\bf{\Phi }}\left( {\bm{u}} \right) \cdot 
		\left( {\begin{array}{*{20}{c}}
				{{\Omega _G}}\\
				{{\Omega _R}}
		\end{array}} \right)
		\left( {{{\bm{x}}_{ij}},{\bm{u}},{t_n} + {s_{ij}}} \right)d{\bm{u}}},
\end{equation}
\begin{equation}\label{equ:ce_w_f_2}
	\begin{aligned}
		&\int {{\bf{\Phi }}\left( {\bm{u}} \right) \cdot \left( {\begin{array}{*{20}{c}}
					G\\
					R
			\end{array}} \right)\left( {{{\bm{x}}_{ij}} - {\bm{u}}{s_{ij}},{\bm{u}},{t_n}} \right)
			d{\bm{u}}} \\ 
		&= {\bm{W}}\left( {{{\bm{x}}_c},{t_n}} \right) + \int {{\bf{\Phi }}
			\left( {\bm{u}} \right) \cdot \left[ {\left( {{{\bm{x}}_{ij}} - {\bm{u}}{s_{ij}} - 
					{{\bm{x}}_c}} \right) \cdot \nabla } \right]\left( {\begin{array}{*{20}{c}}
					G\\
					R
			\end{array}} \right)
			\left( {{{\bm{x}}_{ij}} - {\bm{u}}{s_{ij}},{\bm{u}},{t_n}} \right)d{\bm{u}}}.
	\end{aligned}
\end{equation}
From Eqs.~\ref{equ:ce_w_f_1} and~\ref{equ:ce_w_f_2}, we have
\begin{equation}\label{equ:ce_w_f_12}
	\begin{aligned}
		&{\bm{W}}\left( {{{\bm{x}}_{ij}},{t_n} + {s_{ij}}} \right) - {\bm{W}}\left( 
		{{{\bm{x}}_c},{t_n}} \right)\\
		&= \int {{\bf{\Phi }}\left( {\bm{u}} \right) \cdot \left[ {\left( {{{\bm{x}}_{ij}} 
					- {\bm{u}}{s_{ij}} - {{\bm{x}}_c}} \right) \cdot \nabla } \right]
			\left( {\begin{array}{*{20}{c}}
					G\\
					R
			\end{array}} \right)\left( {{{\bm{x}}_{ij}} - {\bm{u}}{s_{ij}},
				{\bm{u}},{t_n}} \right)d{\bm{u}}}  \\
		&+ {s_{ij}}\int {{\bf{\Phi }}\left( {\bm{u}} \right) 
			\cdot \left( {\begin{array}{*{20}{c}}
					{{\Omega _G}}\\
					{{\Omega _R}}
			\end{array}} \right)
			\left( {{{\bm{x}}_{ij}},{\bm{u}},{t_n} + {s_{ij}}} \right)d{\bm{u}}}.
	\end{aligned}
\end{equation}
Based on the Chapman-Enskog expansion and the Taylor expansion, Eq.~\ref{equ:ce_w_f_12} can be reduced to the following form without regard to high order derivative term,
\begin{equation}\label{equ:ce_w_f_12_simplified}
	\begin{aligned}
		&{\bm{W}}\left( {{{\bm{x}}_{ij}},{t_n} + {s_{ij}}} \right) - {\bm{W}}\left( 
		{{{\bm{x}}_c},{t_n}} \right)\\
		&\approx \int {{\bf{\Phi }}\left( {\bm{u}} \right) \cdot \left[ {\left( {{{\bm{x}}_{ij}}
					- {{\bm{x}}_c}} \right) \cdot \nabla } \right]\left( {\begin{array}{*{20}{c}}
					{{G^*}}\\
					{{R^*}}
			\end{array}} \right)\left( {{{\bm{x}}_c},{\bm{u}},{t_n}} \right)d{\bm{u}}}  + 
		{s_{ij}}\int {{\bf{\Phi }}\left( {\bm{u}} \right) \cdot {\partial _t}\left( 
			{\begin{array}{*{20}{c}}
					{{G^*}}\\
					{{R^*}}
			\end{array}} \right)\left( {{{\bm{x}}_c},{\bm{u}},{t_n}} \right)d{\bm{u}}}.
	\end{aligned}
\end{equation}
Furthermore, the equilibrium state ${\phi ^*}\left( {{{\bm{x}}_{ij}},{{\bm{u}}_k},{t_n} + {s_{ij}}} \right)$ can be approximated as
\begin{equation}\label{equ:ce_feq_w}
	{\phi ^*}\left( {{{\bm{x}}_{ij}},{{\bm{u}}_k},{t_n}
		+ {s_{ij}}} \right)
	\approx {{\phi ^*}\left[ {{{\bm{u}}_k},
			{\bm{W}}\left( {{{\bm{x}}_c},{t_n}} \right)} \right] + \frac{{\partial {\phi ^*}}}
		{{\partial {\bm{W}}}} \cdot \left[ {{\bm{W}}\left( {{{\bm{x}}_{ij}},{t_n} + 
				{s_{ij}}} \right) - {\bm{W}}\left( {{{\bm{x}}_c},{t_n}} \right)} \right]}.
\end{equation}
Therefore, from Eqs.~\ref{equ:ce_w_f_12_simplified} and~\ref{equ:ce_feq_w}, we have
\begin{equation}\label{equ:ce_phi_eq}
	\begin{aligned}
		&\frac{{{s_{ij}}}}{{\tau  + {s_{ij}}}}{\phi ^*}\left( {{{\bm{x}}_{ij}},{{\bm{u}}_k},{t_n} 
			+ {s_{ij}}} \right) \approx \frac{{{s_{ij}}}}{{\tau  + {s_{ij}}}}{\phi ^*}
		\left( {{{\bm{x}}_c},{{\bm{u}}_k},{t_n}} \right)\\
		&
		+ \frac{{{s_{ij}}}}{{\tau  + {s_{ij}}}}\frac{{\partial {\phi ^*}}}{{\partial 
				{\bm{W}}}} \cdot \int {{\bf{\Phi }}\left( {\bm{u}} \right) \cdot \left[
			{\left( {{{\bm{x}}_{ij}} - {{\bm{x}}_c}} \right) \cdot \nabla } \right]
			\left( {\begin{array}{*{20}{c}}
					{{G^*}}\\
					{{R^*}}
			\end{array}} \right)\left( {{{\bm{x}}_c},{\bm{u}},{t_n}} \right)d{\bm{u}}} \\
		&+ \frac{{{s_{ij}}}}{{\tau  + {s_{ij}}}}\frac{{\partial {\phi ^*}}}{{\partial {\bm{W}}}} 
		\cdot {s_{ij}}\int {{\bf{\Phi }}\left( {\bm{u}} \right) \cdot {\partial _t}
			\left( {\begin{array}{*{20}{c}}
					{{G^*}}\\
					{{R^*}}
			\end{array}} \right)\left( {{{\bm{x}}_c},{\bm{u}},{t_n}} \right)d{\bm{u}}}.
	\end{aligned}
\end{equation}
Finally, with the above Eqs.~\ref{equ:ce_phi_f_transport} and~\ref{equ:ce_phi_eq}, the gas distribution function $\phi({{{\bm{x}}_{ij}},{{\bm{u}}_k},{t_n} + s_{ij}})$ used for the flux evaluation in the continuum limit can be expressed as
\begin{equation}\label{equ:ce_phi_f}
	\begin{aligned}
		&\phi \left( {{{\bm{x}}_{ij}},{{\bm{u}}_k},{t_n} + {s_{ij}}} \right) = 
		{\phi ^*}\left( {{{\bm{x}}_c},{{\bm{u}}_k},{t_n}} \right) - \tau {D_t}{\phi ^*}
		\left( {{{\bm{x}}_c},{{\bm{u}}_k},{t_n}} \right)\\
		&+ \frac{\tau }{{\tau  + {s_{ij}}}}\left( {{{\bm{x}}_{ij}} - {{\bm{x}}_c}} \right) 
		\cdot \nabla {\phi ^*}\left( {{{\bm{x}}_c},{{\bm{u}}_k},{t_n}} \right) + 
		\frac{{\tau {s_{ij}}}}{{\tau  + {s_{ij}}}}{\partial _t}{\phi ^*}
		\left( {{{\bm{x}}_c},{{\bm{u}}_k},{t_n}} \right)\\
		&+ \frac{{{s_{ij}}}}{{\tau  + {s_{ij}}}}\frac{{\partial {\phi ^*}}}{{\partial 
				{\bm{W}}}} \cdot \int {{\bf{\Phi }}\left( {\bm{u}} \right) \cdot 
			\left[ {\left( {{{\bm{x}}_{ij}} - {{\bm{x}}_c}} \right) \cdot \nabla } 
			\right]\left( {\begin{array}{*{20}{c}}
					{{G^*}}\\
					{{R^*}}
			\end{array}} \right)\left( {{{\bm{x}}_c},{\bm{u}},{t_n}} \right)d{\bm{u}}} \\
		&+ \frac{{{s_{ij}}}}{{\tau  + {s_{ij}}}}\frac{{\partial {\phi ^*}}}{{\partial 
				{\bm{W}}}} \cdot {s_{ij}}\int {{\bf{\Phi }}\left( {\bm{u}} \right)
			\cdot {\partial _t}\left( {\begin{array}{*{20}{c}}
					{{G^*}}\\
					{{R^*}}
			\end{array}} \right)\left( {{{\bm{x}}_c},{\bm{u}},{t_n}} \right)d{\bm{u}}}. 
	\end{aligned}
\end{equation}
It is noticed that the $\phi({{{\bm{x}}_{ij}},{{\bm{u}}_k},{t_n} + s_{ij}})$ is a precise Chapman-Enskog Navier-Stokes distribution function and the first-order Taylor expansion of $\phi({{{\bm{x}}_{c}},{{\bm{u}}_k},{t_n} })$ with respect to time and space also take into consideration.

\subsection{General framework of the algorithm}\label{sec:framework}
In this section, we make a summary of the whole computation procedure of the present algorithm from time level $t_n$ to $t_{n+1}$ in the following steps:
\\
\textbf{Step 1.} Start the computation with an initial flow field, where the gas distribution functions in each cell are equilibrium state.
\\
\textbf{Step 2.} Compute the micro-flux $F_{ij,k}^n$ across the cell interface of control volumes.

\textbf{(a)} Calculate the distribution function ${\phi}\left({{{\bm{x}}_{ij}}-{{\bm{u}}_k}{s_{ij}},{{\bm{u}}_k},{t_n}}\right)$ according to Eq.~\ref{equ:flux_solve_phi_phi}.

\textbf{(b)} Calculate the macroscopic flow variables $\bm {W}(\bm x_{ij},{t_n} + s_{ij})$ using Eqs.~\ref{equ:halftime_solve_rho}$\sim$\ref{equ:halftime_solve_Erot} and the translational heat flux and rotational heat flux using Eqs.~\ref{equ:halftime_solve_qtra}$\sim$\ref{equ:halftime_solve_qrot}.

\textbf{(c)} Calculate the equilibrium distribution function $\phi^*({{{\bm{x}}_{ij}},{{\bm{u}}_k},{t_n} + s_{ij}})$ from the macroscopic flow variables and heat fluxes.

\textbf{(d)} Calculate the distribution function $\phi({{{\bm{x}}_{ij}},{{\bm{u}}_k},{t_n} + s_{ij}})$ at the cell interface by Eq.~\ref{equ:flux_solve_phi}.

\textbf{(e)} Calculate the micro-flux $F_{ij,k}^n$ using Eq.~\ref{equ:implicit_mic_flux}.
\\
\textbf{Step 3.} Compute the macro-flux ${\bm{F}}_{ij}^{n}$ across the cell interface of control volumes by Eq.~\ref{equ:implicit_mac_equ_mac_flux}.
\\
\textbf{Step 4.} Solve the implicit macroscopic equations Eq.~\ref{equ:implicit_mac_equ_fin} to predict the macroscopic flow variables $\tilde {\bm{W}}_i^{n+1}$.
With the predicted macroscopic flow variables $\tilde {\bm{W}}_i^{n+1}$, the predicted equilibrium distribution function $\tilde \phi _{i,k}^{*,n + 1}$ can be obtained.
\\
\textbf{Step 5.} Solve the implicit microscopic equation Eq.~\ref{equ:implicit_mic_final_equ} to update distribution function $\phi _{i,k}^{n + 1}$.
\\
\textbf{Step 6.} Update the macroscopic flow variables ${\bm{W}}_i^{n+1}$ and heat fluxes $(\bm q_{tr})_i^{n+1}$ and $(\bm q_{rot})_i^{n+1}$ by the integral error compensation technique~\citep{yuan_conservative_2020} according to Eqs.~\ref{equ:solve_rho}$\sim$\ref{equ:solve_qrot}.
\begin{equation}\label{equ:solve_rho}
	\rho _i^{n + 1} = \sum\limits_k {{\omega _k}\left( {G_{i,k}^{n + 1} - \tilde G_{i,k}^{*,n + 1}} 
		\right)}  + \tilde \rho _i^{n + 1},
\end{equation}
\begin{equation}\label{equ:solve_rhoU}
	\left( {\rho {\bm{U}}} \right)_i^{n + 1} = \sum\limits_k {{\omega _k}{{\bm{u}}_k}\left( {G_{i,k}^{n + 1}
			- \tilde G_{i,k}^{*,n + 1}} \right)}  + \left( {\tilde \rho {\bm{\tilde U}}} \right)_i^{n + 1},
\end{equation}
\begin{equation}\label{equ:solve_e}
	\left( {\rho E} \right)_i^{n + 1} = \sum\limits_k {{\omega _k}\left[ {\frac{1}{2}{{\left| {\bm{u}} 
					\right|}^2}\left( {G_{i,k}^{n + 1} - \tilde G_{i,k}^{*,n + 1}} \right) + \left( {R_{i,k}^{n + 1} 
				- \tilde R_{i,k}^{*,n + 1}} \right)} \right]}  + \left( {\tilde \rho \tilde E} \right)_i^{n + 1},
\end{equation}
\begin{equation}\label{equ:solve_erot}
	\left( {\rho {E_{rot}}} \right)_i^{n + 1} = \sum\limits_k {{\omega _k}\left( {R_{i,k}^{n + 1} 
			- \tilde R_{i,k}^{*,n + 1}} \right)}  + \frac{{{K_{rot}}}}{2}\tilde \rho _i^{n + 1}R\left[ 
	{\left( {1 - {1 \mathord{\left/{\vphantom {1 {{Z_{rot}}}}} \right.
					\kern-\nulldelimiterspace} {{Z_{rot}}}}} \right)\left( {{{\tilde T}_{rot}}} \right)_i^{n + 1} 
		+ {{\tilde T_i^{n + 1}} \mathord{\left/
				{\vphantom {{\tilde T_i^{n + 1}} {{Z_{rot}}}}} \right.
				\kern-\nulldelimiterspace} {{Z_{rot}}}}} \right],
\end{equation}
\begin{equation}\label{equ:solve_qtr}
	\left( {{{\bm{q}}_{tr}}} \right)_i^{n + 1} = \frac{1}{2}\sum\limits_k {{\omega _k}{{\bm{c}}_k}
		{{\left| {{{\bm{c}}_k}} \right|}^2}\left( {G_{i,k}^{n + 1} - \tilde G_{i,k}^{*,n + 1}} \right)}  
	+ \left( {{{{\bm{\tilde q}}}_{tr}}} \right)_i^{n + 1}\frac{1}{3}\left( 
	{1 - \frac{{1 - {\omega _0}}}{{{Z_{rot}}}}} \right),
\end{equation}
\begin{equation}\label{equ:solve_qrot}
	\left( {{{\bm{q}}_{rot}}} \right)_i^{n + 1} = \sum\limits_k {{\omega _k}{{\bm{c}}_k}\left( 
		{R_{i,k}^{n + 1} - \tilde R_{i,k}^{*,n + 1}} \right)}  + \left( {{{{\bm{\tilde q}}}_{rot}}} 
	\right)_i^{n + 1}\left( {1 - \delta } \right)\left( {1 - \frac{{1 - {\omega _1}}}{{{Z_{rot}}}}} \right).
\end{equation}
\\
\textbf{Step 7.} Make judgment: If the residuals satisfy the convergent condition, output the flow field and stop the computation. Otherwise, go to \textbf{Step 2}.

\section{Numerical Results and Discussions}\label{sec:cases}
In this section, four three-dimensional test cases are conducted to verify the present method at various Mach (Ma) and Knudsen numbers.
The lid-driven cavity flow is firstly simulated to evaluate the efficiency and accuracy of the present method at different flow regimes.
And then, supersonic and hypersonic flows over a sphere and hypersonic flows over a blunted-cone are performed to validate the present method in the extremely non-equilibrium flows.
Finally, the reentry trajectory of Apollo 6 command module in the range from 200 to 85 km altitude is carried out to assess the ability of the present method for predicting the hypersonic rarefied and near-continuum flows.

In the calculation, dimensionless quantities normalized by the reference length, density, temperature and velocity are introduced.
\begin{equation}\label{equ:ref_quantities}
	{L_{ref}} = {L_c},{\qquad}{\rho _{ref}} = {\rho _\infty },
	{\qquad}{T_{ref}} = {T_\infty },{\qquad}{U_{ref}}=\sqrt {2R{T_{ref}}},
\end{equation}
where $L_c$ is the characteristic length scale of the flow, ${\rho_\infty}$, ${T_\infty}$ are free stream density and temperature, respectively.
By using the reference variables expressed in Eqs.~\ref{equ:ref_quantities} and~\ref{equ:reference}, we are able to obtain a complete dimensionless system.
\begin{equation}\label{equ:reference}
	\begin{array}{*{20}{l}}
		{{t_{ref}} = {L_{ref}}U_{ref}^{ - 1},}&{{R_{ref}} = U_{ref}^2T_{ref}^{ - 1},}&{{E_{ref}} = U_{ref}^2,}\\
		{{p_{ref}} = {\rho _{ref}}U_{ref}^2,}&{{q_{ref}} = {\rho _{ref}}U_{ref}^3,}
		&{{\mu _{ref}} = {\rho _{ref}}{U_{ref}}{L_{ref}}.}
	\end{array}
\end{equation}

\subsection{The lid-driven cavity flow}\label{sec:caseA}
The 3D lid-driven cavity flows at different flow regimes are simulated to investigate the performance of the proposed method.
In all of the simulations, computational domain is a cubic box with the edge length of ${L}$, which is filled with the nitrogen gas.
The VHS molecular model with $\omega = 0.74$ is applied and the initial temperature of gas is $T=273\rm{K}$.
On the wall of the cavity, the diffuse reflection boundary condition with full thermal accommodation is implemented, and the wall temperature $T_{w}=T$.
The Mach number, which is defined by the upper wall velocity and the acoustic velocity, is set as 0.1624.
The rotational collision number keeps a constant value of $Z_{rot}=3$.

\begin{figure}[!htp]
	\centering	
	\subfigure[\label{Fig:CavityFlow_Mesh}]{
		\includegraphics[width=0.45\textwidth]{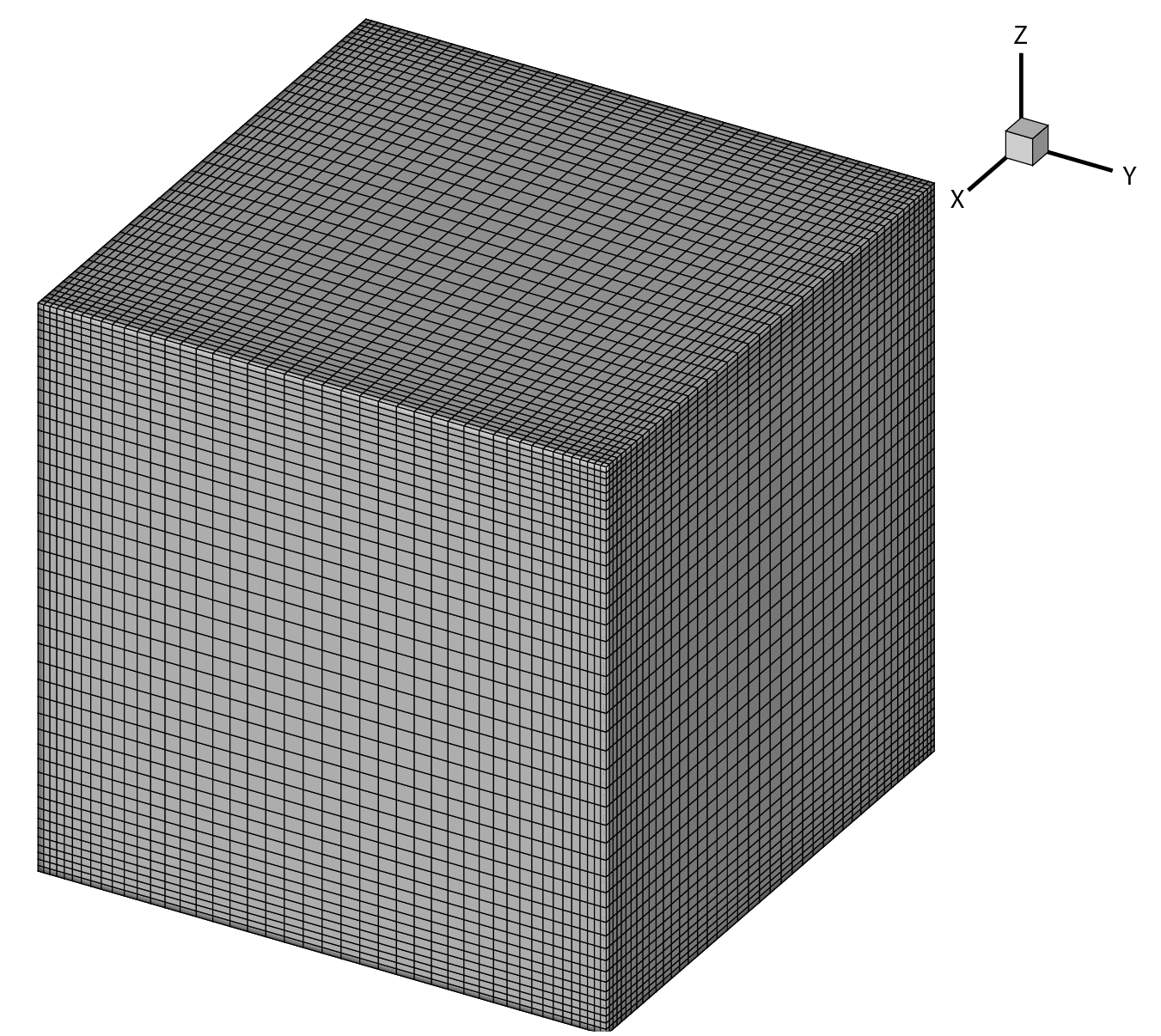}
	}
	\subfigure[\label{Fig:CavityFlow_Micmesh}]{
		\includegraphics[width=0.45\textwidth]{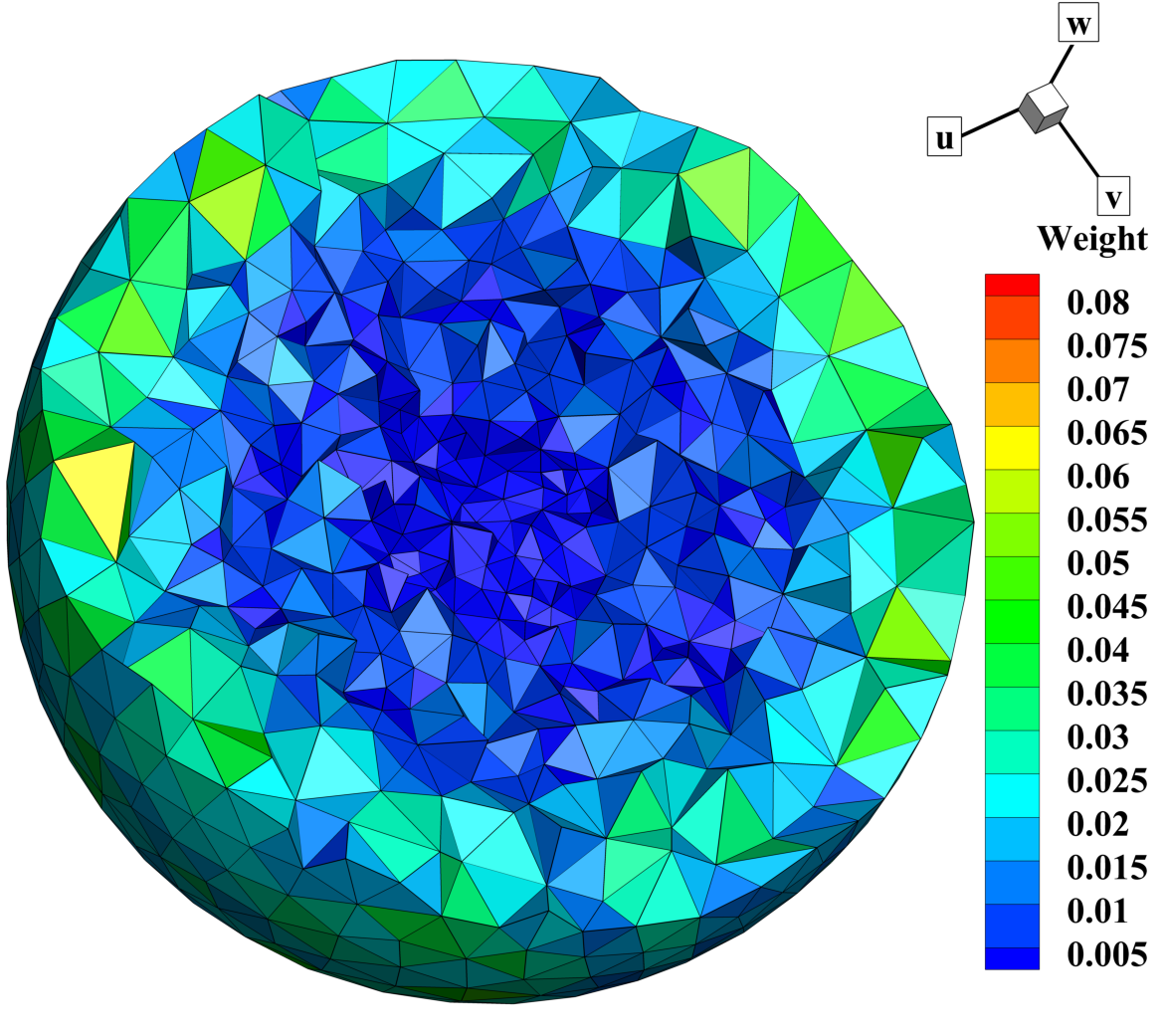}
	}
	\caption{Diagrams of (a) the physical space mesh (85184 cells) and (b) the section view of the unstructured discrete velocity space mesh (total 22710 cells) for the three-dimensional lid-driven cavity flows.}
	\label{Fig:CavityFlow_MeshMicmesh} 
\end{figure}

\begin{figure}[!htp]
	\centering
	\subfigure[\label{Fig:CavityFlow_Kn10_Velocity_Line}]{
		\includegraphics[width=0.45\textwidth]{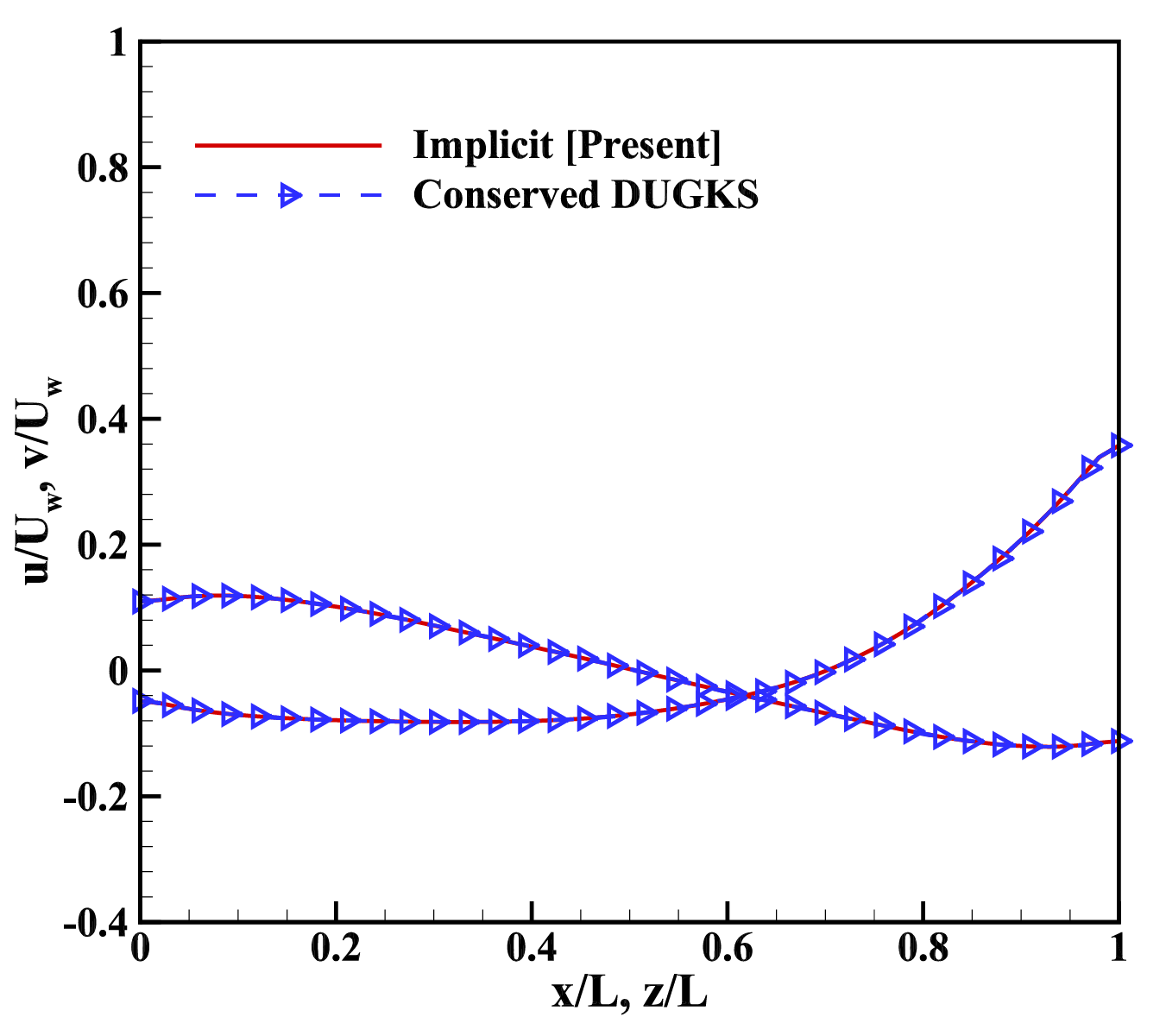}
	}
	\subfigure[\label{Fig:CavityFlow_Kn1_Velocity_Line}]{
		\includegraphics[width=0.45\textwidth]{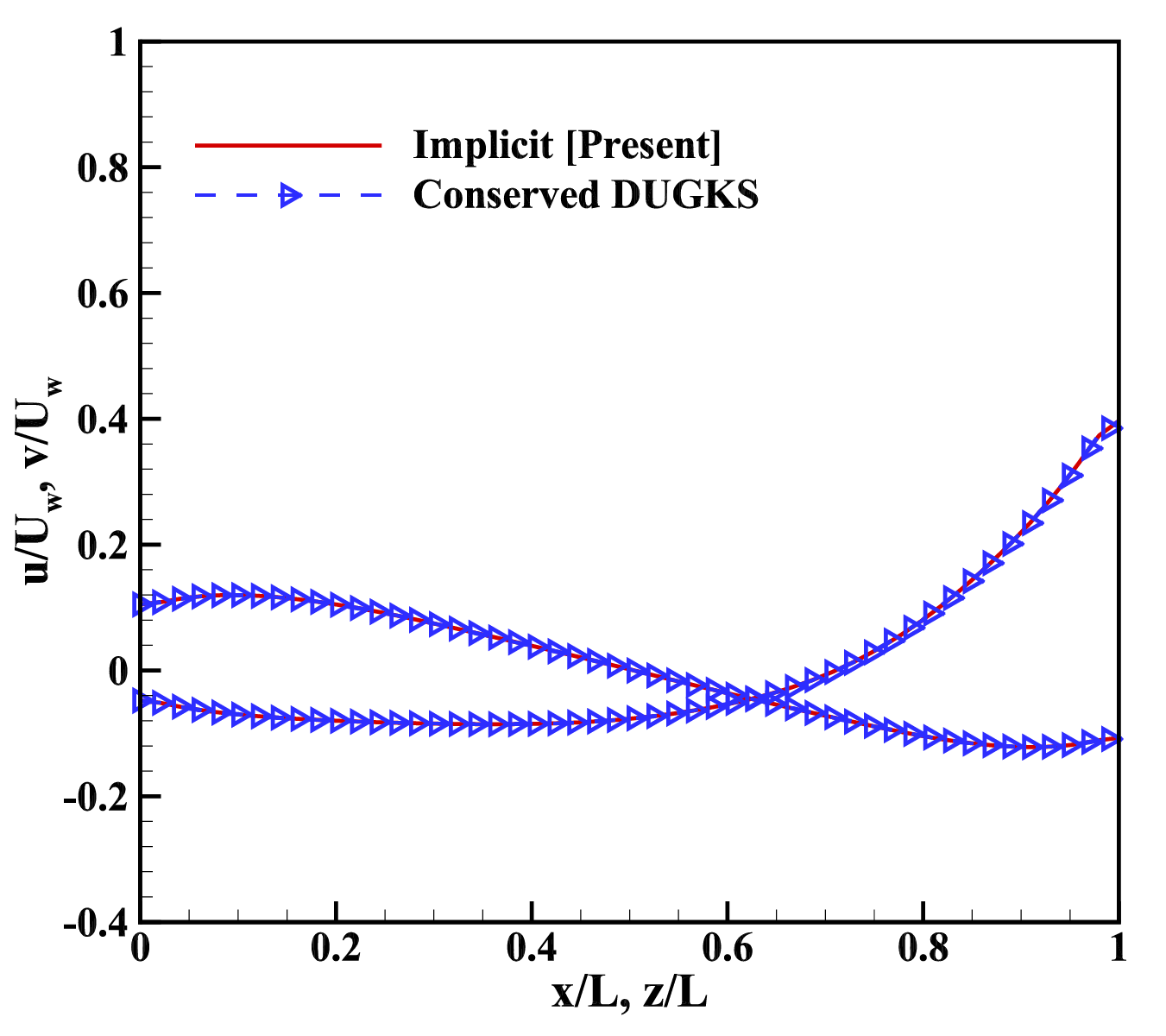}
	}
	\subfigure[\label{Fig:CavityFlow_Kn0.075_Velocity_Line}]{
		\includegraphics[width=0.45\textwidth]{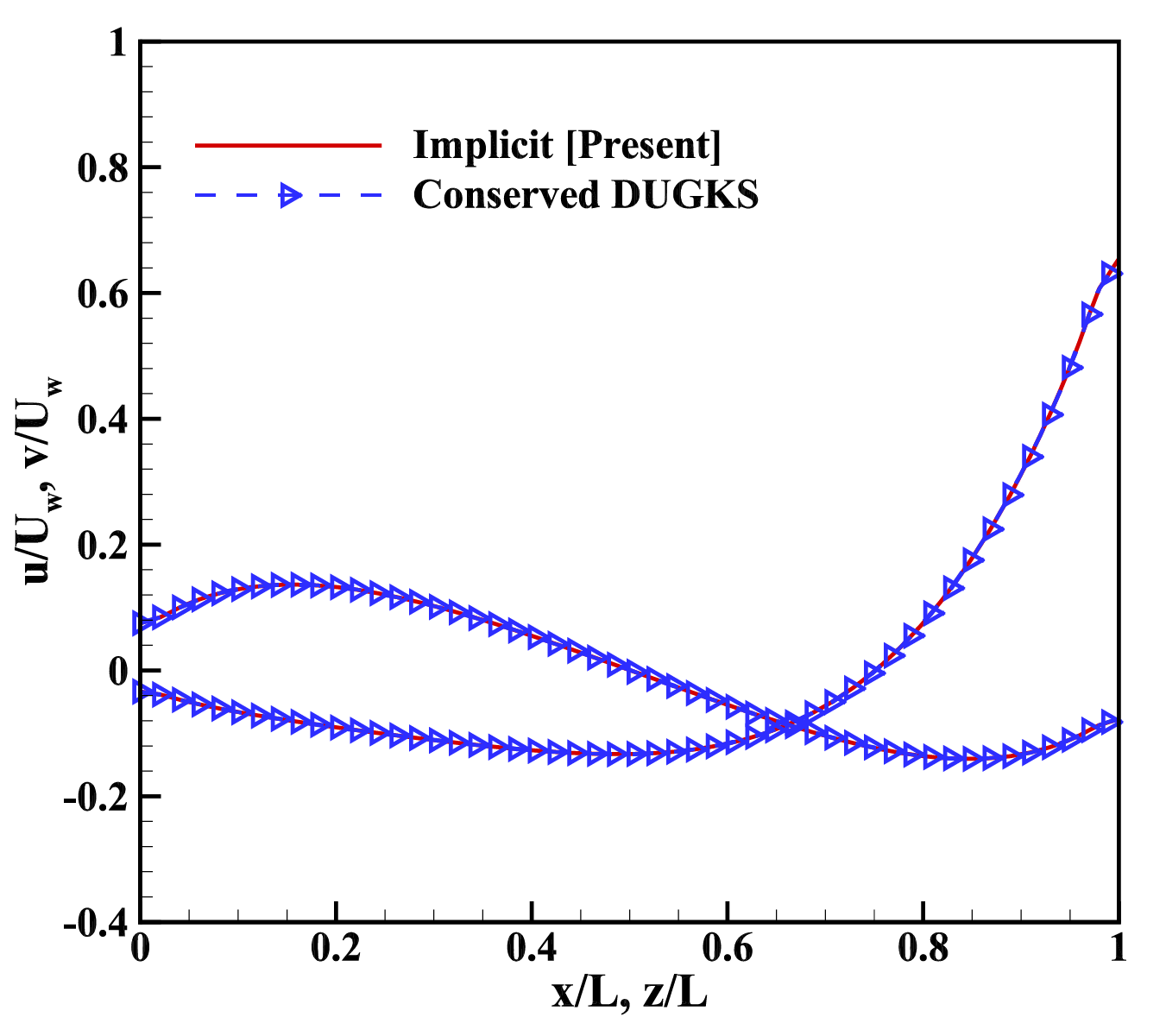}
	}
	\subfigure[\label{Fig:CavityFlow_Re100_Velocity_Line}]{
		\includegraphics[width=0.45\textwidth]{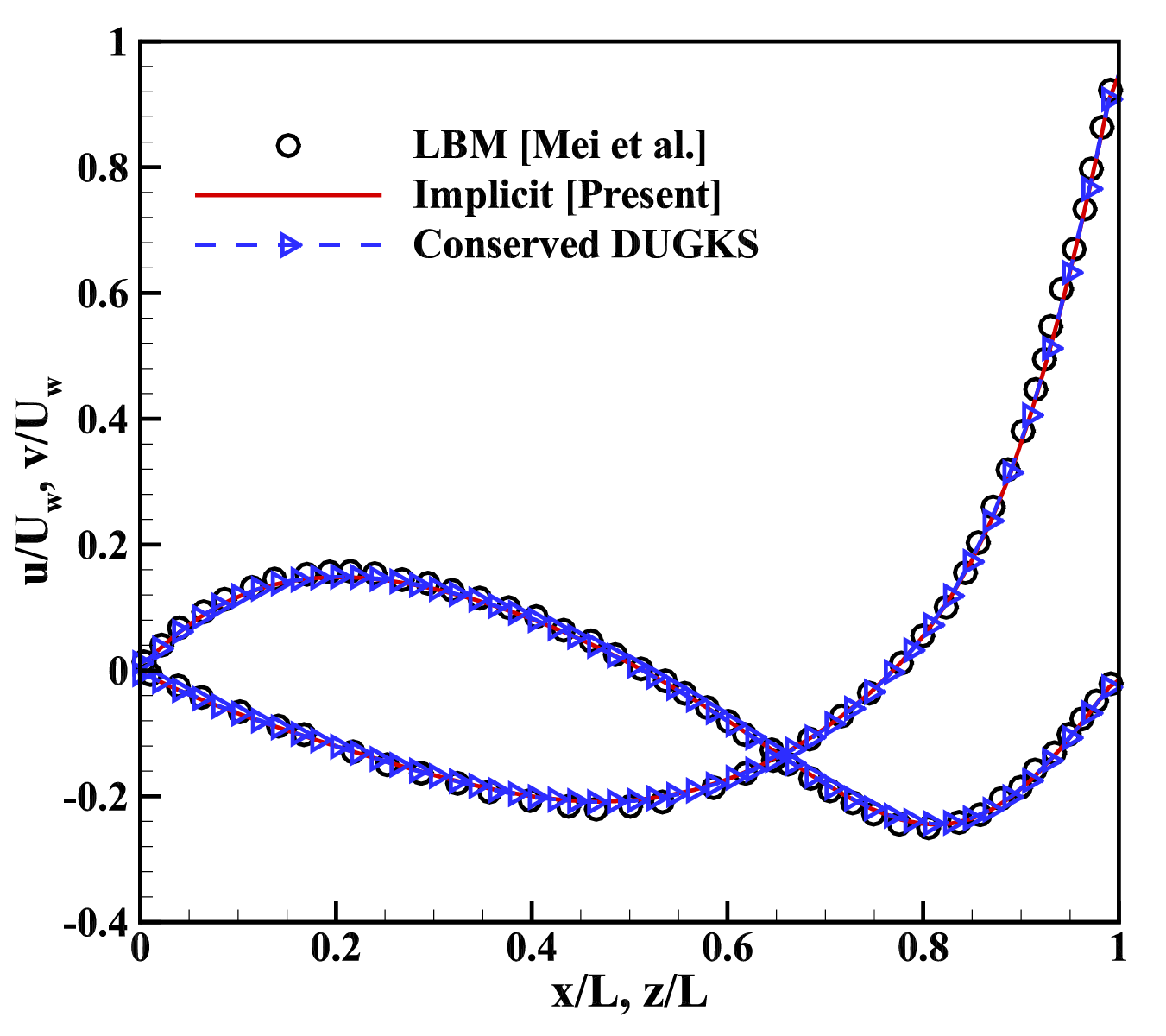}
	}
	\caption{The velocity profiles along the central lines of the three-dimensional lid-driven cavity flows on the plane $Y = 0$ at different Kn numbers. (a) Kn = 10, (b) Kn = 1, (c) Kn = 0.075, (d) Re =100.}
	\label{Fig:CavityFlow_Velocity} 
\end{figure}

\begin{figure}[!htp]
	\centering
	\subfigure[\label{Fig:CavityFlow_Kn10_Res}]{
		\includegraphics[width=0.45\textwidth]{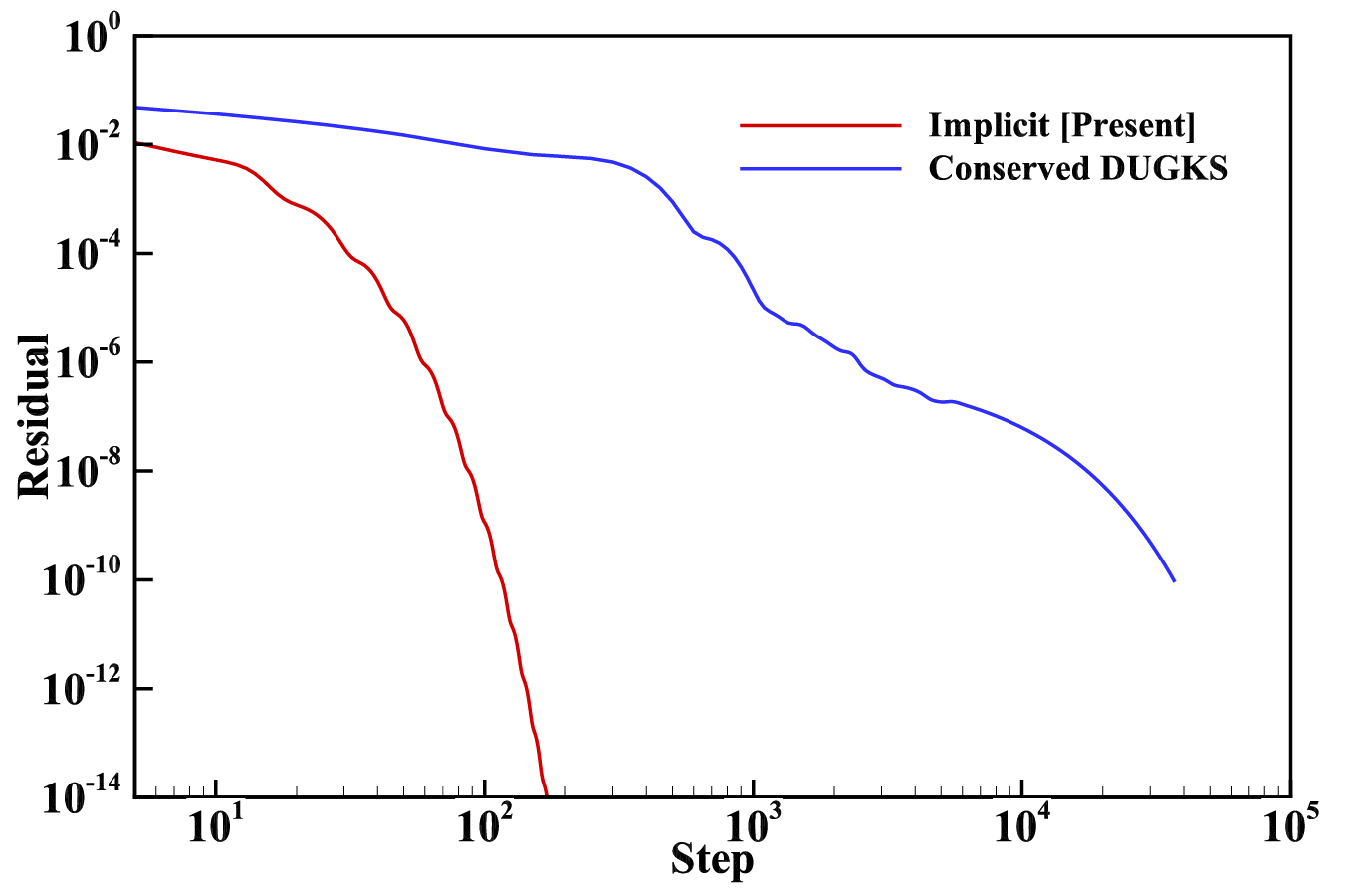}
	}
	\subfigure[\label{Fig:CavityFlow_Kn1_Res}]{
		\includegraphics[width=0.45\textwidth]{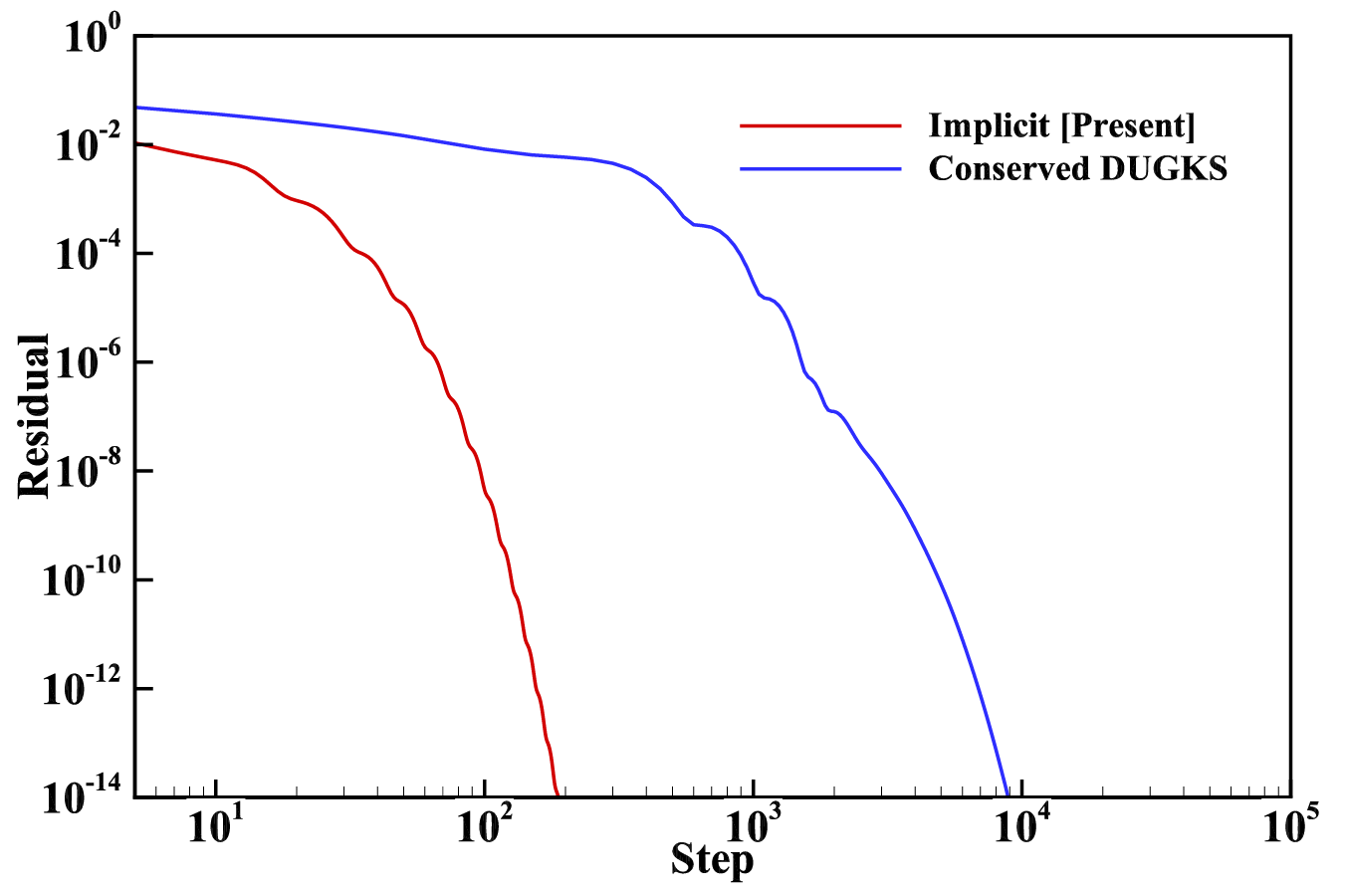}
	}
	\subfigure[\label{Fig:CavityFlow_Kn0.075_Res}]{
		\includegraphics[width=0.45\textwidth]{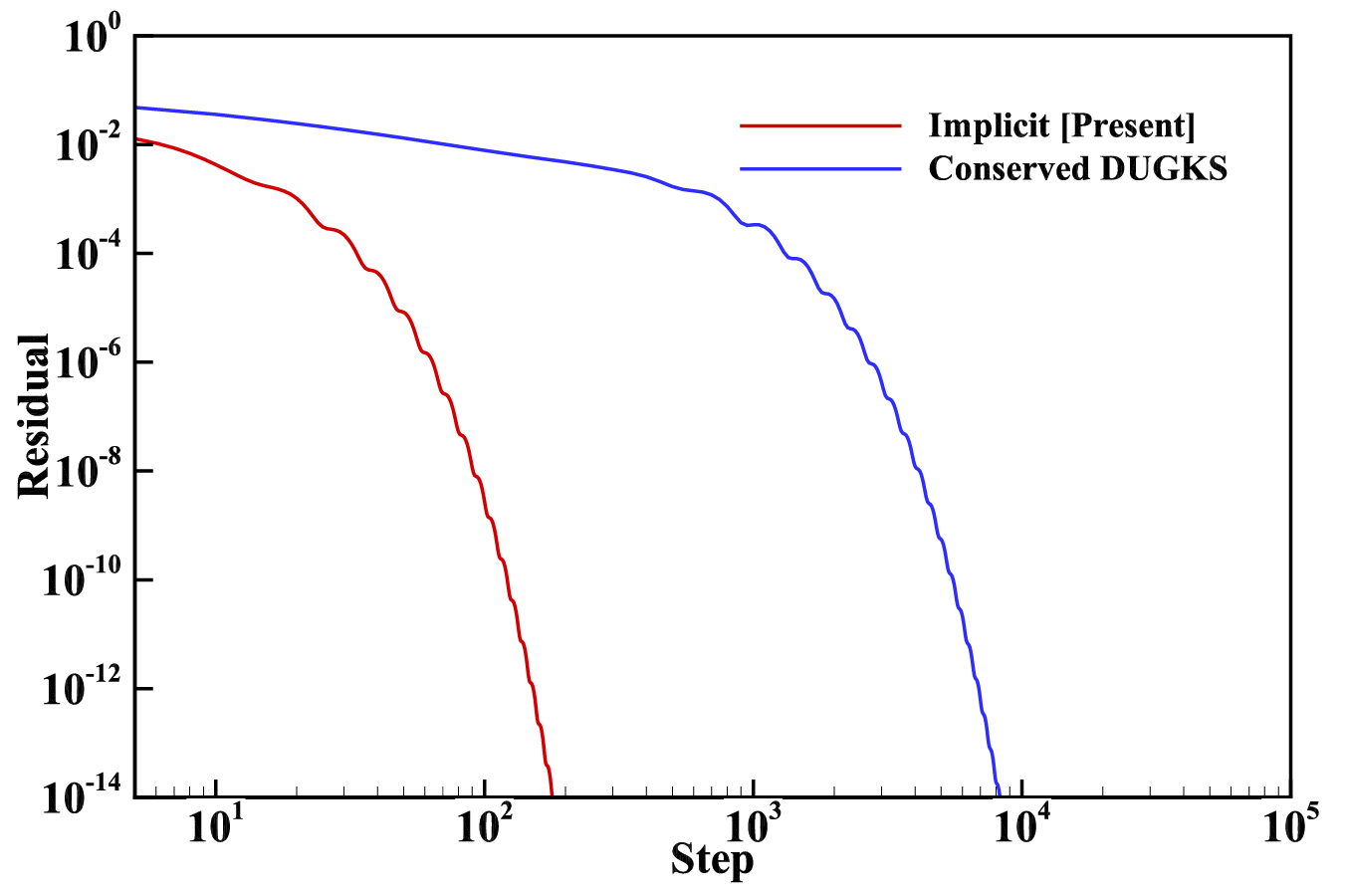}
	}
	\subfigure[\label{Fig:CavityFlow_Re100_Res}]{
		\includegraphics[width=0.45\textwidth]{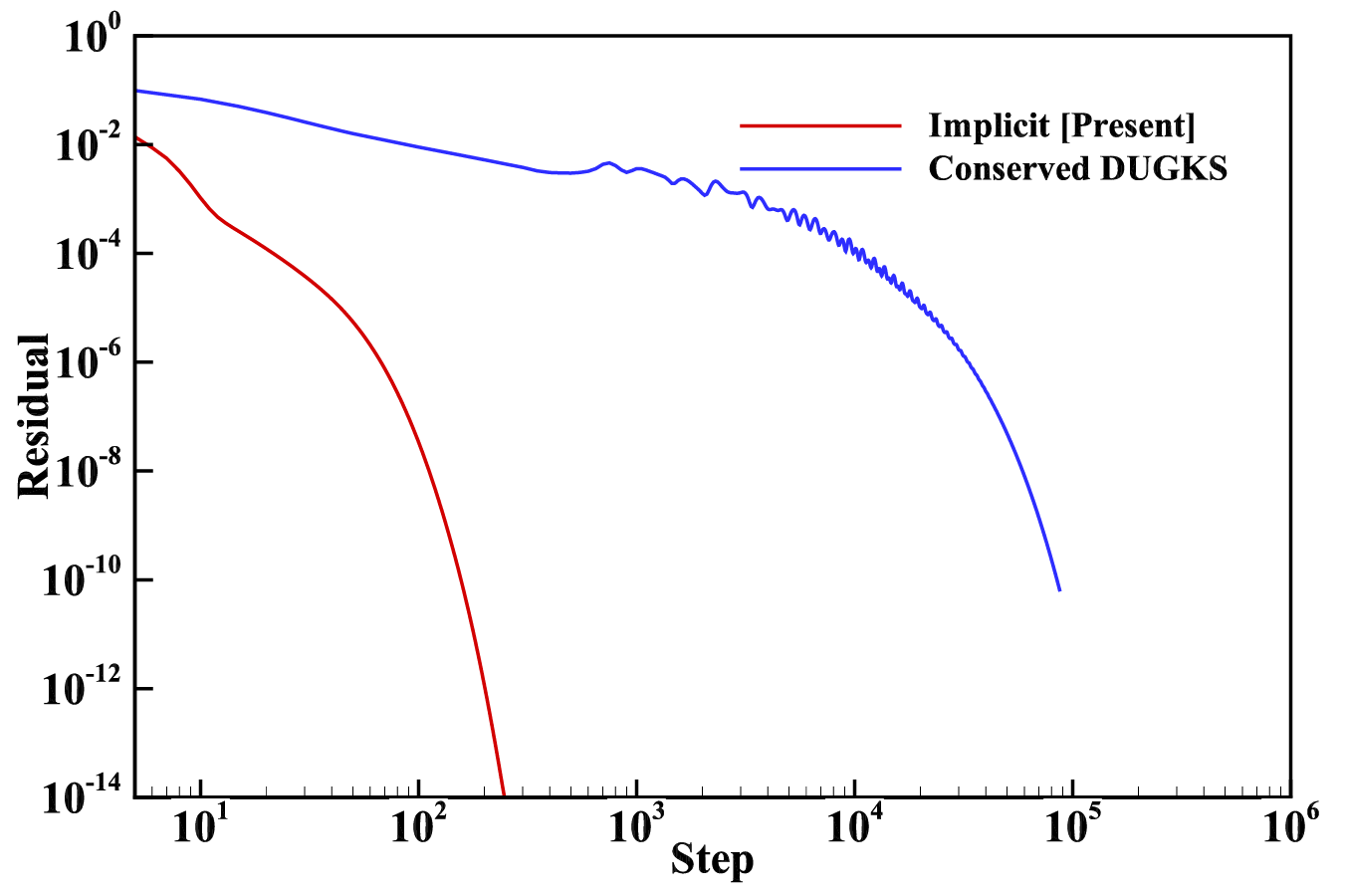}
	}
	\caption{The residual curves of the three-dimensional lid-driven cavity flows at different Kn numbers. (a) Kn = 10, (b) Kn = 1, (c) Kn = 0.075, (d) Re =100.}
	\label{Fig:CavityFlow_Res} 
\end{figure}

In the present study, a uniform mesh (${44^3}$ cells) is used for the cases of Kn = 10, 1, 0.075, while a non-uniform mesh (${44^3}$ cells, Fig~\ref{Fig:CavityFlow_Mesh}) with a mesh size $0.01L$ near the wall is used for the case of Re = 100.
The unstructured DVS mesh with 22710 cells shown in Fig.~\ref{Fig:CavityFlow_Micmesh} is applied for all cases.
The results of velocity profiles along the central lines $(x/L,L/2,L/2)$ and $(L/2,L/2,z/L)$ are shown in Fig.~\ref{Fig:CavityFlow_Velocity}.
It can be seen that there is no much difference in the plots of velocity profiles solved by the present method and the conserved DUGKS in all flow regimes, and the solution of lattice Boltzmann method (LBM) for Re = 100.
Moreover, as reported in Table~\ref{Table:CavityFlow_Speed}, the present method is 1$-$2 orders of magnitude faster than the conserved DUGKS in all flow regimes.
The criterion of convergence is that the global average residual of macroscopic variables is less than ${10^{-10}}$.
Correspondingly, the residual curves of the present method and conserved DUGKS are plotted in Fig.~\ref{Fig:CavityFlow_Res}.
The current implicit algorithm only requires about 200 iteration steps for all cases, and the residual can reach ${10^{-14}}$.

\subsection{Supersonic and hypersonic flows over a sphere}\label{sec:caseB}
\begin{figure}[!htp]
	\centering
	\subfigure[\label{Fig:Sphere_Ma5.45_Mesh}]{
		\includegraphics[width=0.45\textwidth]{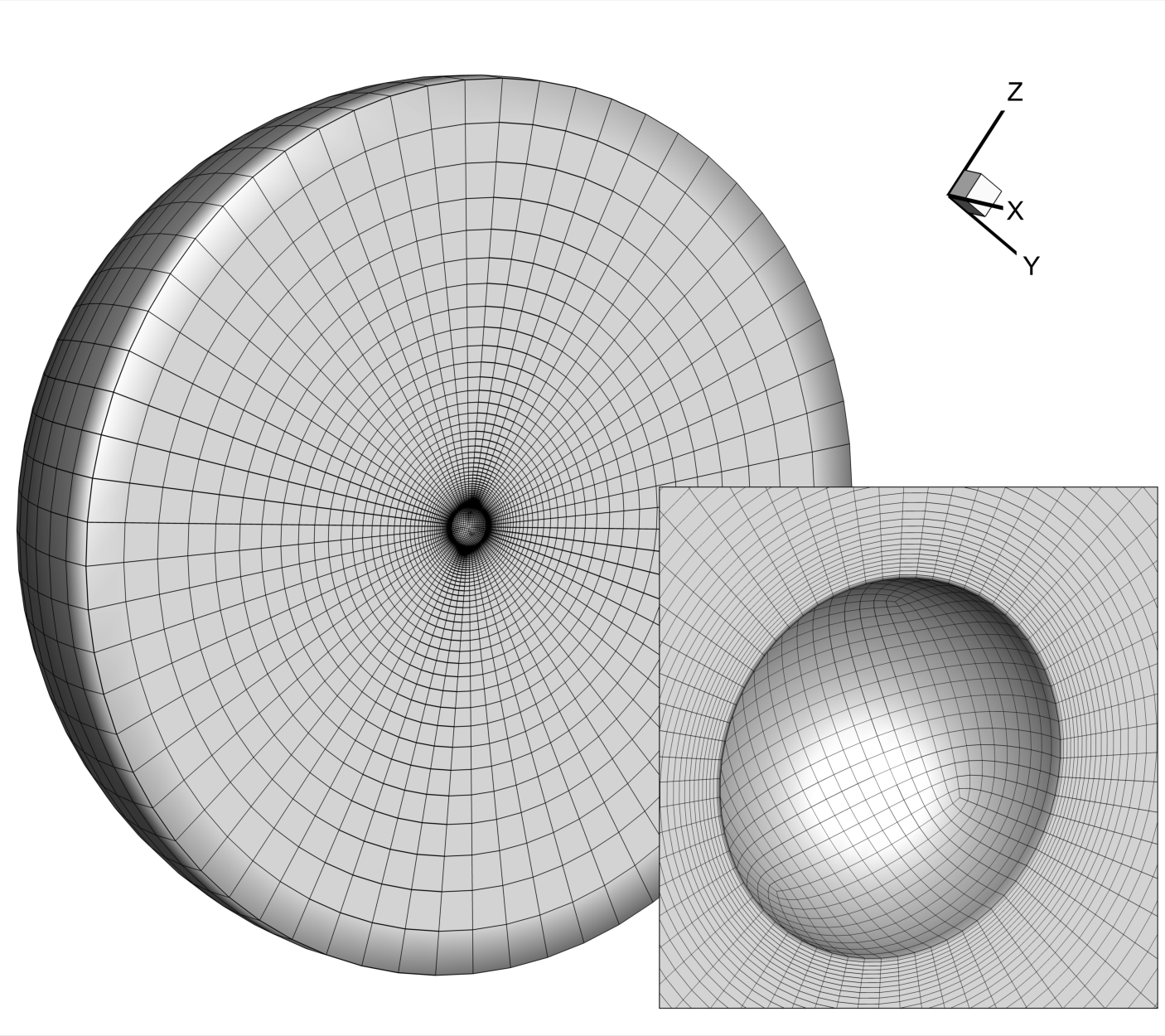}
	}
	\subfigure[\label{Fig:Sphere_Ma5.45_Micmesh}]{
		\includegraphics[width=0.45\textwidth]{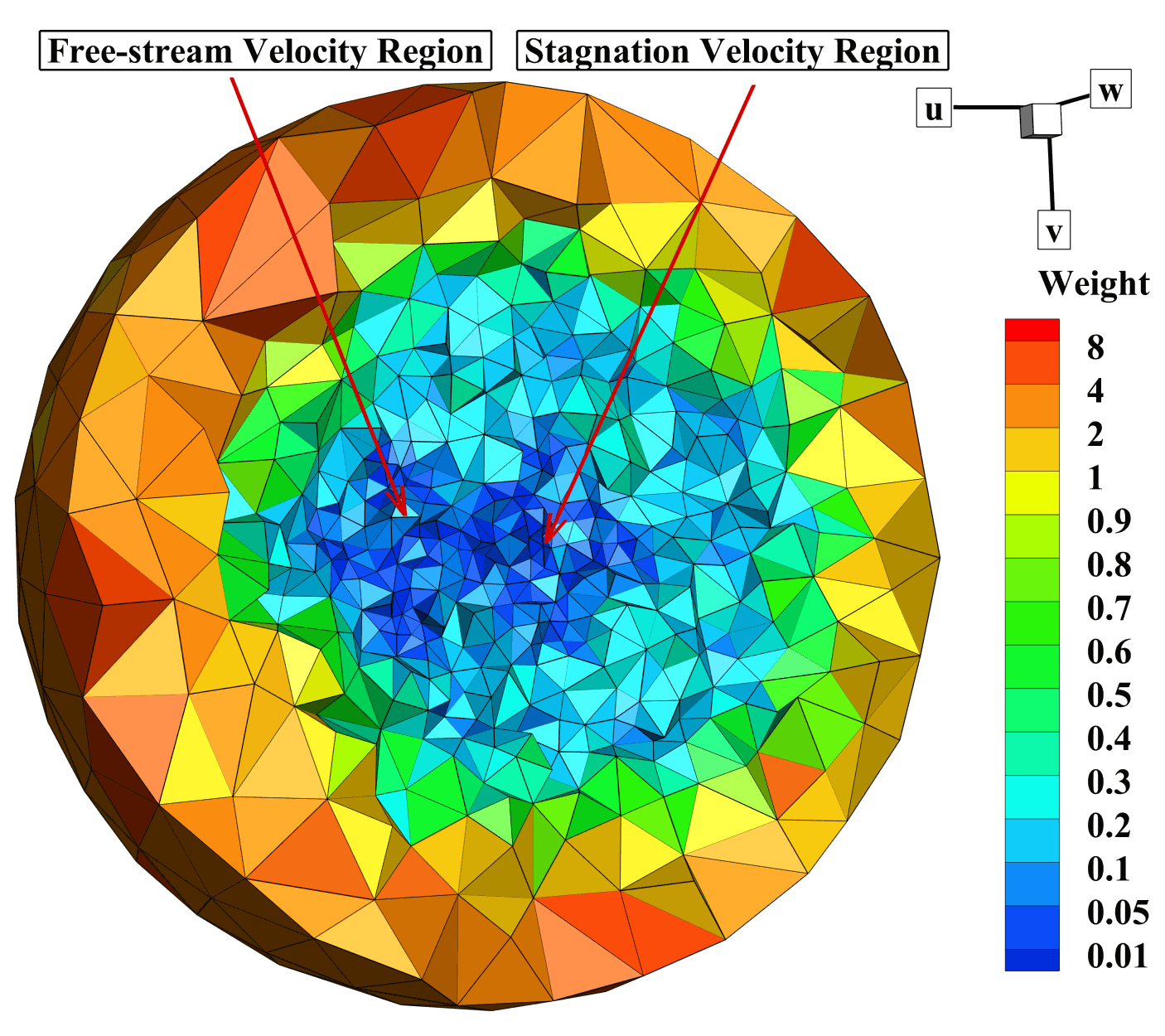}
	}
	\caption{The section views of (a) the physical space mesh (total 83200 cells) and (b) the unstructured discrete velocity space mesh (Ma = 5.45, total 22860 cells) for the hypersonic flow over a sphere.}
	\label{Fig:Sphere_Ma5.45_MeshMicmesh} 
\end{figure}

The experiments of supersonic (Ma = 4.25) and hypersonic (Ma = 5.45) flows over a sphere at different Kn numbers conducted by Wendt~\citep{wendt_drag_1971} are simulated to assess the accuracy and efficiency of the present method for high speed non-equilibrium flows.
The diameter of sphere is $d=2\ \rm{mm}$, Kn and Re numbers depending on the diameter are shown in the third and fourth columns of Table~\ref{Table:Sphere_Drag}.
In all of the simulations, the free-stream total temperature is 300K, and the wall temperature is set as 302K and 315K for Ma = 4.25 and Ma =5.45, respectively.
The dynamic viscosity of the air is calculated using the Sutherland formula~\citep{toro_riemann_2009}, and rotational collision number keep a constant value of 3.
Fig.~\ref{Fig:Sphere_Ma5.45_Mesh} illustrates the section views of the physical space mesh, in which contains 83200 hexahedral cells with a mesh size $0.01d$ near the wall.
A unstructured DVS mesh with 22670 cells for the case of Ma = 4.25 and 22860 cells for the case of Ma = 5.45 (\ref{Fig:Sphere_Ma5.45_Micmesh}) are applied.
As shown in Fig.~\ref{Fig:Sphere_Ma5.45_Micmesh}, the velocity space is discretized into a sphere with center coordinates of $(2, 0, 0)$ and radius of $6\sqrt {R{T_{max}}}$, and $T_{max} = {\bf{max}} ({T_{0},T_{w}})$.
$T_0$ is the total temperature can be predicted from free-stream temperature and Mach number.
In order to ensure the integration accuracy and capture the non-equilibrium distribution, the discrete velocity around the point $(0, 0, 0)$ (solid wall) within the sphere of radius $3\sqrt {R{T_w}}$ and $(4.56, 0, 0)$ (the free-stream velocity is 4.56 in the simulation) within the sphere of radius $3\sqrt {R{T_\infty}}$ are refined.

\begin{figure}[!htp]
	\centering
	\subfigure[\label{Fig:Sphere_Ma4.25Kn0.031_Res}]{
		\includegraphics[width=0.45\textwidth]{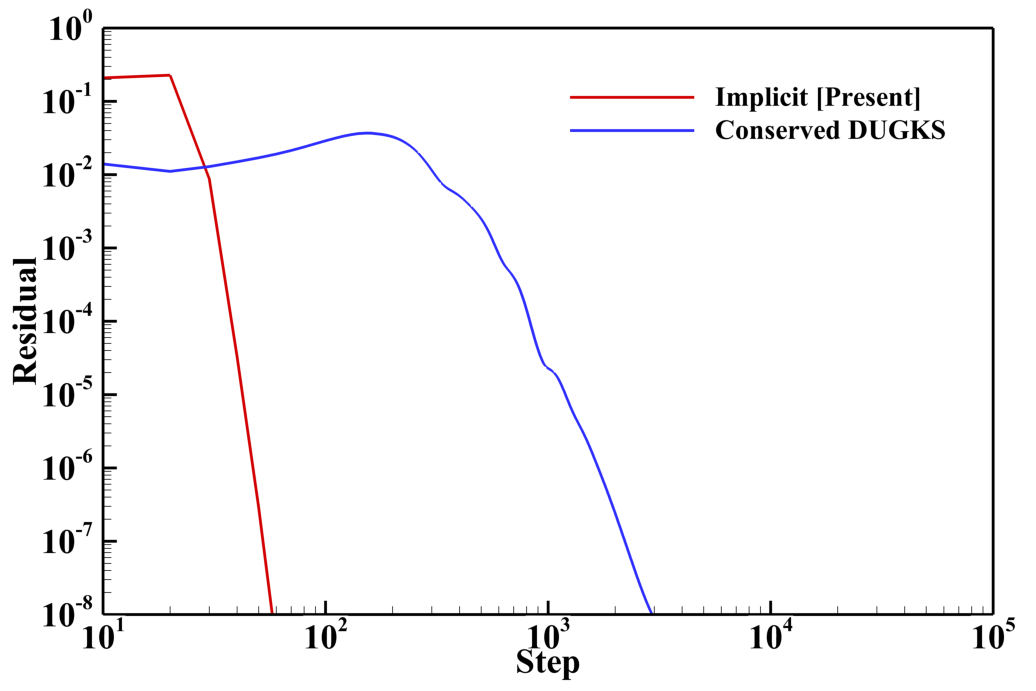}
	}
	\subfigure[\label{Fig:Sphere_Ma5.45Kn1.96_Res}]{
		\includegraphics[width=0.45\textwidth]{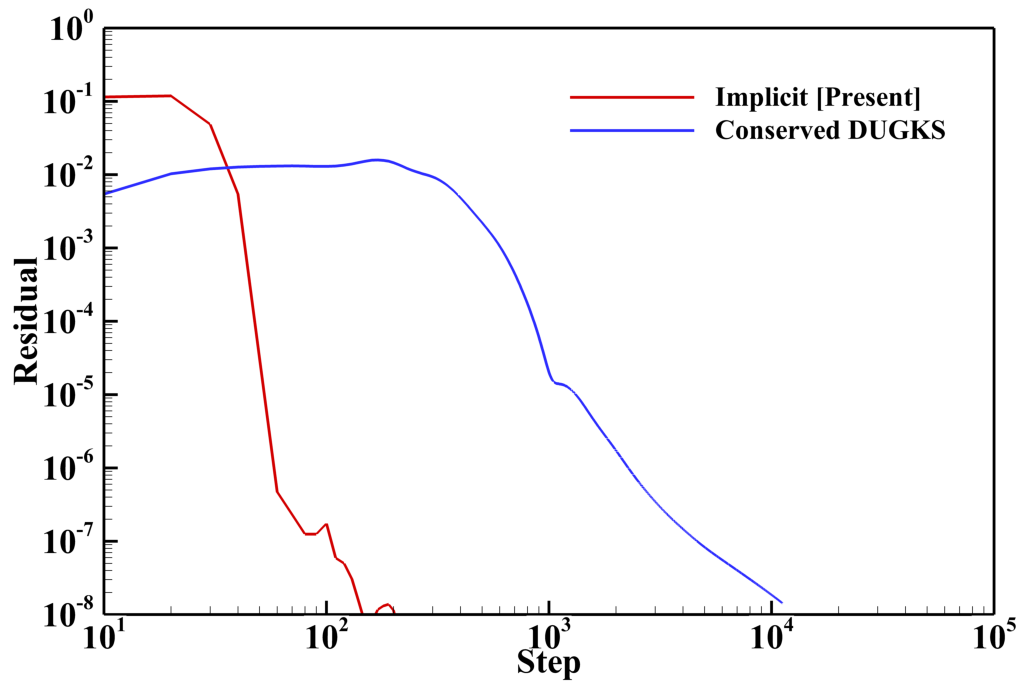}
	}
	\caption{The residual curves of the supersonic and hypersonic flows over a sphere at different Ma and Kn numbers. (a) Ma = 4.25, Kn = 0.031, and (b) Ma = 5.45, Kn = 1.96.}
	\label{Fig:Sphere_Res} 
\end{figure}

The drag coefficient comparisons with the experiments are shown in Table~\ref{Table:Sphere_Drag}.
The reference area is $\pi {\left( {{d \mathord{\left/{\vphantom {d 2}} \right.\kern-\nulldelimiterspace} 2}} \right)^2}$.
Given that the root mean square error of the experiments is about $\pm 2\%$, the present results, where the maximum relative error is less than 2$\%$, can be regarded as excellent.
To further confirm the efficiency and accuracy of the current algorithm, the numerical solutions of the case $\rm {Ma=4.25}$, $\rm {Kn=0.031}$ (Case A4) and $\rm {Ma=5.45}$, $\rm {Kn=1.96}$ (Case B1) will be compared with the results of DS2V code~\citep{bird_molecular_1994}.
In terms of the computational efficiency, detailed results are given in Table~\ref{Table:Sphere_Speed}.
Correspondingly, the residuals of the present method and conserved DUGKS are plotted in Fig.~\ref{Fig:Sphere_Res}.
In comparison with the explicit conserved DUGKS, the present implicit method can accelerate convergence by more than 20 times.

\begin{figure}[!htp]
	\centering
	\subfigure[\label{Fig:Sphere_Ma4.25Kn0.031_Wall_Cp}]{
		\includegraphics[width=0.45\textwidth]{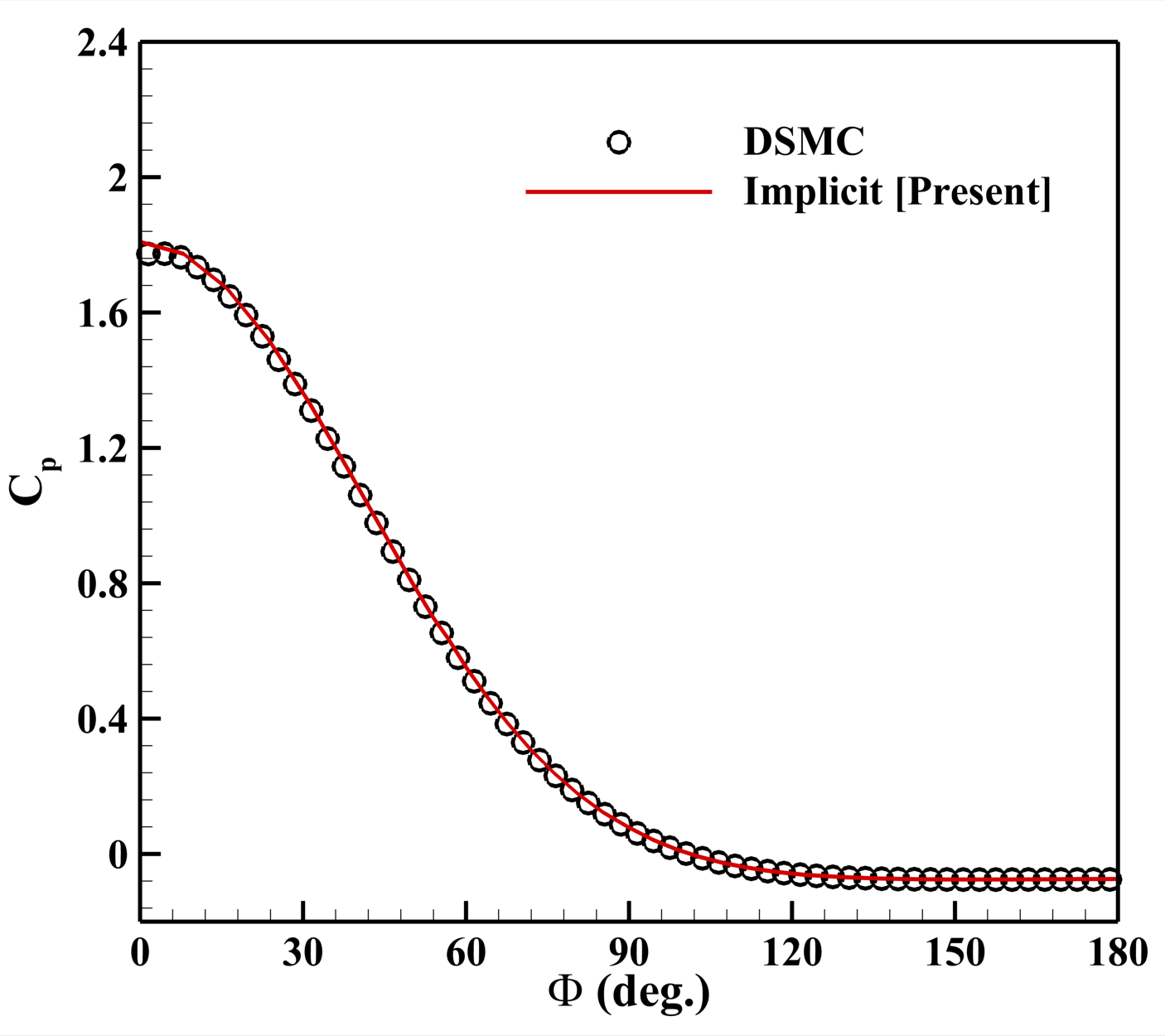}
	}
	\subfigure[\label{Fig:Sphere_Ma4.25Kn0.031_Wall_Cf}]{
		\includegraphics[width=0.45\textwidth]{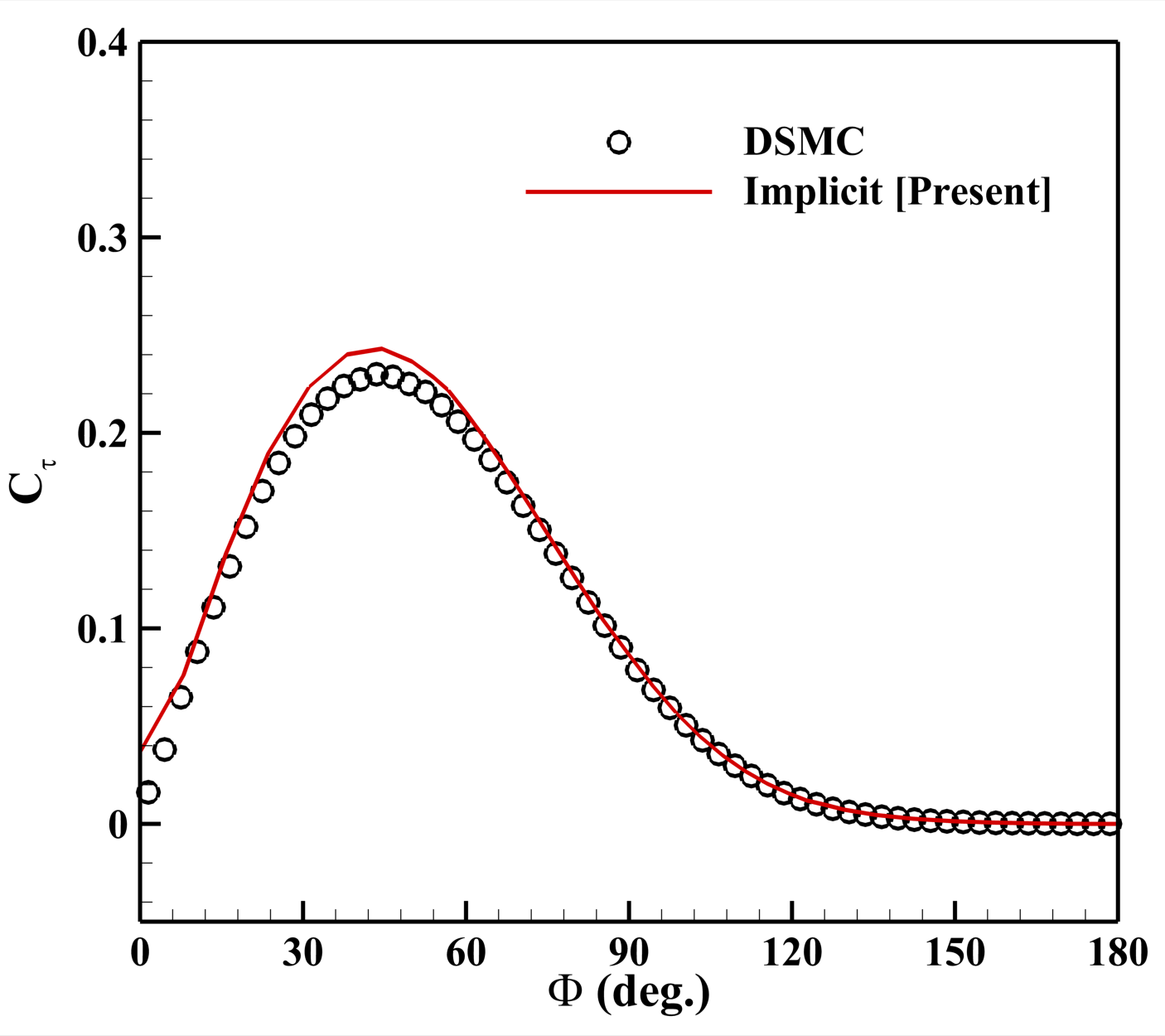}
	}
	\subfigure[\label{Fig:Sphere_Ma4.25Kn0.031_Wall_Ch}]{
		\includegraphics[width=0.45\textwidth]{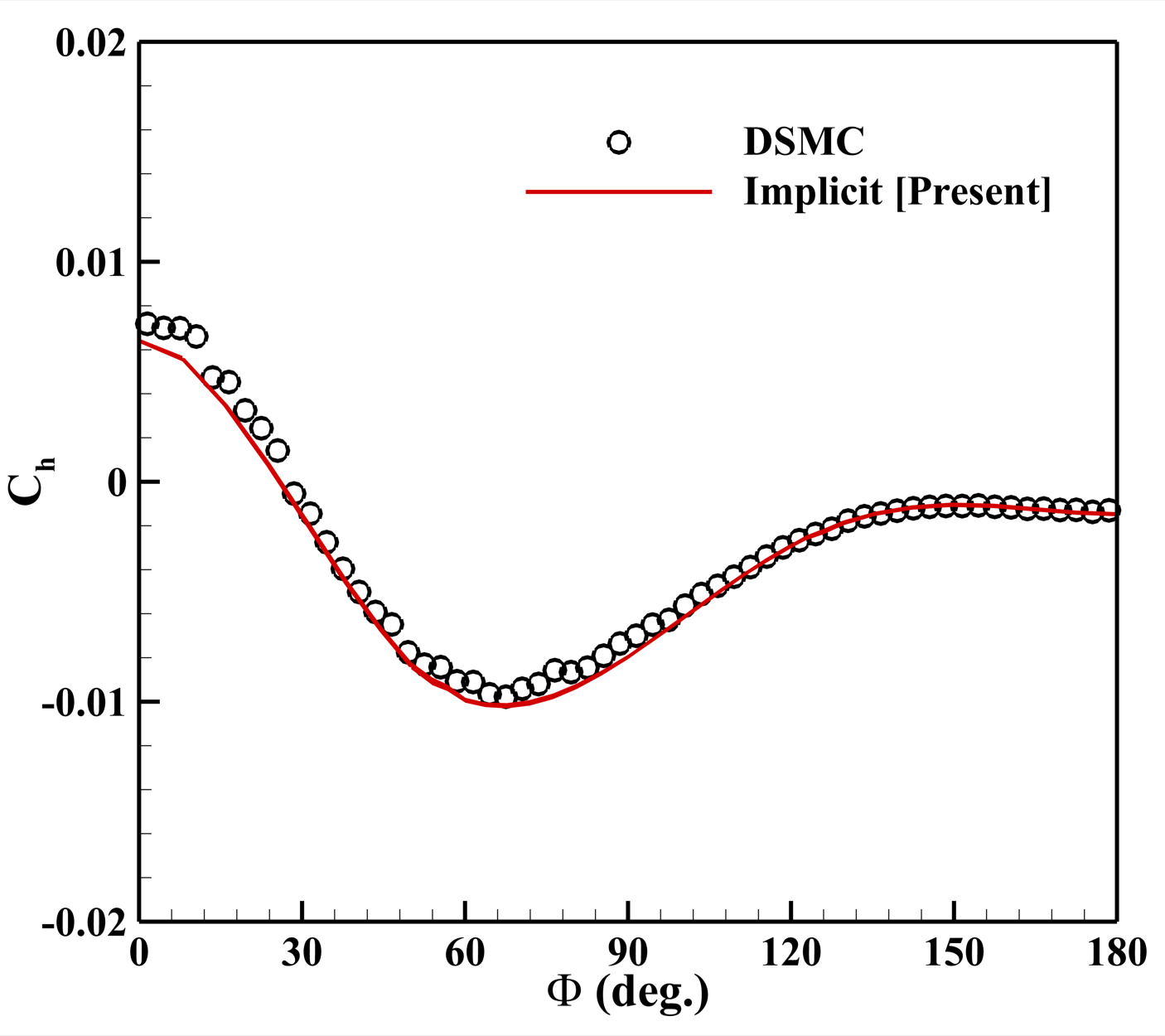}
	}
	\caption{Comparison of DSMC and the present (a) pressure coefficient, (b) shear stress coefficient and (c) heat transfer coefficient for the supersonic flow over a sphere (Ma = 4.25, Kn = 0.031).}
	\label{Fig:Sphere_Ma4.25Kn0.031_Wall}
\end{figure}

\begin{figure}[!htp]
	\centering
	\subfigure[\label{Fig:Sphere_Ma5.45Kn1.96_Wall_Cp}]{
		\includegraphics[width=0.45\textwidth]{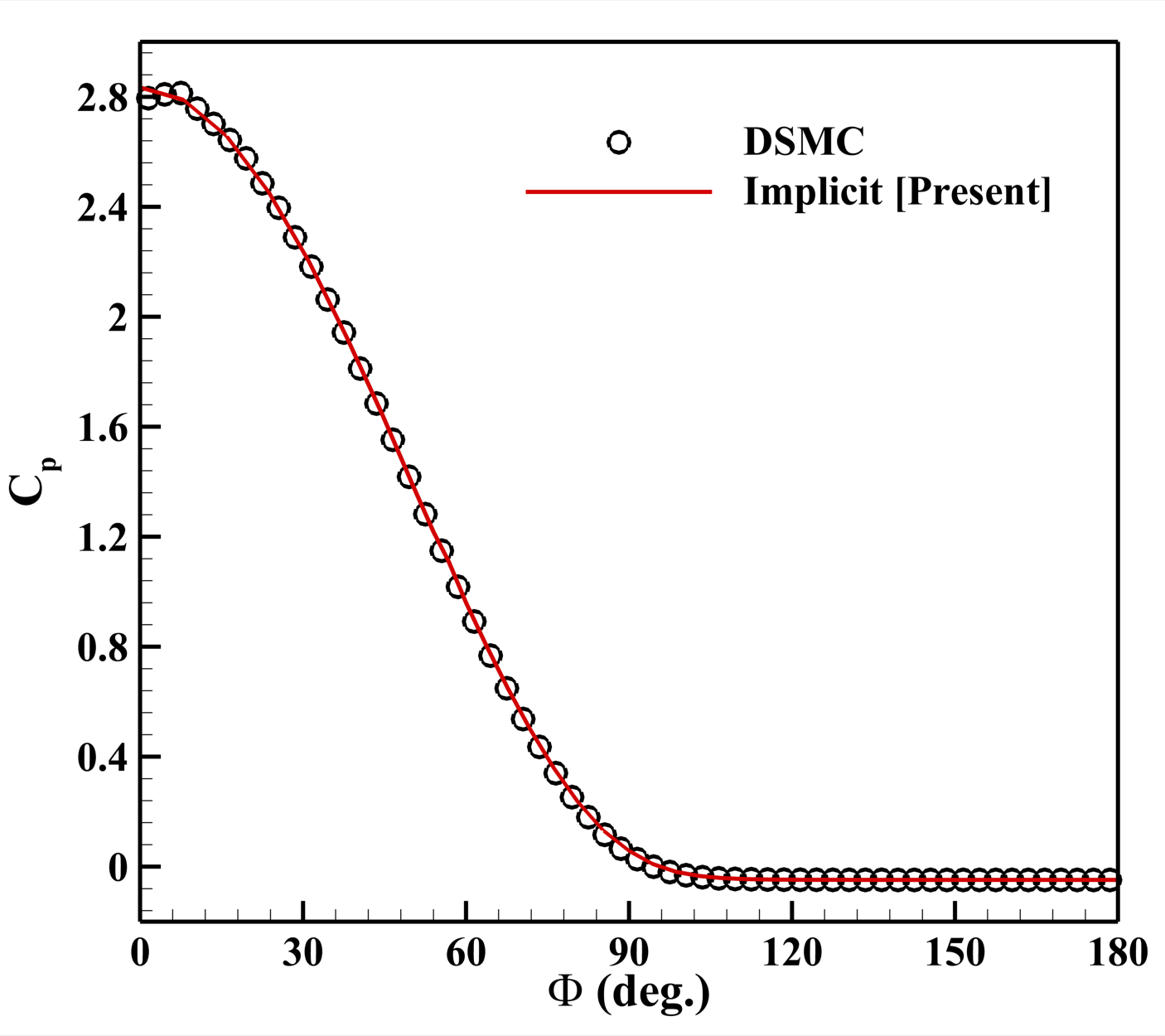}
	}
	\subfigure[\label{Fig:Sphere_Ma5.45Kn1.96_Wall_Cf}]{
		\includegraphics[width=0.45\textwidth]{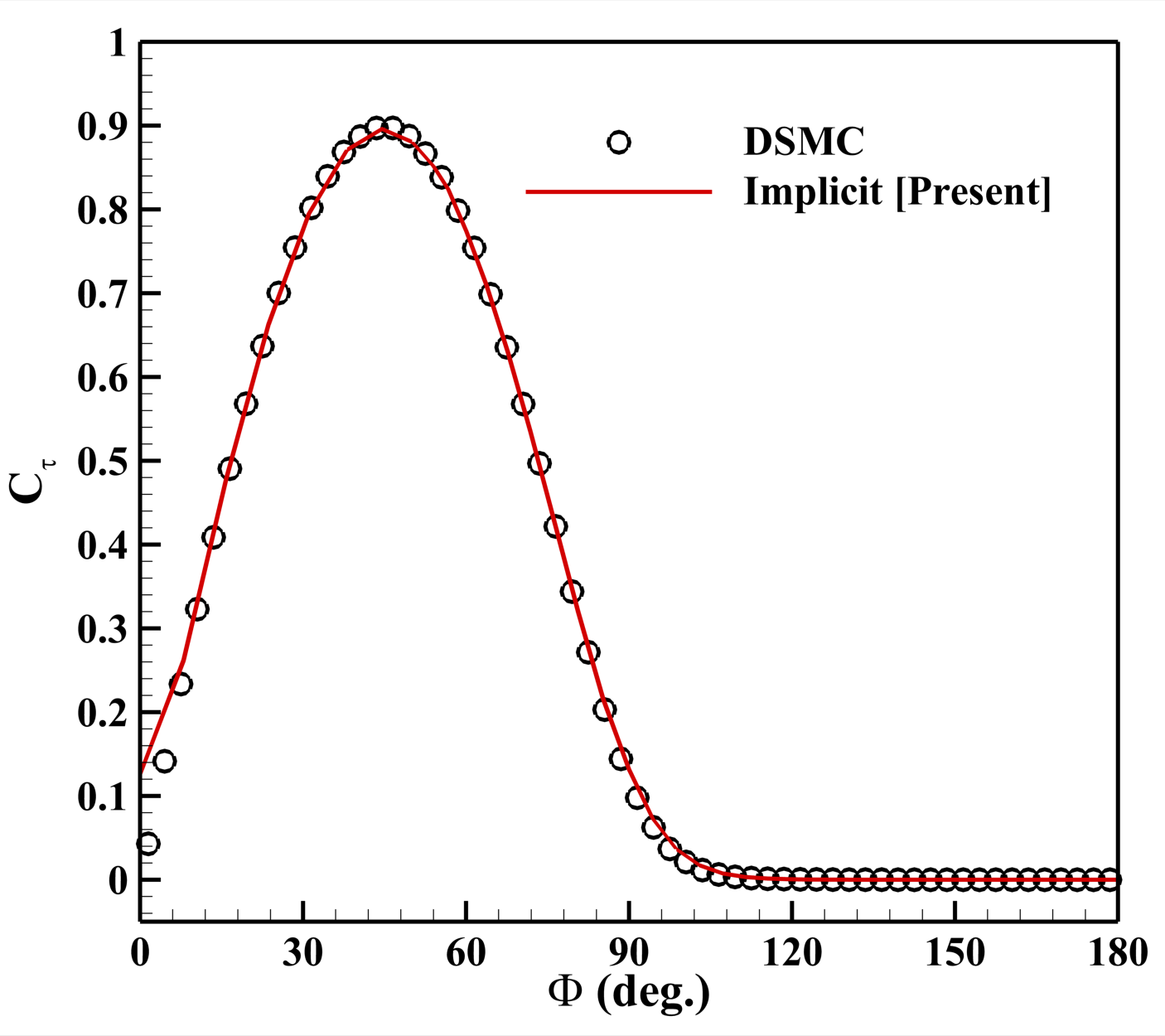}
	}
	\subfigure[\label{Fig:Sphere_Ma5.45Kn1.96_Wall_Ch}]{
		\includegraphics[width=0.45\textwidth]{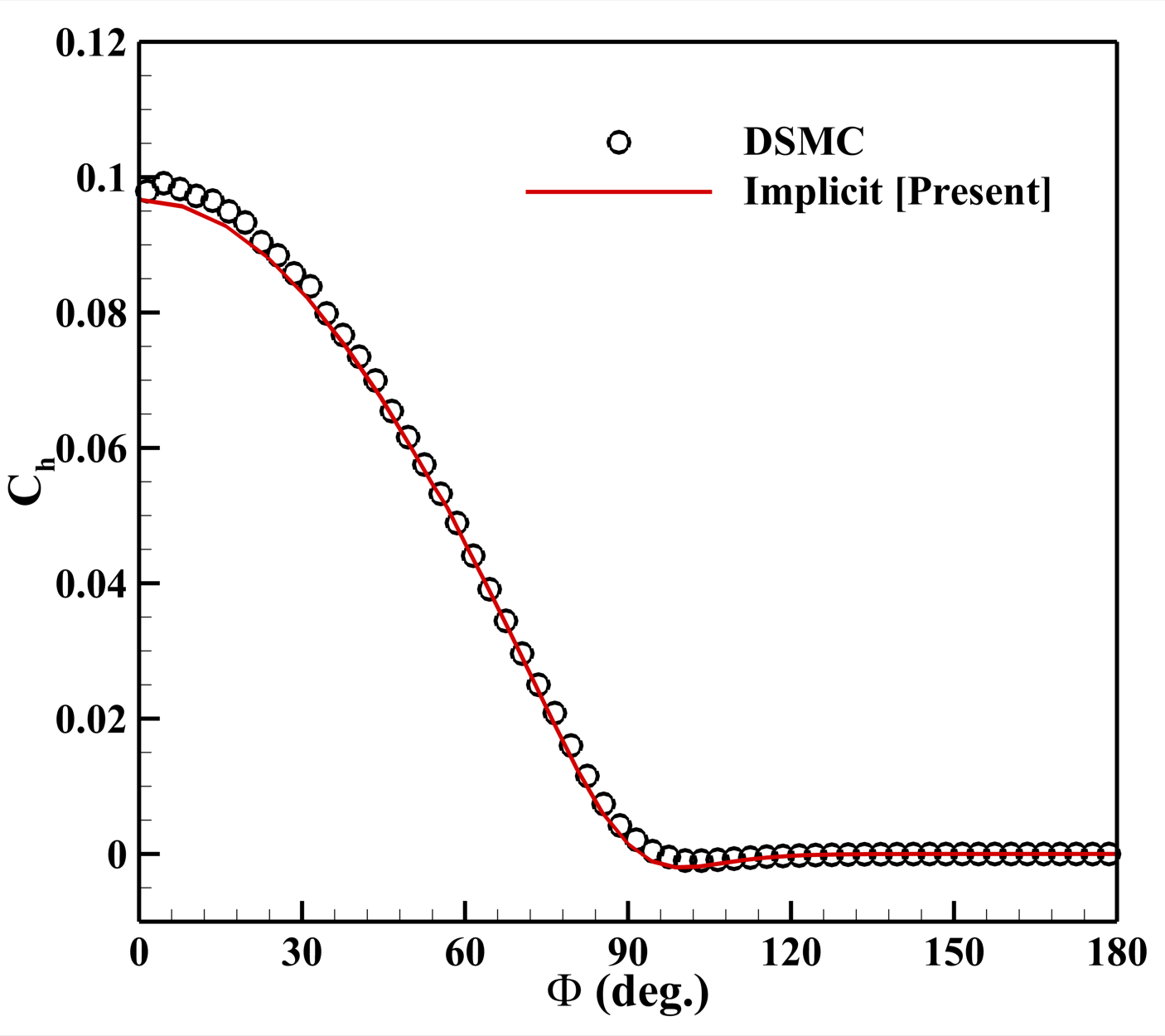}
	}
	\caption{Comparison of DSMC and the present (a) pressure coefficient, (b) shear stress coefficient and (c) heat transfer coefficient for the hypersonic flow over a sphere (Ma = 5.45, Kn = 1.96).}
	\label{Fig:Sphere_Ma5.45Kn1.96_Wall}
\end{figure}

\begin{figure}[!htp]
	\centering
	\subfigure[\label{Fig:Sphere_Ma4.25Kn0.031_StagnationLine_Den}]{
		\includegraphics[width=0.45\textwidth]{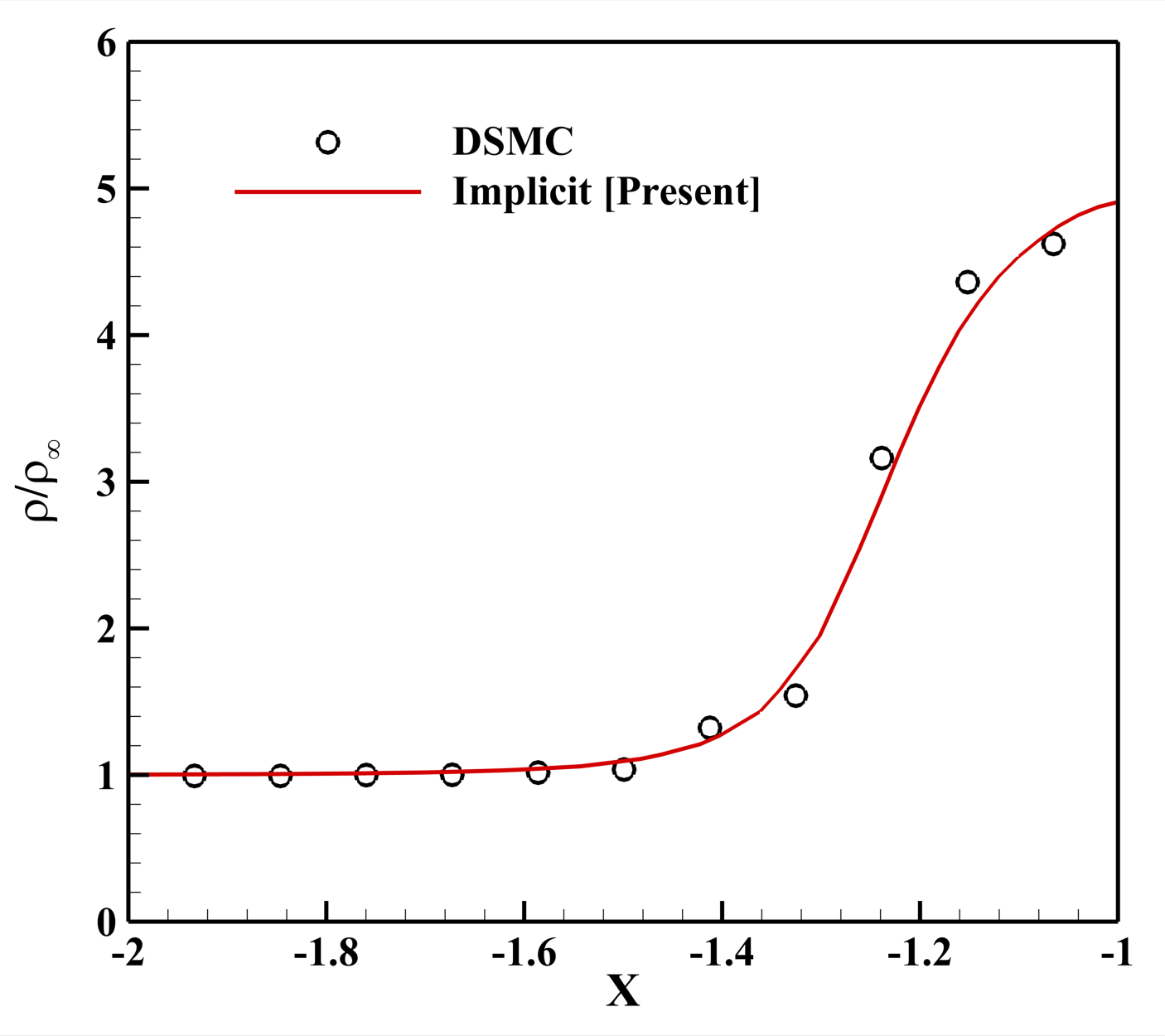}
	}
	\subfigure[\label{Fig:Sphere_Ma4.25Kn0.031_StagnationLine_Pre}]{
	\includegraphics[width=0.45\textwidth]{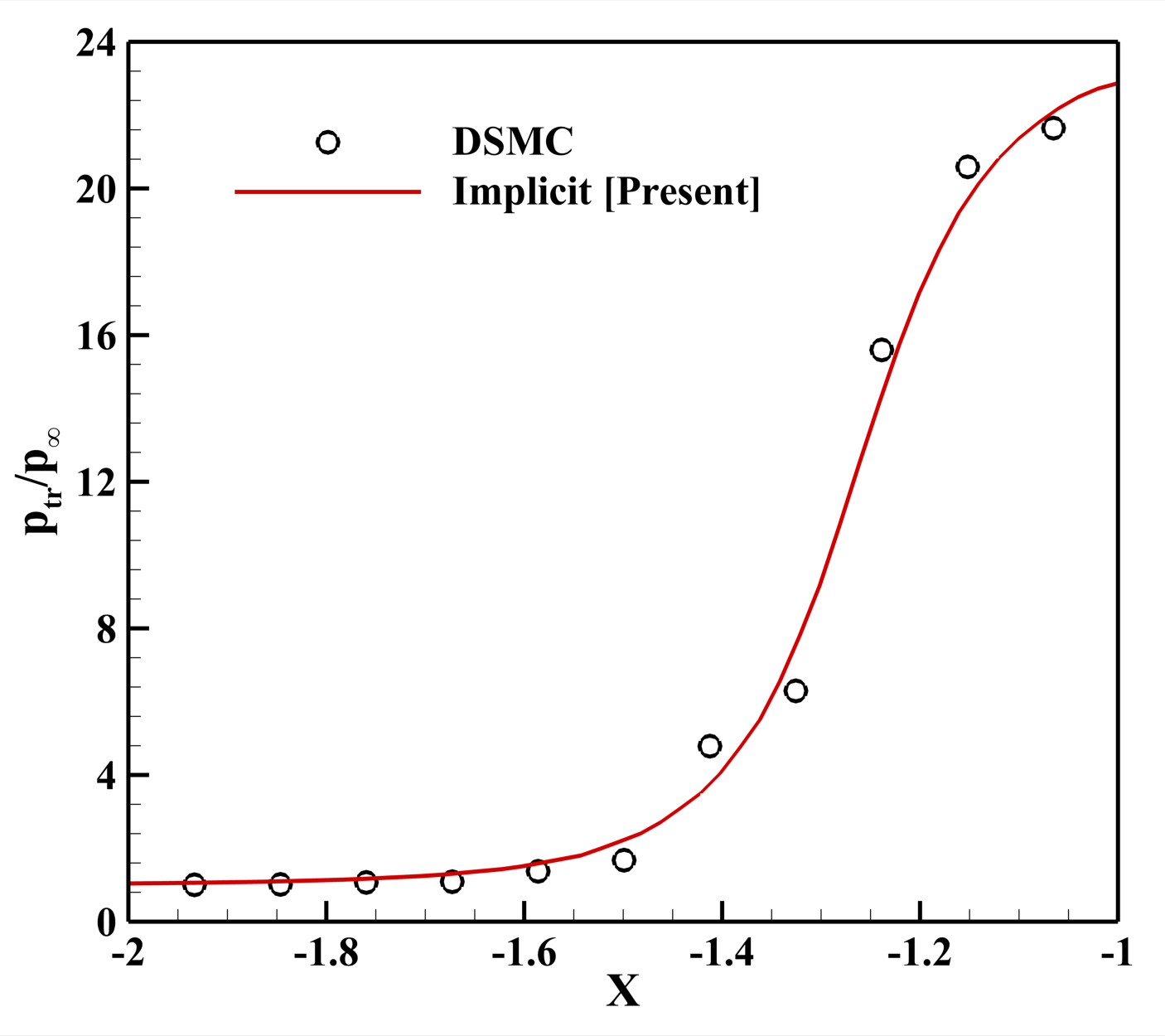}
	}
	\subfigure[\label{Fig:Sphere_Ma4.25Kn0.031_StagnationLine_Vel}]{
		\includegraphics[width=0.45\textwidth]{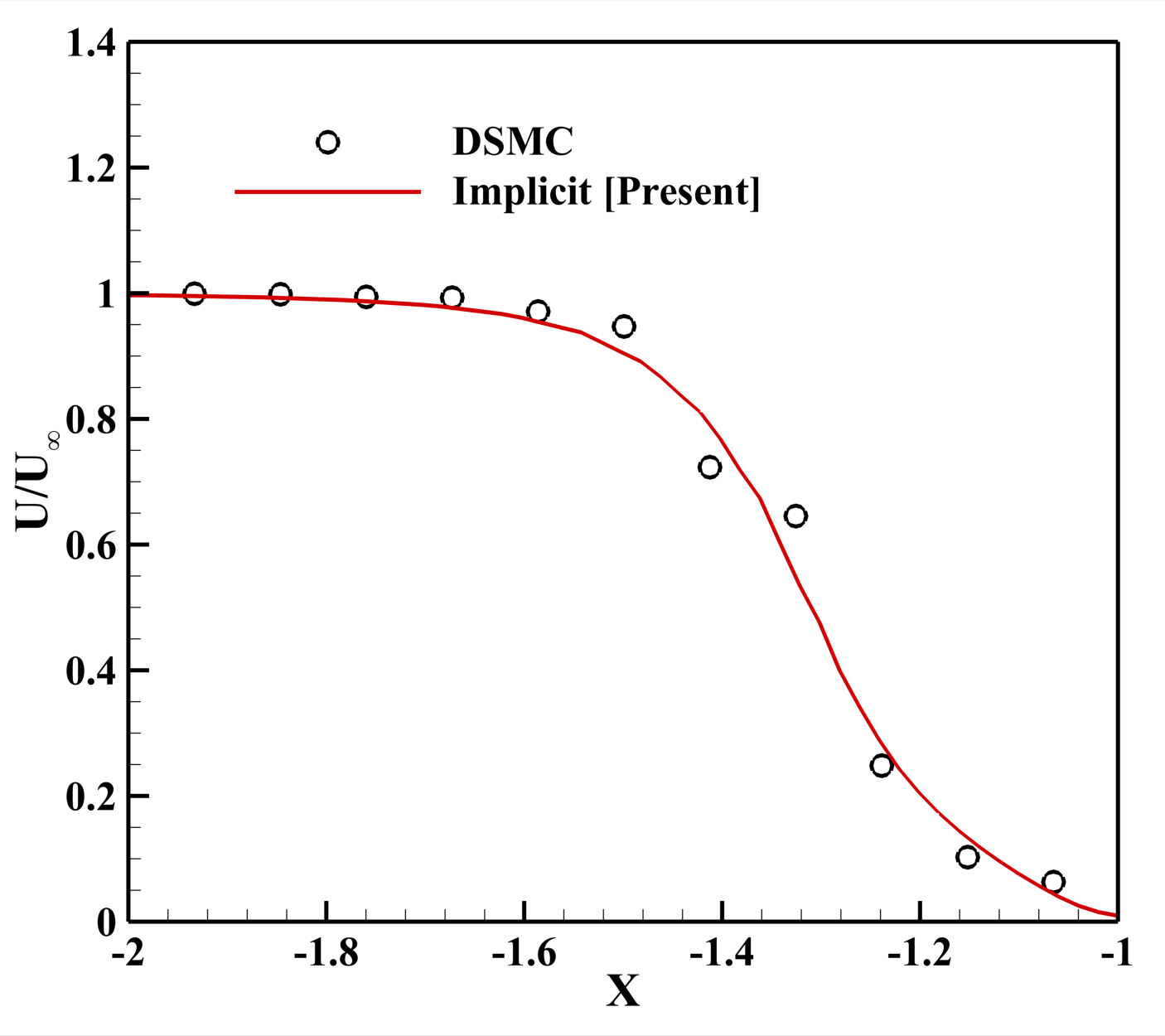}
	}
	\subfigure[\label{Fig:Sphere_Ma4.25Kn0.031_StagnationLine_Tem}]{
		\includegraphics[width=0.45\textwidth]{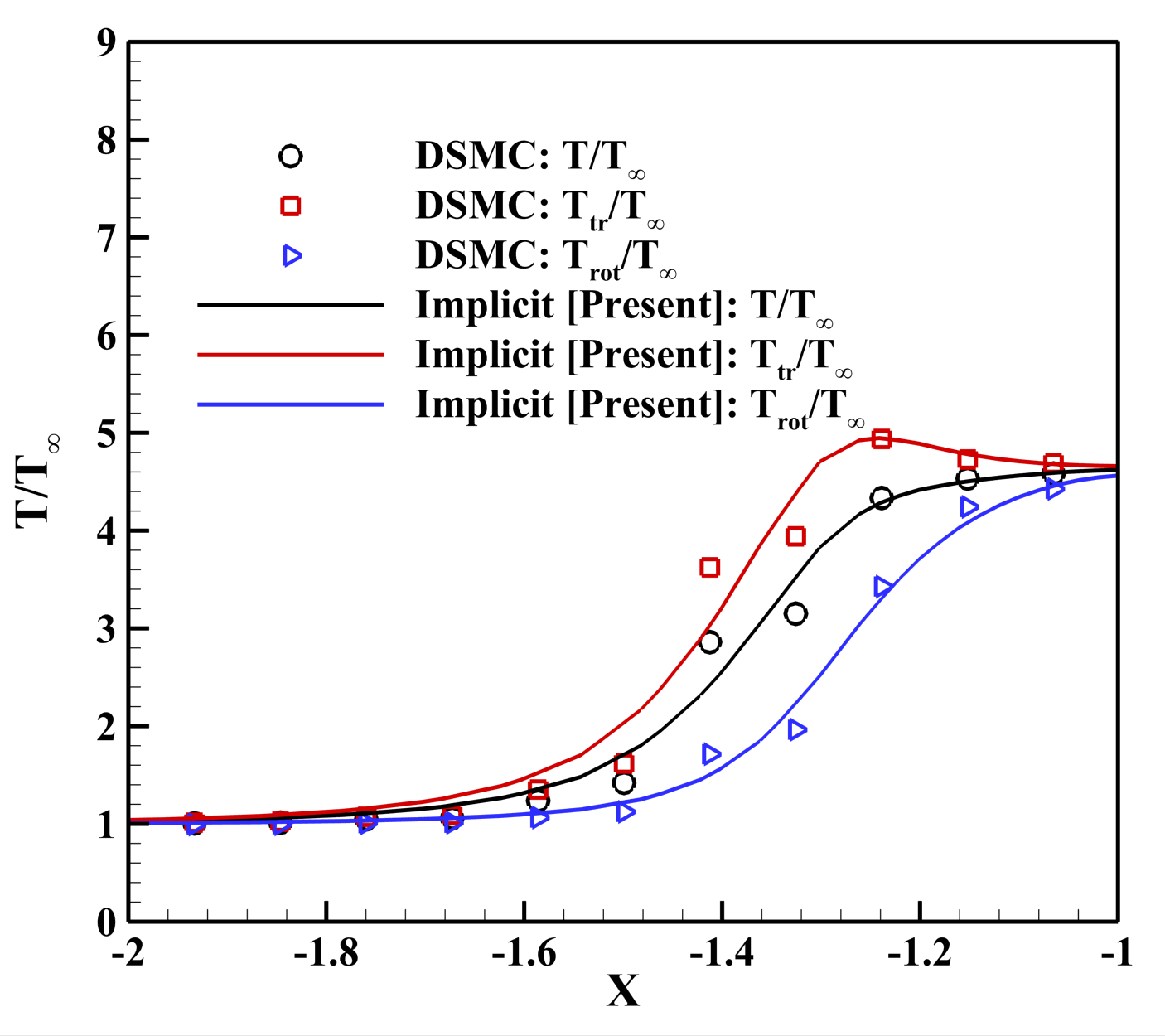}
	}
	\caption{Comparison of the (a) density, (b) pressure, (c) velocity, (d) translational and rotational temperatures along the stagnation line for the supersonic flow over a sphere (Ma = 4.25, Kn = 0.031).}
	\label{Fig:Sphere_Ma4.25Kn0.031_StagnationLine}
\end{figure}

\begin{figure}[!htp]
	\centering
	\subfigure[\label{Fig:Sphere_Ma5.45Kn1.96_StagnationLine_Den}]{
		\includegraphics[width=0.45\textwidth]{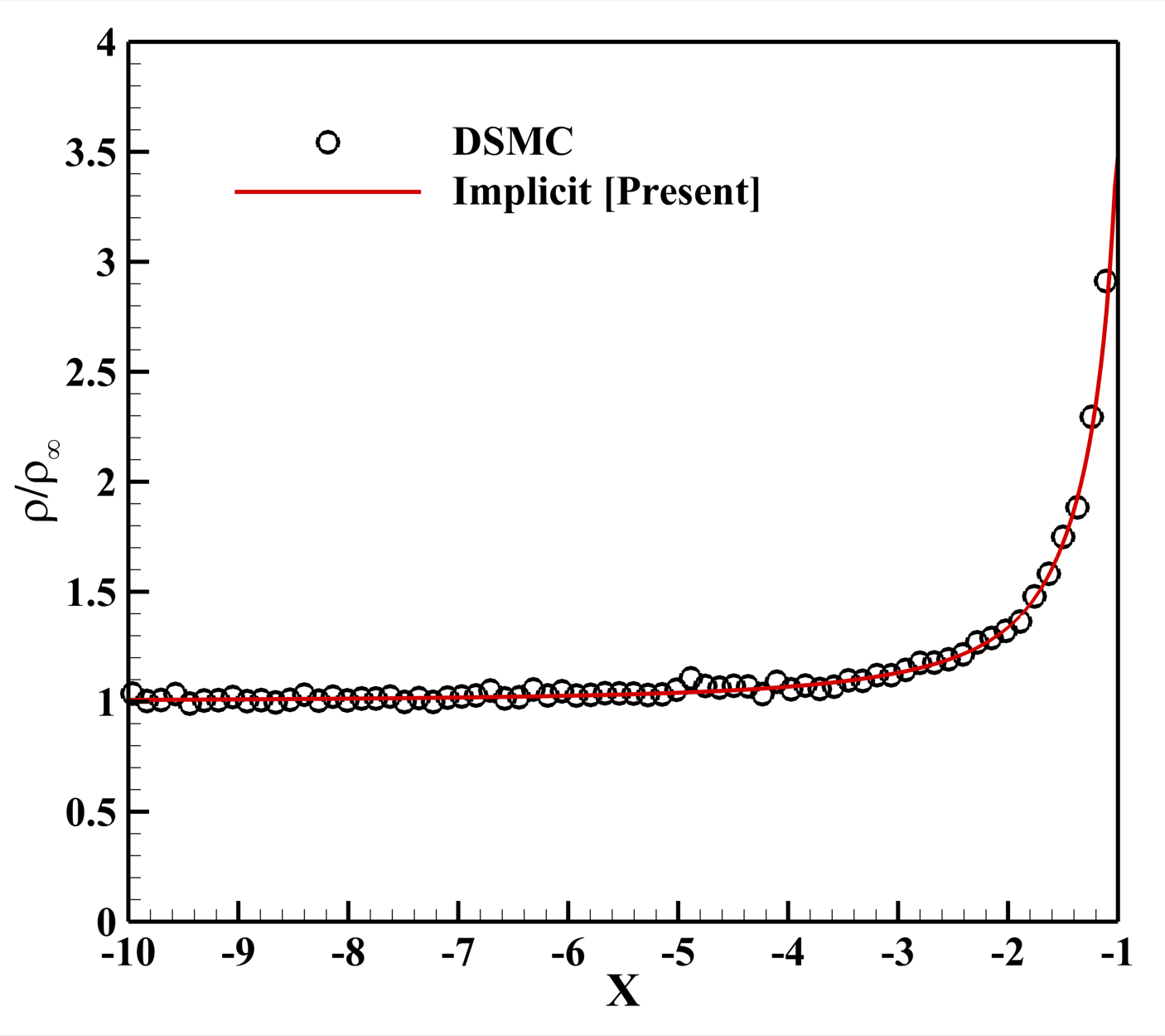}
	}
	\subfigure[\label{Fig:Sphere_Ma5.45Kn1.96_StagnationLine_Pre}]{
	\includegraphics[width=0.45\textwidth]{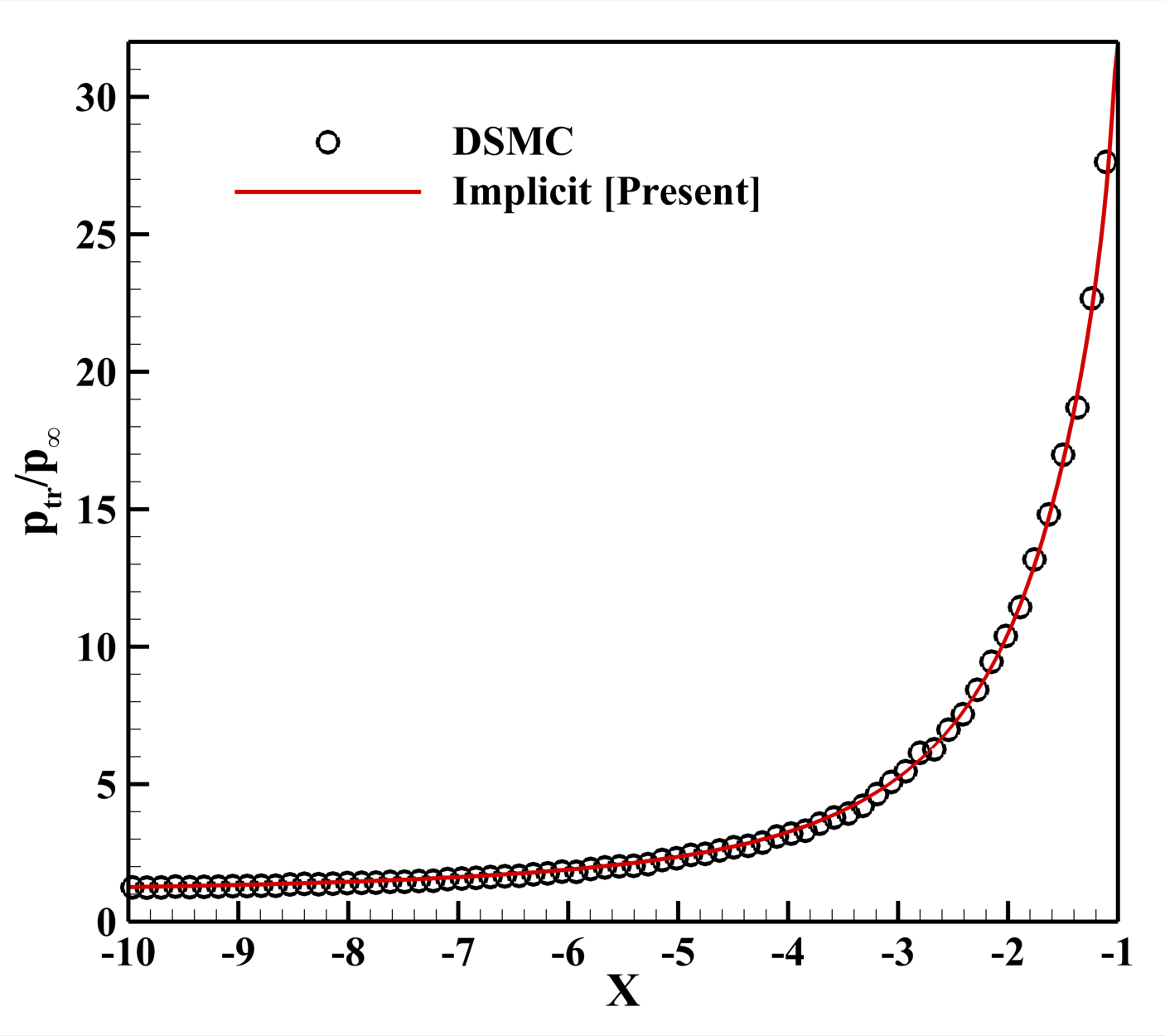}
	}
	\subfigure[\label{Fig:Sphere_Ma5.45Kn1.96_StagnationLine_Vel}]{
		\includegraphics[width=0.45\textwidth]{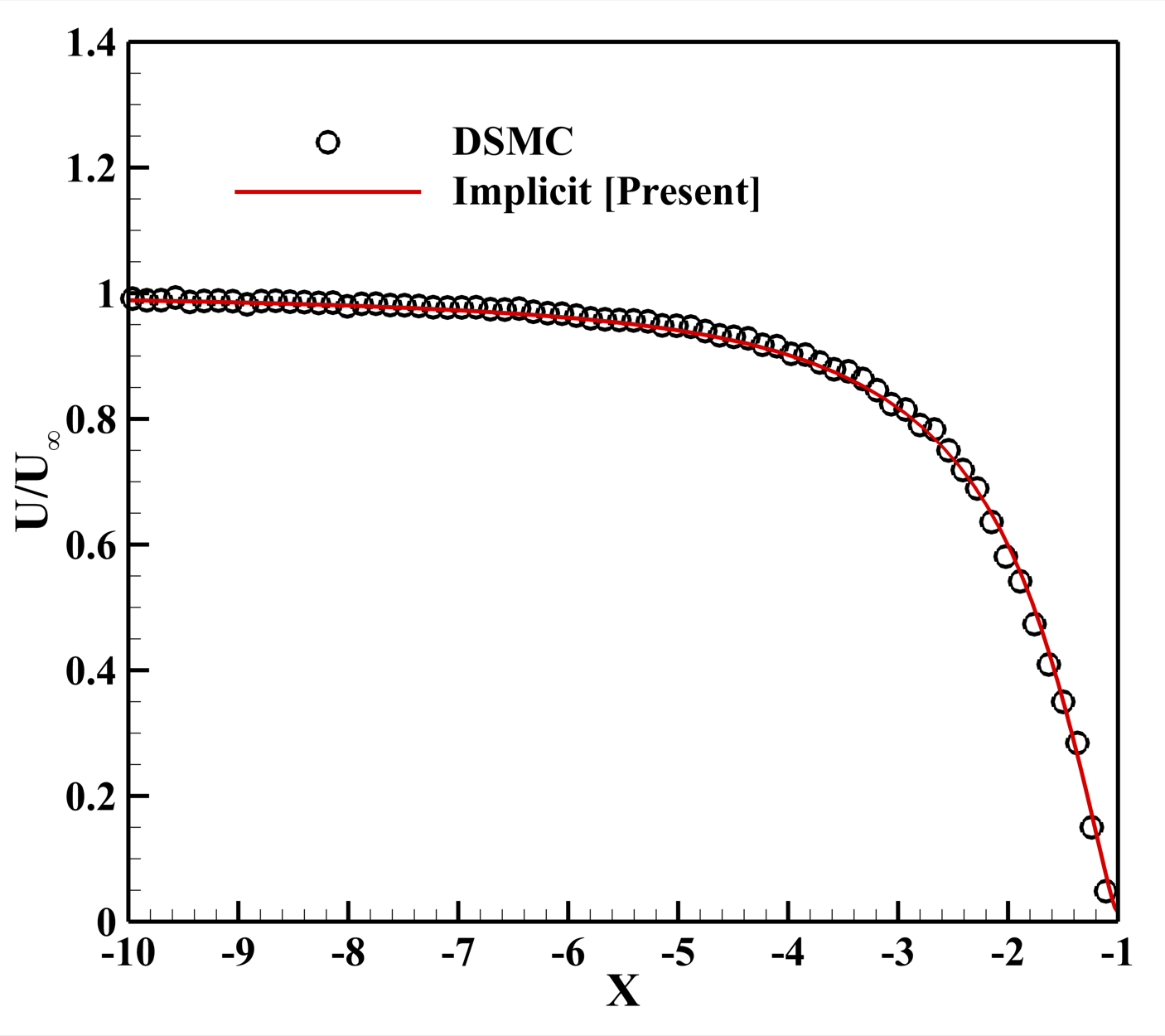}
	}
	\subfigure[\label{Fig:Sphere_Ma5.45Kn1.96_StagnationLine_Tem}]{
		\includegraphics[width=0.45\textwidth]{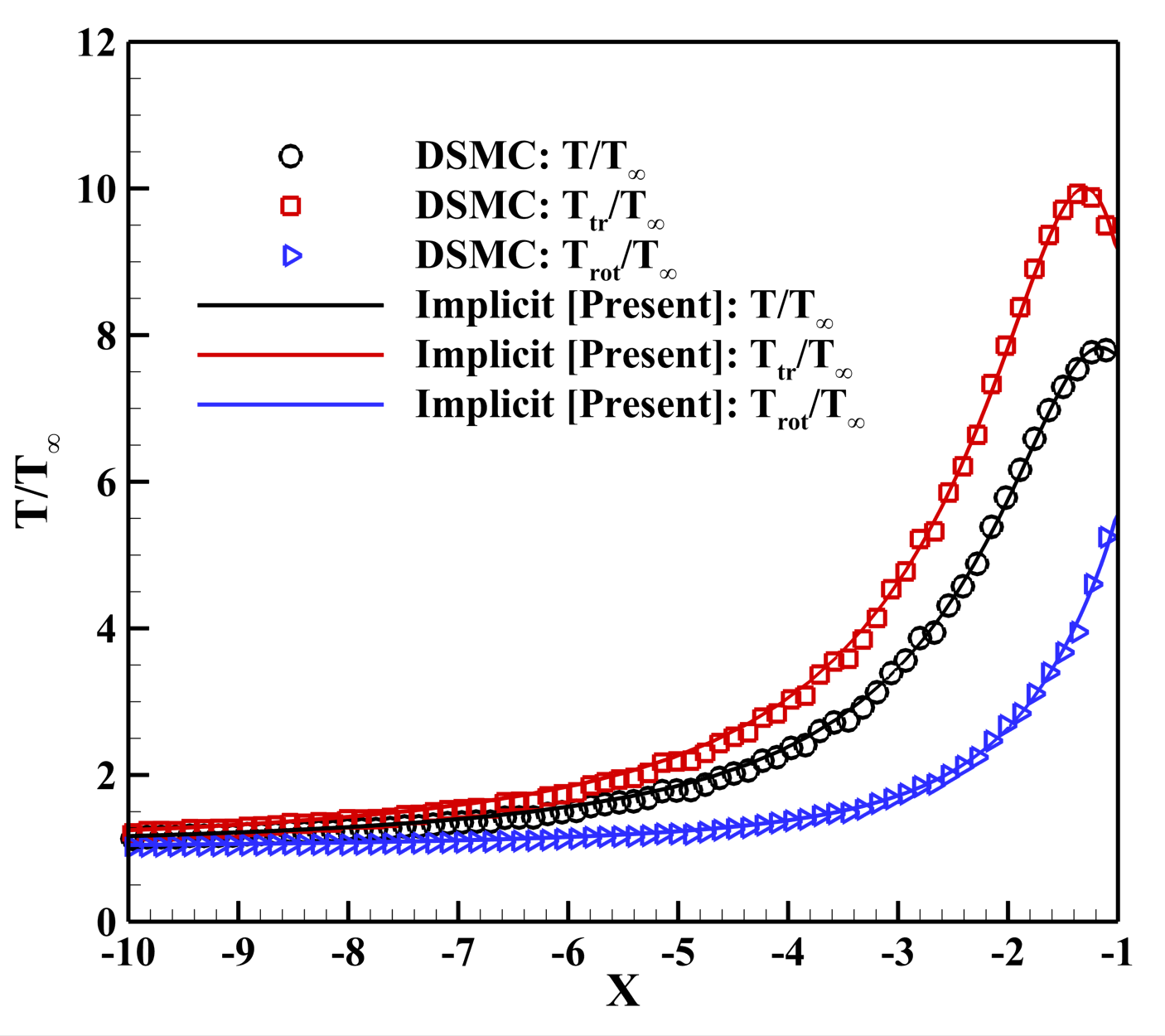}
	}
	\caption{Comparison of the (a) density, (b) pressure, (c) velocity, (d) translational and rotational temperatures along the stagnation line for the hypersonic flow over a sphere (Ma = 5.45, Kn = 1.96).}
	\label{Fig:Sphere_Ma5.45Kn1.96_StagnationLine}
\end{figure}

\begin{figure}[!htp]
	\centering
	\subfigure[\label{Fig:Sphere_Ma4.25Kn0.031_Zsurface_Den}]{
		\includegraphics[width=0.45\textwidth]{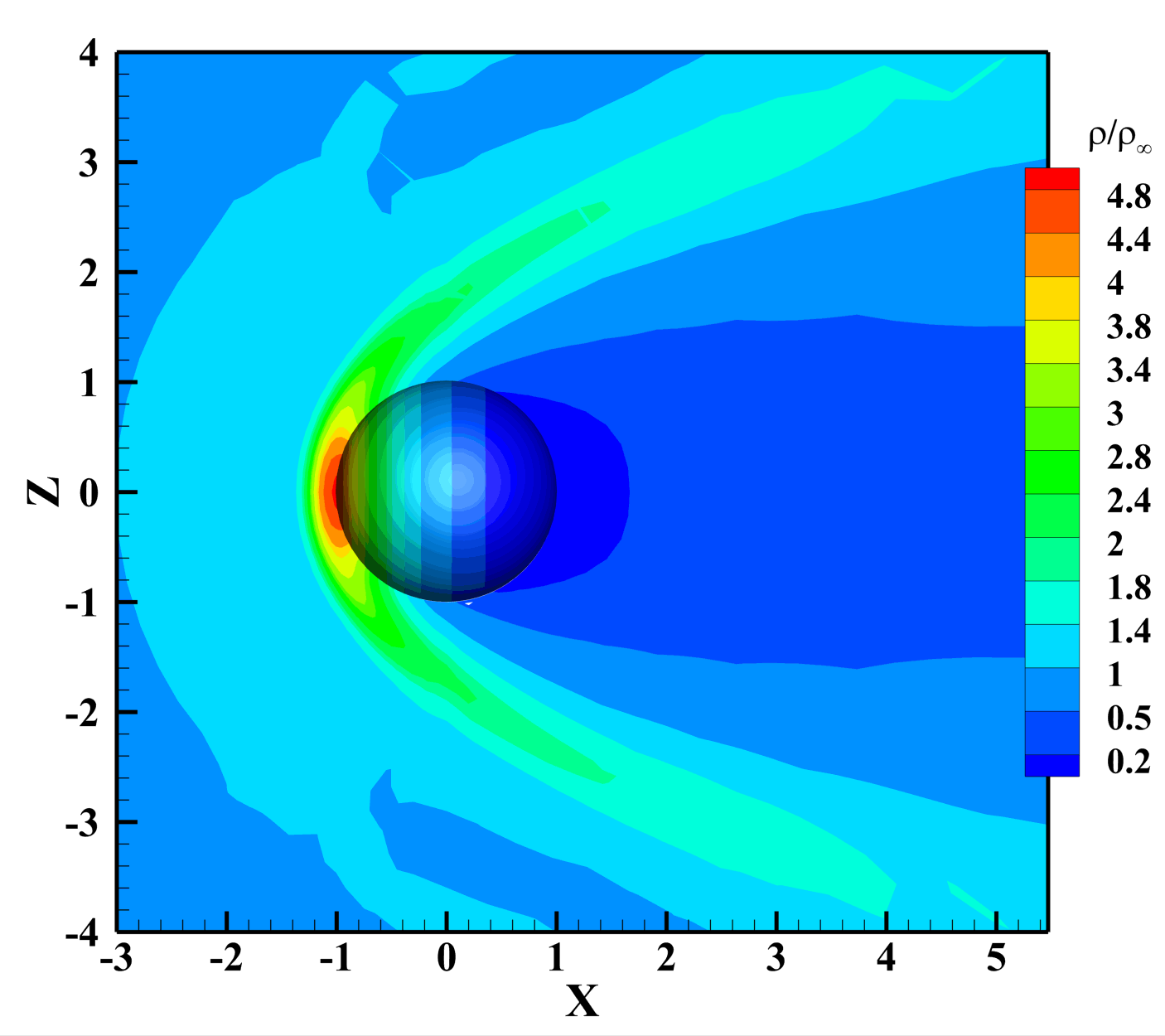}
	}
	\subfigure[\label{Fig:Sphere_Ma4.25Kn0.031_Zsurface_Ma}]{
		\includegraphics[width=0.45\textwidth]{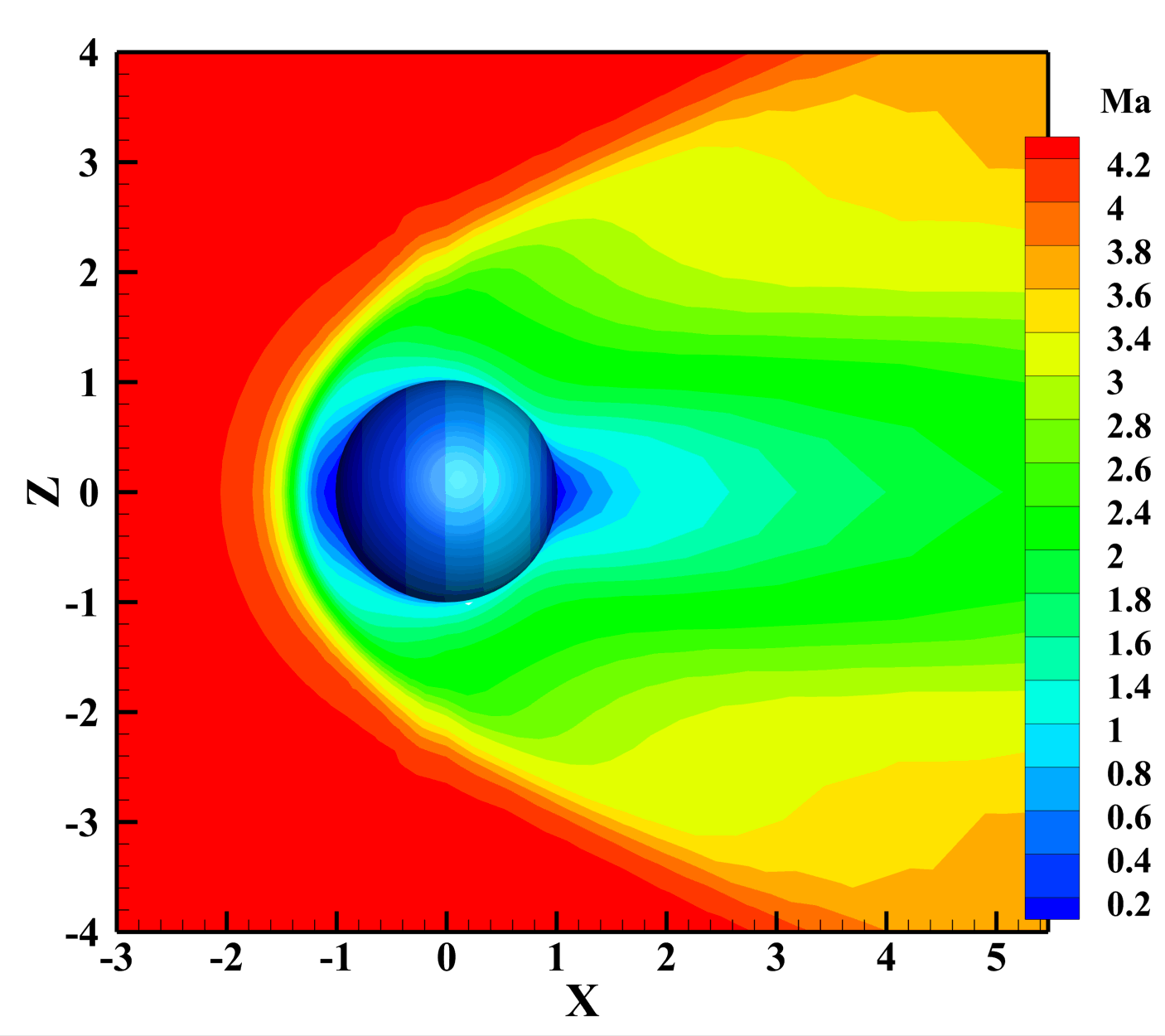}
	}
	\subfigure[\label{Fig:Sphere_Ma4.25Kn0.031_Zsurface_Tem_Tr}]{
		\includegraphics[width=0.45\textwidth]{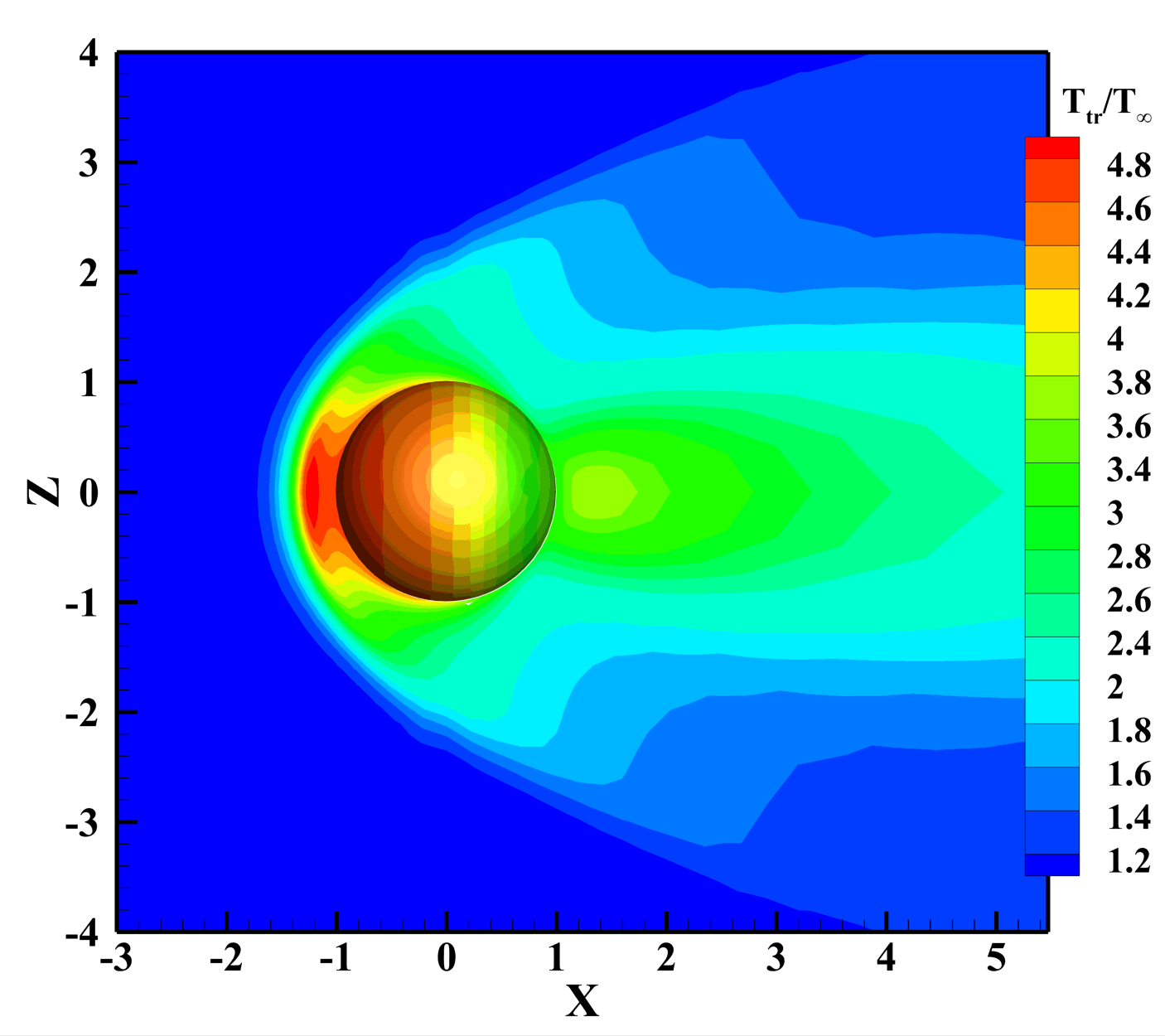}
	}
	\subfigure[\label{Fig:Sphere_Ma4.25Kn0.031_Zsurface_Tem_Rot}]{
		\includegraphics[width=0.45\textwidth]{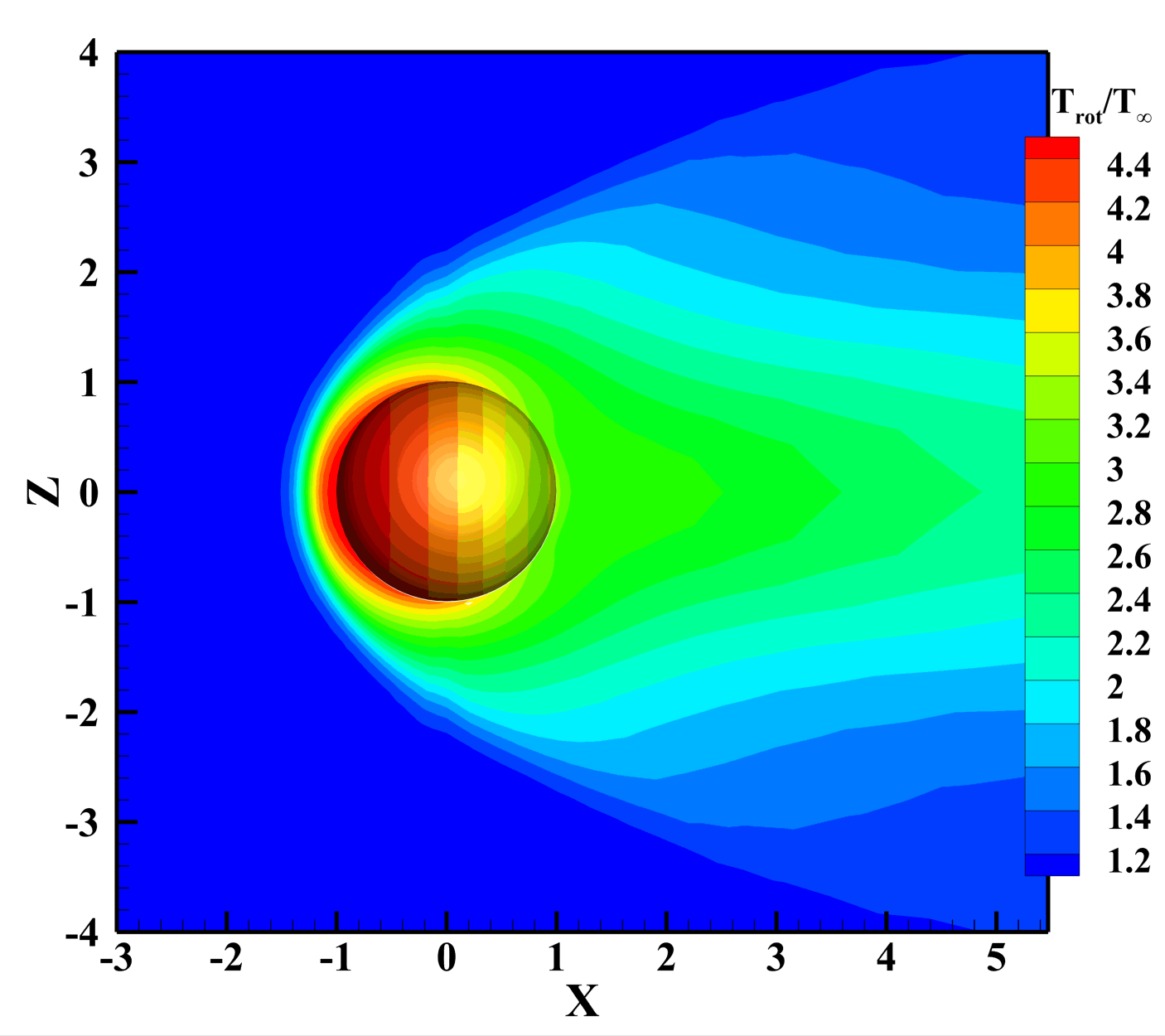}
	}
	\caption{The contour charts of the supersonic flow over a sphere (Ma = 4.25, Kn = 0.031). (a) Density, (b) Ma Number, (c) translational temperature and (d) rotational temperature.}
	\label{Fig:Sphere_Ma4.25Kn0.031_Zsurface}
\end{figure}

\begin{figure}[!htp]
	\centering
	\subfigure[\label{Fig:Sphere_Ma5.45Kn1.96_Zsurface_Den}]{
		\includegraphics[width=0.45\textwidth]{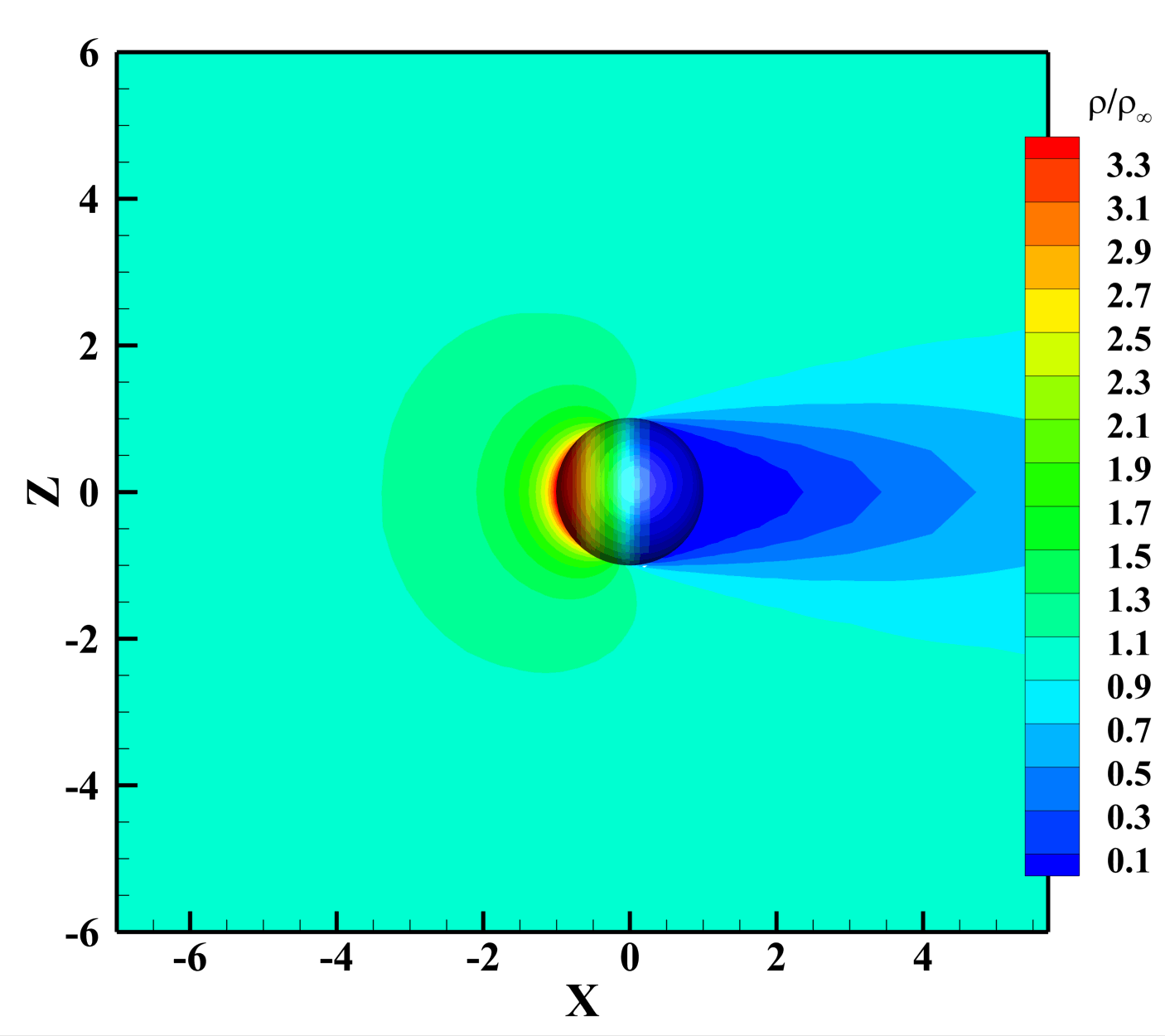}
	}
	\subfigure[\label{Fig:Sphere_Ma5.45Kn1.96_Zsurface_Ma}]{
		\includegraphics[width=0.45\textwidth]{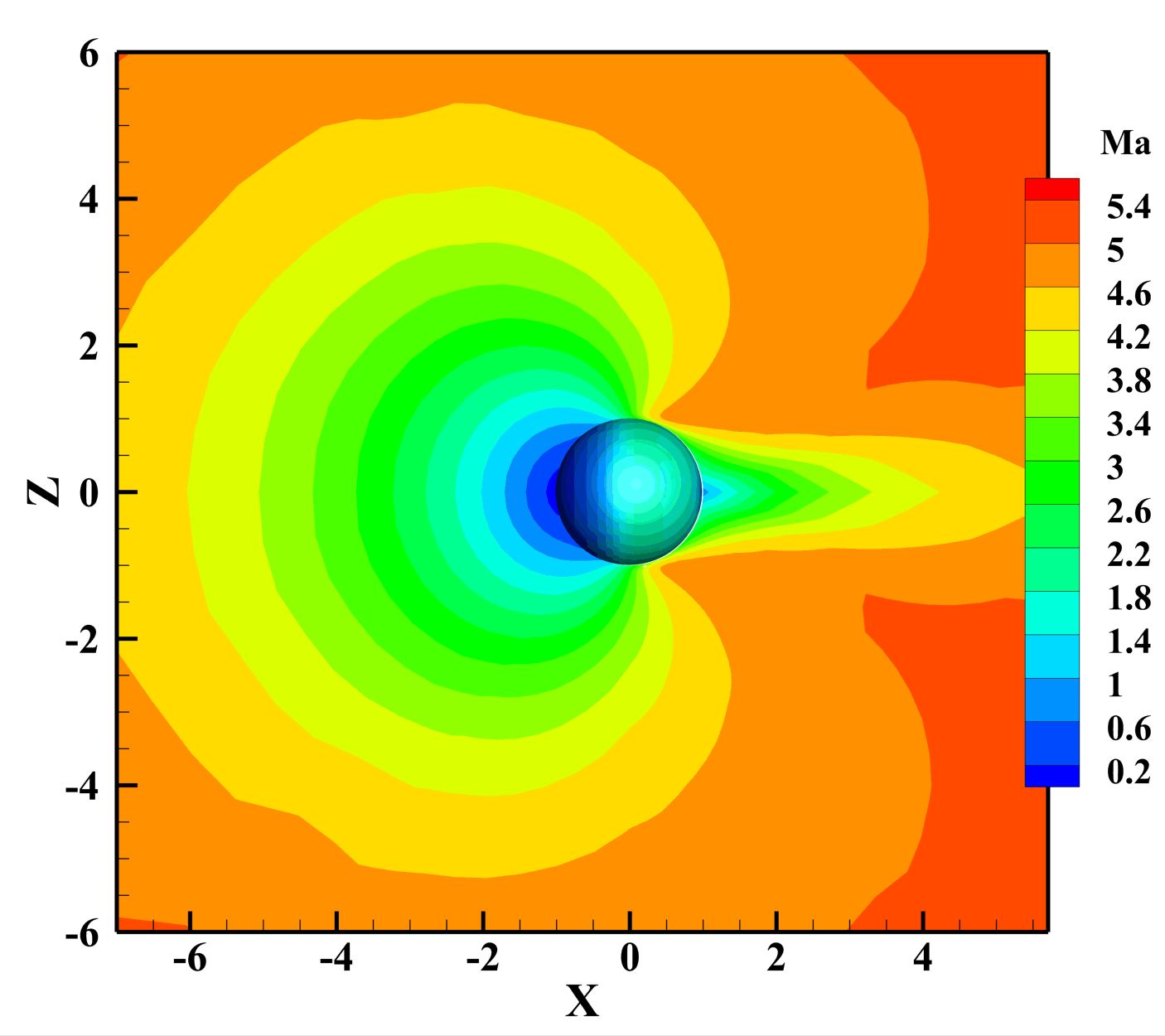}
	}
	\subfigure[\label{Fig:Sphere_Ma5.45Kn1.96_Zsurface_Tem_Tr}]{
		\includegraphics[width=0.45\textwidth]{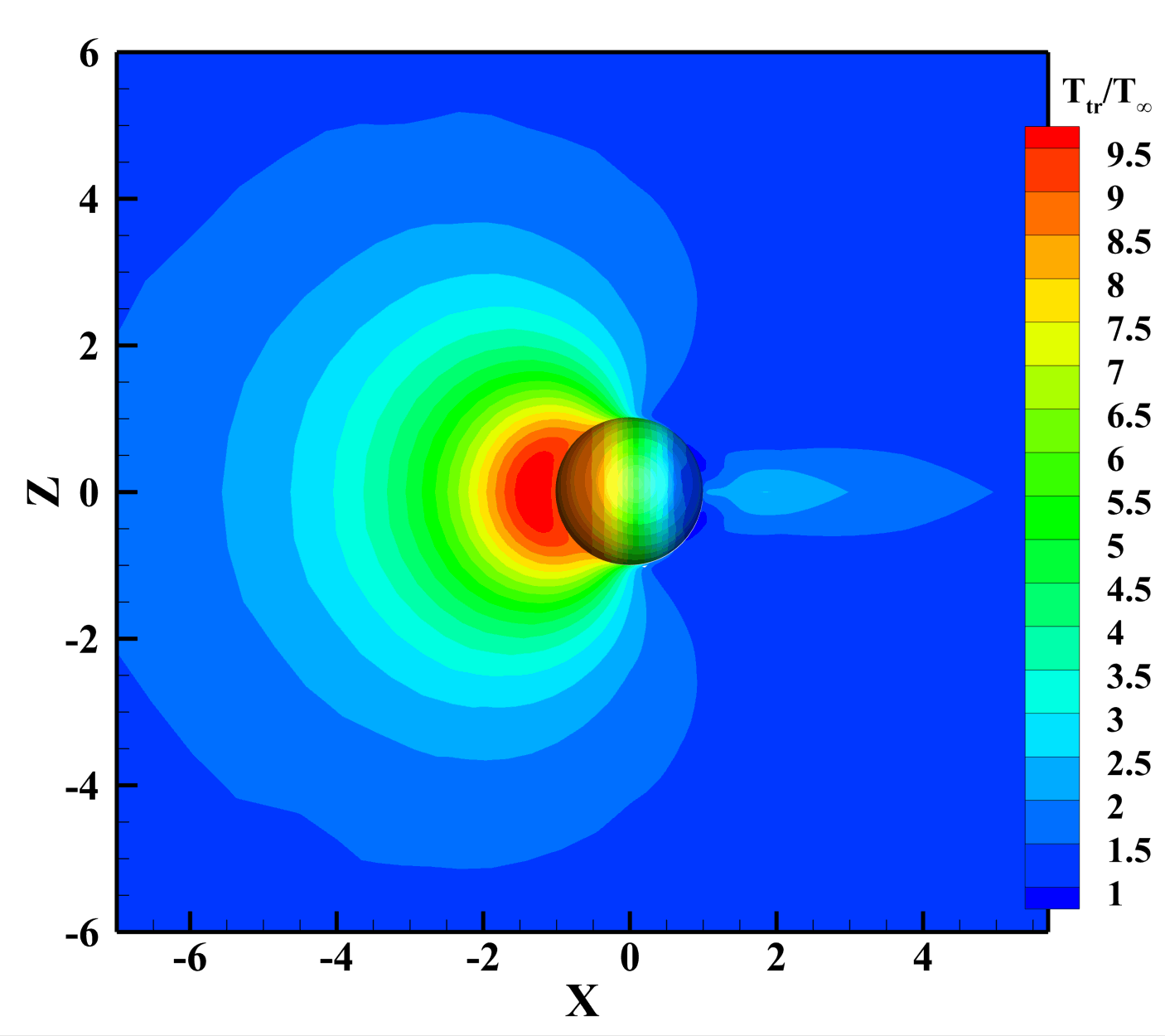}
	}
	\subfigure[\label{Fig:Sphere_Ma5.45Kn1.96_Zsurface_Tem_Rot}]{
		\includegraphics[width=0.45\textwidth]{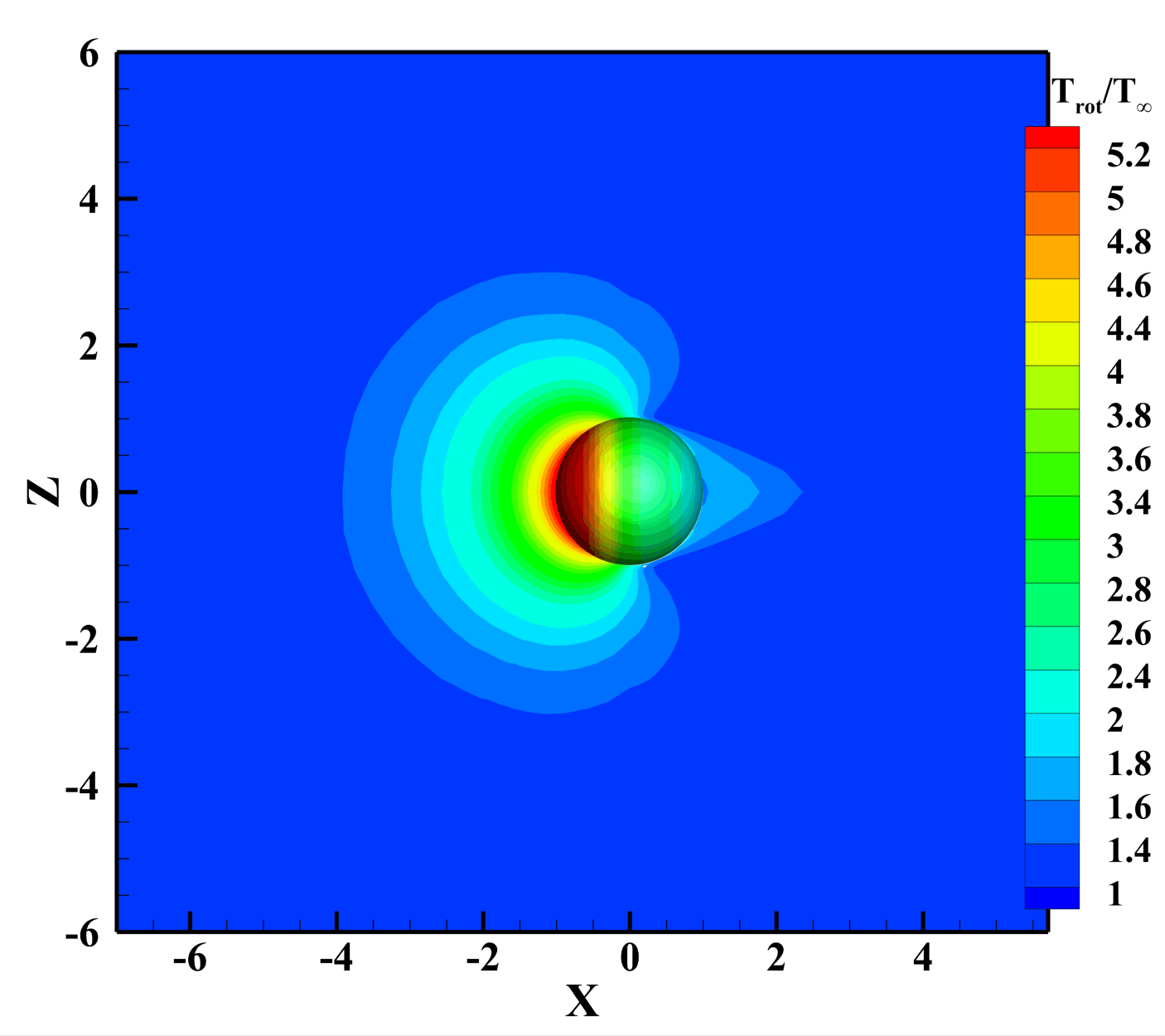}
	}
	\caption{The contour charts of the hypersonic flow over a sphere (Ma = 5.45, Kn = 1.96). (a) Density, (b) Ma Number, (c) translational temperature and (d) rotational temperature.}
	\label{Fig:Sphere_Ma5.45Kn1.96_Zsurface}
\end{figure}

The wall surface pressure coefficient $C_{p}$, shear stress coefficient $C_{\tau}$ and heat transfer coefficient $C_{h}$ of cases A4 and B1 are shown in Figs.~\ref{Fig:Sphere_Ma4.25Kn0.031_Wall} and~\ref{Fig:Sphere_Ma5.45Kn1.96_Wall}, respectively.
Basically, the pressure coefficient match well with the reference data, while the shear stress coefficient is also in good agreement with the reference data, except a little deviation near the stagnation point.
As for heat transfer coefficient, there is a little deviation compared with the reference data for the case A4, while it is consistent with the reference data for the case B1.
In the present work, the $C_{p}$, $C_{\tau}$, and $C_{h}$ are calculated from the formulas,
\begin{equation}\label{equ:cp}
	{C_p} = \frac{{\sum\limits_k {{\omega _k}{{\left( {{{\bm{u}}_k} \cdot {{\bm{n}}_{w}}} \right)}^2}
				{G_{w,k}}}  - {p_\infty }}}{{\frac{1}{2}{\rho _\infty }{{\left| {{{\bm{U}}_\infty }} \right|}^2}}},
\end{equation}
\begin{equation}\label{equ:cf}
	{C_{\tau}} = \frac{{\left| {\sum\limits_k {{\omega _k}\left( {{{\bm{u}}_k} \cdot {{\bm{n}}_{w}}}
					\right){{\bm{u}}_k}{G_{w,k}}}  - {{\bm{n}}_{w}}\sum\limits_k {{\omega _k}{{\left( {{{\bm{u}}_k} 
								\cdot {{\bm{n}}_{w}}} \right)}^2}{G_{w,k}}} } \right|}}{{\frac{1}{2}{\rho _\infty }{{\left|
					{{{\bm{U}}_\infty }} \right|}^2}}},
\end{equation}
\begin{equation}\label{equ:ch}
	{C_h} = \frac{{\sum\limits_k {{\omega _k}\left( {{{\bm{u}}_k} \cdot {{\bm{n}}_{w}}} \right)\left(
				{\frac{1}{2}{{\left| {{{\bm{u}}_k}} \right|}^2}{G_{w,k}} + {R_{w,k}}} \right)} }}{{\frac{1}{2}
			{\rho _\infty }{{\left| {{{\bm{U}}_\infty }} \right|}^3}}},
\end{equation}
where ${p _\infty }$, ${\rho _\infty }$ and ${U _\infty }$ are the free-stream pressure, density and velocity, respectively.
${\bm {n}}_{w}$ is the outward unit vector normal to the wall from the fluid region to the wall.
Figs.~\ref{Fig:Sphere_Ma4.25Kn0.031_StagnationLine} and~\ref{Fig:Sphere_Ma5.45Kn1.96_StagnationLine} show the quantitative comparison of the density, pressure, velocity, and temperatures along the stagnation line for cases A4 and B1, respectively.
The satisfactory agreements are achieved between these two methods.
Furthermore, Figs.~\ref{Fig:Sphere_Ma4.25Kn0.031_Zsurface} and~\ref{Fig:Sphere_Ma5.45Kn1.96_Zsurface} illustrate the contours of the density, Ma number, and the translational, rotational temperatures in the middle of the $Y$ plane for those two cases, respectively.
As the Kn number increases, the compressive wave becomes weaker in front of the sphere.

\subsection{Hypersonic rarefied flow over a blunted-cone}\label{sec:caseC}
\begin{figure}[!htp]
	\centering
	\subfigure[\label{Fig:Cone_Ma10.15_Mesh}]{
		\includegraphics[width=0.45\textwidth]{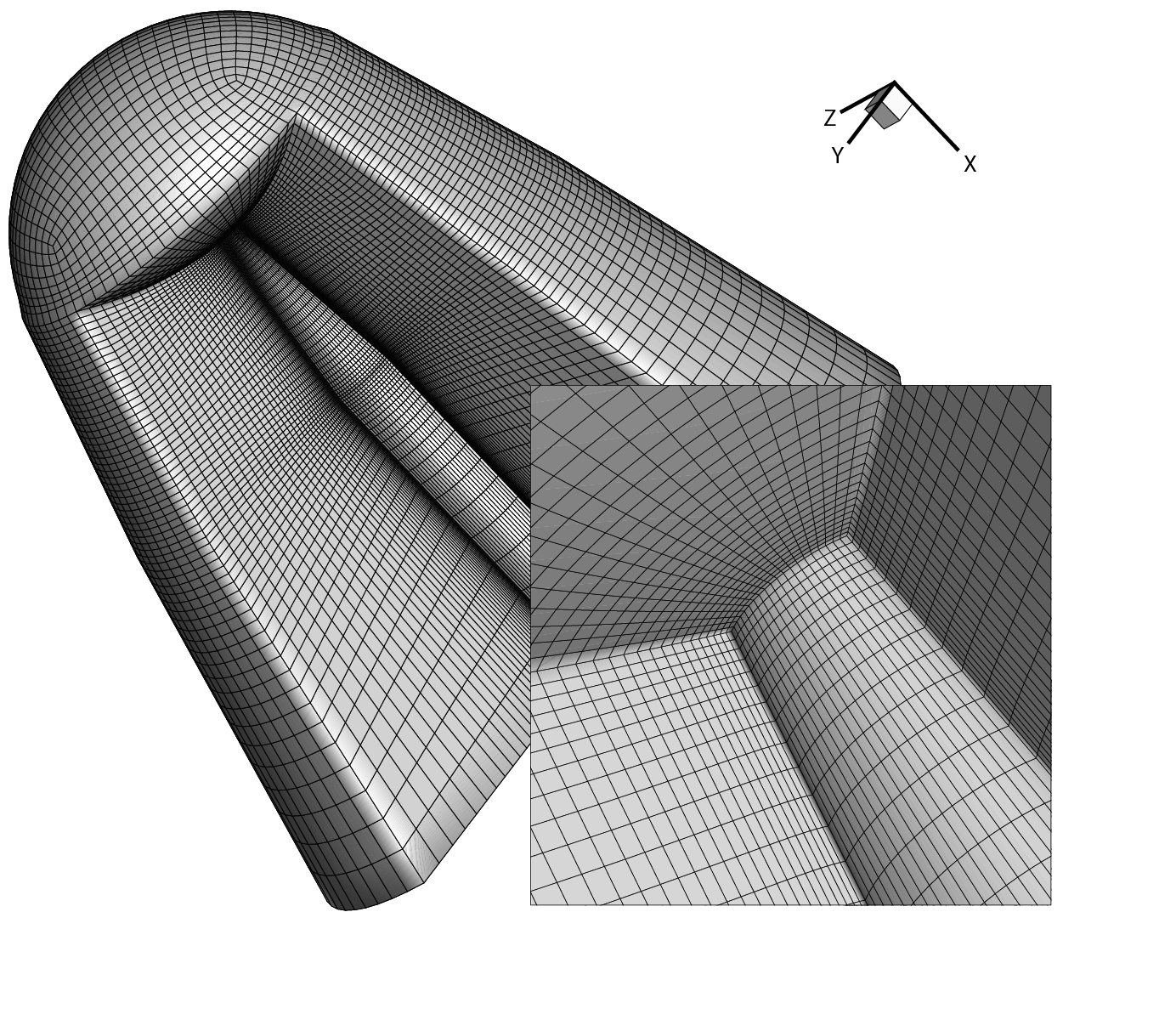}
	}
	\subfigure[\label{Fig:Cone_Ma10.15_Micmesh}]{
		\includegraphics[width=0.45\textwidth]{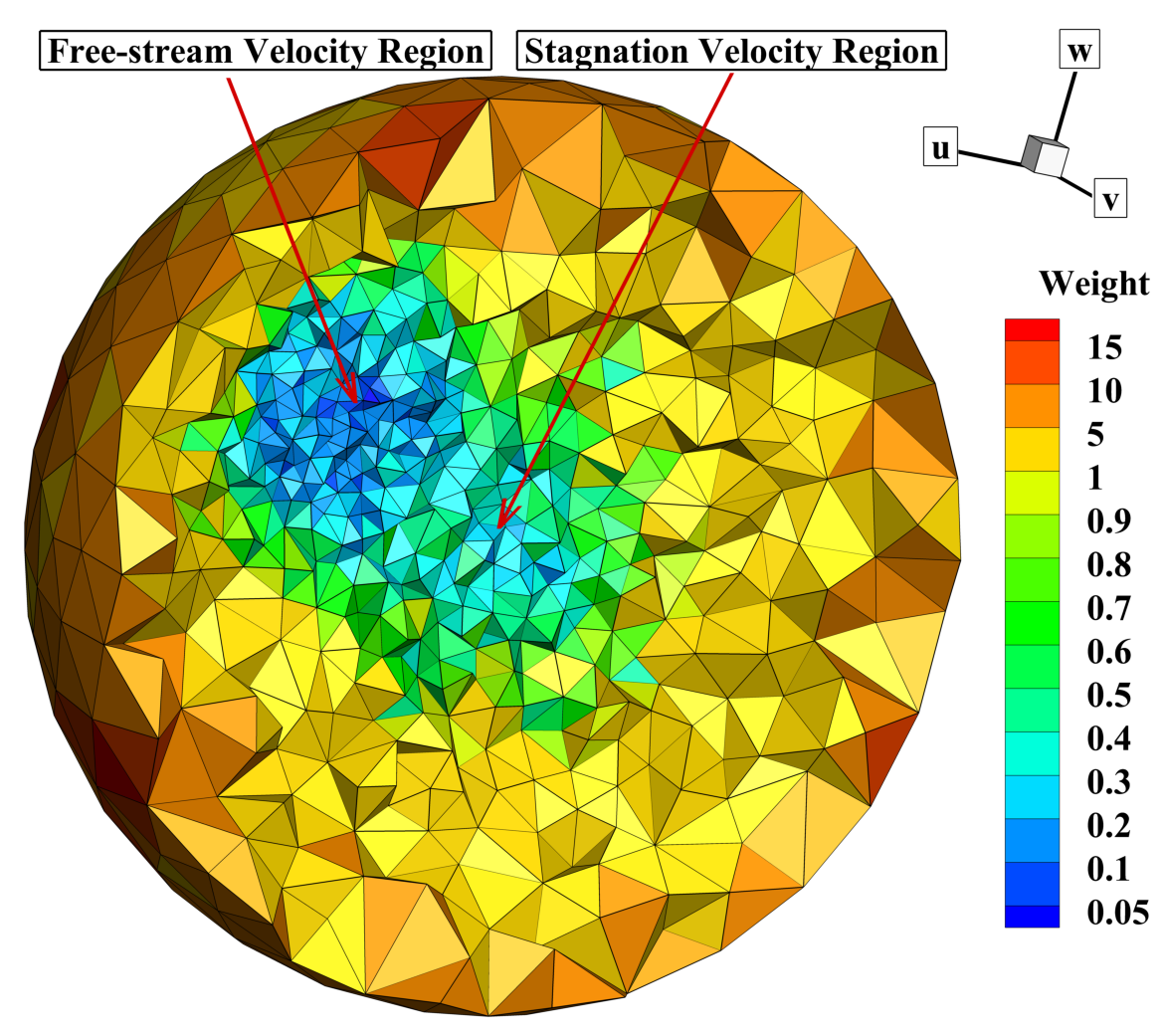}
	}
	\caption{The section views of (a) the physical space mesh (total 260640 cells) and (b) the unstructured discrete velocity space mesh (AOA = 25$^ \circ$, total 22620 cells) for the hypersonic rarefied flow over a blunted-cone.}
	\label{Fig:Cone_Ma10.15_MeshMicmesh} 
\end{figure}

In this study, computer simulations are presented using the present method for hypersonic rarefied flow over a blunted-cone at 0, 10, 20 and 25 degrees angle-of-attack (AOA), and compared to the wind tunnel data~\citep{boylan_aerodynamics_1967} and DSMC result~\citep{padilla_assessment_2006}.
Geometry of the blunted-cone for wind tunnel test in the von Karman Gas Dynamics Facilitycan (VKF) Tunnel L can be found in the work of Boylan~\citep{boylan_aerodynamics_1967}, and flow conditions are shown in Table~\ref{Table:Cone_Parameter}.
In the simulation, Sutherland formula is utilized to calculate the dynamic viscosity. The diffuse reflection boundary condition with full thermodynamic accommodation at the wall surface is applied.
Fig.~\ref{Fig:Cone_Ma10.15_Mesh} illustrates the physical space mesh, in which contains 260640 hexahedral cells.
The unstructured DVS mesh with 22620 cells is shown in Fig.~\ref{Fig:Cone_Ma10.15_Micmesh}.
All of the simulations are run on Xi'an Future Artificial Intelligence Computing Center.
Each compute node has two Intel Xeon 6248R CPUs at 3.0GHz, and 192GB memory.
It takes about three hours of calculation to obtain a stable aerodynamic coefficient using the discrete particle velocity based MPI parallel strategy with 500 cores.

\begin{figure}[!htp]
	\centering
	\subfigure[\label{Fig:Cone_Ma10.15_Cd}]{
		\includegraphics[width=0.45\textwidth]{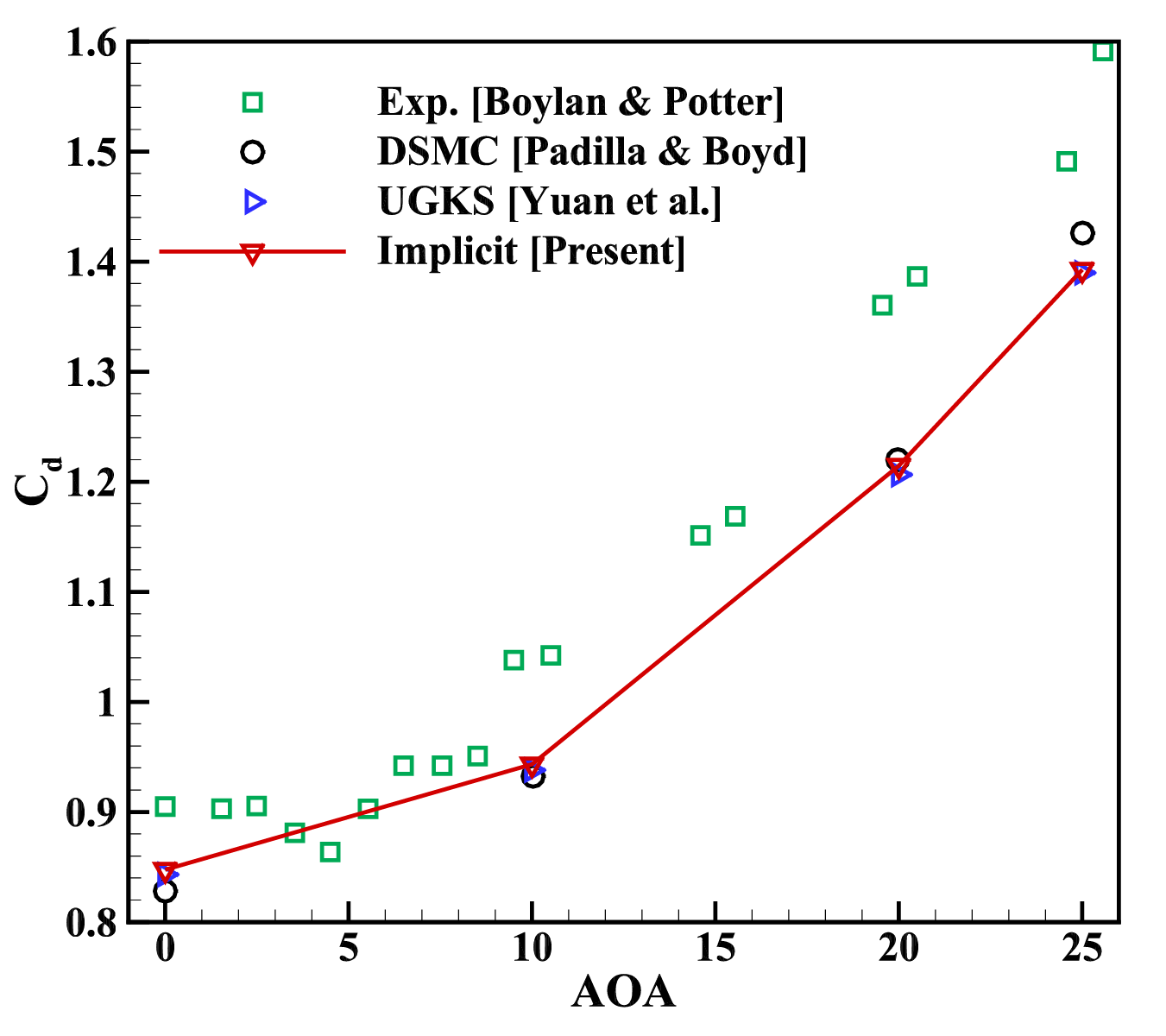}
	}
	\subfigure[\label{Fig:Cone_Ma10.15_Cl}]{
		\includegraphics[width=0.45\textwidth]{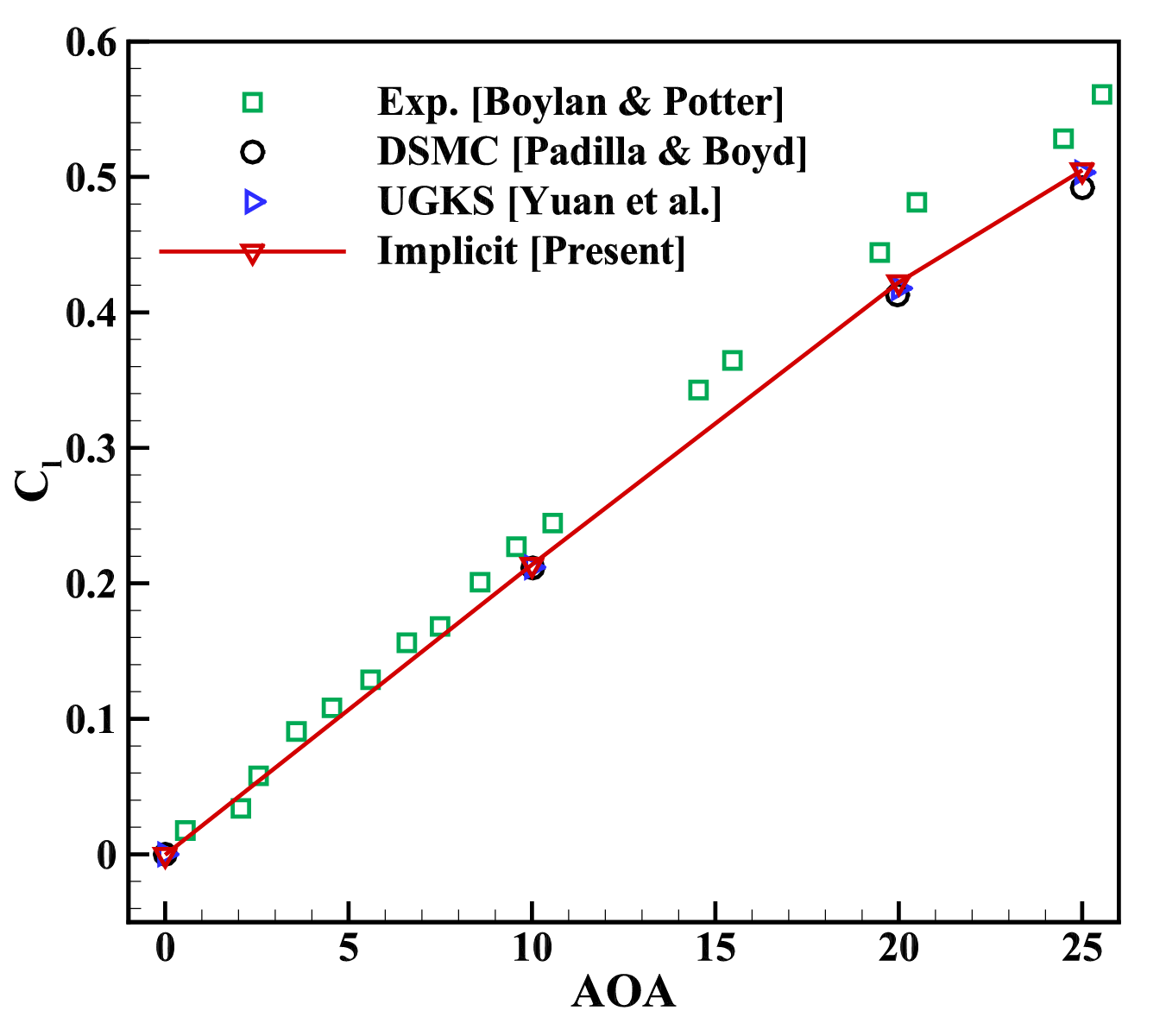}
	}
	\caption{Variation of (a) the drag coefficient and (b) the lift coefficient with AOA for the hypersonic rarefied flow over a blunted-cone.}
	\label{Fig:Cone_Ma10.15_ClCd} 
\end{figure}

\begin{figure}[!htp]
	\centering
	\subfigure[\label{Fig:Cone_Ma10.15_LD}]{
		\includegraphics[width=0.45\textwidth]{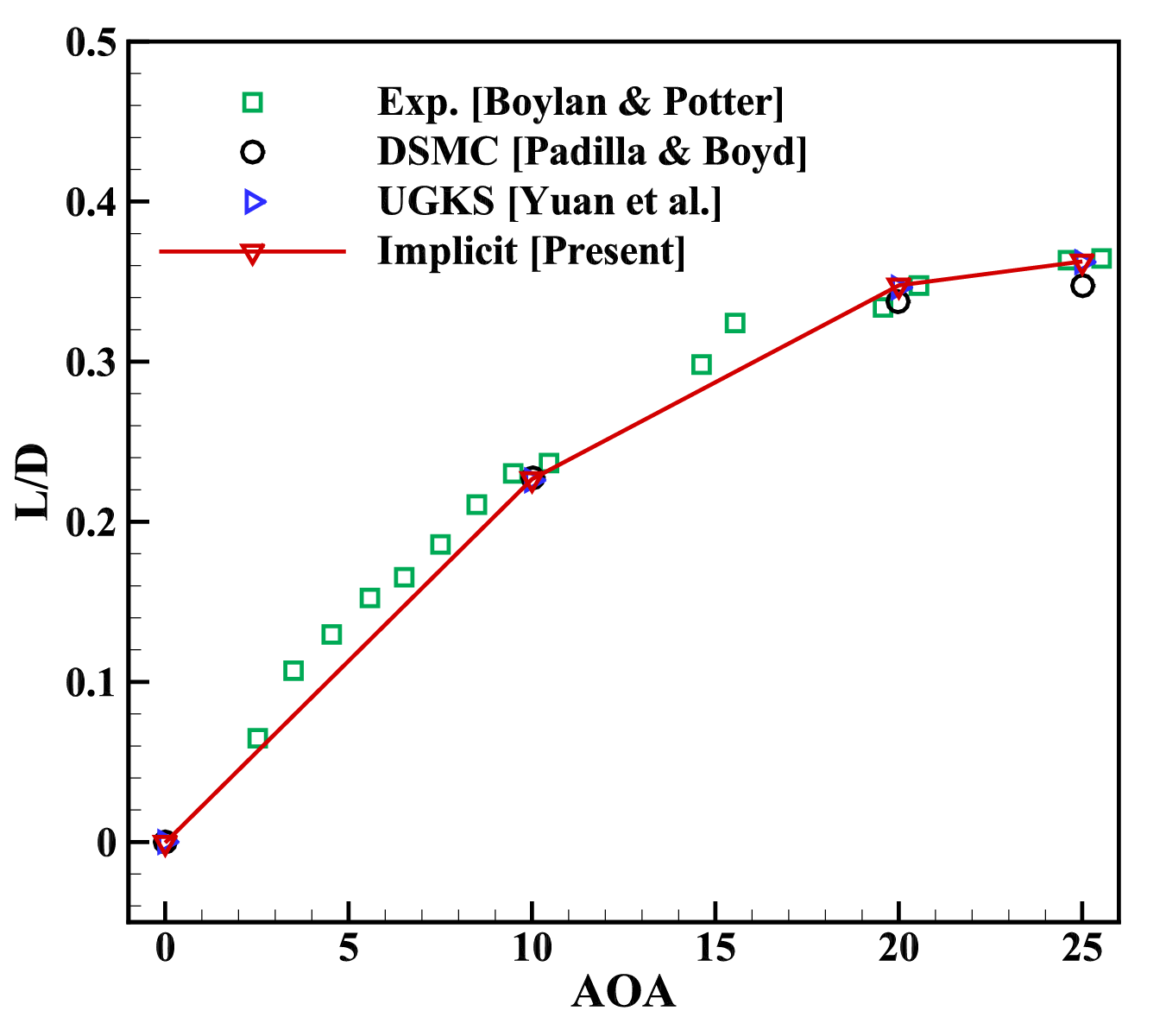}
	}
	\subfigure[\label{Fig:Cone_Ma10.15_Cm}]{
		\includegraphics[width=0.45\textwidth]{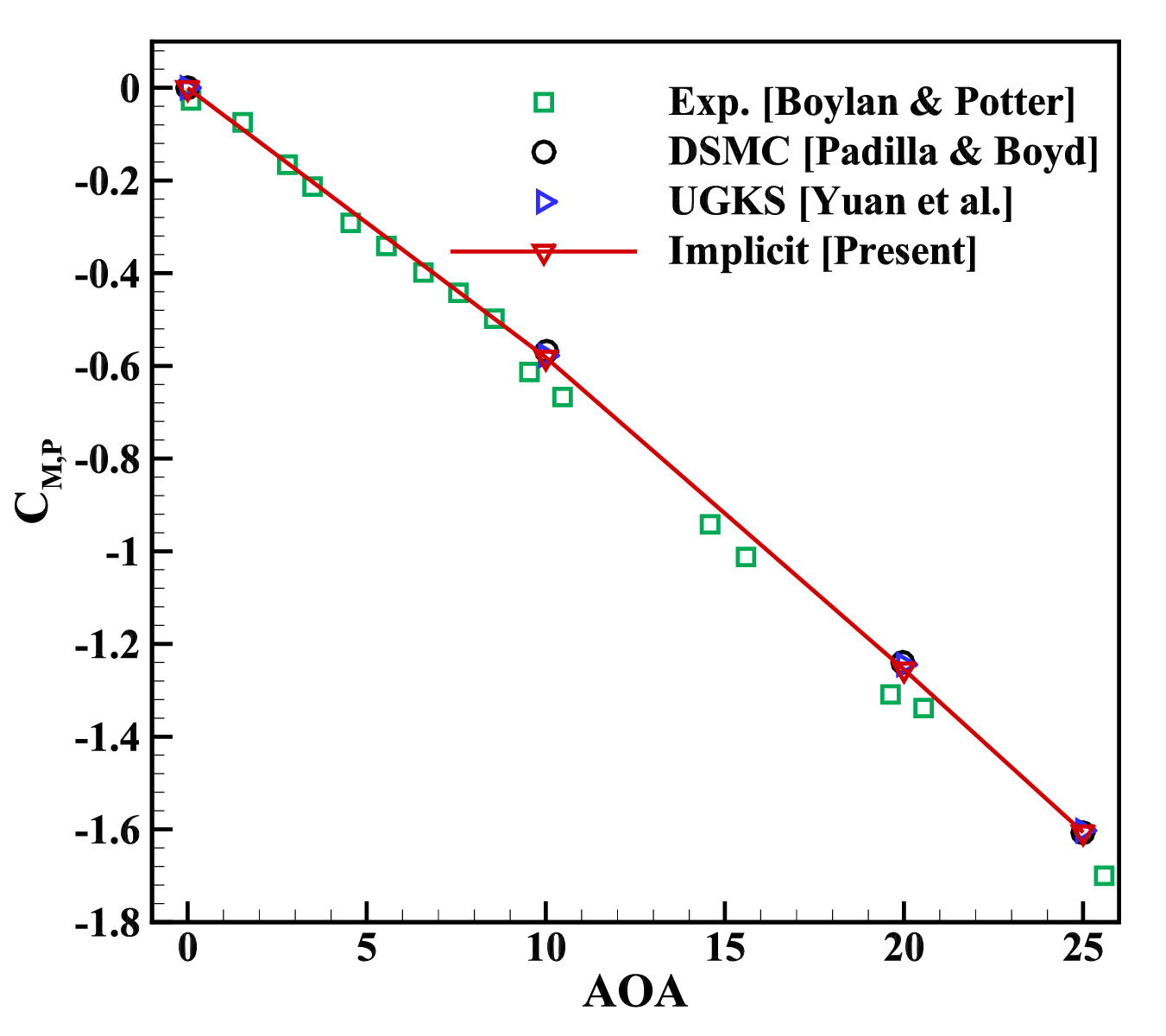}
	}
	\caption{Variation of (a) the lift-to-drag ratio and (b) the pitching moment coefficient with AOA for the hypersonic rarefied flow over a blunted-cone.}
	\label{Fig:Cone_Ma10.15_LDCm}
\end{figure}

The lift and drag coefficients at various angles-of-attack are shown in Fig.~\ref{Fig:Cone_Ma10.15_ClCd}.
The trends of lift and drag coefficients agree well with the wind tunnel data.
The differences of drag coefficient solved by the present method and DSMC are about 2.35$\%$, 1.17$\%$, -0.45$\%$, and -2.33$\%$ at AOA $= 0^ \circ$, 10$^ \circ$, 20$^ \circ$ and 25$^ \circ$, respectively.
As for lift coefficient, the percentage differences between the present and DSMC data are about 1.14$\%$, 2.24$\%$ and 2.62$\%$ at AOA $= 10^ \circ$, 20$^ \circ$ and 25$^ \circ$, respectively.
Fig.~\ref{Fig:Cone_Ma10.15_LD} compares the lift-to-drag ratio at various angles of attack.
The present solution, DSMC data and wind tunnel data demonstrate good agreement, and the maximum relative error in comparison with DSMC is about 4.35$\%$ at AOA $= 25^ \circ$.
The comparisons of pitching moment coefficient are shown in Fig.~\ref{Fig:Cone_Ma10.15_Cm}.
Once again, there is no significant difference between the present and DSMC result, as well as wind tunnel data.
The relative errors of pitching moment coefficient between the present method and DSMC are 2.33$\%$, 1.24$\%$, and -0.06$\%$ at AOA = 10$^ \circ$, 20$^ \circ$ and 25$^ \circ$, respectively.
In conclusion, the aerodynamic coefficients predicted by the present method are consistent with those of DSMC, and they both reasonably agree well with the wind tunnel data.

\begin{figure}[!htp]
	\centering
	\subfigure[\label{Fig:Cone_Ma10.15_Line_AOA0_Cp}]{
		\includegraphics[width=0.45\textwidth]{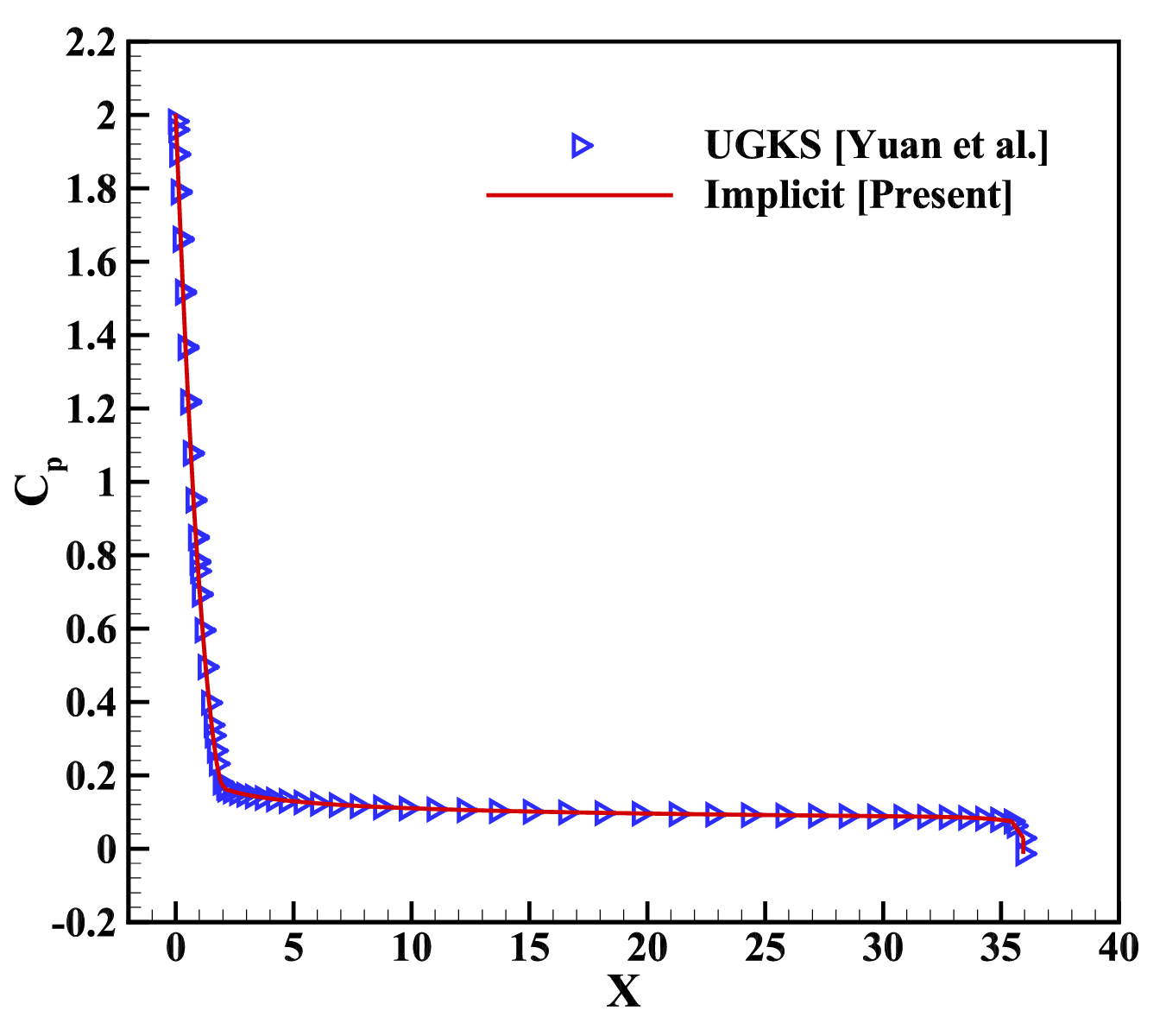}
	}
	\subfigure[\label{Fig:Cone_Ma10.15_Line_AOA0_Ch}]{
		\includegraphics[width=0.45\textwidth]{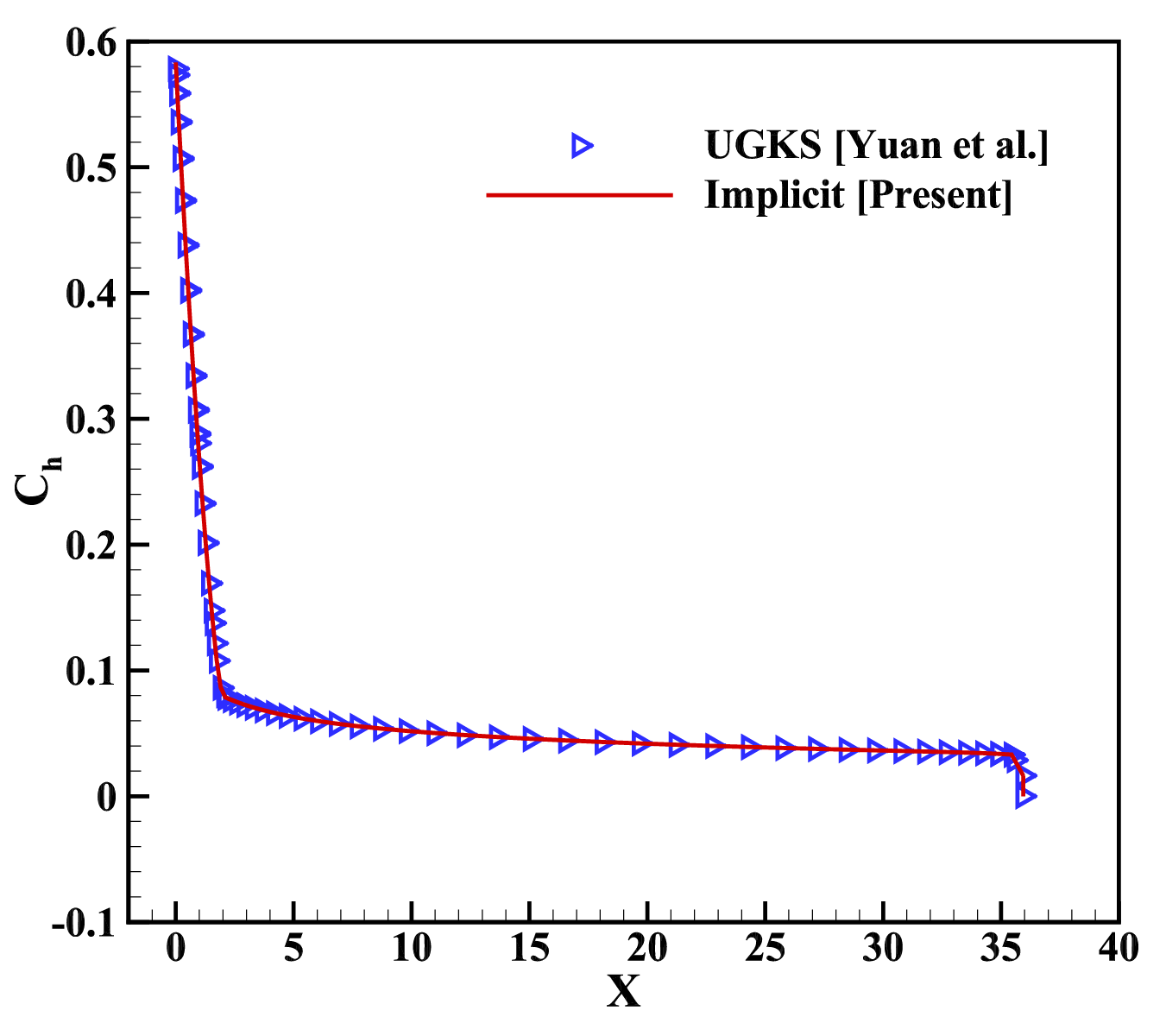}
	}
	\caption{Comparison of UGKS and the present (a) pressure coefficient and (b) heat transfer coefficient for the hypersonic rarefied flow over a blunted-cone at AOA = 0$^ \circ$.}
	\label{Fig:Cone_Ma10.15_Line_AOA0_CpCh}
\end{figure}

\begin{figure}[!htp]
	\centering
	\subfigure[\label{Fig:Cone_Ma10.15_Line_AOA25_Cp}]{
		\includegraphics[width=0.45\textwidth]{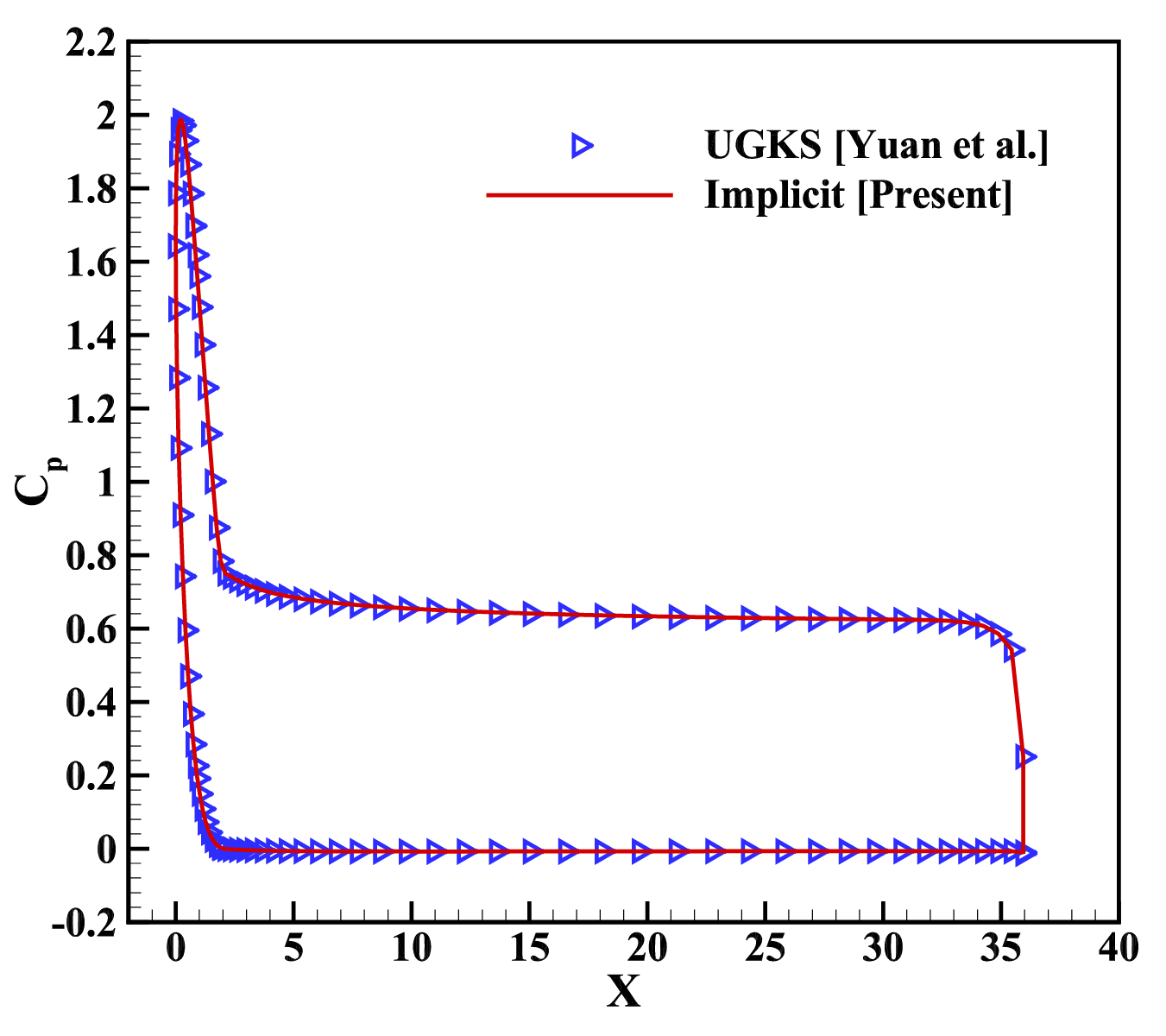}
	}
	\subfigure[\label{Fig:Cone_Ma10.15_Line_AOA25_Ch}]{
		\includegraphics[width=0.45\textwidth]{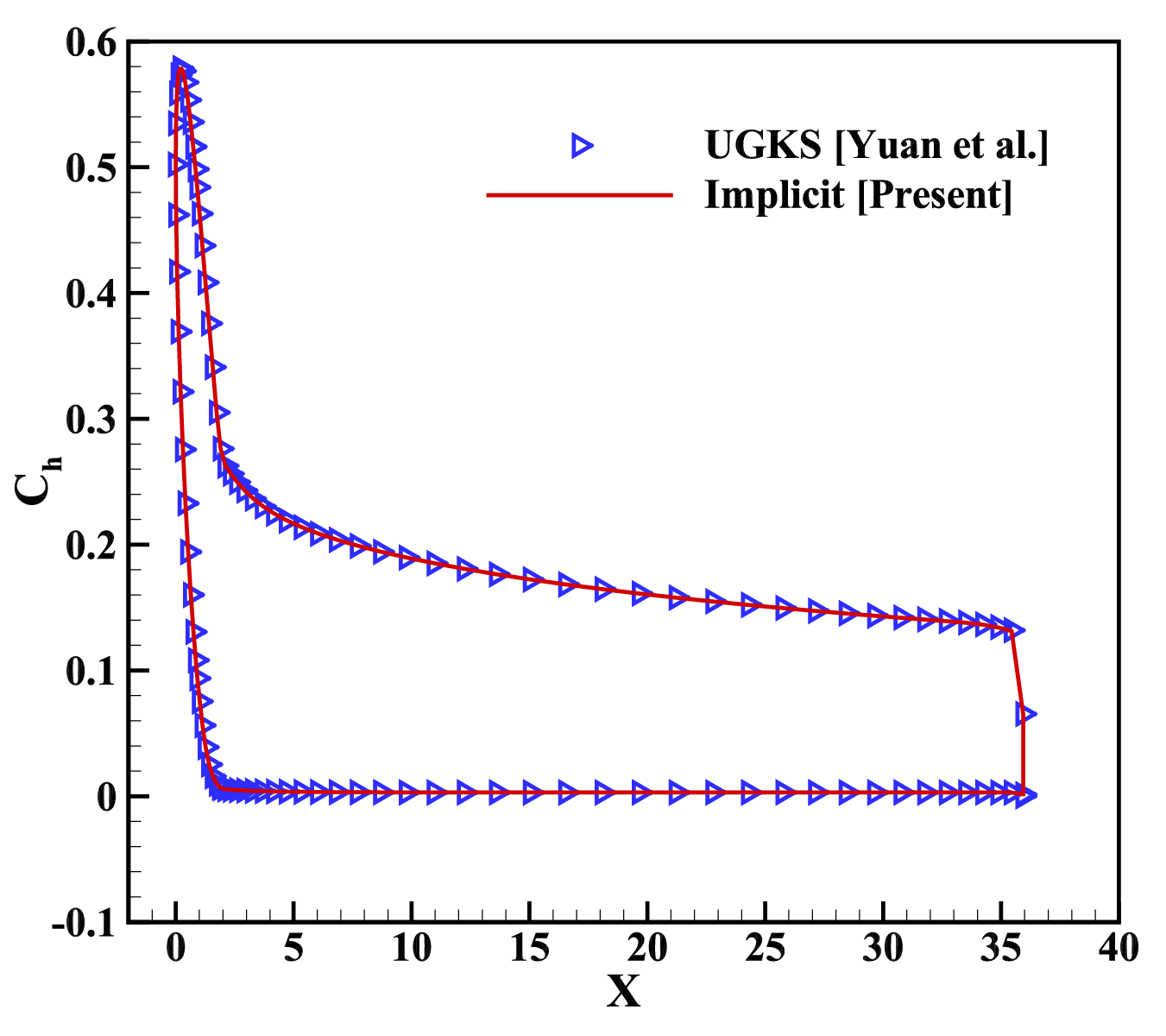}
	}
	\caption{Comparison of UGKS and the present (a) pressure coefficient and (b) heat transfer coefficient for the hypersonic rarefied flow over a blunted-cone at AOA = 25$^ \circ$.}
	\label{Fig:Cone_Ma10.15_Line_AOA25_CpCh}
\end{figure}

\begin{figure}[!htp]
	\centering
	\subfigure[\label{Fig:Cone_Ma10.15_Mac_Ma_AOA20}]{
		\includegraphics[width=0.45\textwidth]{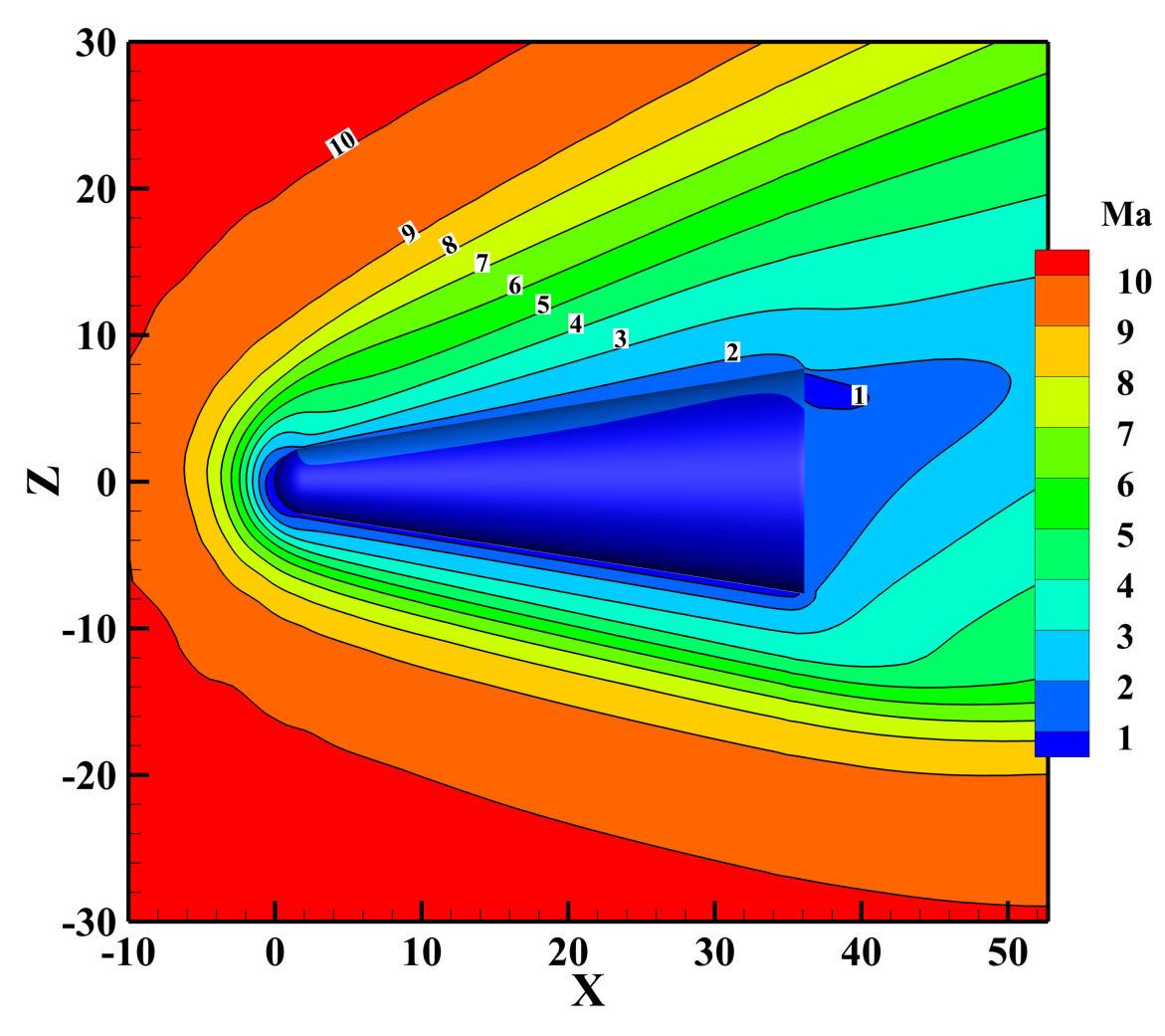}
	}
	\subfigure[\label{Fig:Cone_Ma10.15_Mac_T_AOA20}]{
		\includegraphics[width=0.45\textwidth]{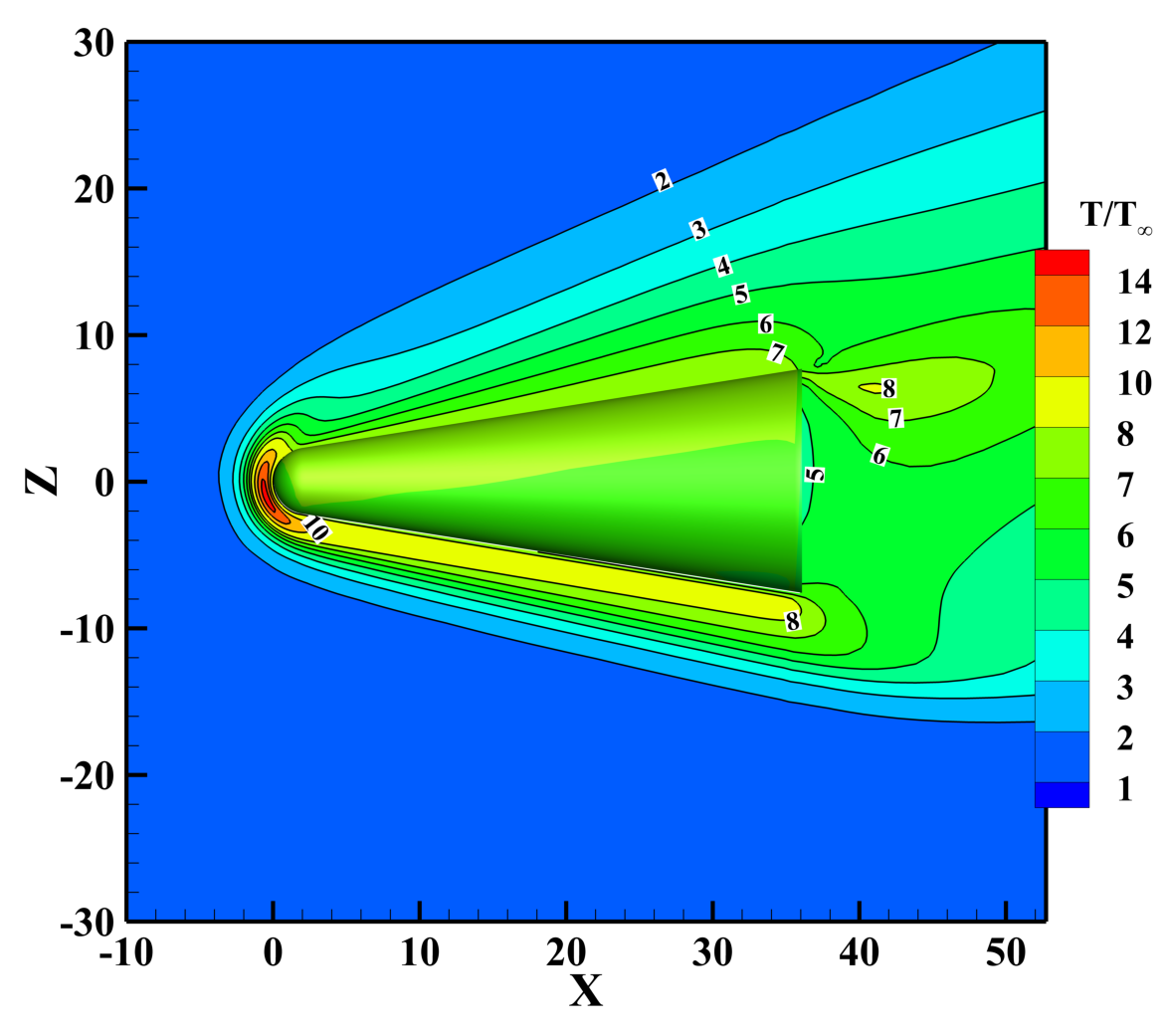}
	}
	\subfigure[\label{Fig:Cone_Ma10.15_Mac_Ttr_AOA20}]{
		\includegraphics[width=0.45\textwidth]{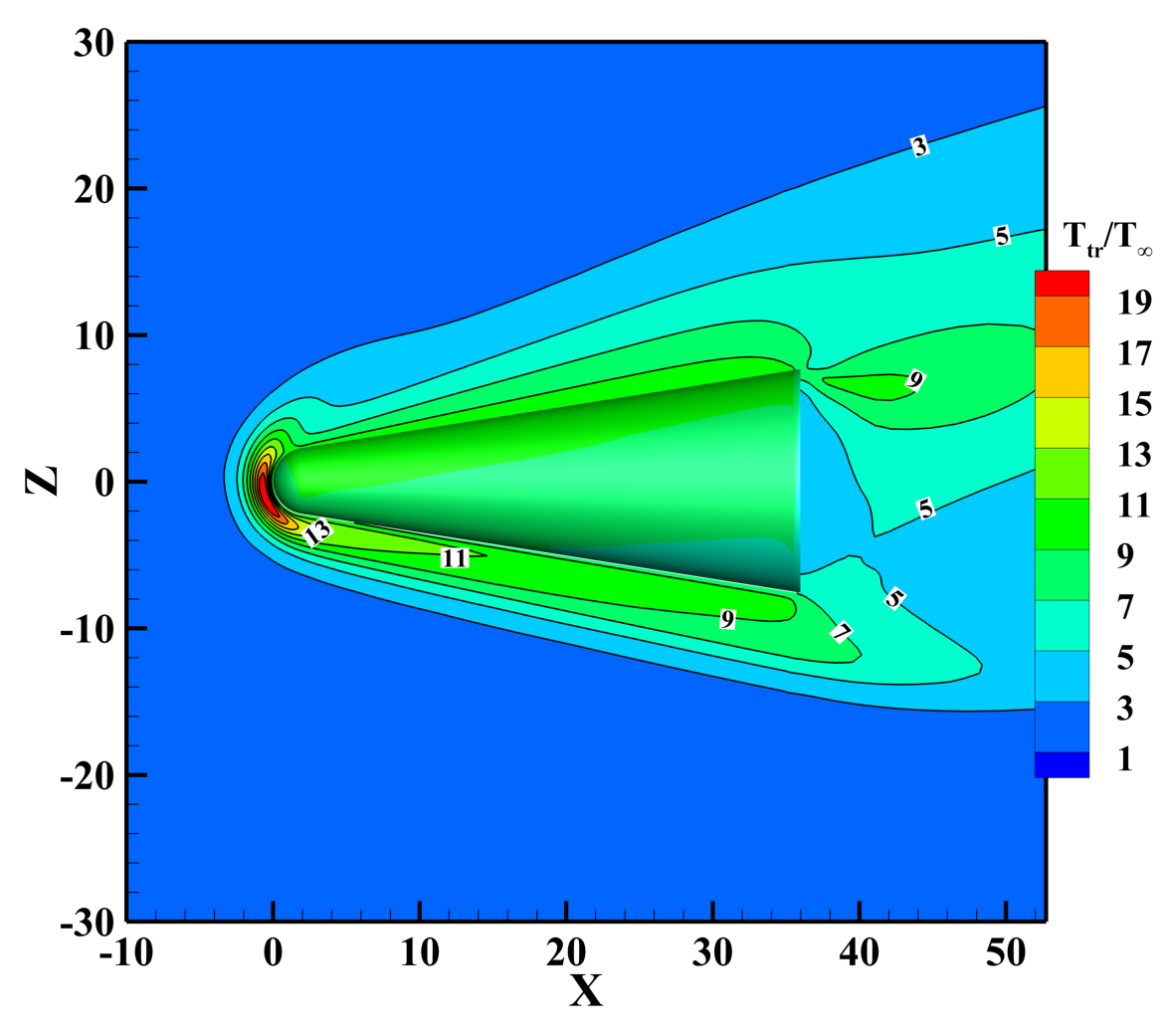}
	}
	\subfigure[\label{Fig:Cone_Ma10.15_Mac_Trot_AOA20}]{
		\includegraphics[width=0.45\textwidth]{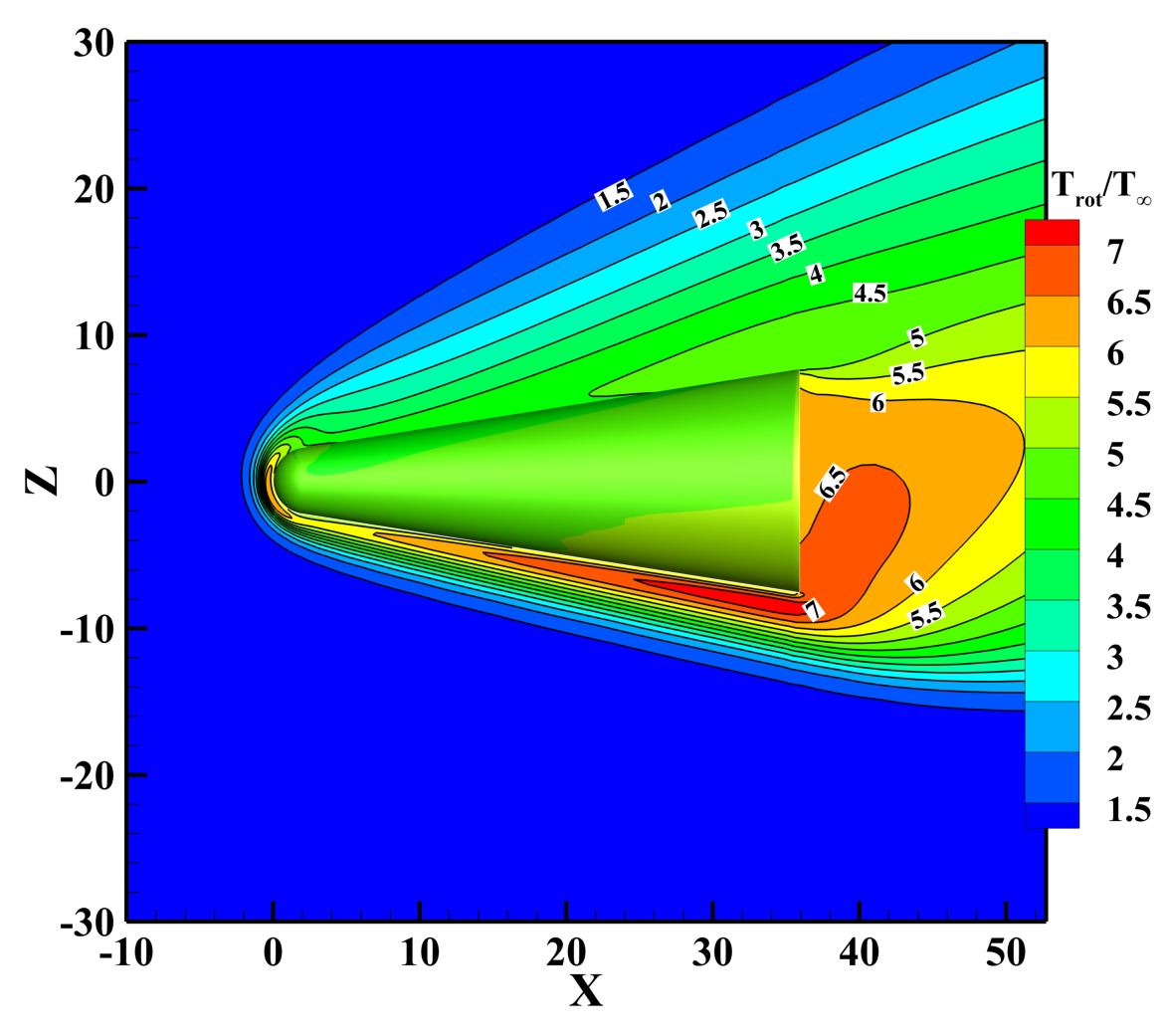}
	}
	\caption{The contour charts of the hypersonic rarefied flow over a blunted-cone at AOA = 20$^ \circ$. (a) Mach number, (b) temperature, (c) translational temperature and (d) rotational temperature.}
	\label{Fig:Cone_Ma10.15_MacField}
\end{figure}

The wall pressure and heat transfer coefficients at AOA $=$ 0, 10, 20 and 25 degrees are shown in Figs.~\ref{Fig:Cone_Ma10.15_Line_AOA0_CpCh}$\sim$\ref{Fig:Cone_Ma10.15_Line_AOA25_CpCh}, respectively.
Both the pressure and heat transfer coefficients are consistent with those of UGKS~\citep{liu_unified_2014}, because they both employ the similar concept of direct modeling to construct multi-scale numerical flux.
Fig.~\ref{Fig:Cone_Ma10.15_MacField} illustrates the contours of Ma number and temperatures of blunted-cone at AOA $= 20^ \circ$.
The translational temperature reaches the maximum value near the stagnation point while the maximum rotational temperature comes later along the upwind side around the blunted-cone tail.
This demonstrates the energy transfer process from translational degrees of freedom to rotational degrees of freedom along the flow direction on the upwind side of the blunted-cone.

\subsection{Apollo 6 command module}\label{sec:caseD}
\begin{figure}[!htp]
	\centering
	\subfigure[\label{Fig:Apollo_Mesh}]{
		\includegraphics[width=0.45\textwidth]{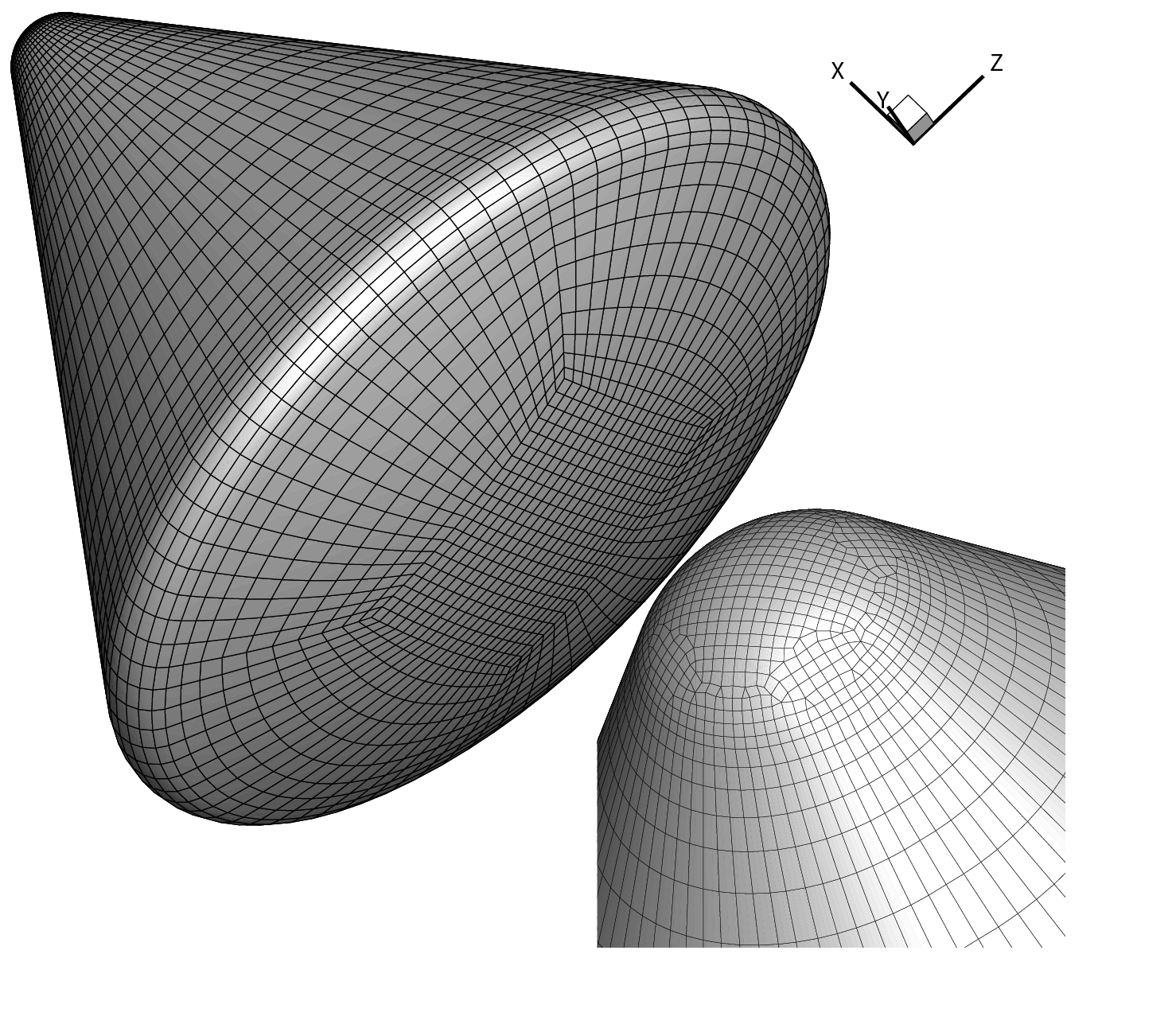}
	}
	\subfigure[\label{Fig:Apollo_Micmesh}]{
		\includegraphics[width=0.45\textwidth]{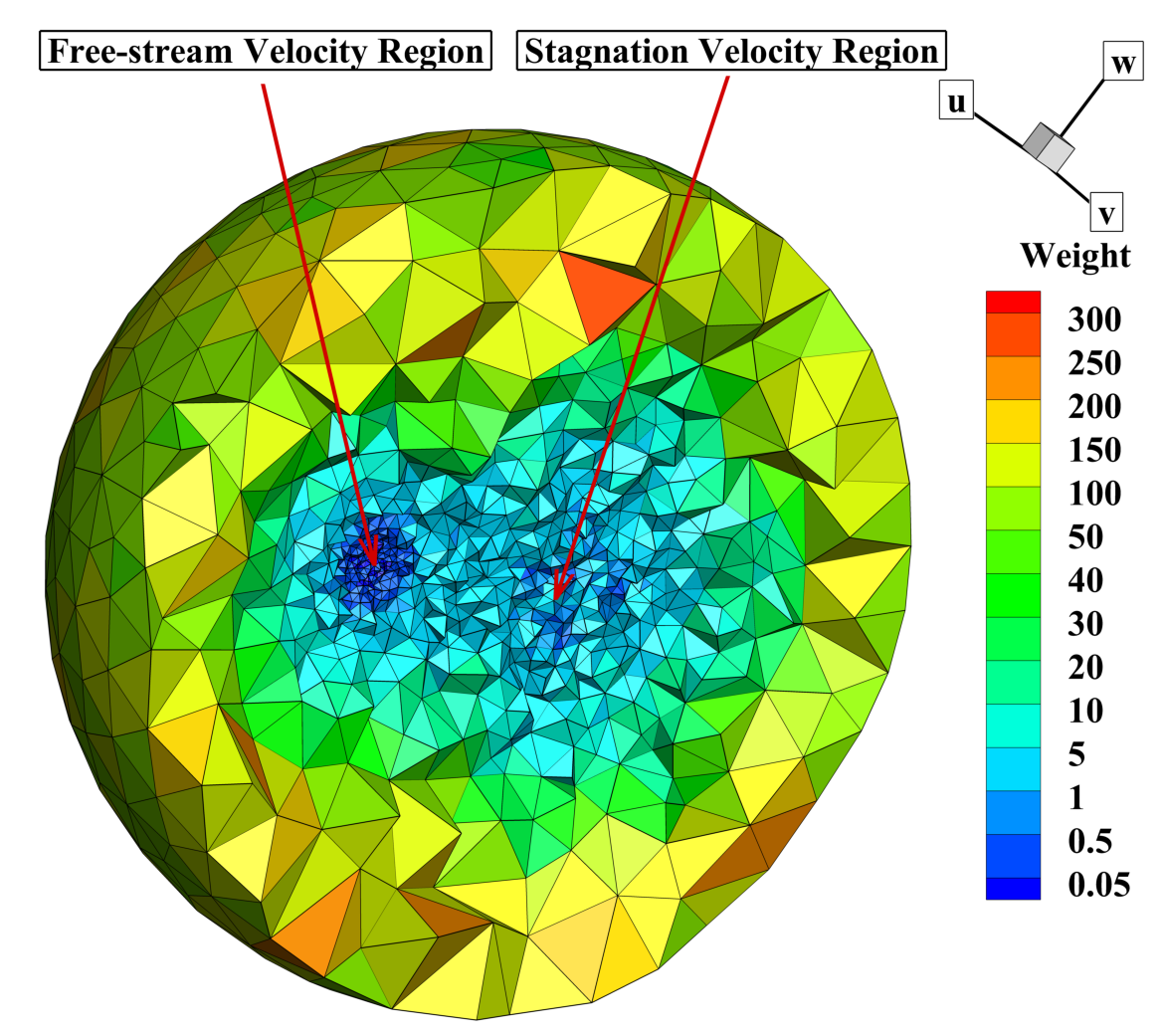}
	}
	\caption{(a) The body mesh for the Apollo 6 command module (4420 cells for surface mesh and 154700 cells for volume mesh). (b) The section view of the unstructured discrete velocity space mesh (H = 85km, total 32510 cells) for the Apollo 6 command module.}
	\label{Fig:Apollo_MeshMicmesh} 
\end{figure}

The objective of this subsection focuses on the reentry trajectory of Apollo 6 command module between 200 and 85 km by maintaining a constant velocity ${U_\infty}=9.6\ \rm{km/s}$ and AOA $=-25^\circ$.
In the work of Moss et. al.~\citep{moss_dsmc_2006}, the rarefied portion of the trajectory was simulated with the DS3V code~\citep{bird_molecular_1994} by using a five-species reacting air gas model.
For the continuum portion of the trajectory, CFD simulations were made by using LAURA (Langley Aerothermodynamic Upwind Relaxation Algorithm) Navier-Stokes code~\citep{gnoffo_conservation_1989}.
The atmospheric conditions are given in Table~\ref{Table:Apollo_Parameter}.
The single-specie nitrogen gas and VHS molecular model is used.
As shown in Fig.~\ref{Fig:Apollo_Mesh}, an unstructured body mesh consisting of 4420 grid cells is used to define the wall surface, and the computational domain has total 154700 cells.
The unstructured DVS mesh with 34560, 32110, 32900, 30220 and 32510 cells are used for altitude of 200, 150, 120, 100 and 85 km, respectively.
Fig.~\ref{Fig:Apollo_Micmesh} illustrates the section view of the unstructured DVS mesh for altitude of 85km.
All simulations take a rotational collision number of 5 and use the diffuse reflection boundary condition with full thermodynamic accommodation at the wall surface.

\begin{figure}[!htp]
	\centering
	\subfigure[\label{Fig:Apollo_Cd}]{
		\includegraphics[width=0.45\textwidth]{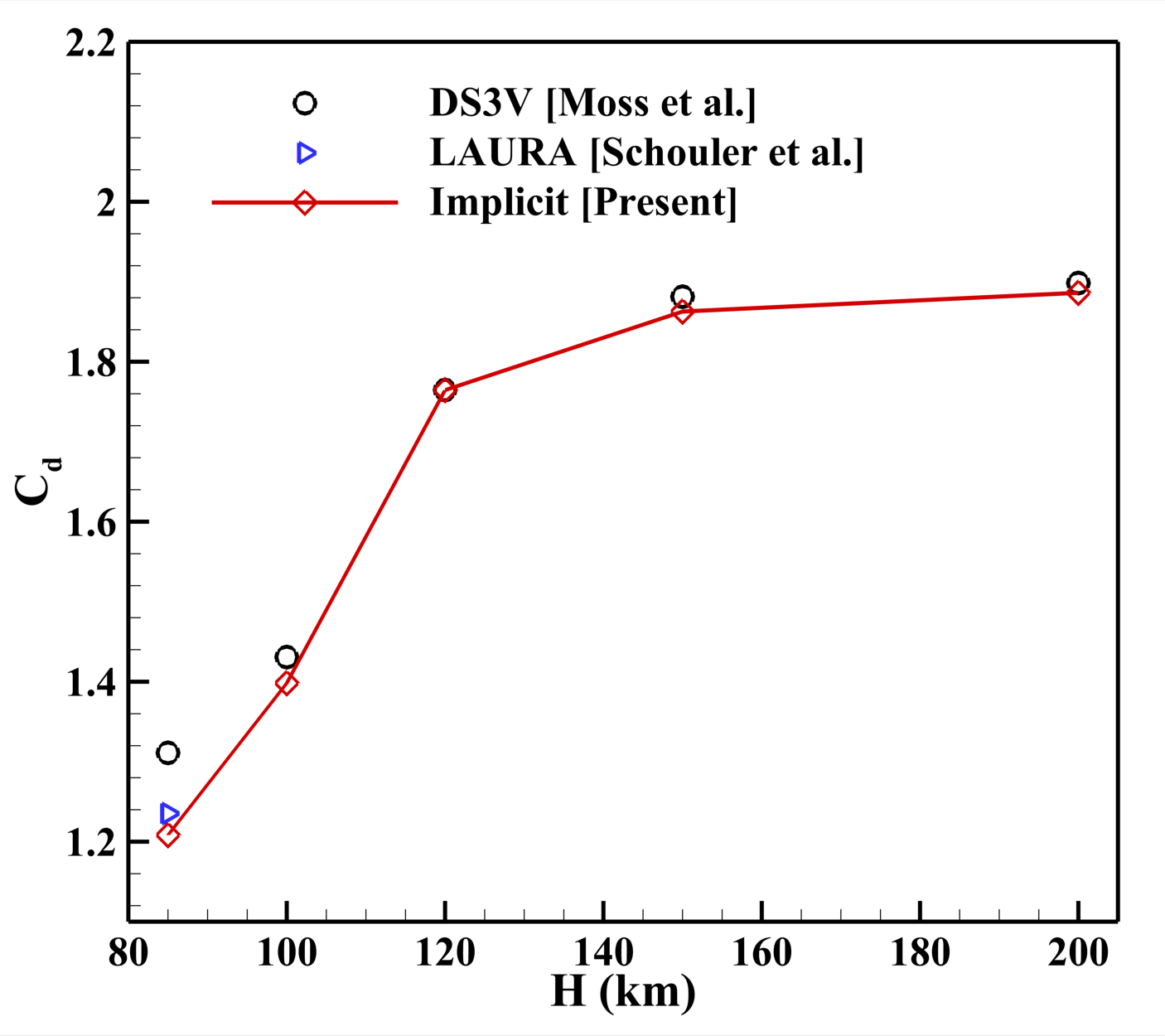}
	}
	\subfigure[\label{Fig:Apollo_Cl}]{
		\includegraphics[width=0.45\textwidth]{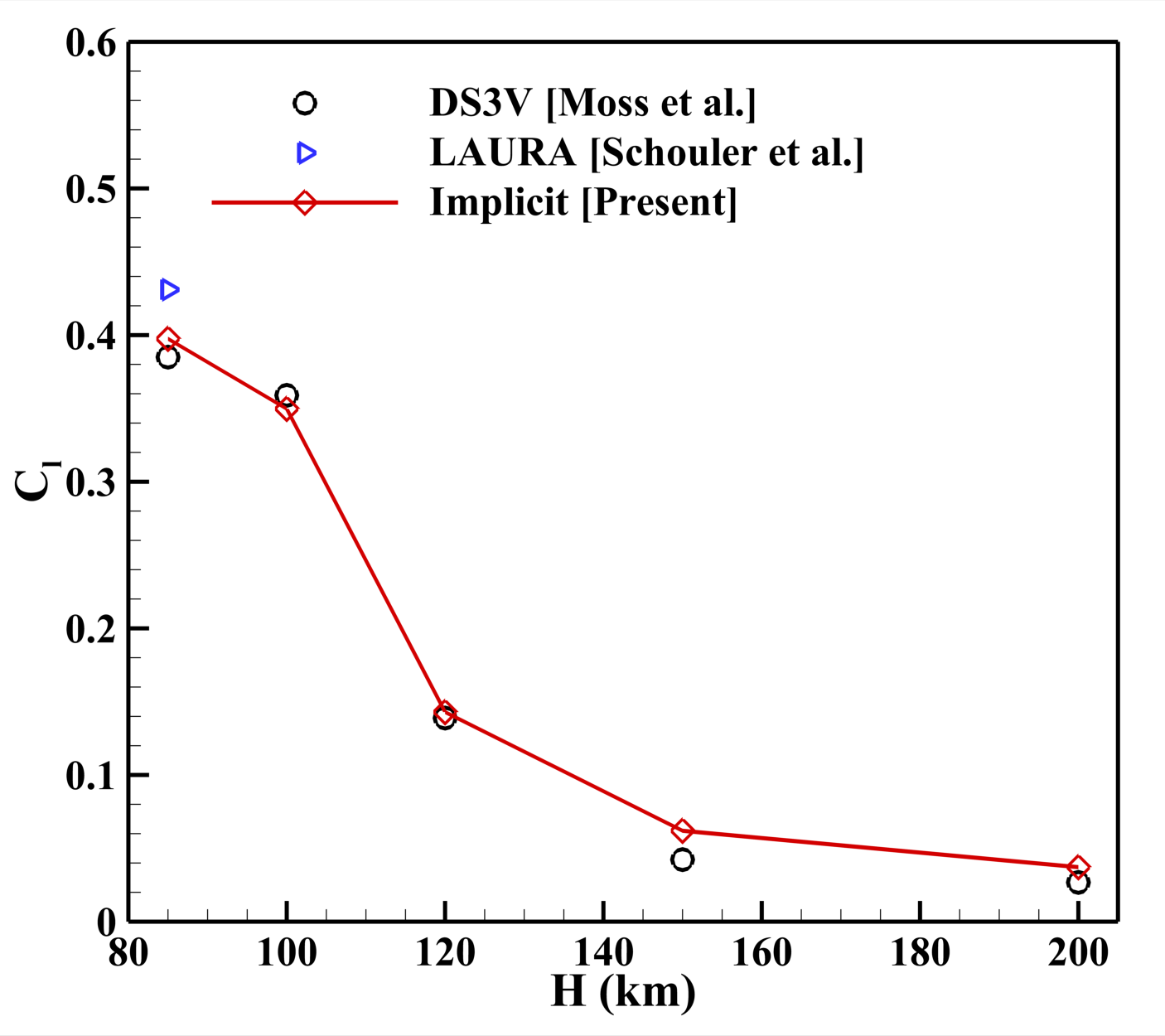}
	}
	\caption{Evolution of (a) the drag coefficient and (b) the lift coefficient of the Apollo 6 command module as a function of the altitude between 85 and 200 km.}
	\label{Fig:Apollo_ClCd} 
\end{figure}

\begin{figure}[!htp]
	\centering
	\subfigure[\label{Fig:Apollo_LD}]{
		\includegraphics[width=0.45\textwidth]{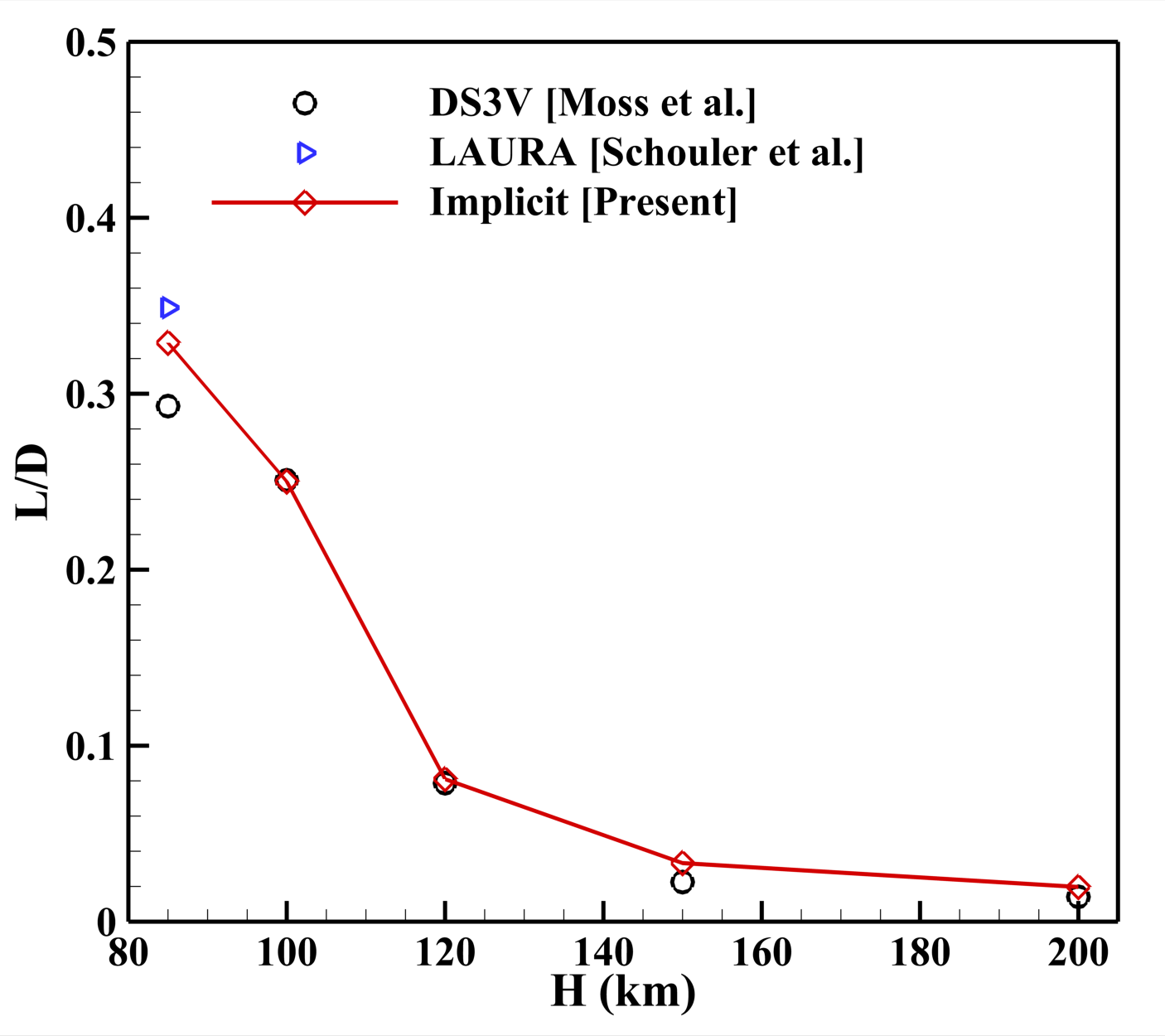}
	}
	\subfigure[\label{Fig:Apollo_Cm}]{
		\includegraphics[width=0.45\textwidth]{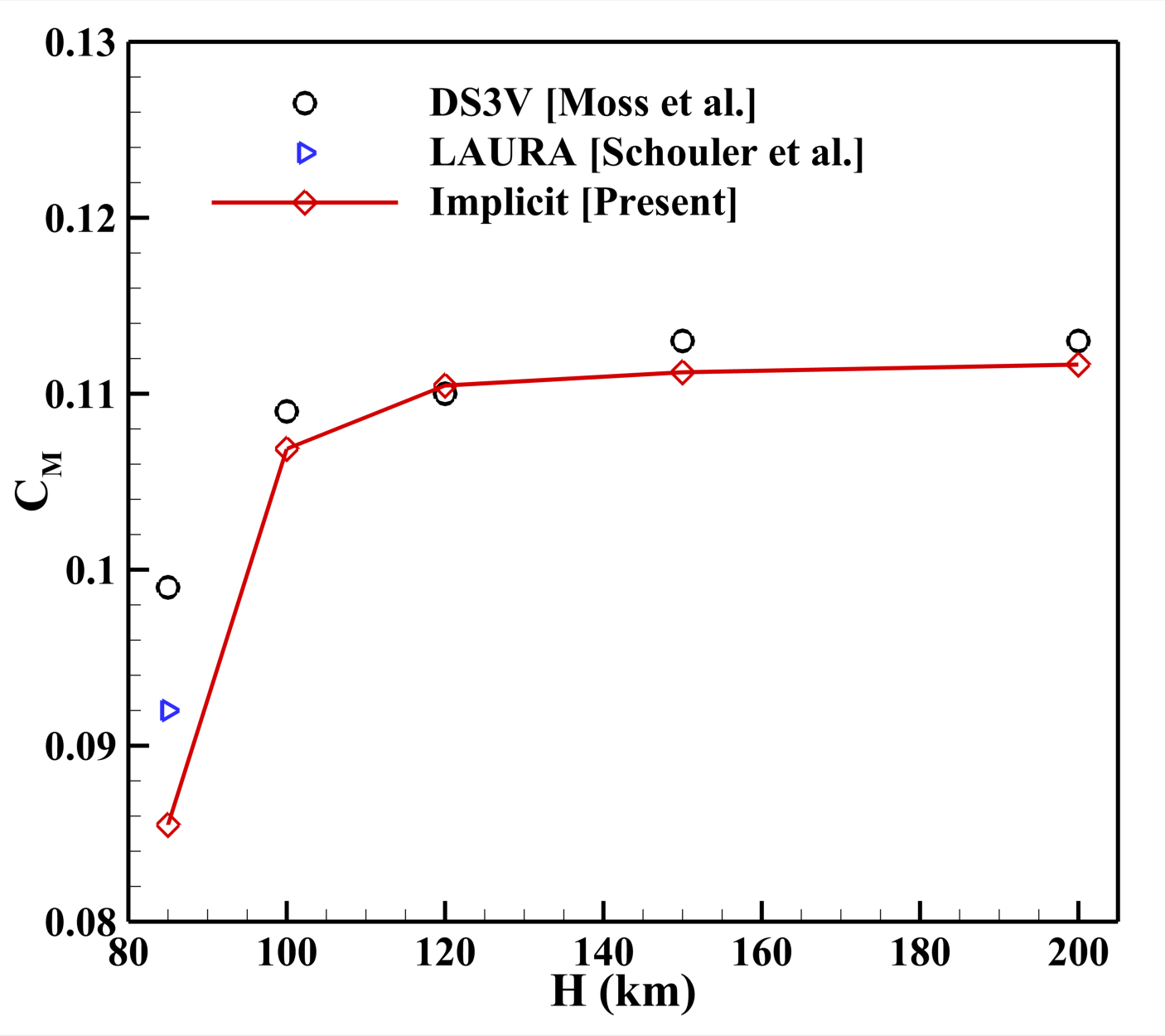}
	}
	\caption{Evolution of (a) the lift-to-drag ratio and (b) the pitching moment coefficient of Apollo 6 command module as a function of the altitude between 85 and 200 km.}
	\label{Fig:Apollo_LDCm}
\end{figure}

\begin{figure}[!htp]
	\centering
	\subfigure[\label{Fig:Apollo_VelSlip}]{
		\includegraphics[width=0.45\textwidth]{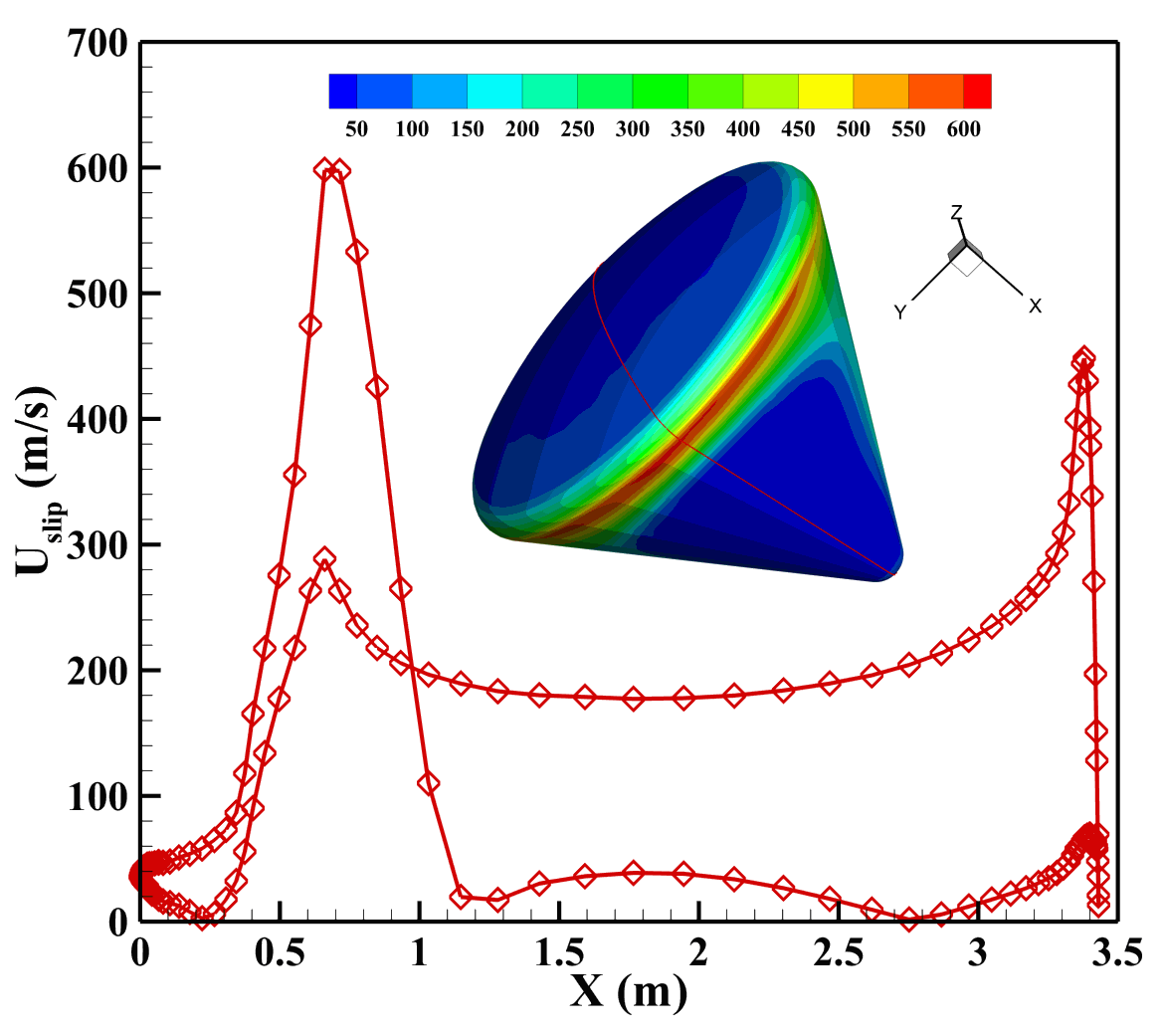}
	}
	\subfigure[\label{Fig:Apollo_TemJump}]{
		\includegraphics[width=0.45\textwidth]{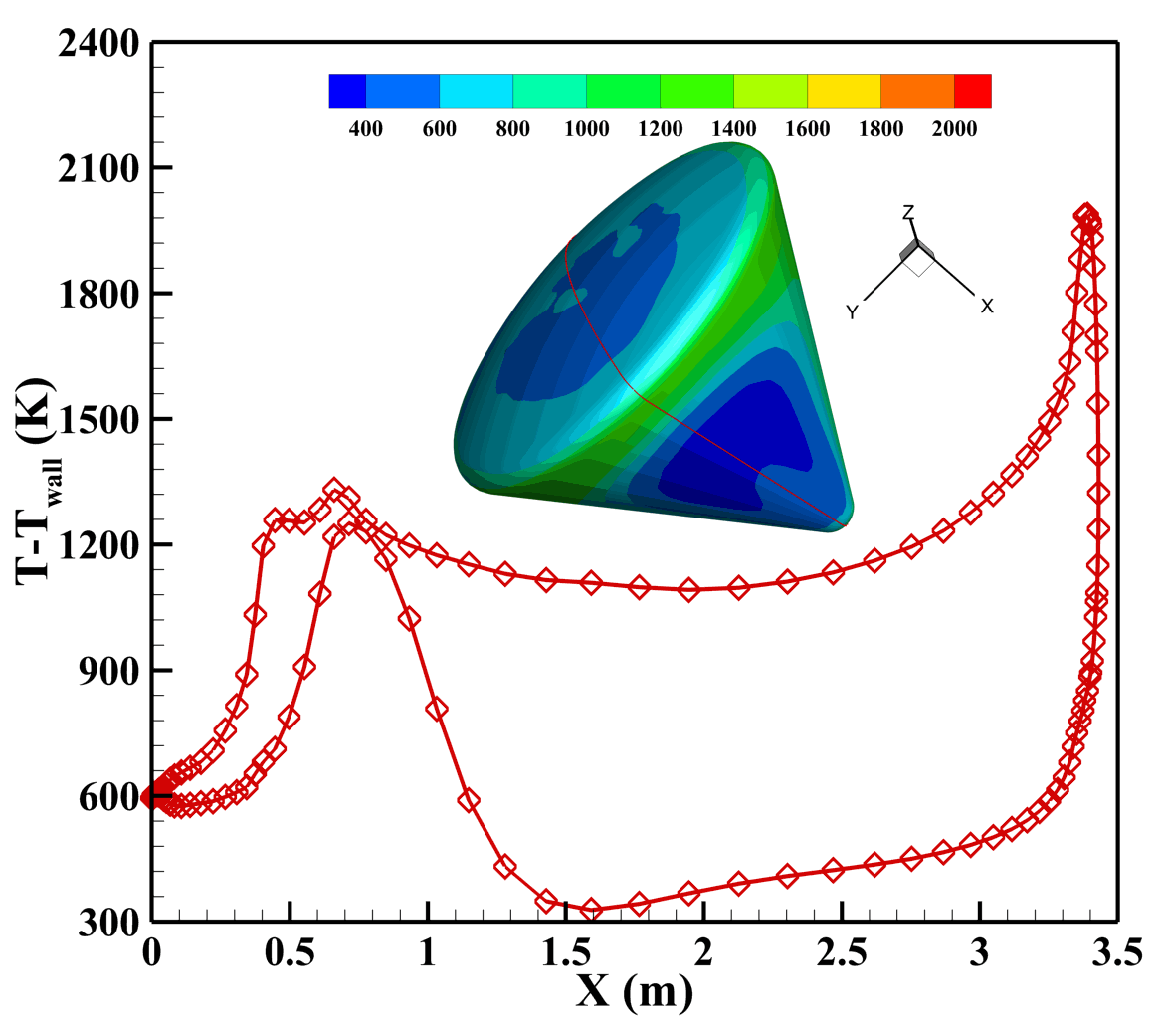}
	}
	\caption{(a) Velocity slip and (b) temperature jump on the wall surface of the Apollo 6 command module at altitude of 85 km.}
	\label{Fig:Apollo_Vel_Tem}
\end{figure}

\begin{figure}[!htp]
	\centering
	\subfigure[\label{Fig:Apollo_SurZ0_H100}]{
		\includegraphics[width=0.45\textwidth]{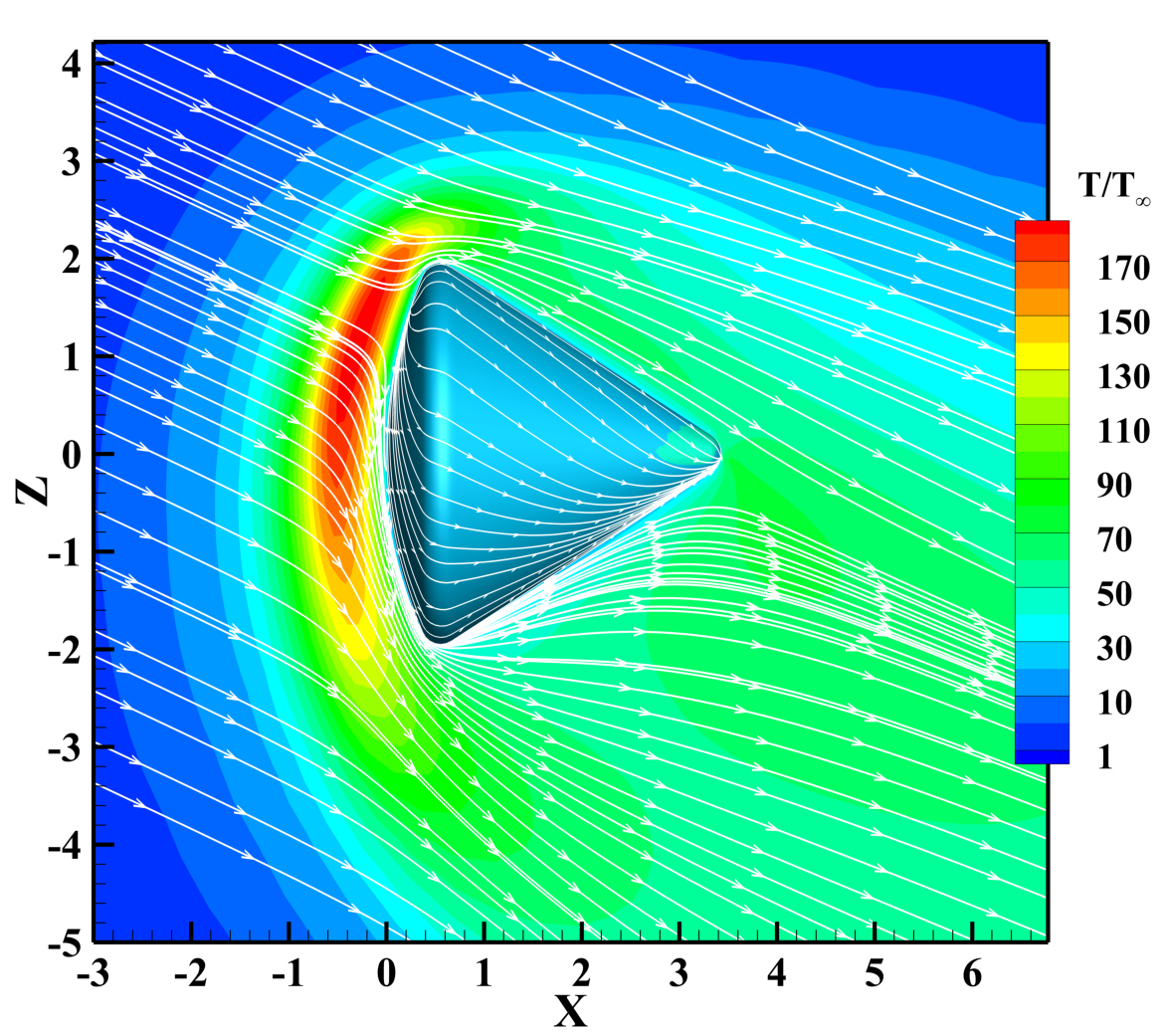}
	}
	\subfigure[\label{Fig:Apollo_SurZ0_H85}]{
		\includegraphics[width=0.45\textwidth]{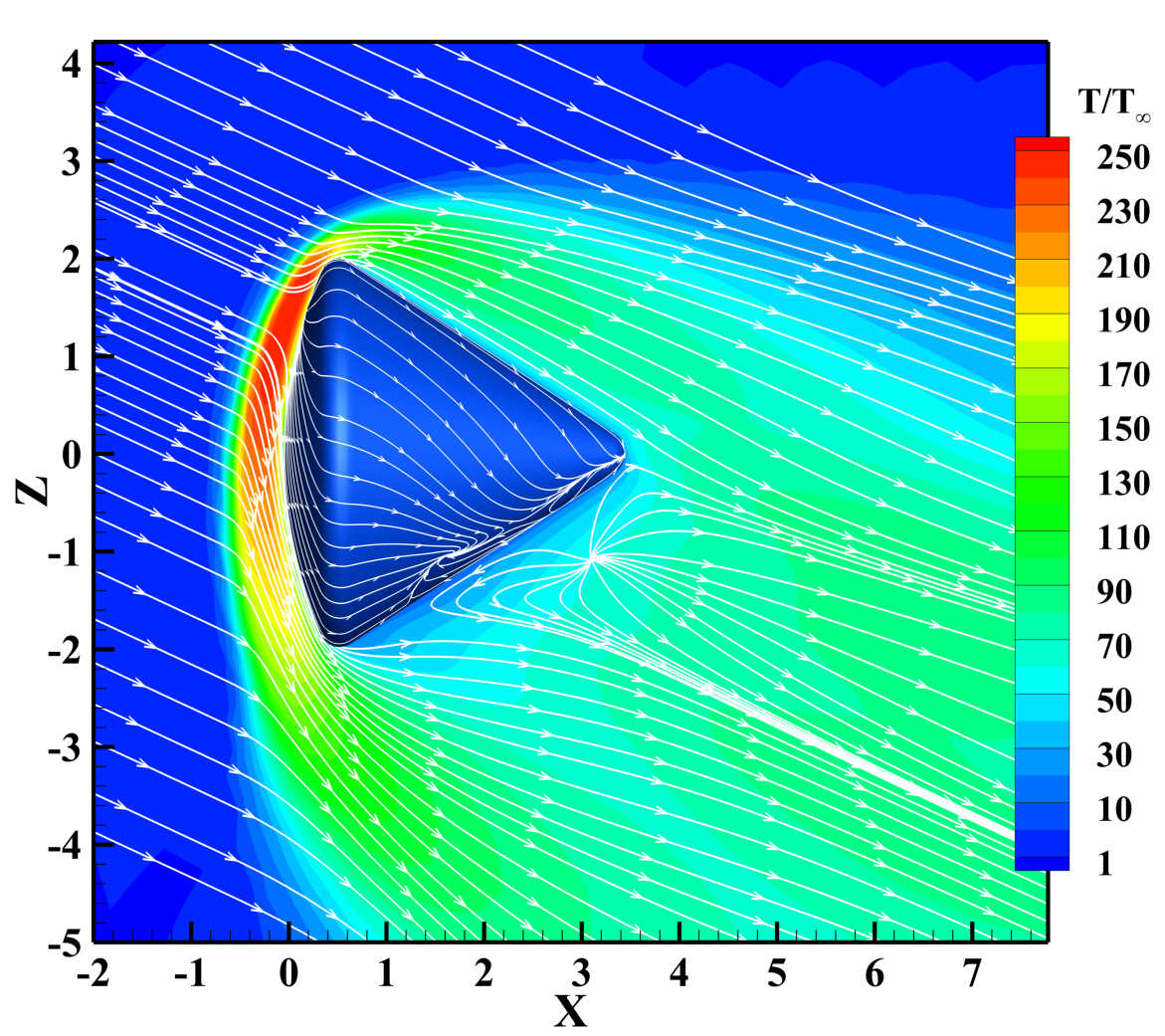}
	}
	\caption{The temperature contour charts and streamlines at altitude of 100 and 85 km for the Apollo 6 command module. (a) Altitude of 100km, Ma = 33.96, Kn = 0.0338. (b) Altitude of 85km, Ma = 35.59, Kn = 0.0024.}
	\label{Fig:Apollo_SurZ0}
\end{figure}

Aerodynamic results of the numerical simulations are presented in Figs.~\ref{Fig:Apollo_ClCd} and~\ref{Fig:Apollo_LDCm}.
The aerodynamic coefficients are shown to be very sensitive to rarefied gas effect.
The lift coefficient and lift-to-drag ratio increase with a decrease in altitude, while the drag coefficient and pitching moment coefficient decrease with decreasing altitude.
The aerodynamic coefficients predicted by the present method excellently agree well with those of DSMC at altitude between 200 and 100 km.
At altitude of 100 and 120 km, the differences of the aerodynamic coefficients between the present results and those from DS3V are less than 3$\%$.
At altitude of 150 and 200 km, the differences of the drag coefficient and pitching moment coefficient between the present and DS3V results are less than 2$\%$.
Meanwhile, we observed that the present results, at an altitude of 85 km, clearly diverge from those of the DS3V code and LAURA N-S code.
In fact, we consider the present results are more reliable without considering the influence of species.
In the work of Moss et. al.~\citep{moss_dsmc_2006}, DS3V simulations were made with a global mean mcs/mfp of 3.980 and 2.450, but a grid resolved DS3V simulation requires a mcs/mfp value that is of order 0.1.
We noted that the results of DS3V gradually approach our results as the mean mcs/mfp decreases.
On the other hand, the velocity slip and temperature jump have been observed as shown in Fig.~\ref{Fig:Apollo_Vel_Tem}, and they have a significant impact on the prediction of aerothermodynamic properties.
However, LAURA simulations conducted by Moss et. al.~\citep{moss_dsmc_2006} did not consider the velocity slip and temperature jump boundary conditions.
Fig.~\ref{Fig:Apollo_SurZ0} illustrate the attached and detached streamlines and temperature contour at altitude of 100 and 85 km.

In order to generate an appropriate unstructured DVS and reduce the number of cells as few as possible, we explore the skills and criteria for generating 3D unstructured DVS based on the Gaussian distribution and numerical experiments of Apollo 6 command module.
The present analysis is based on the dimensionless system, and the relevant dimensionless velocity and temperature at altitude of 150, 120, 100 and 85 km are shown in Table.~\ref{Table:Apollo_DimensionlessParameter}.
In the present study, the velocity space is dispersed in a spherical region according to $3 \sigma$ criterion of the Gaussian distribution.
All test cases are shown in Table.~\ref{Table:Apollo_DVS}.
The spherical center coordinate is set as $0.4({U_{\infty}},{V_{\infty}},{W_{\infty}})$ for all unstructured DVS, and the radius is $5\sqrt {R{T_0}}$ for cases No. M1 $\sim$ No. M5, $4\sqrt {R{T_0}}$ for case No. M6 and $3\sqrt {R{T_0}}$ for case No. M7.
The total temperature ${T_0}$ is estimated to use formula
\begin{equation}\label{equ:solve_t0}
	{T_0} = {T_\infty }\left( {1 + \frac{{\gamma - 1}}{2}{\rm{Ma}^2}} \right).
\end{equation}
Furthermore, the discrete velocity mesh is refined in the spherical regions of the free-stream velocity point and the zero velocity point according to the free-stream temperature and the wall temperature.
The spherical center coordinates are set as $({U_{\infty}},{V_{\infty}},{W_{\infty}})$ and $(0,0,0)$, corresponding radii are $3\sqrt {R{T_{\infty}}}$ and $3\sqrt {R{T_{w}}}$,  respectively.

\begin{figure}[!htp]
	\centering
	\subfigure[\label{Fig:Apollo_Cd_Micmesh}]{
		\includegraphics[width=0.45\textwidth]{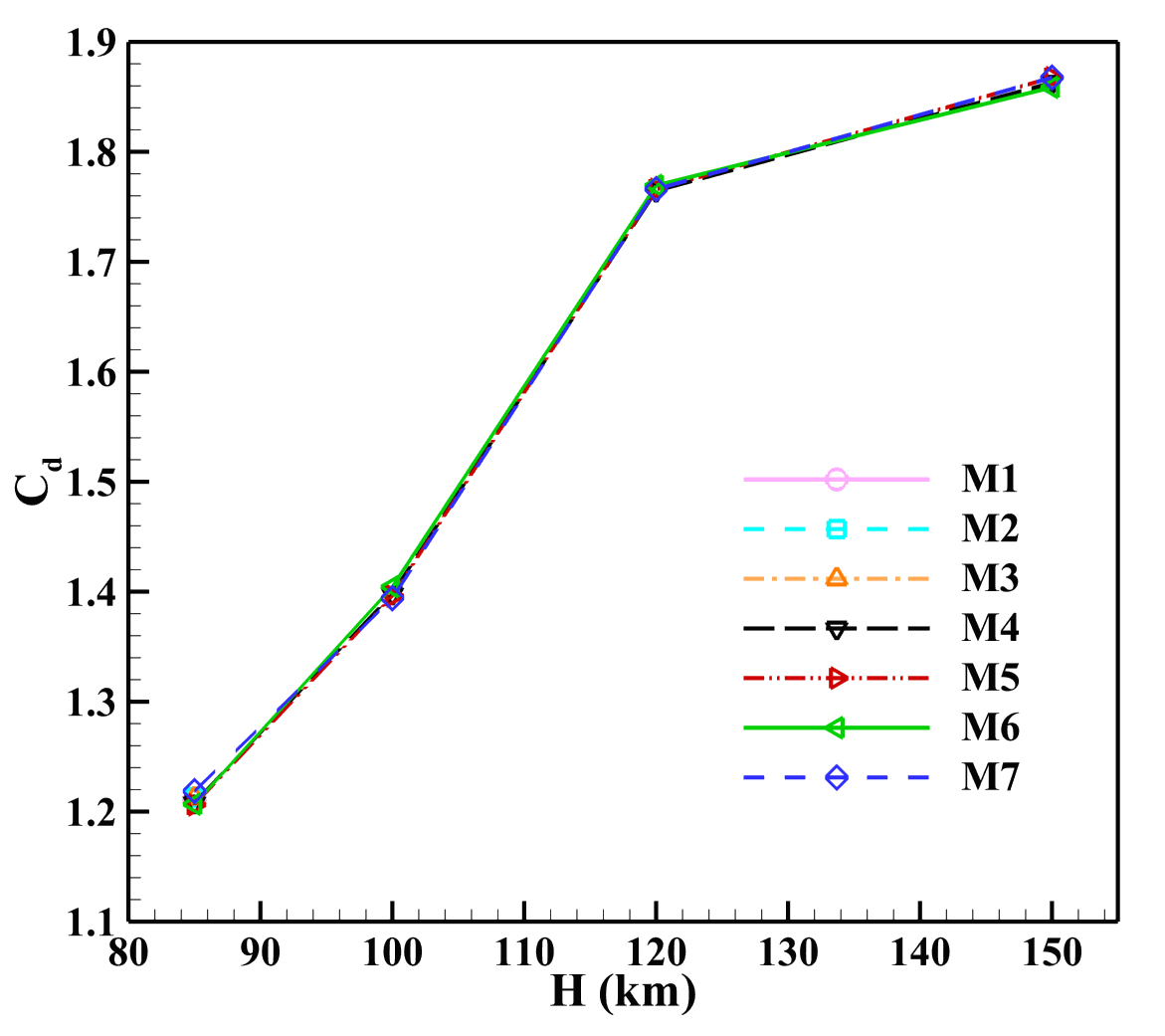}
	}
	\subfigure[\label{Fig:Apollo_Cl_Micmesh}]{
		\includegraphics[width=0.45\textwidth]{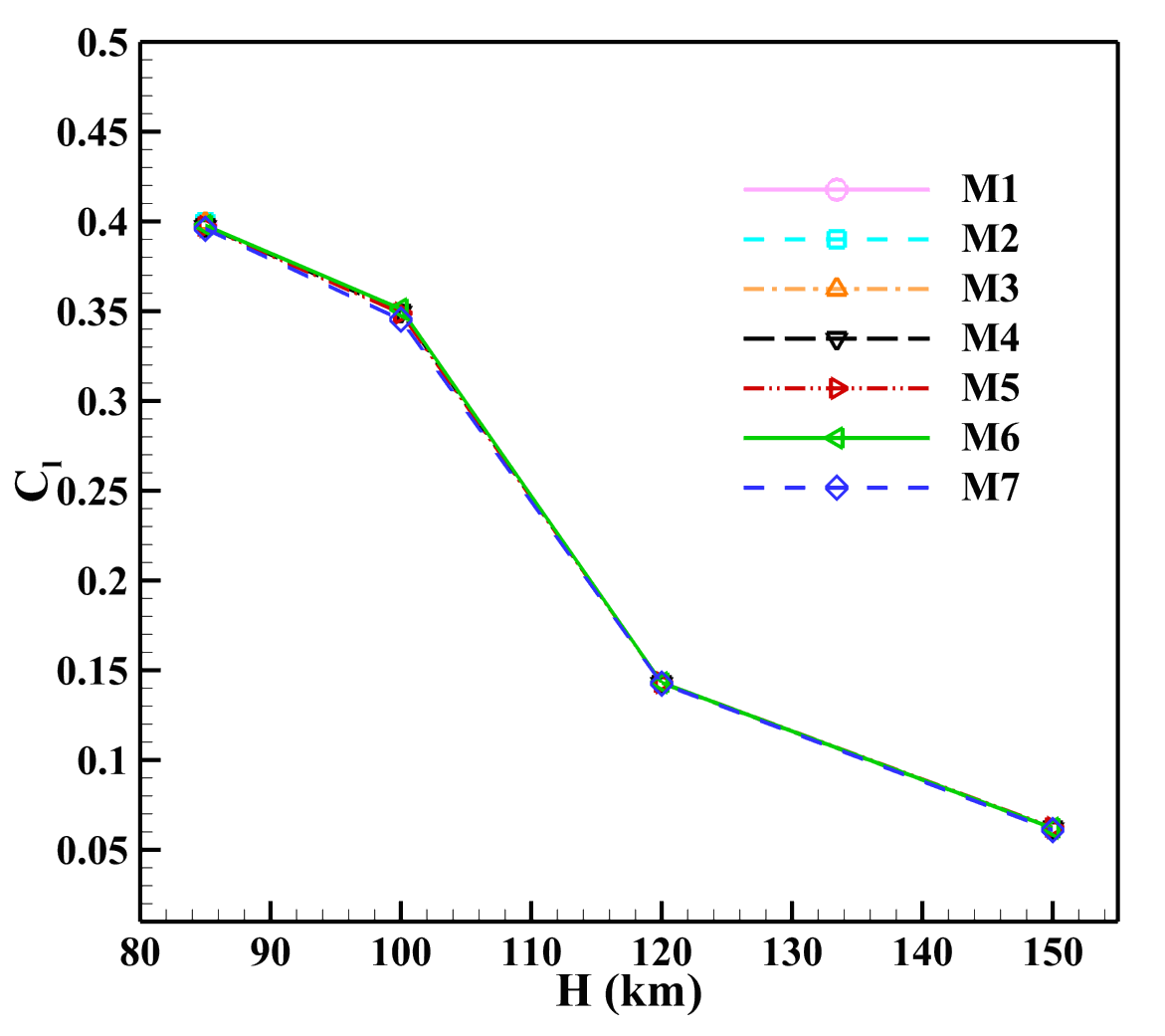}
	}
	\subfigure[\label{Fig:Apollo_LD_Micmesh}]{
		\includegraphics[width=0.45\textwidth]{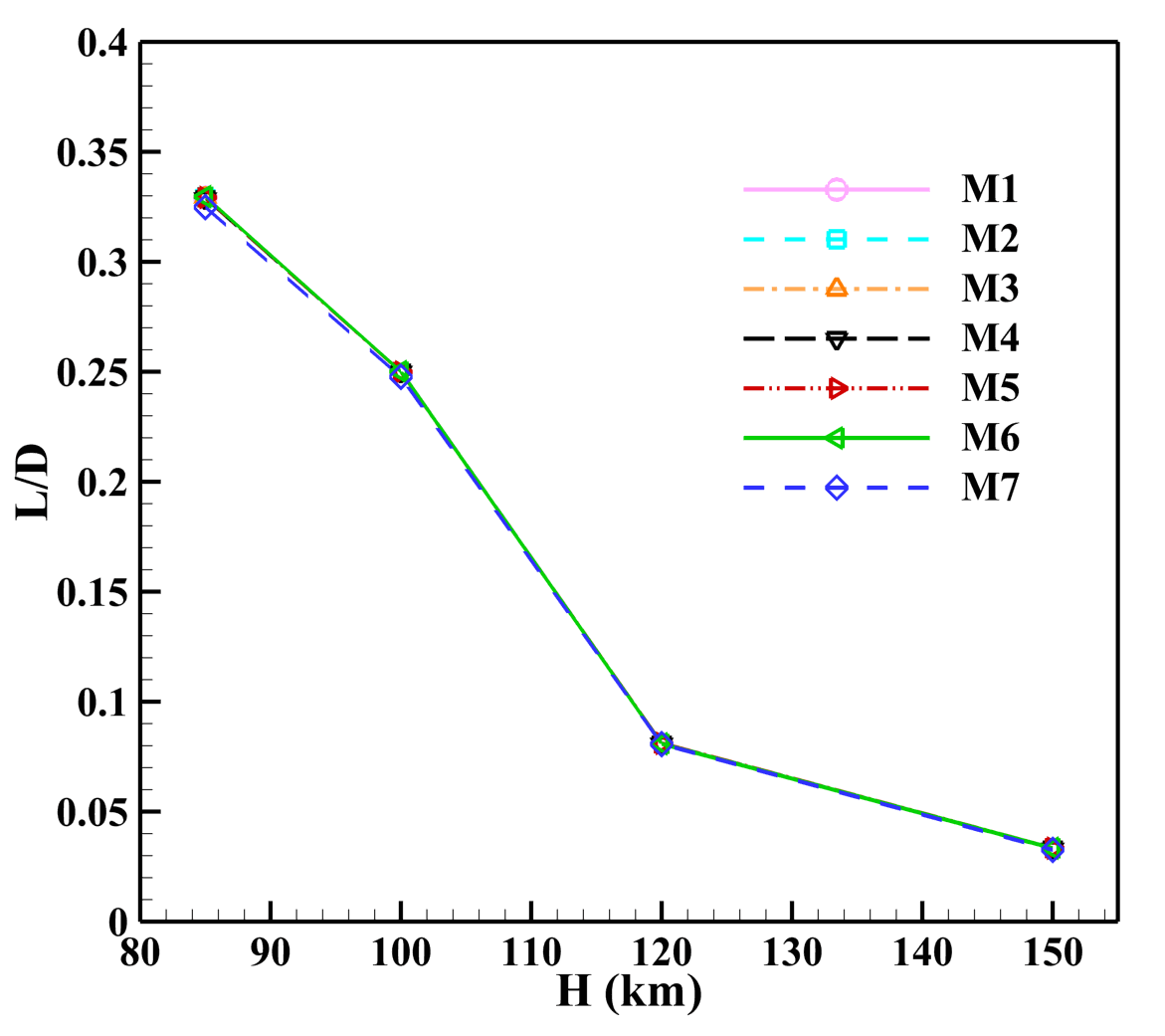}
	}
	\subfigure[\label{Fig:Apollo_Cm_Micmesh}]{
		\includegraphics[width=0.45\textwidth]{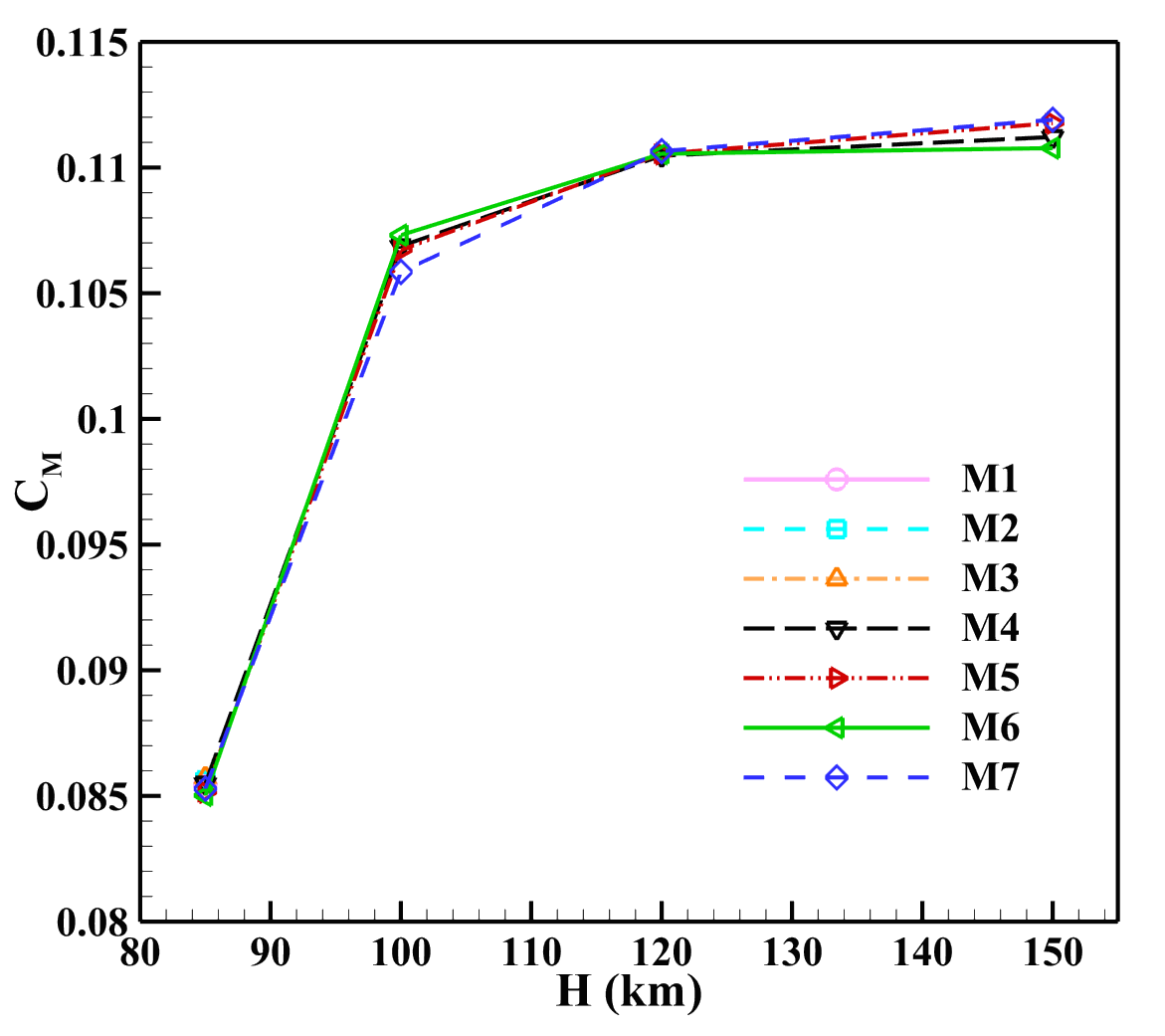}
	}
	\caption{Evolution of (a) the drag coefficient, (b) the lift coefficient, (c) the lift-to-drag ratio and (d) the pitching moment coefficient of the Apollo 6 command module as a function of the altitude between 85 and 150 km using different unstructured DVS.}
	\label{Fig:Apollo_ClCdCmLD_Micmesh} 
\end{figure}

\begin{figure}[!htp]
	\centering
	\subfigure[\label{Fig:Apollo_H85_Cp}]{
		\includegraphics[width=0.3\textwidth]{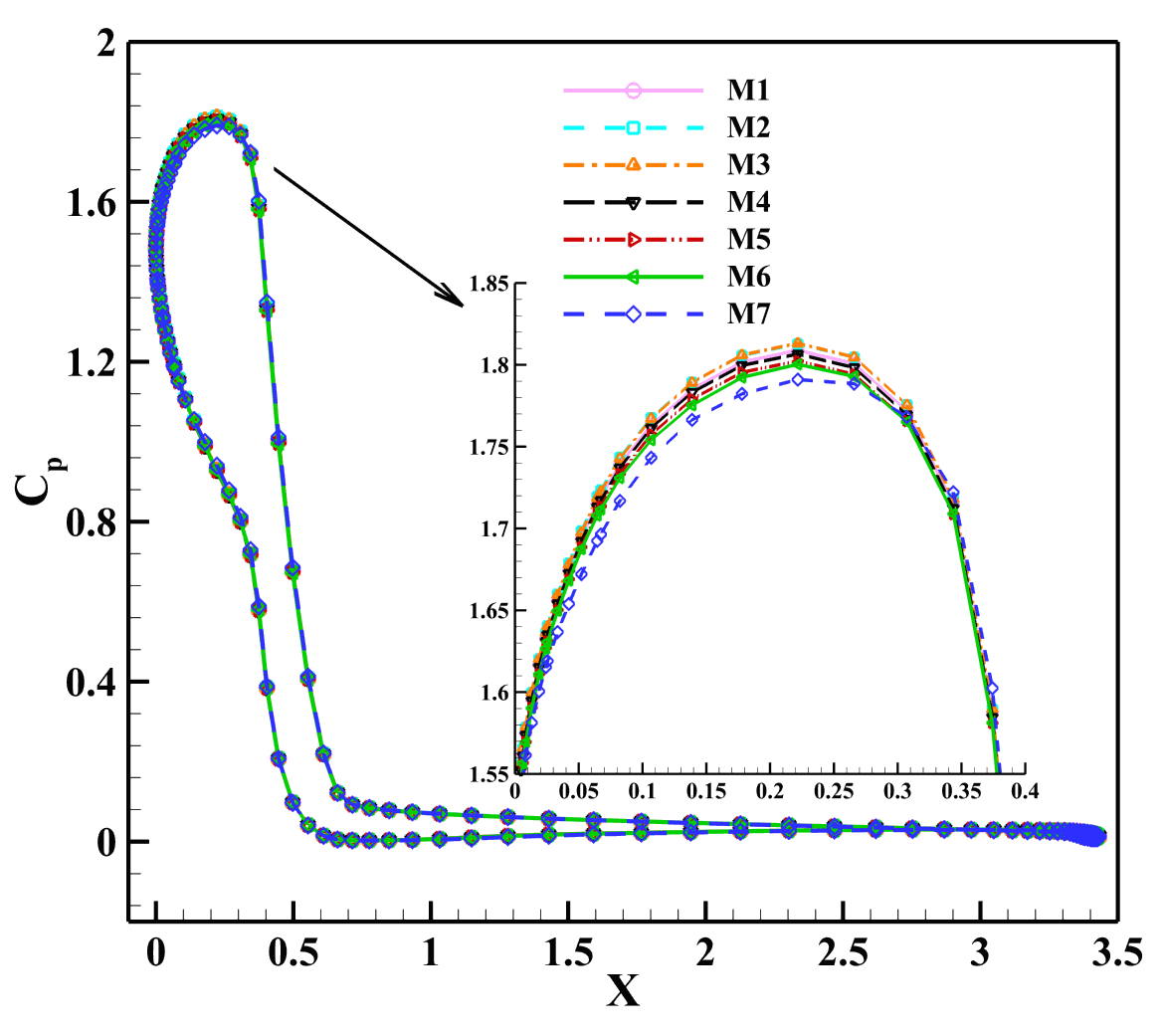}
	}
	\subfigure[\label{Fig:Apollo_H85_Cf}]{
		\includegraphics[width=0.3\textwidth]{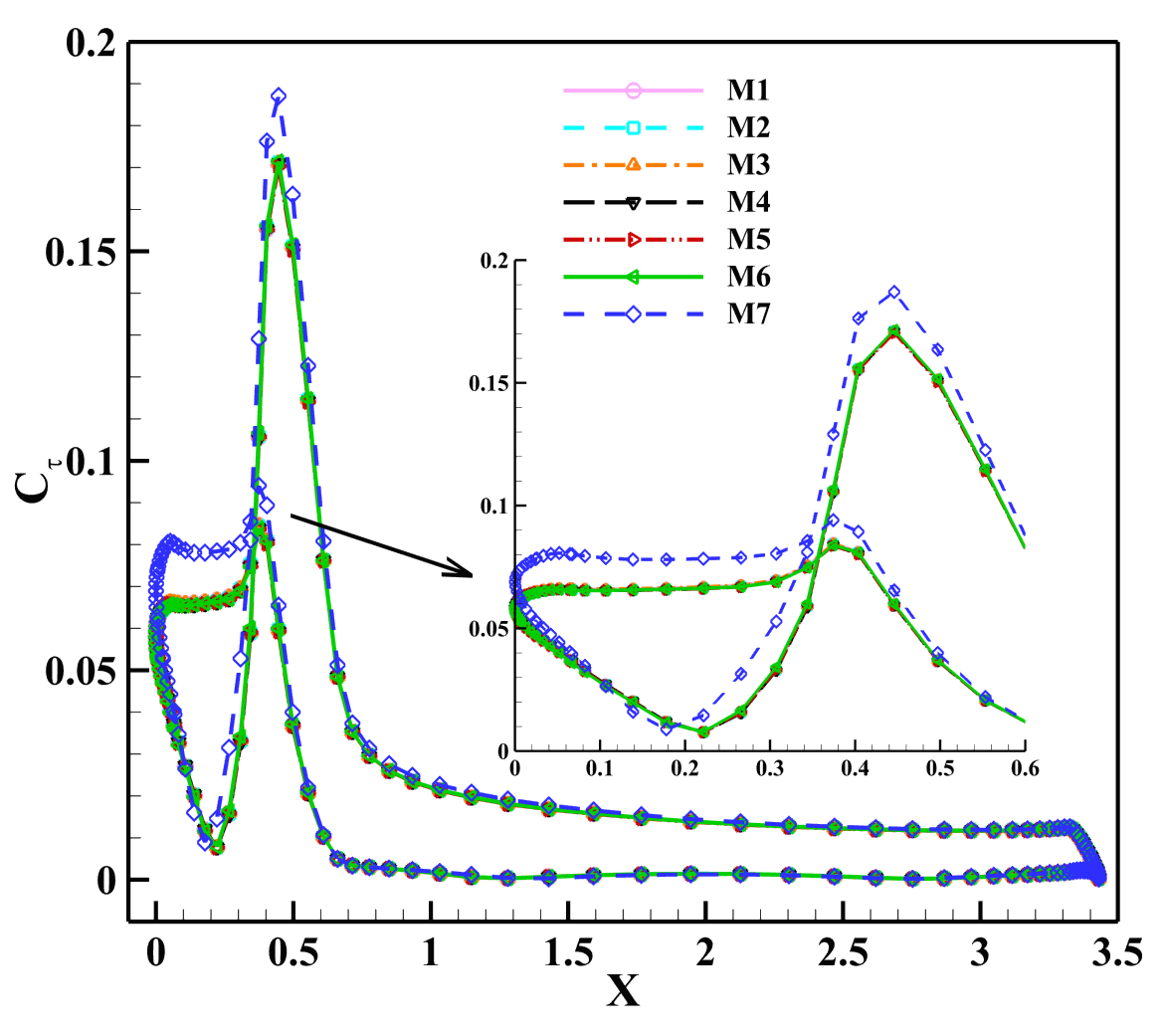}
	}
	\subfigure[\label{Fig:Apollo_H85_Ch}]{
		\includegraphics[width=0.3\textwidth]{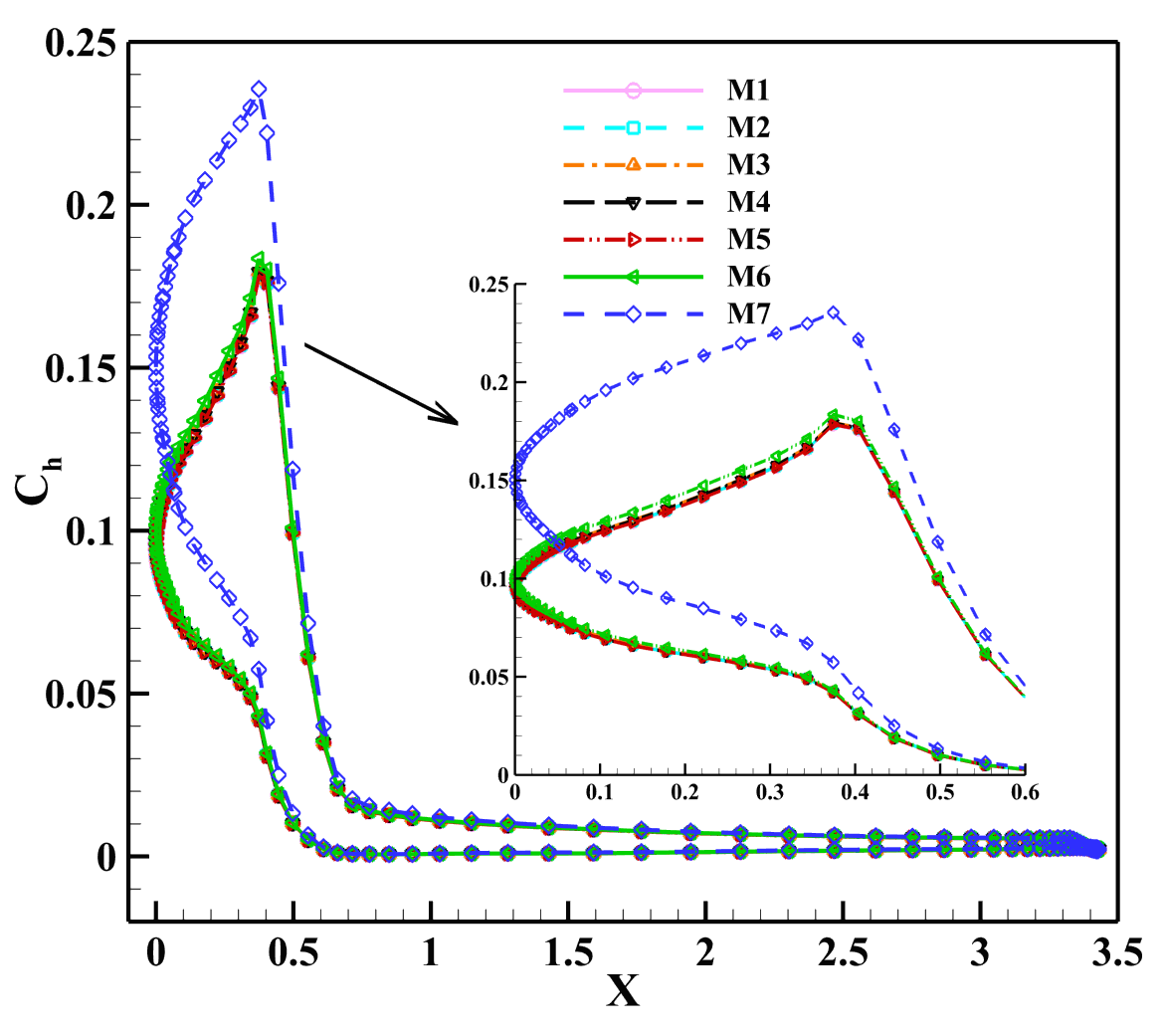}
	}
	\caption{Comparison of the present (a) pressure coefficient, (b) shear stress coefficient and (c) heat transfer coefficient for the Apollo 6 command module using different unstructured DVS (Altitude of 85km).}
	\label{Fig:Apollo_H85_CpCfCh}
\end{figure}

\begin{figure}[!htp]
	\centering
	\subfigure[\label{Fig:Apollo_H150_Cp}]{
		\includegraphics[width=0.3\textwidth]{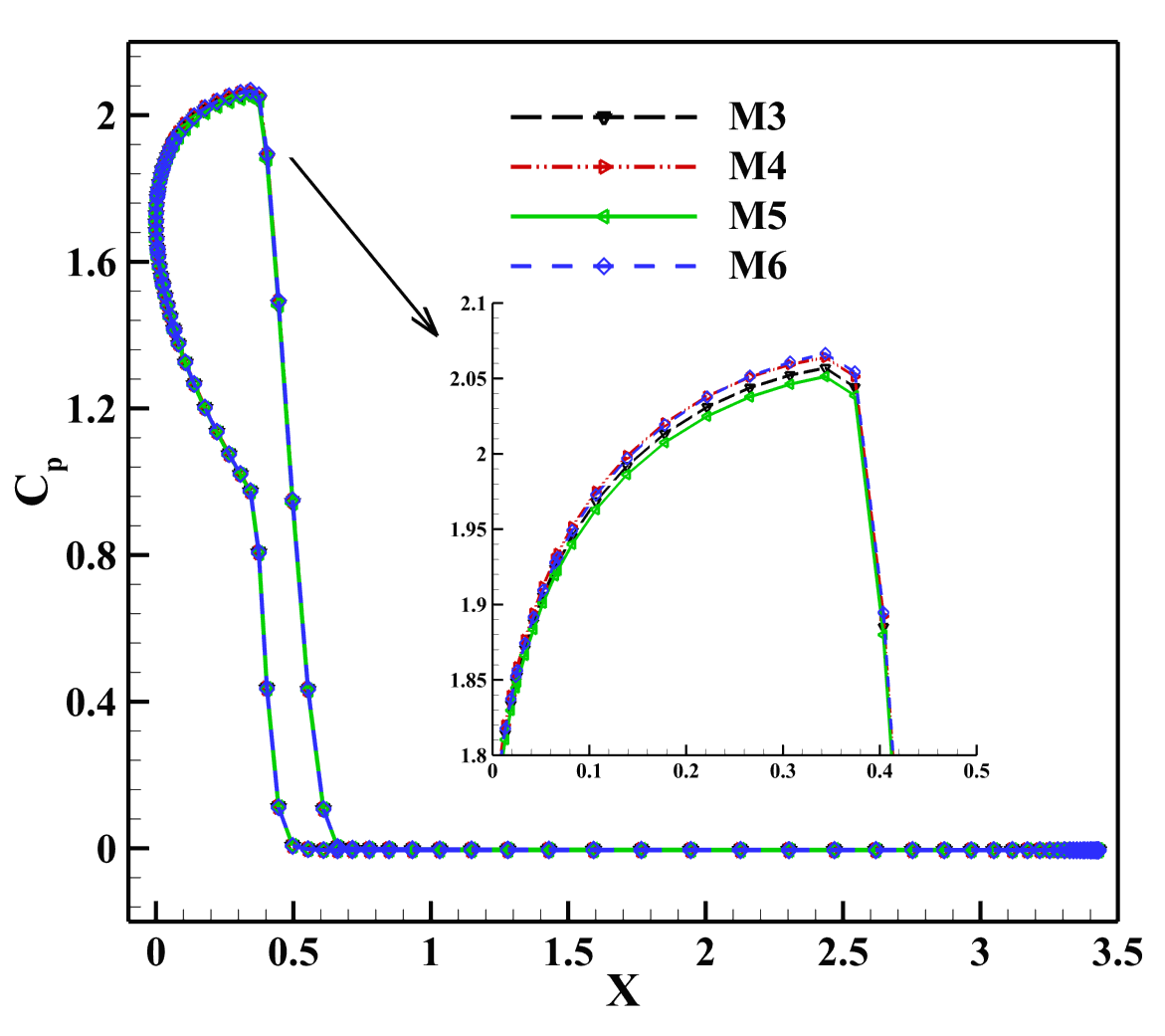}
	}
	\subfigure[\label{Fig:Apollo_H150_Cf}]{
		\includegraphics[width=0.3\textwidth]{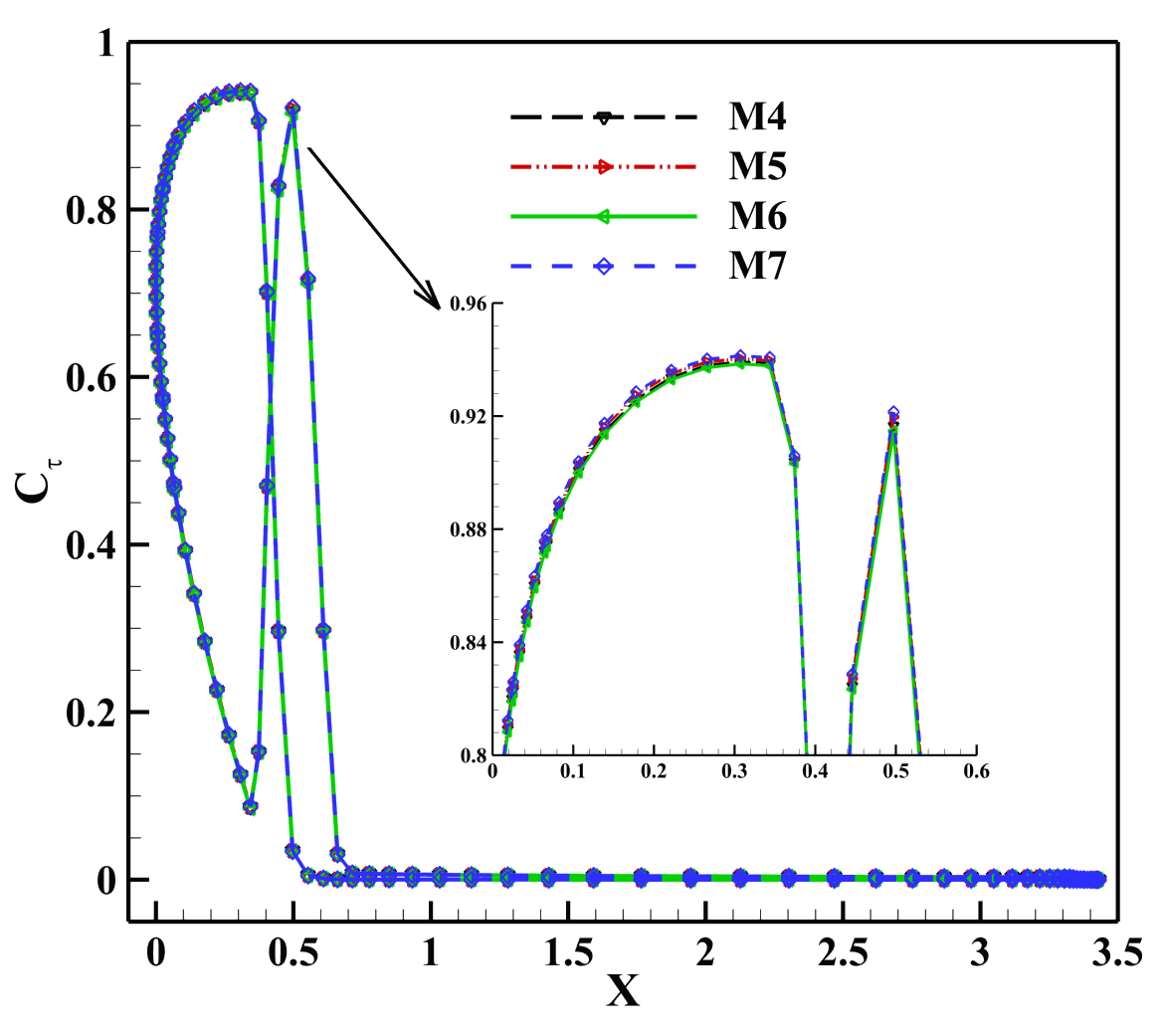}
	}
	\subfigure[\label{Fig:Apollo_H150_Ch}]{
		\includegraphics[width=0.3\textwidth]{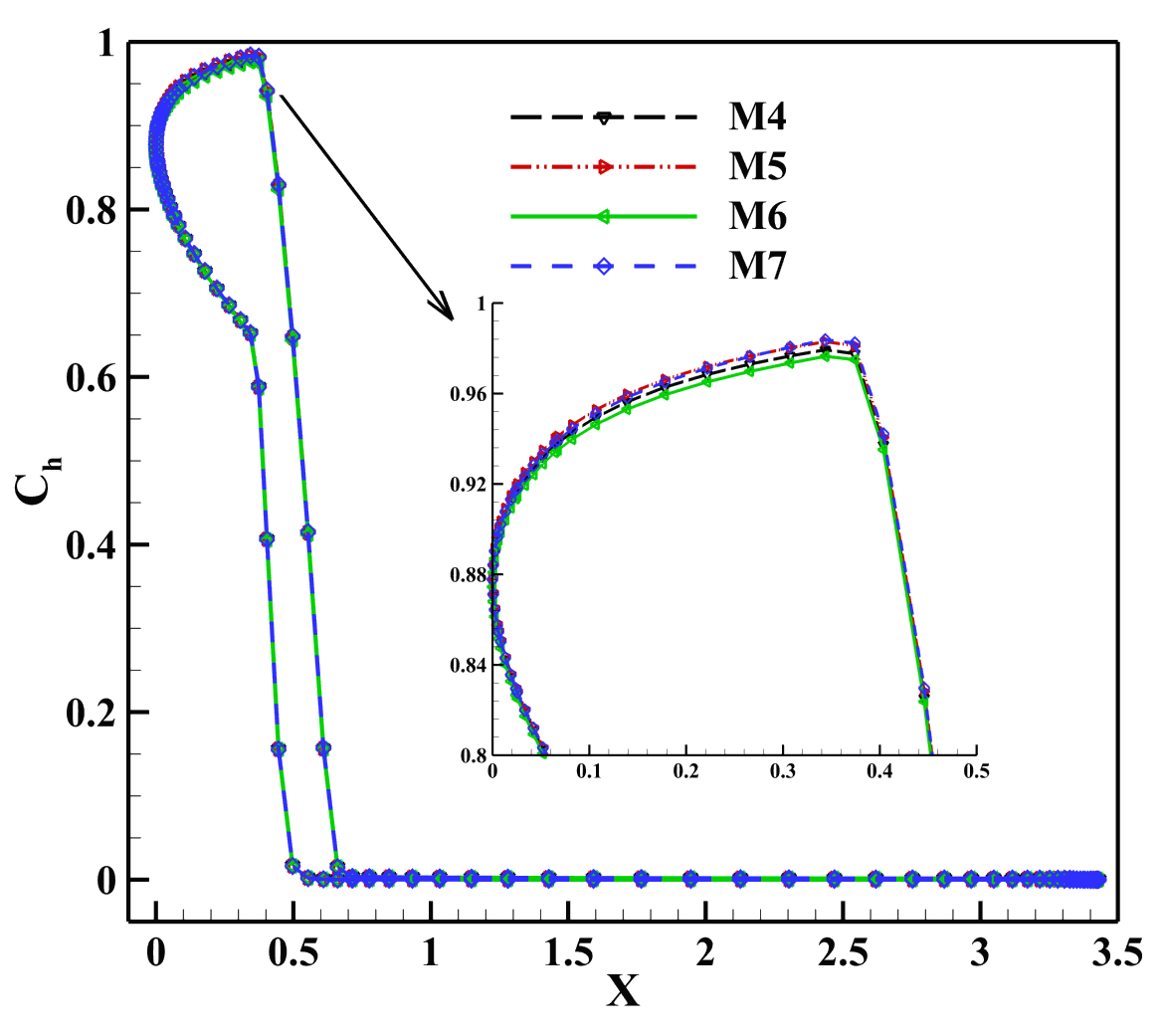}
	}
	\caption{Comparison of the present (a) pressure coefficient, (b) shear stress coefficient and (c) heat transfer coefficient for the Apollo 6 command module using different unstructured DVS (Altitude of 150km).}
	\label{Fig:Apollo_H150_CpCfCh}
\end{figure}

Aerodynamic coefficients calculated by using different DVS are shown in Fig.~\ref{Fig:Apollo_ClCdCmLD_Micmesh}.
The solutions solved by different DVS demonstrate good agreement, and the reasonable aerodynamic coefficients can be obtained by using the DVS of No. M7.
The maximum relative errors in comparison with the case No. M1 (85km) and case No. M4 (100, 120, 150 km) are about 1.30$\%$, -1.26$\%$, -0.37$\%$ and -1.96$\%$, respectively.
The aerodynamic coefficients predicted by the refined DVS of No. M6 excellently agree well with results of the case No. M1 (85km) and case No. M4 (100, 120, 150 km), the maximum relative error is less than 0.62$\%$.
In general, the unstructured DVS with about 15000 cells can provide acceptable and reasonable aerodynamic coefficients for Apollo 6 command module, and a spherical region with the radius of $3\sqrt {R{T_0}}$ is enough to capture the aerodynamic coefficient.
The surface pressure coefficient $C_{p}$, shear stress coefficient $C_{\tau}$ and heat transfer coefficient $C_{h}$ at altitude of 85, 100, 120 and 150 km are shown in Figs.~\ref{Fig:Apollo_H85_CpCfCh} and~\ref{Fig:Apollo_H150_CpCfCh}, respectively.
At altitude of 85 km, it is needed a unstructured DVS with about 25000 cells (No. M5) to capture the $C_{h}$.
The number of cells in unstructured DVS required for solving heat transfer coefficient decreases with the increase of Kn number.
The total temperature $T_0$ calculated according to Eq.~\ref{equ:solve_t0} approach the real maximum temperature $T_{max}$ of the flow field in the near-continuum and continuum flows.
On the other hand, the total temperature $T_0$ calculated according to Eq.~\ref{equ:solve_t0} overestimate the real maximum temperature of the flow field at large Kn number, and the discrete range of DVS determined according to this temperature is correspondingly larger.
Therefore, a spherical region with the radius of $5\sqrt {R{T_{max}}}$ is enough to capture the heat transfer coefficient.
At altitude of 150 km, the radius of discrete range solved by $3\sqrt {R{T_{0}}}$ and $5\sqrt {R{T_{max}}}$ are very approximate.
In general, the discrete range of the unstructured DVS for hypersonic rarefied flows can be determined according to the empirical criteria.
The spherical center coordinate is set as $0.4({U_{\infty}},{V_{\infty}},{W_{\infty}})$, and the radius is $5\sqrt {R{T_0}}$ for near-continuum flow, and $3\sqrt {R{T_0}}$ for transitional flow.

\begin{figure}[!htp]
	\centering
	\includegraphics[width=0.45\textwidth]{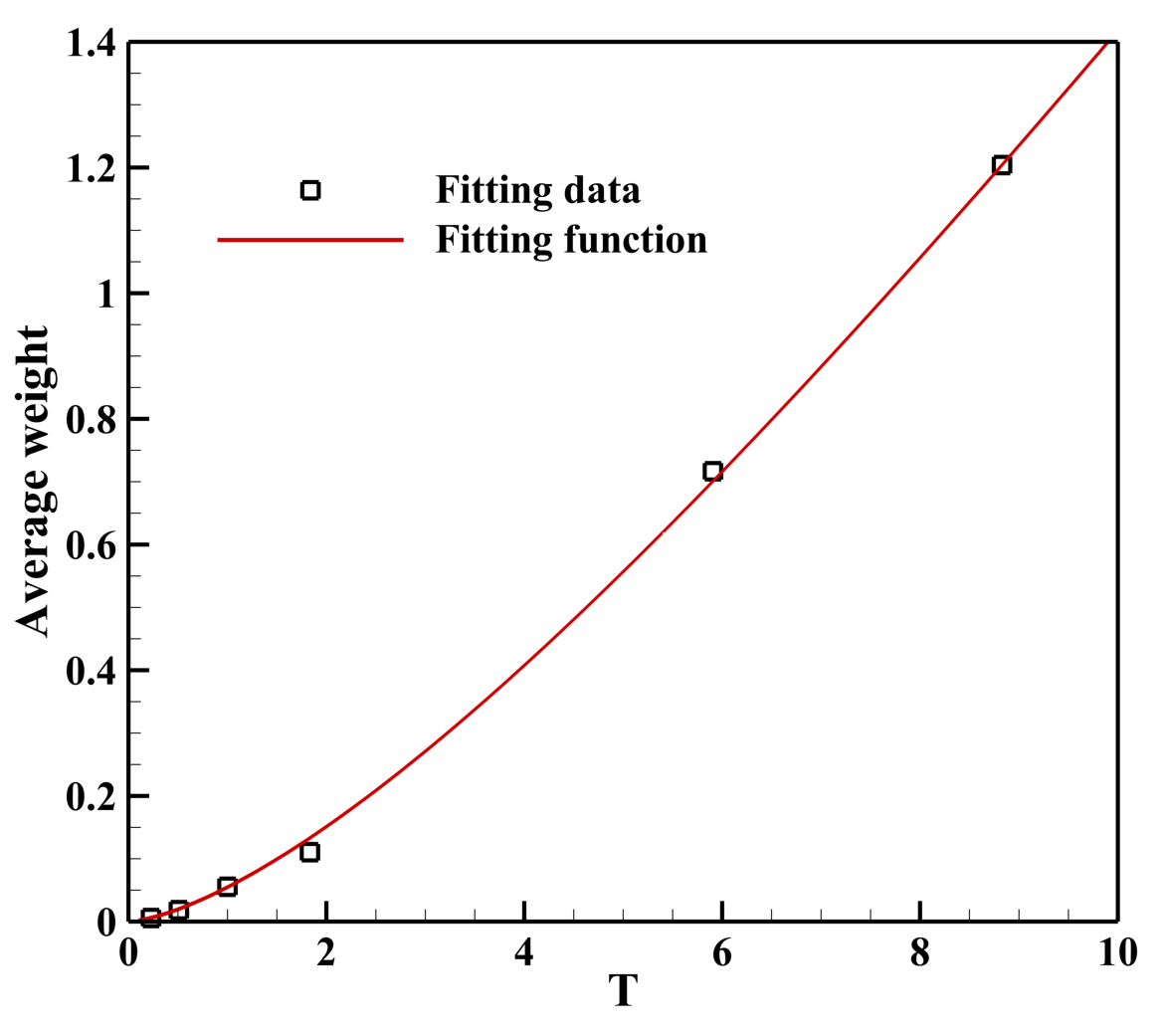}
	\caption{Evolution of the average weight as a function of the temperature (free-stream temperature and wall temperature).}
	\label{Fig:Apollo_Temperature_Weight}
\end{figure}

The center coordinates and radius of the refined spherical regions have been determined according to $3 \sigma$ criterion.
The number of cells in the refined spherical regions should also be examined.
In the refined spherical regions of the free-stream velocity point and the zero velocity point, the number of cells mainly related to free-stream and wall temperatures, respectively.
Fig.~\ref{Fig:Apollo_Temperature_Weight} illustrates the evolution of the average weight (case No. M5) as a function of the temperature in the refined spherical regions.
The average weight is defined as the volume of the sphere divided by the number of cells in the refined spherical region.
In order to determine the average weight expediently, a fitting function shown in Eq.~\ref{equ:solve_weight} is constructed according to the existing data.
\begin{equation}\label{equ:solve_weight}
	\omega = \frac{{\frac{4}{3}\pi {r^3}}}{{730\left[ {\arctan \left( {\frac{{T - 1}}{{40}}} \right) + 1} \right]}},
\end{equation}
where $r = 3\sqrt {RT}$, $T$ is the free-stream temperature or wall temperature.
The range and average weight in the refined spherical regions can be approximately determined according to the above skills or criteria.

\section{Conclusion}\label{sec:conclusion}
In this paper, a conservative implicit scheme is further developed for three-dimensional steady flows of diatomic gases in all flow regimes.
In the present method, the implicit macroscopic equations are simultaneously solved with the gas kinetic equation based on the Rykov model in a finite volume framework.
The use of coupled implicit macroscopic and microscopic iterative equations markedly increases the efficiency of the present method, especially for the highly non-equilibrium and near-continuum flows.
In order to capture the multi-scale solution accurately and efficiently, a difference scheme of the model equation is used to evolve the initial distribution function inside the cell to the interface with a local physical time step, which takes into account the particle transport and collision process.
Then an asymptotic analysis of the gas distribution function used for the flux evaluation in the continuum limit is accomplished.
In addition, the use of unstructured DVS and MPI parallel in DVS further speed up the calculation.
The unstructured DVS is more flexible than Cartesian DVS, and can greatly reduce the number of discrete velocity points, especially in the three-dimensional hypersonic flow simulations.
Furthermore, based on numerical experiments of Apollo 6 command module, an empirical generation criterion for three-dimensional unstructured DVS is proposed to generate the reasonable unstructured DVS more quickly and easily.

In the numerical tests, the efficiency and accuracy of the present method are verified by a series of three-dimensional low-speed, supersonic and hypersonic flows in all flow regimes.
The present implicit method can improve the convergence rate by one or two orders of magnitude compared to explicit conserved DUGKS.
Moreover, the test cases of supersonic and hypersonic flows are performed, in which shows good accuracy in the prediction of aerothermodynamic properties comparing with the results of DSMC and experiment.
In addition, benefited from the ability to simulate multi-scale flows, the present method are more reliable and accurate compared with single-scale N-S solver in the slip and transitional flow regimes.
And it is more efficient than DSMC method in the near-continuum flow regime.
In conclusion, the present method is efficient and accurate for finding steady-state solutions of the gas kinetic equations for diatomic gas flow in all flow regimes.
It is promising to develop an accurate and efficient multi-scale flow solver for three dimensional complex flow simulations, especially the hypersonic flows in the slip and transitional flow regimes.

\section*{Acknowledgments}
The authors thank Prof. Kun Xu in Hong Kong University of Science and Technology for discussions of the direct modeling of multi-scale flows.
Rui Zhang thanks Dr. Ruifeng Yuan at Southern University of Science and Technology for providing the meshes of blunted-cone.
Rui Zhang thanks Mr. Junzhe Cao at Northwestern Polytechnical University for providing the physical space mesh of Apollo 6.
This work is supported by the high performance computing power and technical support provided by Xi'an Future Artificial Intelligence Computing Center.
The present work is supported by the National Natural Science Foundation of China (Grants No. 12172301, No. 11902266, No. 12072283 and No. 11902264) and the 111 Project of China (No. B17037).


\bibliography{CMAME_ImplicitRykov3D}

\newpage

\begin{table}
	\centering
	\caption{\label{Table:CavityFlow_Speed} Comparison of the efficiency between the explicit conserved DUGKS and the present method for the three-dimensional lid-driven cavity flows in all flow regimes (MPI parallel with 240 cores).}
	\begin{tabular}{p{0.17\columnwidth}<{\raggedright} p{0.13\columnwidth}<{\raggedleft}
			p{0.17\columnwidth}<{\raggedleft} p{0.13\columnwidth}<{\raggedleft} 
			p{0.17\columnwidth}<{\raggedleft} p{0.12\columnwidth}<{\raggedleft} }
		\bottomrule[1.4pt]
		\hline
		\toprule
		\multirow{1}{*}{Case}
		&\multicolumn{2}{p{0.3\columnwidth}<{\centering}}{Explicit conserved DUGKS} 
		&\multicolumn{2}{p{0.3\columnwidth}<{\centering}}{Present}
		&\multicolumn{1}{p{0.18\columnwidth}<{\centering}}{Speedup}  \\
		\cline{2-3} \cline{4-5}		
		&Steps &Time (min) &Steps &Time (min)\\
		\hline
		Kn = 10	    &36600  &5639.7		&116	&46.6	&121.1 \\
		Kn = 1	    &4900	&709.5		&125	&50.9	&14.0 \\
		Kn = 0.075	&5500	&799.4		&121	&46.6	&17.2 \\
		Re = 100	&84700 	&12531.2	&157	&64.6	&194.0 \\		
		\bottomrule[1.4pt]
		\hline
		\toprule
	\end{tabular}
\end{table}

\begin{table}
	\centering
	\caption{\label{Table:Sphere_Drag} Comparison of the drag coefficients for the supersonic and hypersonic flows over a sphere.}
	\begin{tabular}{p{0.1\columnwidth}<{\centering} p{0.1\columnwidth}<{\centering} 
			p{0.1\columnwidth}<{\centering} p{0.12\columnwidth}<{\centering} 
			p{0.16\columnwidth}<{\centering} p{0.17\columnwidth}<{\centering} p{0.19\columnwidth}<{\centering} }
		\bottomrule[1.4pt]
		\hline
		\toprule
		No. &Ma	&Kn	&Re &Exp. (Air) &Present ($\rm{N_2}$) &Relative error \\
		\hline
		A1	&4.25	&0.121	&53.0	&1.69	&1.670	&-1.18$\%$	\\
		A2	&4.25	&0.080	&80.5	&1.53	&1.539	& 0.59$\%$	\\
		A3	&4.25	&0.043	&150.0	&1.37	&1.396	& 1.90$\%$	\\
		A4	&4.25	&0.031	&210.0	&1.35	&1.342	&-0.59$\%$	\\
		B1	&5.45	&1.960	&4.2	&2.60	&2.582	&-0.69$\%$ \\
		B2	&5.45	&0.957	&8.6	&2.44	&2.433	&-0.29$\%$	\\
		B3	&5.45	&0.490	&16.8	&2.28	&2.235	&-1.97$\%$	\\
		B4	&5.45	&0.256	&32.1	&2.04	&2.000	&-1.96$\%$	\\
		\bottomrule[1.4pt]
		\hline		
		\toprule
	\end{tabular}
\end{table}

\begin{table}
	\centering
	\caption{\label{Table:Sphere_Speed} Comparison of the efficiency between the explicit conserved DUGKS and the present method for the supersonic and hypersonic flows over a sphere (MPI parallel with 280 cores).}
	\begin{tabular}{p{0.3\columnwidth}<{\centering} p{0.14\columnwidth}<{\raggedleft} 
			p{0.15\columnwidth}<{\raggedleft} 
			p{0.09\columnwidth}<{\raggedleft} p{0.15\columnwidth}<{\raggedleft} 
			p{0.09\columnwidth}<{\raggedleft} }
		\bottomrule[1.4pt]
		\hline
		\toprule
		\multirow{1}{*}{Case}
		&\multicolumn{2}{p{0.30\columnwidth}<{\centering}}{Explicit conserved DUGKS} 
		&\multicolumn{2}{p{0.24\columnwidth}<{\centering}}{Present}
		&\multicolumn{1}{p{0.12\columnwidth}<{\centering}}{Speedup}  \\
		\cline{2-3} \cline{4-5}		
		&Steps &Time (min) &Steps &Time (min)\\
		\hline
		Ma = 4.25, Kn = 0.031	    &2900   &553	&60		&27		&20.5 \\
		Ma = 5.45, Kn = 1.960	    &7500	&1421	&80		&36		&39.5 \\
		\bottomrule[1.4pt]
		\hline		
		\toprule
	\end{tabular}
\end{table}

\begin{table}
	\centering
	\caption{\label{Table:Cone_Parameter} Conditions for the hypersonic rarefied flow over a blunted-cone.}
	\begin{tabular}{p{0.25\columnwidth}<{\centering} p{0.25\columnwidth}<{\centering}}
		\bottomrule[1.4pt]
		\hline
		\toprule
		Property &Value \\
		\hline
		Gas							&$\rm{N_2}$			\\
		Base diameter $\rm d$		&15.24mm			\\
		Ma						    &10.15				\\
		$\rm{Kn_{HS}}$				&0.065				\\
		$\rm{Re_d}$					&232.8				\\
		$\rm{T_{\infty}}$			&143.5K				\\
		$\rm{T_{w}}$				&600K				\\
		$\rm{Z_{rot}}$				&4.24				\\
		\bottomrule[1.4pt]
		\hline		
		\toprule
	\end{tabular}
\end{table}

\begin{table}
	\centering
	\caption{\label{Table:Apollo_Parameter} Conditions for the Apollo 6 command module.}
	\begin{tabular}{p{0.10\columnwidth}<{\centering} p{0.09\columnwidth}<{\centering} 
			p{0.09\columnwidth}<{\centering} p{0.18\columnwidth}<{\centering} p{0.1\columnwidth}<{\centering} 
			p{0.12\columnwidth}<{\centering} p{0.19\columnwidth}<{\centering} p{0.09\columnwidth}<{\centering}}
		\bottomrule[1.4pt]
		\hline
		\toprule
		Altitude	&Ma	&Kn	&$\rm{\rho_{\infty}}$ ($\rm{Kg/m^3}$)	
		&$\rm{T_{\infty}}$ (K)	&$\rm{T_{w}}$ (K)		
		&R ($\rm{J/{(Kg \cdot K)}}$)	&$\rm{\omega}$ \\
		\hline
		200		&13.02	&44.740		&$3.2829 \times 10^{-10}$	&1026	&234	&378.448	&0.7718	\\	
		150		&16.19	&7.5900		&$2.1383 \times 10^{-9}$	&733	&373	&342.541	&0.7614	\\		
		120		&23.72	&0.7730		&$2.2642 \times 10^{-8}$	&368	&675	&317.845	&0.7535	\\	
		100		&33.96	&0.0338		&$5.5824 \times 10^{-7}$	&194	&1146	&294.235	&0.7476	\\		
		85		&35.59	&0.0024		&$7.9550 \times 10^{-6}$	&181	&1598	&287.103	&0.7471	\\
		\bottomrule[1.4pt]
		\hline		
		\toprule
	\end{tabular}
\end{table}

\begin{table}
	\centering
	\caption{\label{Table:Apollo_DimensionlessParameter} The dimensionless velocity and temperature for the Apollo 6 command module at altitude of 150, 120, 100 and 85 km.}
	\begin{tabular}{p{0.1\columnwidth}<{\centering} p{0.1\columnwidth}<{\centering} 
			p{0.1\columnwidth}<{\centering} p{0.29\columnwidth}<{\centering} p{0.08\columnwidth}<{\centering} 
			p{0.08\columnwidth}<{\centering} p{0.20\columnwidth}<{\centering}}
		\bottomrule[1.4pt]
		\hline
		\toprule
		Altitude	&Ma	&Kn	&($\rm{U_{\infty}}$,$\rm{V_{\infty}}$,$\rm{W_{\infty}}$) 
		&$\rm{T_{\infty}}$ &$\rm{T_{w}}$ 
		&$\rm{T_0}$\\
		\hline
		150		&16.19	&7.5900		&(12.2779, 0, -5.7253)	&1.0	&0.5089	&53.4232	\\		
		120		&23.72	&0.7730		&(17.9887, 0, -8.3883)	&1.0	&1.8342	&113.5277	\\	
		100		&33.96	&0.0338		&(25.7504, 0, -12.0076)	&1.0	&5.9072	&231.6563	\\		
		85		&35.59	&0.0024		&(26.9882, 0, -12.5848)	&1.0	&8.8287	&254.3296	\\
		\bottomrule[1.4pt]
		\hline		
		\toprule
	\end{tabular}
\end{table}

\begin{table}
	\centering
	\caption{\label{Table:Apollo_DVS} The number of discrete velocity cells (M1$\sim$M7 are case No.) for the Apollo 6 command module in the tests of unstructured DVS.}
	\begin{tabular}{p{0.11\columnwidth}<{\centering} p{0.11\columnwidth}<{\centering} 
			p{0.11\columnwidth}<{\centering} p{0.11\columnwidth}<{\centering} p{0.11\columnwidth}<{\centering} 
			p{0.11\columnwidth}<{\centering} p{0.11\columnwidth}<{\centering} p{0.11\columnwidth}<{\centering}}
		\bottomrule[1.4pt]
		\hline
		\toprule
		Altitude	&M1	&M2	&M3	&M4	&M5 &M6	&M7 \\
		\hline
		150	&-		&-		&-		&32110	&24930	&20560	&14480	\\
		120	&-		&-		&-		&32900	&24650	&20030	&14480	\\
		100	&-		&-		&-		&30220	&24900	&19660	&14580	\\
		85	&66680	&54960	&43940	&32510	&24920	&19320	&13900	\\
		\bottomrule[1.4pt]
		\hline		
		\toprule
	\end{tabular}
\end{table}

\begin{figure}[!htp]
	\centering
	\subfigure[\label{Fig:Cone_Ma10.15_Line_AOA10_Cp}]{
		\includegraphics[width=0.45\textwidth]{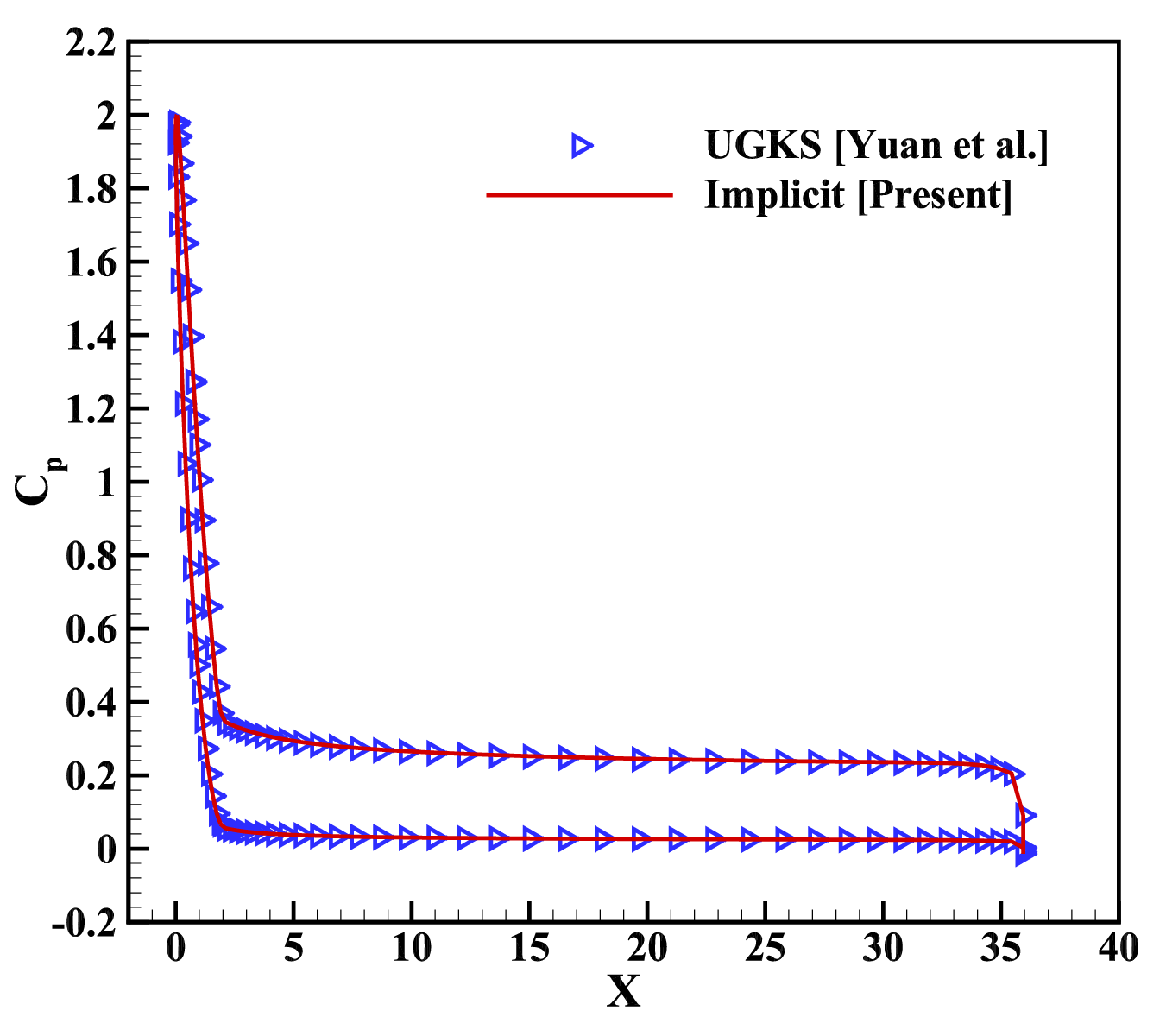}
	}
	\subfigure[\label{Fig:Cone_Ma10.15_Line_AOA10_Ch}]{
		\includegraphics[width=0.45\textwidth]{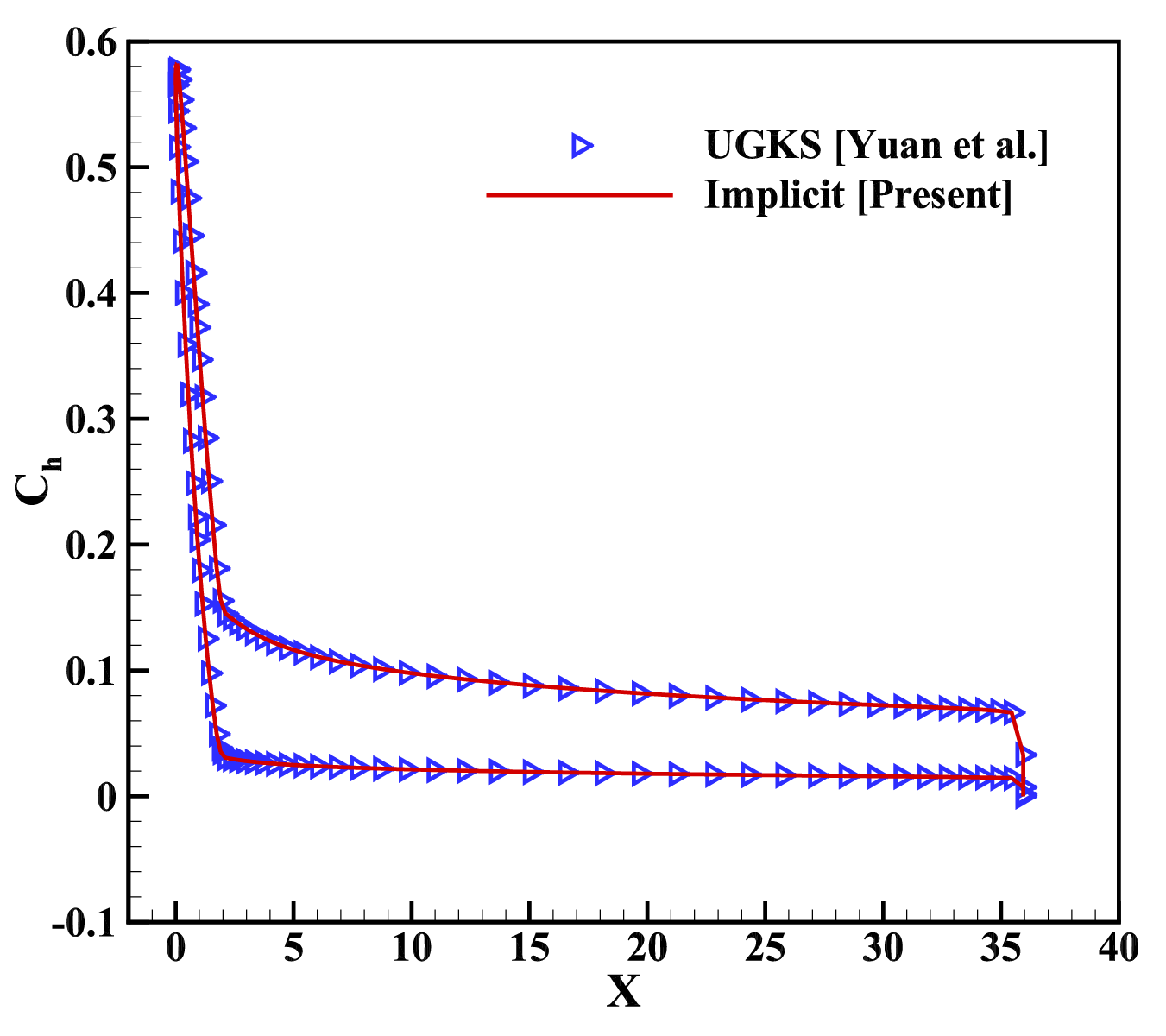}
	}
	\caption{Comparison of UGKS and the present (a) pressure coefficient and (b) heat transfer coefficient for the hypersonic rarefied flow over a blunted-cone at AOA = 10$^ \circ$.}
	\label{Fig:Cone_Ma10.15_Line_AOA10_CpCh}
\end{figure}

\begin{figure}[!htp]
	\centering
	\subfigure[\label{Fig:Cone_Ma10.15_Line_AOA20_Cp}]{
		\includegraphics[width=0.45\textwidth]{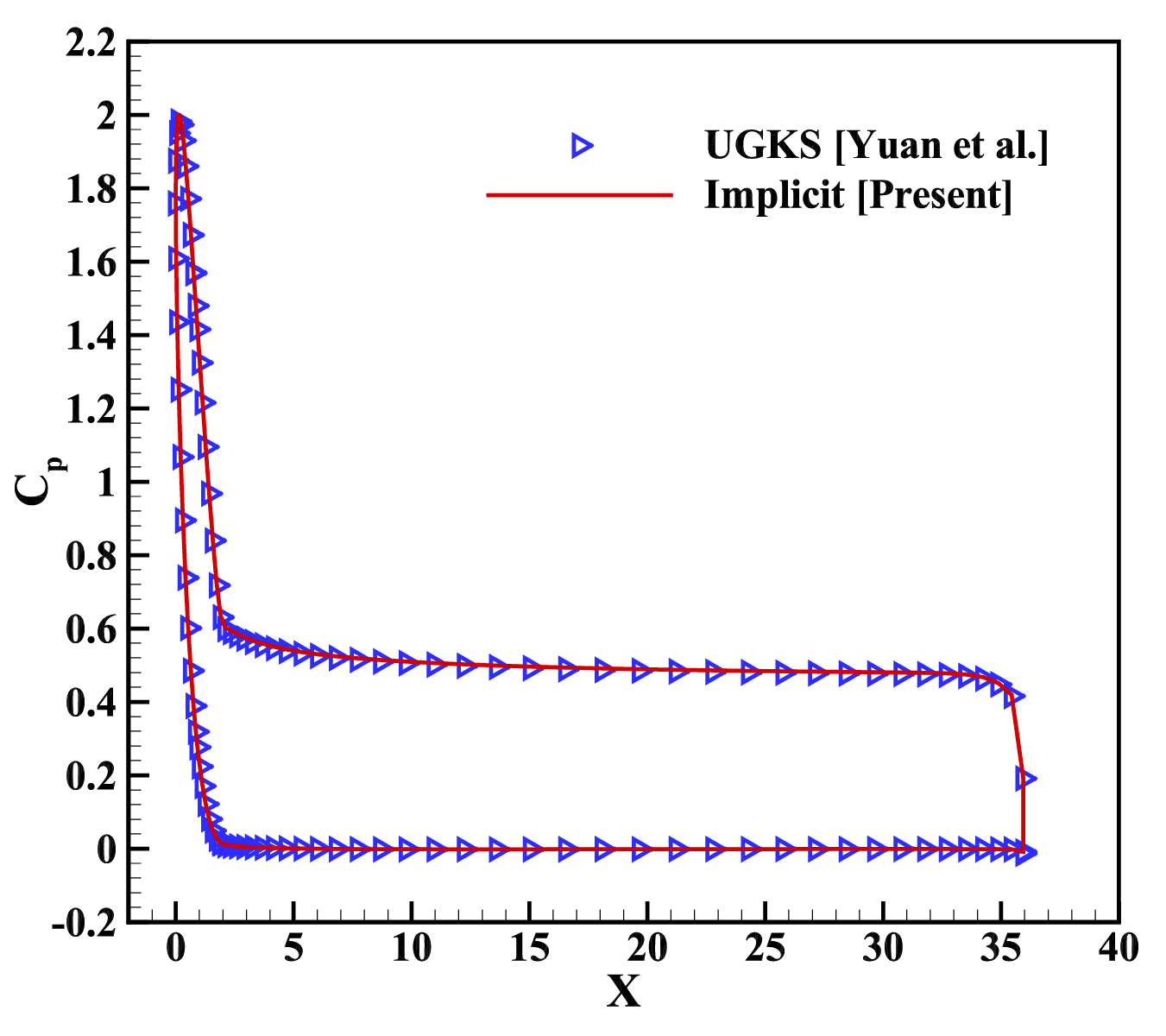}
	}
	\subfigure[\label{Fig:Cone_Ma10.15_Line_AOA20_Ch}]{
		\includegraphics[width=0.45\textwidth]{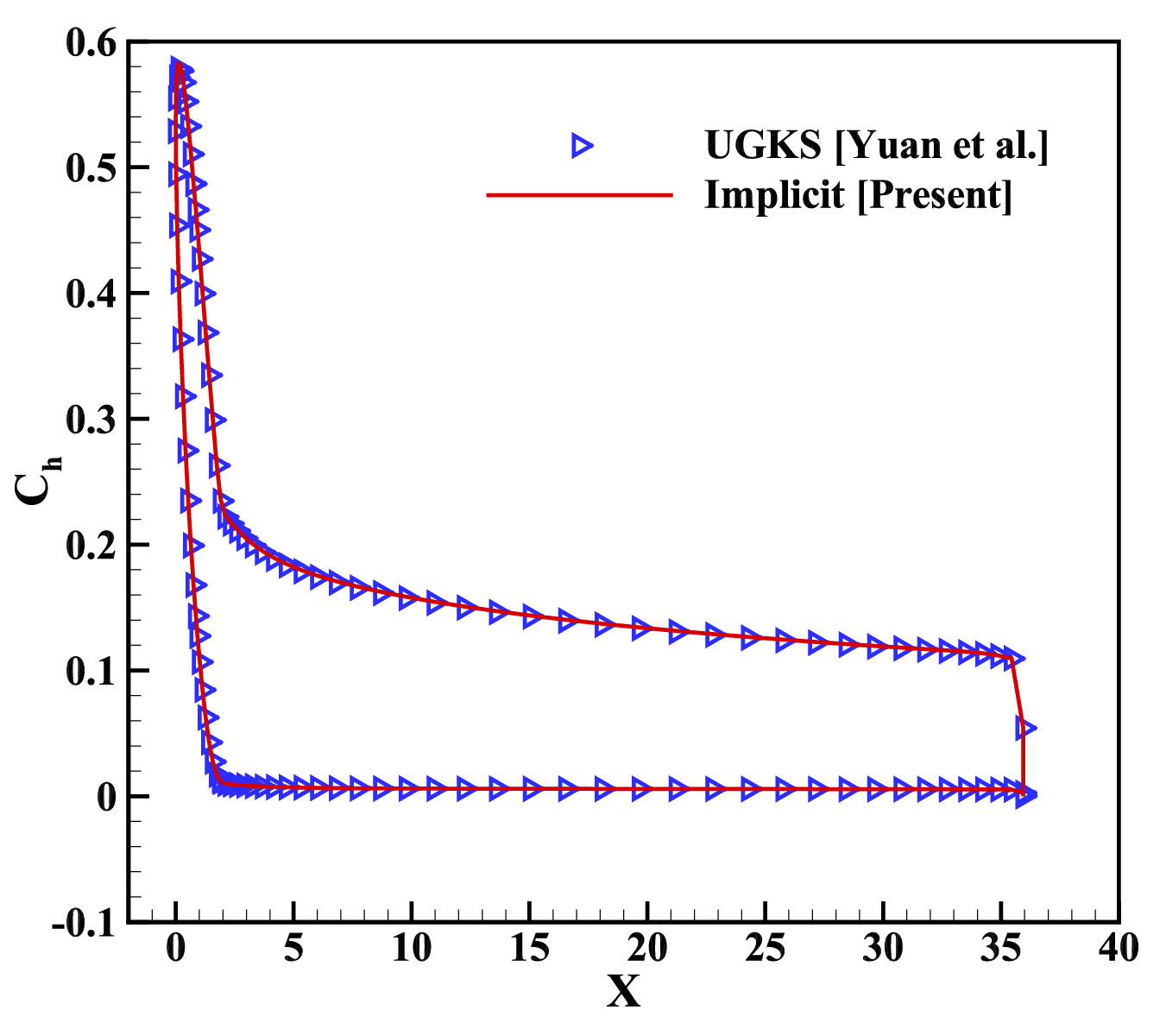}
	}
	\caption{Comparison of UGKS and the present (a) pressure coefficient and (b) heat transfer coefficient for the hypersonic rarefied flow over a blunted-cone at AOA = 20$^ \circ$.}
	\label{Fig:Cone_Ma10.15_Line_AOA20_CpCh}
\end{figure}

\begin{figure}[!htp]
	\centering
	\subfigure[\label{Fig:Apollo_H100_Cp}]{
		\includegraphics[width=0.3\textwidth]{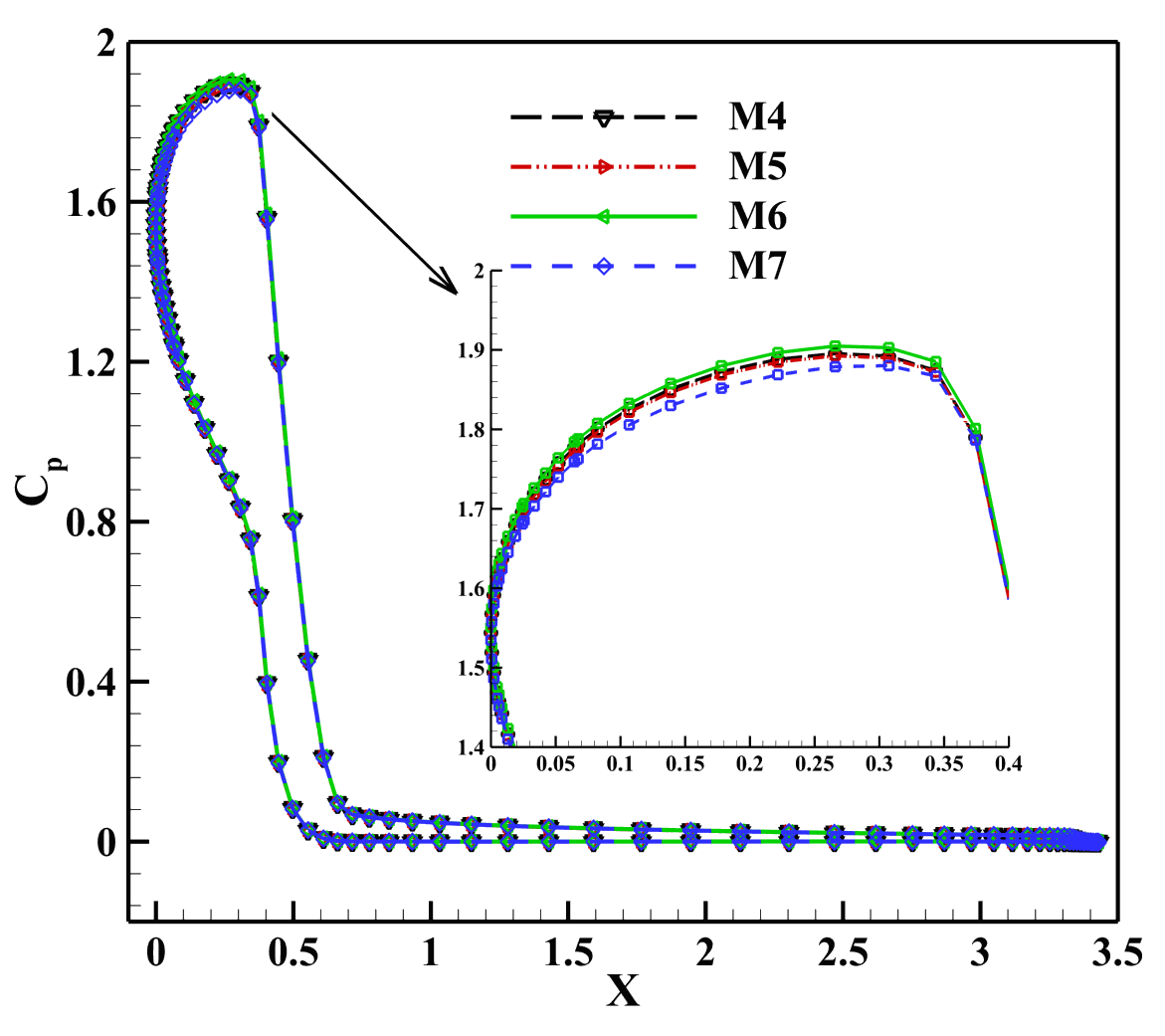}
	}
	\subfigure[\label{Fig:Apollo_H100_Cf}]{
		\includegraphics[width=0.3\textwidth]{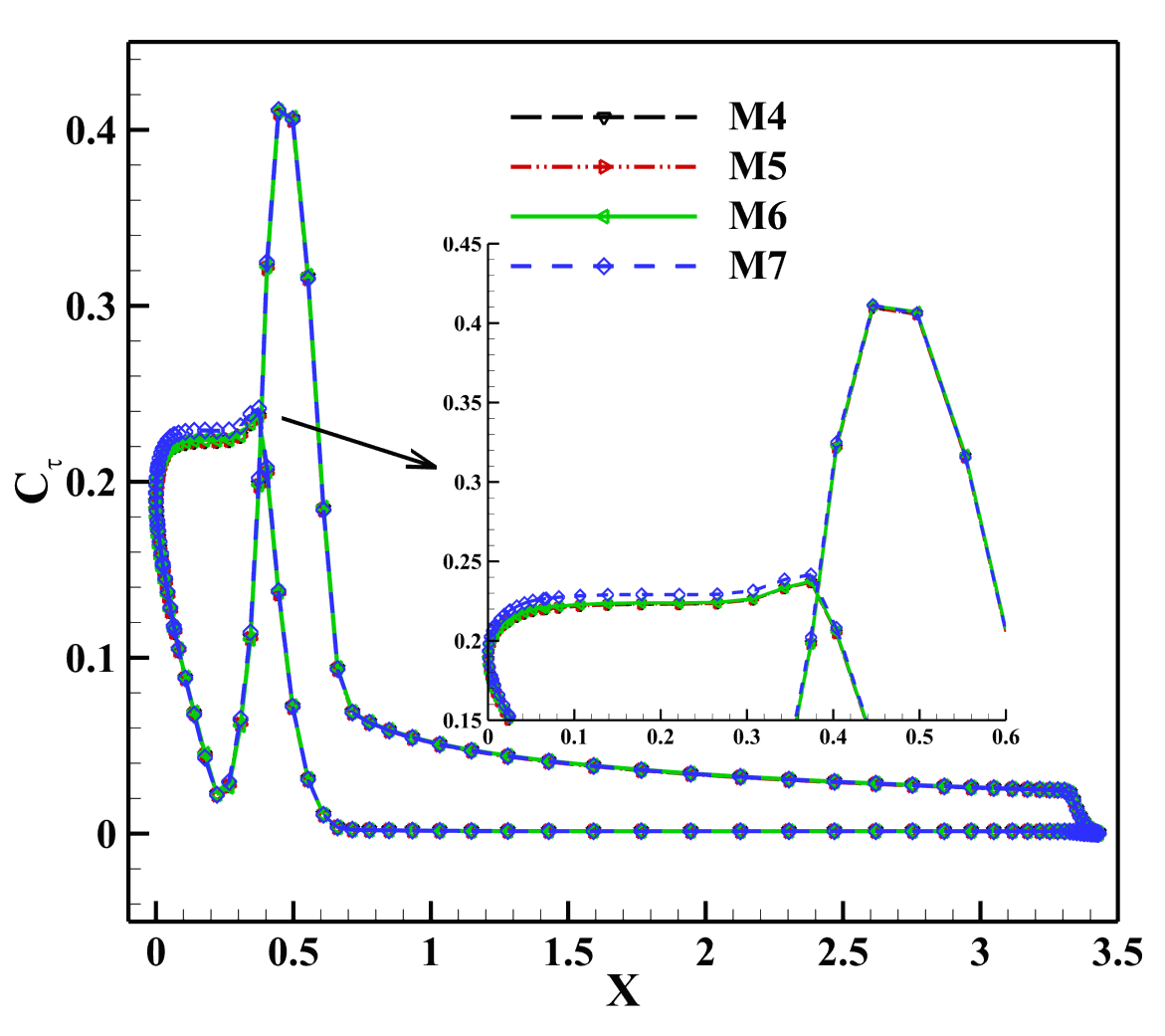}
	}
	\subfigure[\label{Fig:Apollo_H100_Ch}]{
		\includegraphics[width=0.3\textwidth]{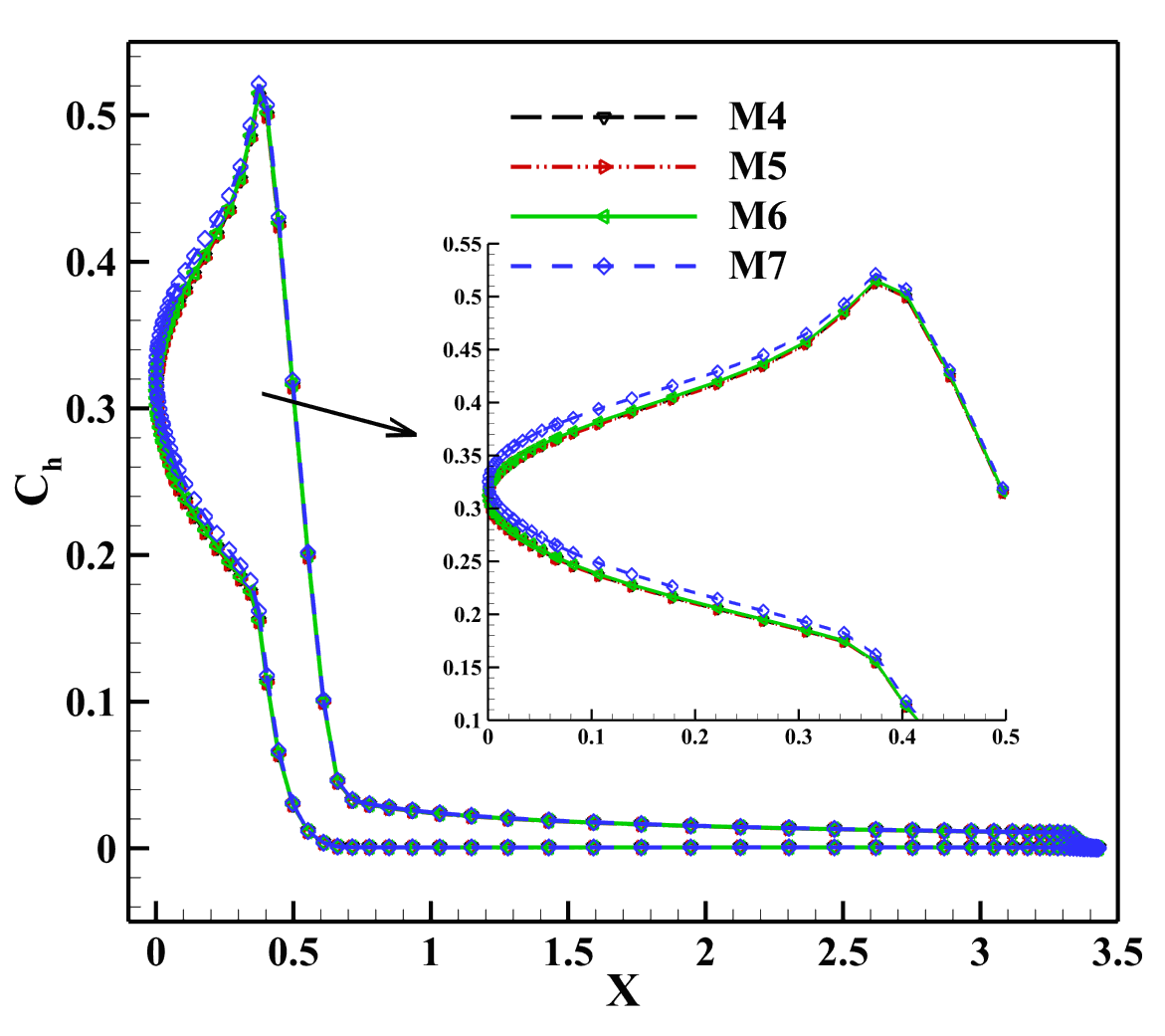}
	}
	\caption{Comparison of the present (a) pressure coefficient, (b) shear stress coefficient and (c) heat transfer coefficient for the Apollo 6 command module using different unstructured DVS (Altitude of 100km).}
	\label{Fig:Apollo_H100_CpCfCh}
\end{figure}

\begin{figure}[!htp]
	\centering
	\subfigure[\label{Fig:Apollo_H120_Cp}]{
		\includegraphics[width=0.3\textwidth]{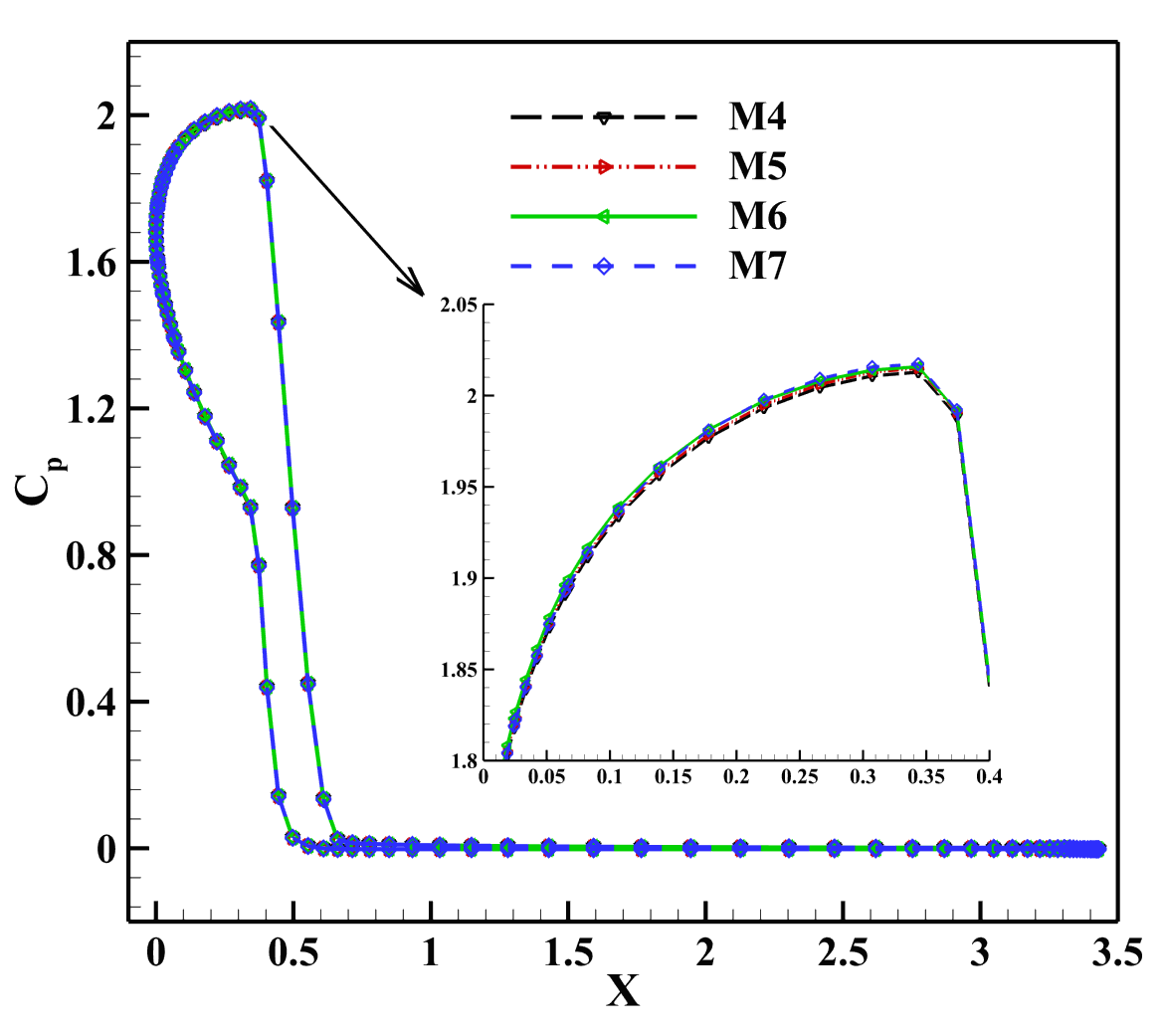}
	}
	\subfigure[\label{Fig:Apollo_H120_Cf}]{
		\includegraphics[width=0.3\textwidth]{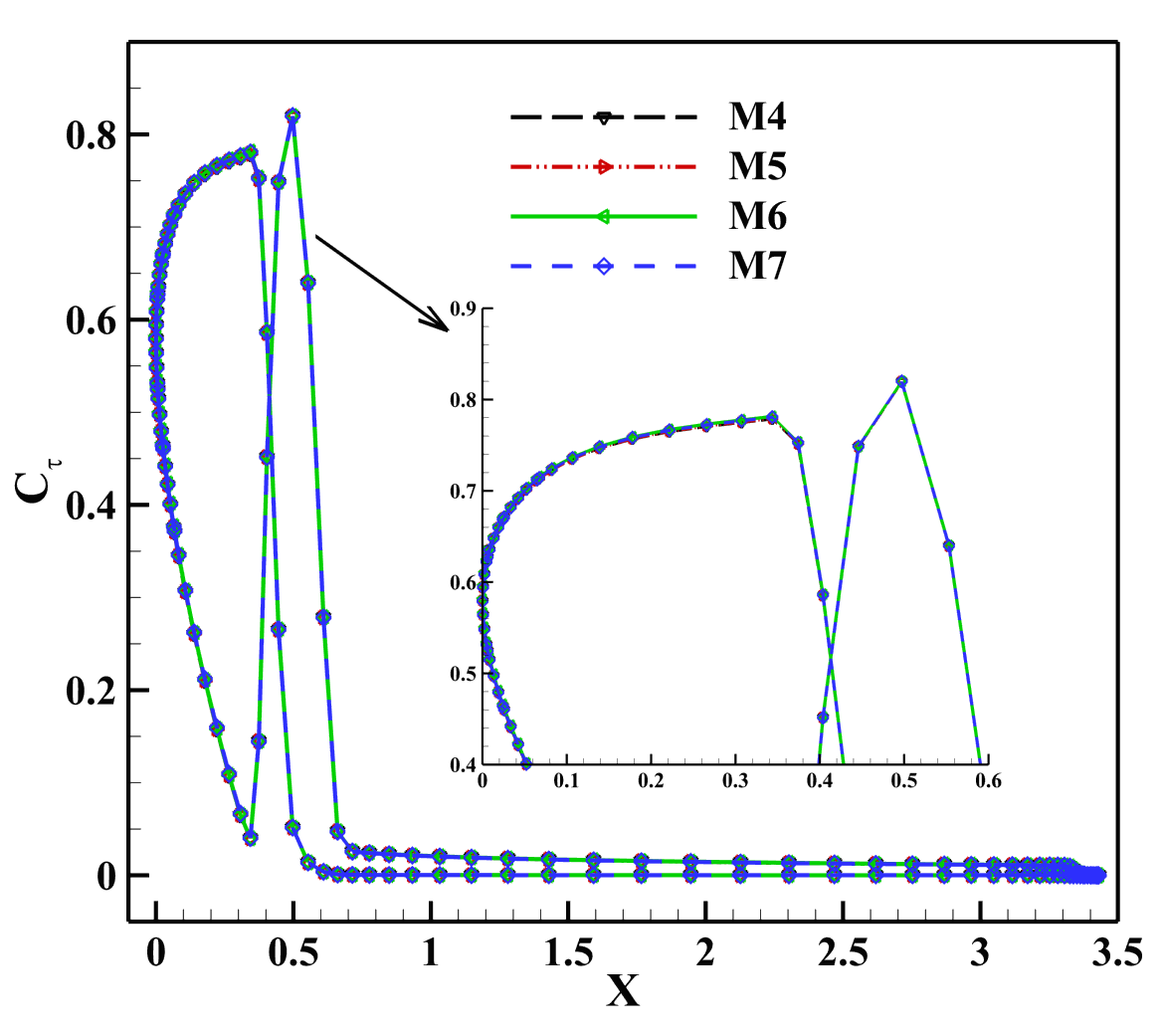}
	}
	\subfigure[\label{Fig:Apollo_H120_Ch}]{
		\includegraphics[width=0.3\textwidth]{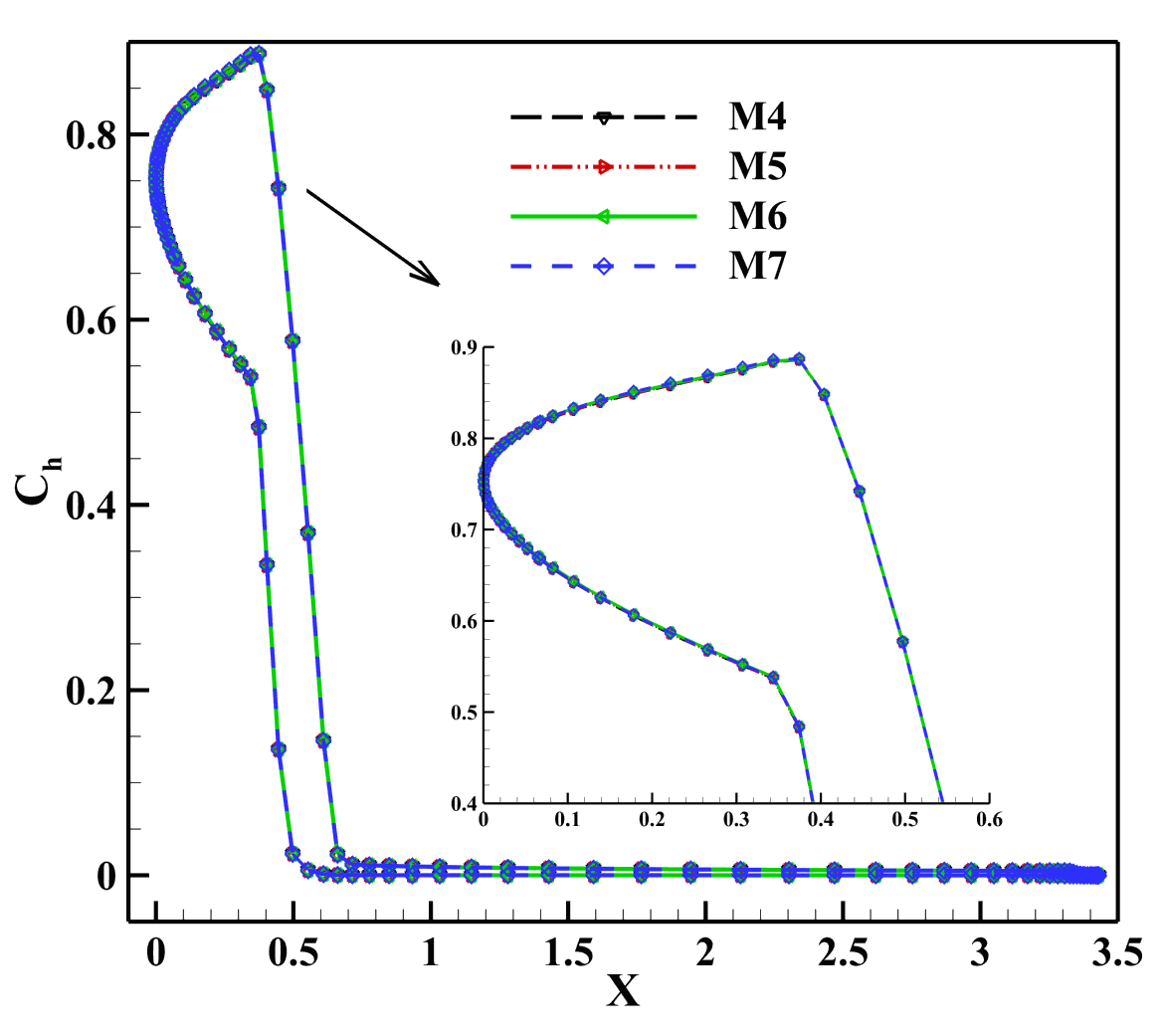}
	}
	\caption{Comparison of the present (a) pressure coefficient, (b) shear stress coefficient and (c) heat transfer coefficient for the Apollo 6 command module using different unstructured DVS (Altitude of 120km).}
	\label{Fig:Apollo_H120_CpCfCh}
\end{figure}

\end{document}